# THE COMETS DISCOVERED FROM NEW ZEALAND, AND THE ASTRONOMERS WHO FOUND THEM

**John Drummond**
*University of Southern Queensland, West Street, Toowoomba, QLD 4350, Australia.*
E-mail: kiwiastronomer@gmail.com

**Wayne Orchiston**
*University of Science and Technology of China, Hefei, Anhui, China; and University of Southern Queensland, West Street, Toowoomba, QLD 4350, Australia.*
E-mail: wayne.orchiston@gmail.com

**and**

**Carolyn Brown and Jonathan Horner**
*University of Southern Queensland, West Street, Toowoomba, QLD 4350, Australia.*
E-mails: Carolyn.Brown@unisq.edu.au
jonti.horner@unisq.edu.au

**Abstract: Eleven comets were discovered by six New Zealanders from New Zealand shores. New Zealand's important geographical position south of the Equator is highlighted. This location allows observers to discover and observe comets that would be difficult to see from the Northern Hemisphere. The lives of these comet discoverers are explored, as well as the discovery circumstances and an analysis of the morphological changes of their comets. We find that numerous New Zealand observers made many observations of these comets, however, relatively few were included in international publications, this possibly being due to papers about them not being submitted to overseas journals. We also note an interesting trend in that the number of New Zealand newspaper articles relating to these comet discoveries plummeted after the 1946 discovery by Albert Jones. We speculate as to why this may have happened.**

## 1 INTRODUCTION

New Zealand has had a rich astronomical heritage since the Māori first settled in Aotearoa / New Zealand (for simplicity, henceforth 'NZ') around 800 years ago (see Orchiston 2016: 33–88; Orchiston and Orchiston, 2017). The history of 'modern' astronomical observations began in NZ when those accompanying James Cook landed at various locations on the coast of the North and South Islands between 1769 and 1777, and the astronomers on the vessels were able to carry out numerous astronomical observations (Orchiston, 1998; 2016: 107–203). In the decades that followed, growing European settlement and especially the presence of surveyors with the New Zealand Company saw ever-burgeoning astronomical activity. NZ was deemed an important geographical and astronomical location as it filled a void in the Southern Hemisphere, effectively plugging the gap between South America and Australia. Occasionally, astronomical events could only be observed from NZ due to its geographical isolation and the time-constraints when these events happened.

Comets are an example of a transient object that is sometimes only visible for a short period of time, and which can strongly favour observations in one hemisphere or the other. A famous example is the apparition of Comet C/2006 P1 (McNaught), which moved in an orbit that made it hard to see from the Northern Hemisphere when at its best, whilst being a spectacular sight for Southern observers (Drummond, 2007). Several NZ astronomers (both amateur and professional) have paid close attention to these temporary visitors and observed their astrophysical changes. This research paper investigates one aspect of NZ cometary astronomy, namely comets officially discovered by New Zealanders from NZ.[1] To date, only six New Zealanders have achieved this distinction (Orchiston, 2016: 500, Table 17.3). Of those six, two, namely John Grigg (1838–1920) and Rodney Austin (b. 1945) discovered three comets each, whilst the other discoverers are known for single or double discoveries. New Zealand-born rocket scientist William Ashley (Bill) Bradfield (1927–2014) discovered all 18 of his comets while he was living in Australia, so he will not be included in this paper.

In this work, we present a detailed descript-

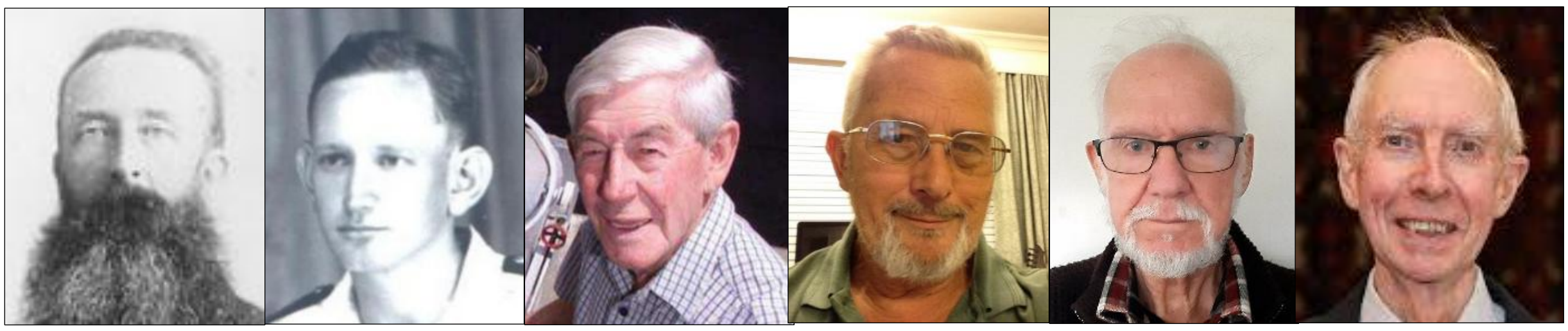

Figure 1: The six New Zealanders who discovered comets from New Zealand shores. Left to right: John Grigg, photographed in 1877 (https://nzmusicalnotables.com/); Murray Geddes, around 1940 (Orchiston Collection); Albert Jones, around 2000 (after Toone, 2016), Michael (Mike) Clark, in 2015 (Orchiston Collection); Rodney (Rod) Austin, in 2025 (courtesy: Rodney Austin); and Alan Gilmore, in 2025 (Wikipedia). Four of the six (Jones, Clark, Austin and Gilmore) are or were known personally to the first two authors of this paper.

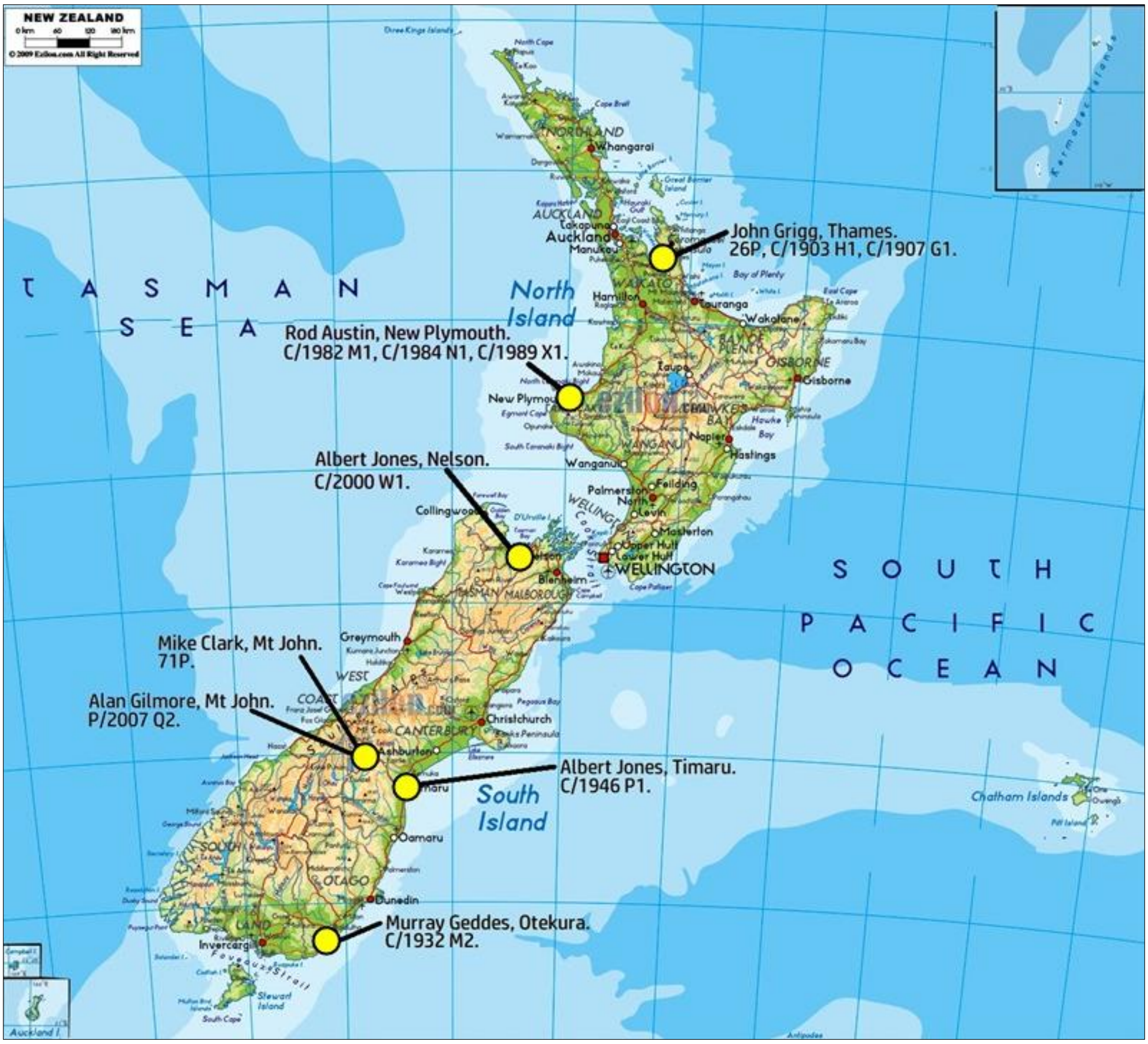

Figure 2: Map of New Zealand showing the five locations from which the six New Zealand comet discoverers discovered their eleven comets. The locations are denoted by the yellow circles (base map: Fotolip; modifications: John Drummond and Wayne Orchiston).

tion of the comets that were discovered by these six men (see Figure 1), with a particular focus on the observations of those comets that were carried out from New Zealand. Briefly stated, six comets were discovered from NZ's North Island and five from the South Island (see Figure 2). Even at their best, most were fairly faint, with the brightest being C/1989 X1 (Austin), which reached an apparent visual magnitude of 4 in August 1982 (Kronk and Meyer, 2010: 759–760). Three of these comets are now known to be periodic comets, with or-

Table 1: Comets discovered from New Zealand shores by New Zealand astronomers.

| No. | Comet's Name | Comet's Discoverer(s) | Discovery Date | Location of Discoverer | Decl'n of Comet (°) | Maximum Magnitude |
|---|---|---|---|---|---|---|
| 1 | 1902 O1 (26P/Grigg-Skjellerup) | John Grigg and J.F. Skjellerup | 23 Jul 1902 | Thames (NIs) | +07 | "Faint" (Jul 1902) |
| 2 | C/1903 H1 (Grigg) | John Grigg | 17 Apr 1903 | Thames (NIs) | –11 | "Very faint" (May 1903) |
| 3 | C/1907 G1 (Grigg-Mellish) | John Grigg and J.E. Mellish | 8 Apr 1907 | Thames (NIs) | –44 | 8(?) (Apr 1907) |
| 4 | C?1932 M2 (Geddes) | Murray Geddes | 22 Jun 1932 | Otekura (SIs) | –84 | 9 (Jun 1932) |
| 5 | C/1946 P1 (Jones) | Albert Jones | 6 Aug 1946 | Timaru (SIs) | –13 | 7 (Oct 1946) |
| 6 | 1973 L1 71P/Clark | Mike Clark | 9 Jun 1973 | Tekapo (SIs) | –31 | 12 (Jul 1973) |
| 7 | C/1982 M1 (Austin) | Rod Austin | 18 Jun 1982 | New Plymouth (NIs) | –40 | 4 (Aug 1982) |
| 8 | C/1984 N1 (Austin) | Rod Austin | 8 Jul 1984 | New Plymouth (NIs) | –39 | 4.8 (Aug 1984) |
| 9 | C/1989 X1 (Austin) | Rod Austin | 6 Dec 1989 | New Plymouth (NIs) | –62 | 4.3 (Apr 1990) |
| 10 | C/2000 W1 (Utsunomiya-Jones) | Albert Jones and S. Utsunomiya | 26 Nov 2000 | Nelson (SIs) | –77 | 6 (Dec 2000) |
| 11 | P/2007 Q2 (Gilmore) | Alan Gilmore | 20 Aug 2007 | Tekapo (SIs) | –01 | 18 (Aug 2007) |

Key: Discovery Dates are in Universal Time (UT); Location of Discoverer: NIs = North Island; SIs = South Island; both 'Tekapo' discoveries were made at nearby Mount John Observatory; Maximum Magnitude column lists the brightest apparent visual magnitude the comet reached during the apparition, and the month and year when this occurred. This Table is based on data in Kronk, 2007; 2009; Kronk and Meyer, 2010; Kronk, Meyer and Sergeant, 2017; Orchiston, 2016: 500; Vsekhsvyatskii, 1964; the NASA/JPL Horizons webpage, and GUIDE.

bital periods of less than 200 years. Of these eleven discoveries, 10 (91 %) were found in the Southern sky, that is, had a negative declination. NZ's isolated Southern Hemisphere location and its relatively small population and low levels of light pollution helped with these discoveries. Nine of the NZ discoveries were made visually and two photographically. Table 1 provides an overview of the comet-discoverers and their comets.

A desired outcome of this work is also to identify observations of these 'New Zealand comets' made by fellow New Zealanders, as well as the discoverers, and to gauge how many (or how few) appeared in international publications.

## 2 METHODOLOGY

Key search vehicles for this research were the NZ newspaper search-engine Papers Past (https://paperspast.natlib.govt.nz/newspapers), interviews with the surviving discoverers, and the Smithsonian Astrophysical Observatory Astrophysics Data System searches (ADS: https://ui.adsabs.harvard.edu/). Papers Past is an online digital archival and research tool produced by the National Library of NZ. It enables users to digitally hunt through a database of archived newspaper articles published across NZ from 1839 until 1998, with a large catalogue of digitised historical resources available to users. In addition, information on the comets themselves and the broader history of cometary astronomy was obtained through the study of a wide variety of peer-reviewed journal papers, magazine and newsletter articles and monographs. We also utilised the planetarium software, GUIDE 9.1 (2020) (henceforth GUIDE) to determine cometary paths in the sky, solar elongation, etc.

Our work has identified many newspaper articles that provide additional information on the circumstances of the discovery of the comets in question, or that detail additional follow-up observations made by many New Zealanders. This research paper covers observations made between 1902 (New Zealand's first official comet discovery) and 2007 (the most recent discovery made). Of the eleven comets discovered by New Zealanders in NZ, nine were found in the twentieth century, and two in the twenty-first. We also include a number of photographs and sketches of these 'New Zealand' comets made by New Zealanders in the course of this work. We found that many of these observations and photographs had never previously seen the 'light of day' in international journals. Indeed, relatively little material from NZ has been included in the most authoritative sources on cometary astronomy (Vsekhsvyatskii, 1964; Kronk, 1999; 2003; 2007; 2009; Kronk and Meyer, 2010; and Kronk et al. 2017). This research paper will attempt to shed light on some of the cometary astronomy done from NZ and how the results of that work can be used to supplement observations from overseas nations.

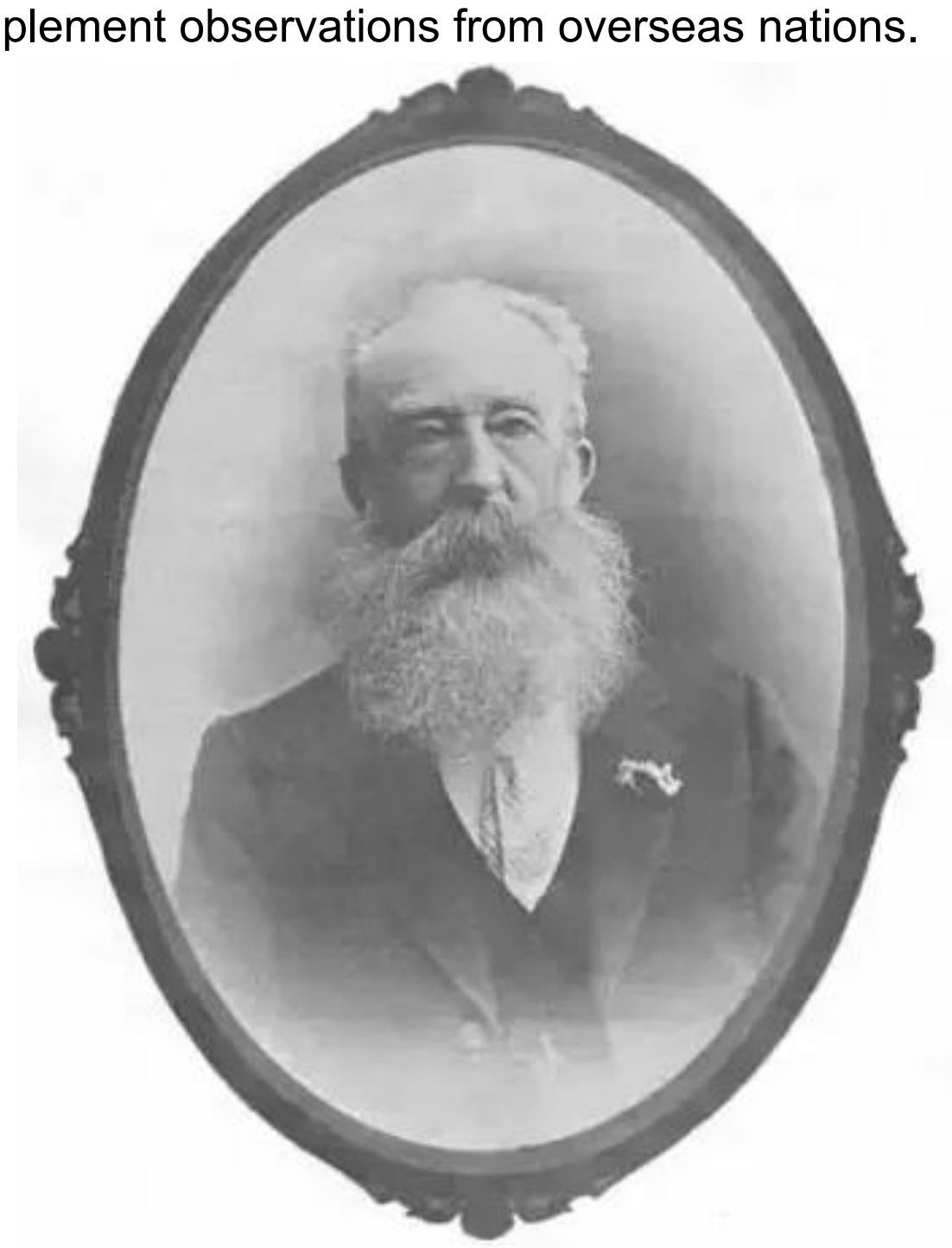

Figure 3: John Grigg later in life (after Grigg, 1970).

The following is an investigation of the six NZ comet discoverers who discovered comets from NZ and a look at 'their' comets. They are presented mainly in chronological order based on the comet discovery dates. Section 3 is about John Grigg and his three comets. Section 4 covers Murray Geddes and his comet. Section 5 investigates Albert Jones and how he set two world records with his two comet discoveries. Mike Clark's first comet discovery by photography in NZ is discussed in Section 6. In Section 7 we look at Rod Austin, who, along with John Grigg, has discovered the most comets (three) from NZ. Section 8 focuses on the most recent person to discover a comet from NZ, Alan Gilmore. Additional observations and photographs of these comets made by fellow NZ astronomers are presented together with a discussion of the astrophysical cometary changes that were observed in the comets.

## 3 JOHN GRIGG

John Grigg (1838–1920; Orchiston, 2016) discovered three comets (Orchiston, 2016: 500), and after Archdeacon Stock in 1881 (Drummond, 2023: 376; Orchiston, 2016: 254; Orchiston and Orchiston, 2026) was the first NZ astronomer person to officially find an unknown comet from NZ. Two of Grigg's comets, 1902 O1 (26P/Grigg-Skjellerup) and C/1907 G1 (Grigg-Mellish), were shared discoveries, whilst the other, C/1903 H1 (Grigg), solely bears his name. All three comets were discovered telescopically, two were first observed at declinations south of the Celestial Equator (see Table 1). None reached naked eye visibility. Orchiston (2016: 481) comprehensively covers the life and astronomical exploits of Grigg, calling him "... New Zealand's leading amateur astronomer during the first decade of the twentieth century ..." As Grigg's life and other astronomical achievements are covered in great detail in that work, we focus here primarily on the comets Grigg discovered, and direct the interested reader to Orchiston's earlier work for a detailed overview of Grigg's life.

John Grigg was born in London (although an Obituary [1920: 6] claims the 'Isle of Thanet') on 4 June 1838 (Orchiston, 2016: 272), and he had no siblings. A visit to the Royal Observatory, Greenwich, London (Grigg, 2020) at the age of 15 (Williams, 2016) initiated an interest in astronomy, a pursuit that would lead to several discoveries and contributions to the international astronomical community.

At the age of 25 Grigg migrated to NZ with his wife (Emma) in 1863 (Mackrell, 1985: 75), but unfortunately Emma passed away four years later. John Grigg then moved to the Coromandel Peninsula in 1868, lured there by gold strikes which occurred several years earlier (Isdale, 1967). Rather than prospecting for gold, Grigg established a furnishing and musical instruments business and also taught singing. He married for a second time at age 33, however his second wife, Sarah (née Allaway), died suddenly in 1874. Undeterred, Grigg married for the final time in 1887, to Jane Henderson. Between his three wives, they had six sons, three daughters, and an adopted son. His portrait in later life is shown in Figure 3.

Grigg arranged the construction of his first observatory (Figure 4), the 'Thames Observatory', in 1885, and it was built by the skilled carpenter Mr P. Sinclair. It was located in Pollen Street, near the centre of Thames, "... at the rear of [his] business premises ..." (Mr Grigg's Observatory, 1885: 3). This was used to house "... a new and expensive telescope and transit instrument ..." (*ibid.*). The main portion of the observatory was circular with a diameter of 2.4 metres (7' 11"), with walls 1.9 meters (6' 2") high. A rotating dome sat atop the walls. The dome slot provided a 30° field of view—presumably as seen from the telescope. If so, this would allow the telescope to track for approx.-imately two hours before the dome would need to be rotated if the observer wanted continuous observations of the same target. The top of the slot extended 15° past the zenith—which was very handy for equatorial mountings as, at times, the telescope position on the mount was beyond the vertical point above it. The dome was "... constructed of paper-mache, on a framework of pohutukawa ribs." (*ibid.*). One wonders how it fared in rain and wind storms. The telescope pier was a concrete pillar 1.4 metres (4' 7") high on which stood a 0.6 metre (2 foot) metal pedestal that carried the equatorial mount and telescope. A lean-to extension to the observatory housed the transit telescope for timekeeping. The transit slit was only 7.5 cm (3 inches) wide, but with a 160° field along the meridian it offered access to a range of 'clock stars' to choose from for determining time (*ibid.*).

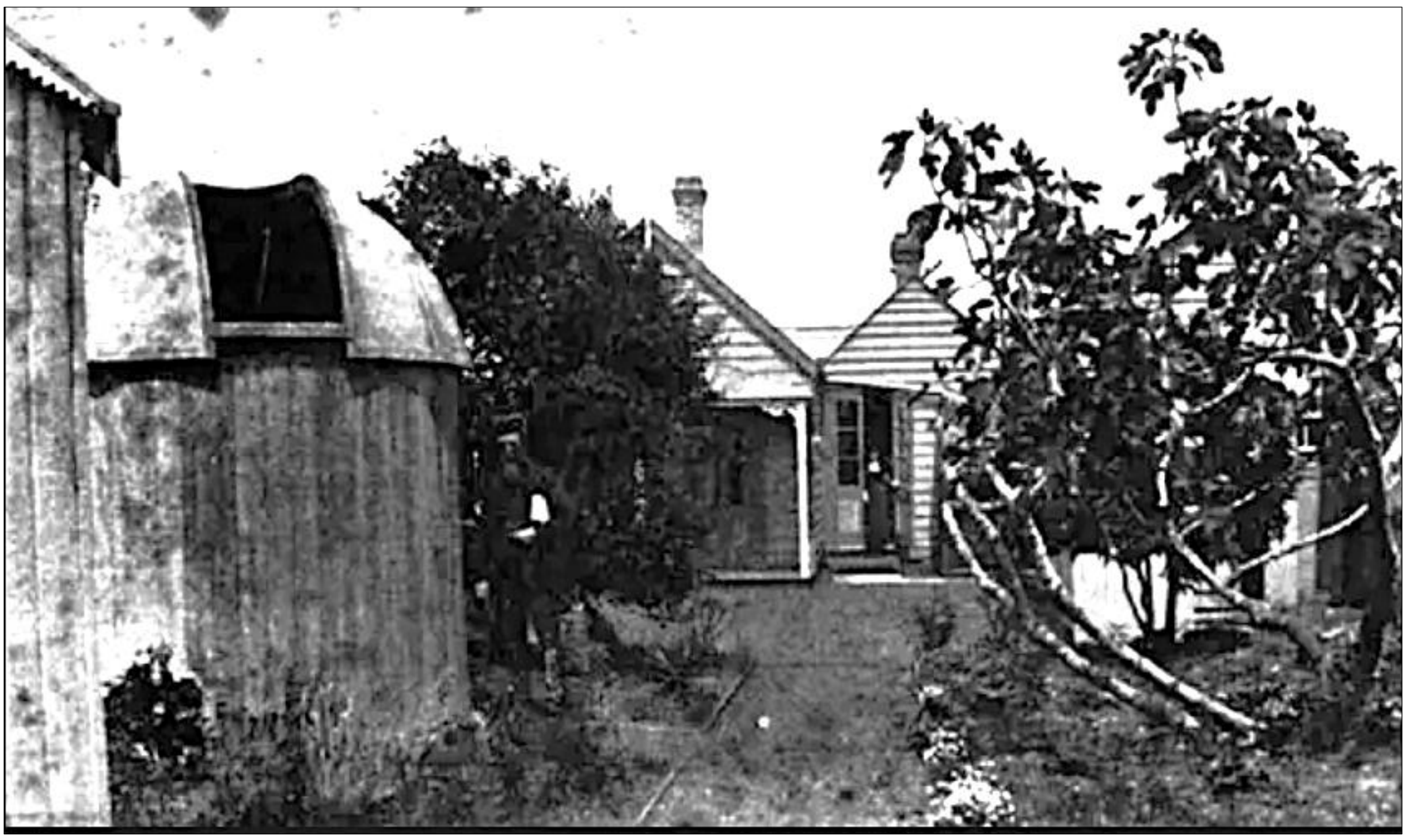

Figure 4: Grigg's Pollen Street observatory, near the centre of Thames, NZ. The building was to the rear of his business premises. We are unsure who the figure in the door is (after Grigg, 1970).

The principal telescope was a 3.5-inch (8.9-cm) Wray refractor (Figure 5), which had a focal length of 1,372 mm (Mr Grigg's Observatory, 1885: 3), and therefore a focal ratio of f/15.4 (focal length divided by aperture). The complement of eyepieces permitted magnifications from 60 to 480 times, which equates to eyepieces with focal lengths ranging from 22 mm to approximately 3 mm for Grigg's main instrument. The equatorial mount had Right Ascension (RA henceforth—unless in a quotation) and Declination (Dec henceforth) circles of 15 cm (6 inch) diameter which thanks to a vernier scale allowed positional readings down to "... 10s in right ascension and 3' in declination …" (Orchiston, 2016: 275). Grigg made his own drive with a weight and small spring clock for the equatorial mount in 1893, based on a design by D.H. Sparling published in the *English Mechanic* on 27 January 1893 (Grigg, 1902a: 126).

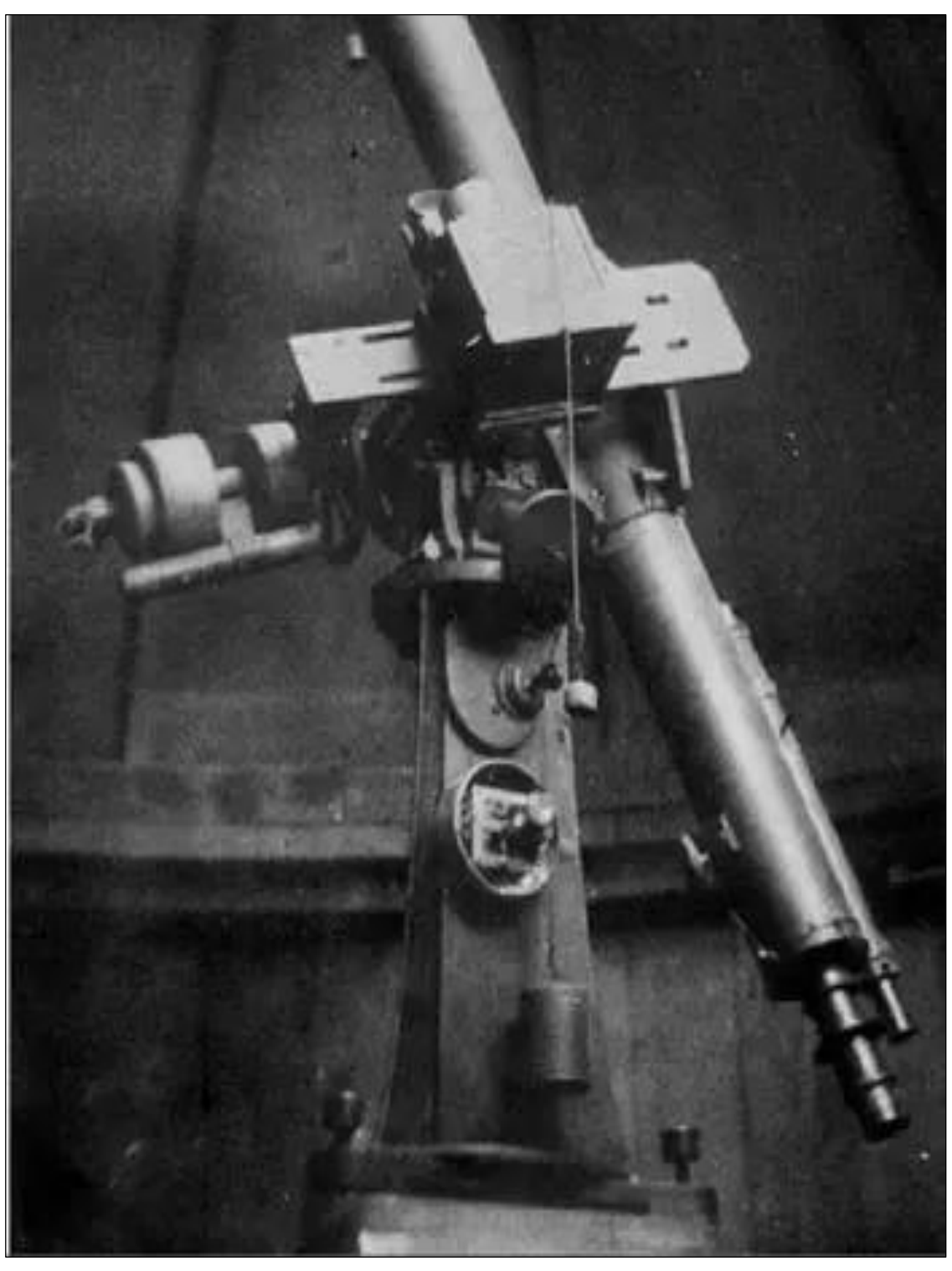

Figure 5: Grigg's 3.5-inch f/15.4 Wray refractor on an equatorial mounting, with which he discovered his three comets (photograph: courtesy Wanganui Observatory archives).

The transit telescope was manufactured by A.J. Frost, and had a focal length of around 330 mm. Grigg wrote that it worked "... very accurately, and will time a star to a fraction of a second if skillfully manipulated." Grigg kept a regulator clock (as well as other time pieces) "... with dead beat escapement …" in his business premises for "... the convenience of the public …" to admire and to allow "... his fellow townsmen in regulating their time-pieces …" (Mr Grigg's Observatory, 1885: 3). It was set to NZ time, which at that stage was 11 hours and 30 minutes ahead of Greenwich time (GMT).[2] Grigg also had a 3-inch (75-mm) altazimuth-mounted portable refractor on loan to take on trips or to view an object not visible from his observatory.

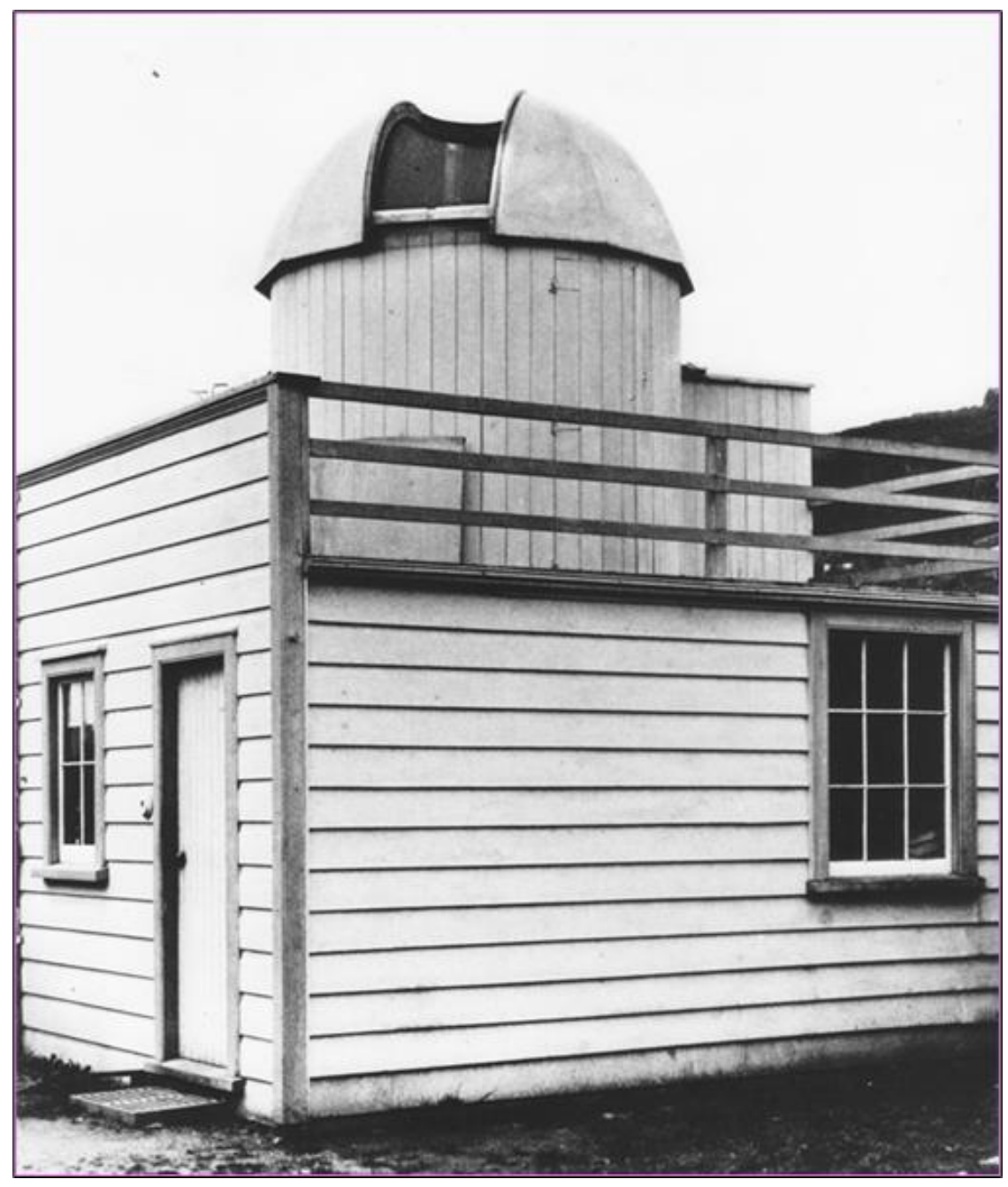

Figure 6: Grigg's second observatory was located at his residence in Queen Street, Thames, where he resided later in his life. Note the main observatory and transit room on the top floor and his office and workshop below (photograph: Angus Collection).

With help from the American 1882 transit of Venus team who determined precise positions for Mount Eden (Auckland) he was able to determine an observatory latitude of 37° 08′ 34.06″ South and longitude 175° 32′ 50.41″ East. Using *Google Earth*, these coordinates place it 300 metres east of Pollen Street (the stated location). The positional precision back then was undoubtedly not as accurate as today!

According to modern weather records, Thames has clear skies for 30% of the year, mostly clear skies for 15%, is partly cloudy for 15%, mostly cloudy for 10% and has total cloud or rain for 30% (Weatherspark). Having clear or mostly clear skies for ~45% of the time would have permitted Grigg to carry out a reasonable number of observations under suitable conditions in any given year.

Grigg retired from full-time work in 1894 when he was 56 in order to "... give more time to his scientific inclinations." (Orchiston, 2016: 279). He then resided in Queen Street (Thames), which ran parallel to Pollen Street where he formerly lived. At Queen Street he constructed a larger, two-storied observatory with the dome and transit room upstairs (simply the relocated original observatory building), while the ground floor served as an office and workshop (Orchiston, 2016: 279), as shown in Figure 6.

Orchiston (2016: 484) points out that Grigg's astronomical interests were varied. His principal interest was comets, both the observation of known ones and the discovery of new ones. He ran a search program from 1886 or 1887 until approximately 1907 (i.e. for about 21 years).

Grigg was also a pioneer in NZ of astrophotography, and he photographed the Sun, the Moon, star fields and comets (Orchiston, 1995; 2016: 597–607). Figure 7 shows one of his photographs of Comet C/1901 G1 (the 'Great Comet') in Orion, which he included with a letter he sent the famous Australian astronomer John Tebbut in 1901. This photograph is now in the Mitchell Library in Sydney, and is currently considered to be the earliest-known successful photograph of a comet taken from NZ, as three purported images of the Great Comet of 1882 shown over Mount Egmont are believed to be fakes (Orchiston et al, 2020: 633–636).

The presence of the 'Belt' in Orion in Figure 7, helps determine that the tail of Comet C/1901 G1 (as recorded in the photograph) was ~7° long at this time (on 12 May). Also note the slightly shorter and fainter ion tail on the left of the dust tail. Contrary to the claim by the journalist Mackrell (1985: 78), for this fairly wide field-of-view (~11° × 11°), Grigg must have used the 'star camera' attached to his Wray telescope and conspicuous in Figure 5 and not the prime-focus camera that he could attach to the eyepiece end of the Wray telescope (as shown in a photograph in Orchiston, 2016: 600). It appears that the above comet photograph is actually a cropped version of the original, which likely covered a wider field of view. The faintest stars recorded are around magnitude 7.5. In his obituary notice (Obituary, 1920), it was claimed that

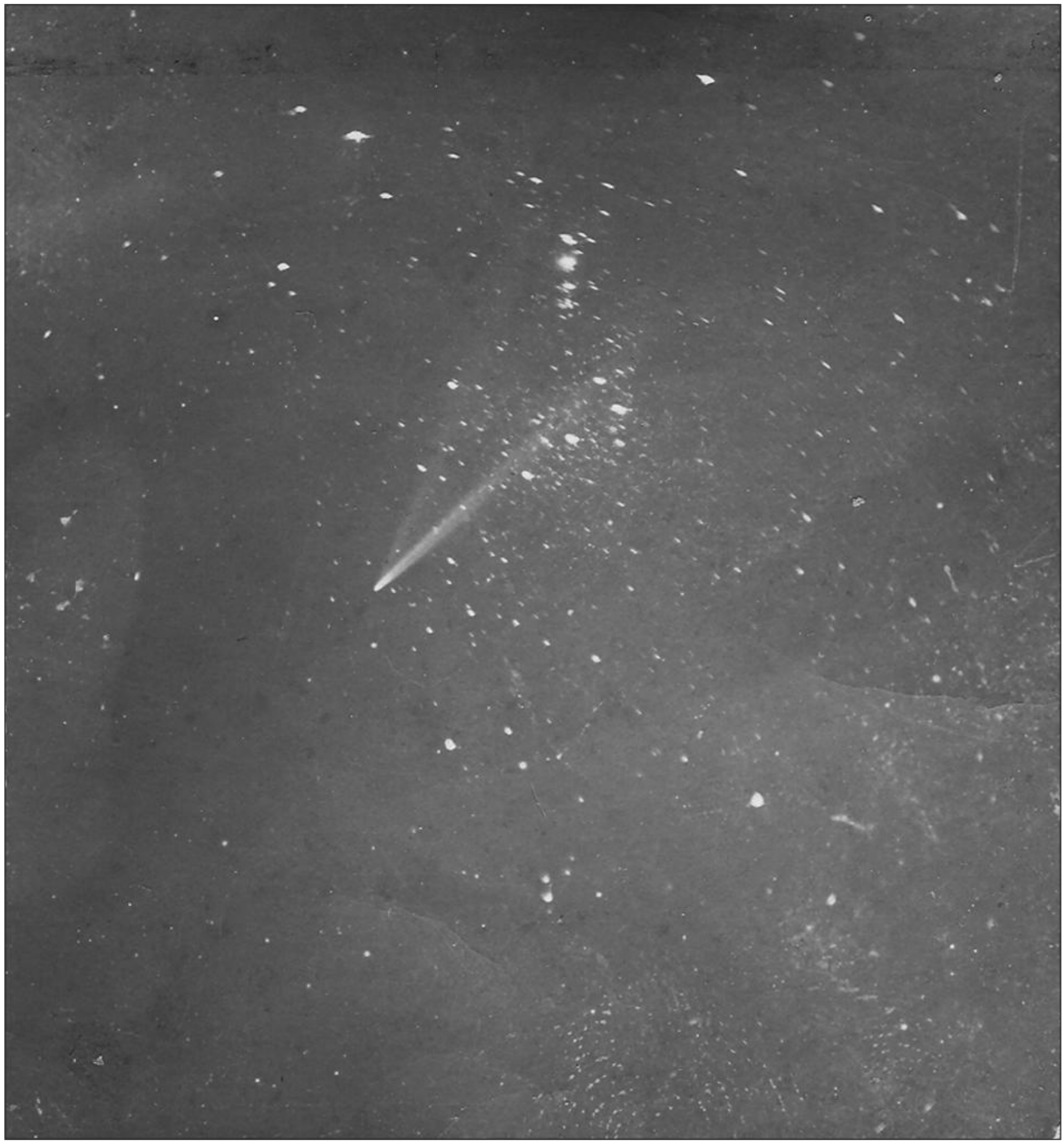

Figure 7: C/1901 G1 in Orion as photographed by John Grigg on 11 May 1901 (UT) at approximately 7 pm local time (based on GUIDE) (photograph courtesy Mitchell Library, Sydney).

Grigg "... secured the only successful photograph in the world of the 1901 comet …", however, Kronk (2007: 12) reveals that Charles Dillon Perrine (1867–1951) "... obtained four photographs of the comet with the Pierson-Dallmeyer camera …" on 6 May 1901 (UT) from Padang, Indonesia. This was during the Lick Observatory's solar eclipse expedition to what was then the Dutch East Indies (see Pearson and Orchiston, 2011).

As previously stated, Grigg and Austin are NZ's leading discoverers of comets, with three apiece named after them. Following is an overview of the comets Grigg discovered.

### 3.1 1902 O1 (26P/Grigg-Skjellerup), Initially 1902 II = 1902c

Initial details of this comet are listed below in Table 2, its orbit is illustrated in Figure 8 and its path through the sky is plotted in Figure 9. For further information see Kronk (2007: 18).

Grigg's first comet discovery (C/1902 O1) occurred on "... July 22d 18h 30m Gr. M. T. ..." 1902 (Greenwich Mean Time) when he was conducting his "... monthly survey …" for undiscovered comets (Grigg, 1902c: 389; Kronk, 2007: 18–19). The comet, initially 1902 II = 1902c (Kronk, 2007: 19) was discovered tele-scopically using the 3.5-inch Wray refractor at 25× magnification and appeared as a "... faint nebula … about twice the diameter of Jupiter …", that is, approximately 1.5′ across (Grigg, 1902c: 389; Kronk, 2007: 19; Orchiston, 2016: 280). The RA was 11 hours and 35 minutes and the Dec was +07° 00′; it was in Virgo, very close to the border with Leo (Grigg, 1902c: 389). Grigg identified "... the daily motion is south 22min, east 1 ¼ degrees." (*Poverty Bay Herald*, 1902: 2). On 6 August 1902 Grigg (1902c: 390–391) wrote to the *Astronomische Nachrichten* stating that he had found a "... nebulous object …" which was not located in his own list of nebulae nor in Webb's or Proctor's Atlases. He never got to confirm any cometary movement as a fire broke out near his premises and ended his observing for the night (Grigg, 1902c: 389). Grigg confirmed the comet's movement four nights later on 26 July (NZST). Grigg observed the comet for a total of 14 times on six nights from discovery (22 July) until 3 August 1902 (Kronk, 1984: 255; Orchiston, 2016: 485). With this crude astrometry, Grigg tried to determine basic orbital elements (A New Comet, 1902: 3; Kronk, 2007: 19; Orchiston, 2016: 280). Based on these provisional elements he confirmed that it was not "... Comet Brooks 2a or Tempel 3-Swift …" (Grigg, 1902c: 390). Grigg's observations were published in the *Astronomische Nachrichten* (Barachi and Grigg, 1903: 213).

Table 2: An overview of Grigg's first comet. Note that for the eleven overview tables for the New Zealand comet discoveries, we have endeavoured to utilise information primarily from the Minor Planet Center (MPC) (MPC Search). This database has been chosen as a standard for this paper as it allows the calculation of the orbit of the comet at the apparition at which it was discovered, rather than that of the current epoch (modern day)—a critical distinction for a paper talking about these comets in their historical context. If MPC material was unavailable, we then used NASA/JPL material (NASA/JPL Horizons), if this also was not published, we used information from Kronk's *Cometography* volumes. In addition, papers in research journal were used. Also, specific to New Zealand's geographic location, GUIDE planetarium software was accessed for comet visibility times. Whilst the elements from the database(s) are given to more significant figures than presented here, in these orbital element tables we have chosen to round all values to a meaningful number of significant figures for internal consistency.

| | |
|---|---|
| Discovery Date | 22 July 1902, ~6pm NZST (22 July 1902 UT) |
| Discovery Magnitude | "… extremely faint …" (Kronk. 2007, 19) |
| Discovery Declination | +07° (Leo) |
| Perihelion date (orbital elements not stated in MPC for 1902. Used NASA/JPL, or Kronk if needed). | 03.503 July 1902 (NASA/JPL) |
| Perihelion distance | 0.753 au (NASA/JPL) |
| Perigee | 10 August 1902 UT (Kronk, 2007: 18) |
| Perigee distance | 0.7389 au (Kronk, 2007: 18) |
| Brightest | 'Faint' (Kronk, 2007: 19) |
| Visible from NZ | 23 July (discovery) to late October 1902 (GUIDE) |
| Last observed (1902 apparition) | 3 August 1902 UT (Kronk, 2007: 19) |
| Eccentricity of the orbit (e) | 0.73629 (Marsden, 1972b) |
| Semi-Major axis (a) | 2.799 au (GUIDE) |
| Aphelion distance (Q) | 4.83 au (Marsden, 1972b) |
| Inclination (i) | 8.294° (Marsden, 1972b) |
| Epoch | 20 June 1902 (NASA/JPL) |
| Period | 5.26 years (NASA/JPL – 2012 epoch) |
| Observations in Kronk (2007) | 6, all by Grigg (including discovery) |
| Observations in MPC | 827 (1922 05 22.93 – 2024 10 07.26) |
| Observations in COBS | Not listed |
| Papers Past newspaper articles | Relating to Grigg or 26P: 35 (July 1902) |

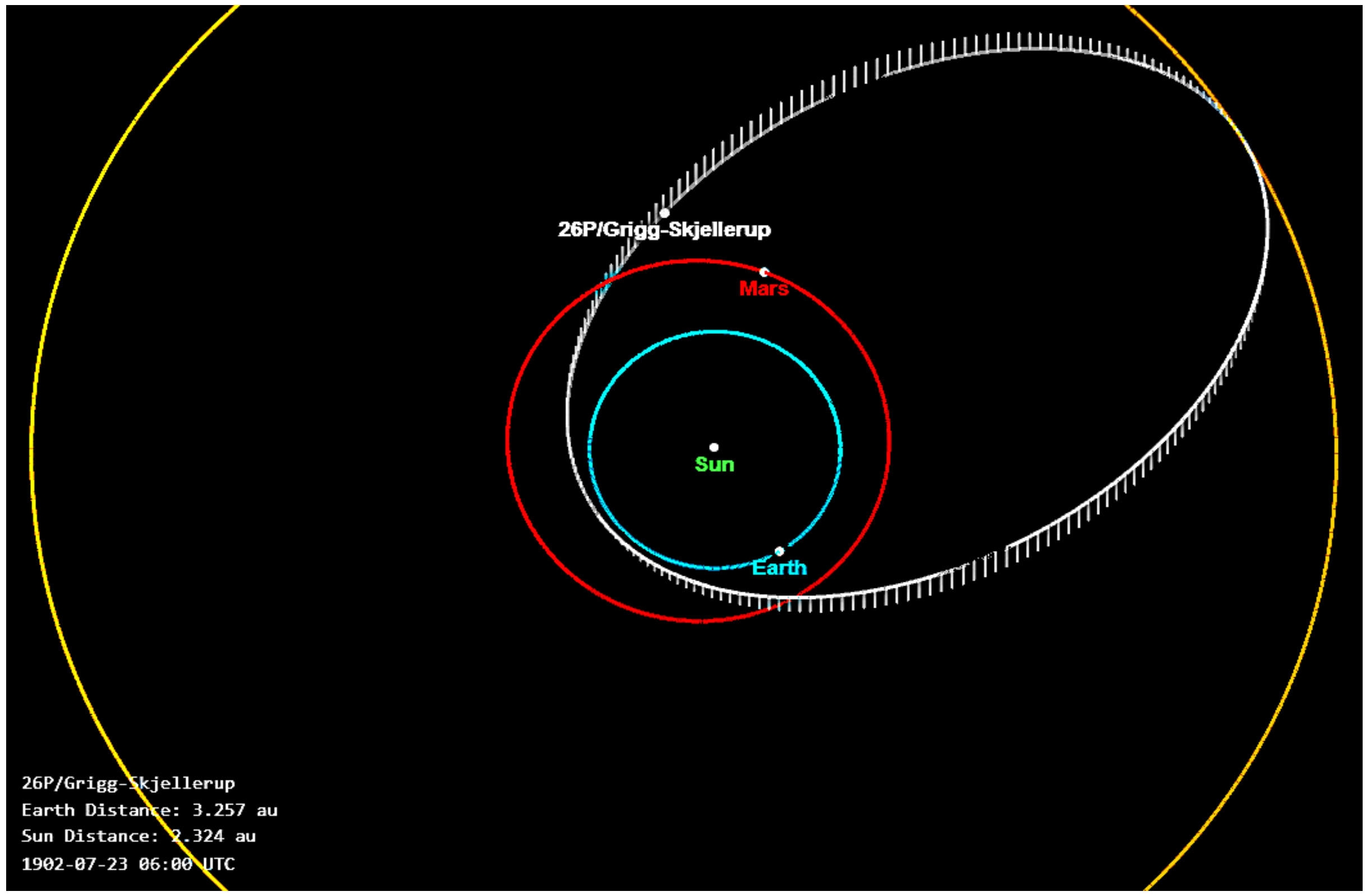

Figure 8: The orbit of Comet 26P/Grigg-Skjellerup at the time of discovery on 23 July 1902 UT. Note how the aphelion reaches Jupiter's orbit. The comet orbits the Sun in an anti-clockwise direction as seen from above. The vertical lines reveal where the comet's orbit would place it above and below the ecliptic plane. The orbital period of the comet at this epoch is 5.26 years. From NASA/Jet Propulsion Laboratory (JPL) Small-Body Database Lookup (https://ssd.jpl.nasa.gov/).

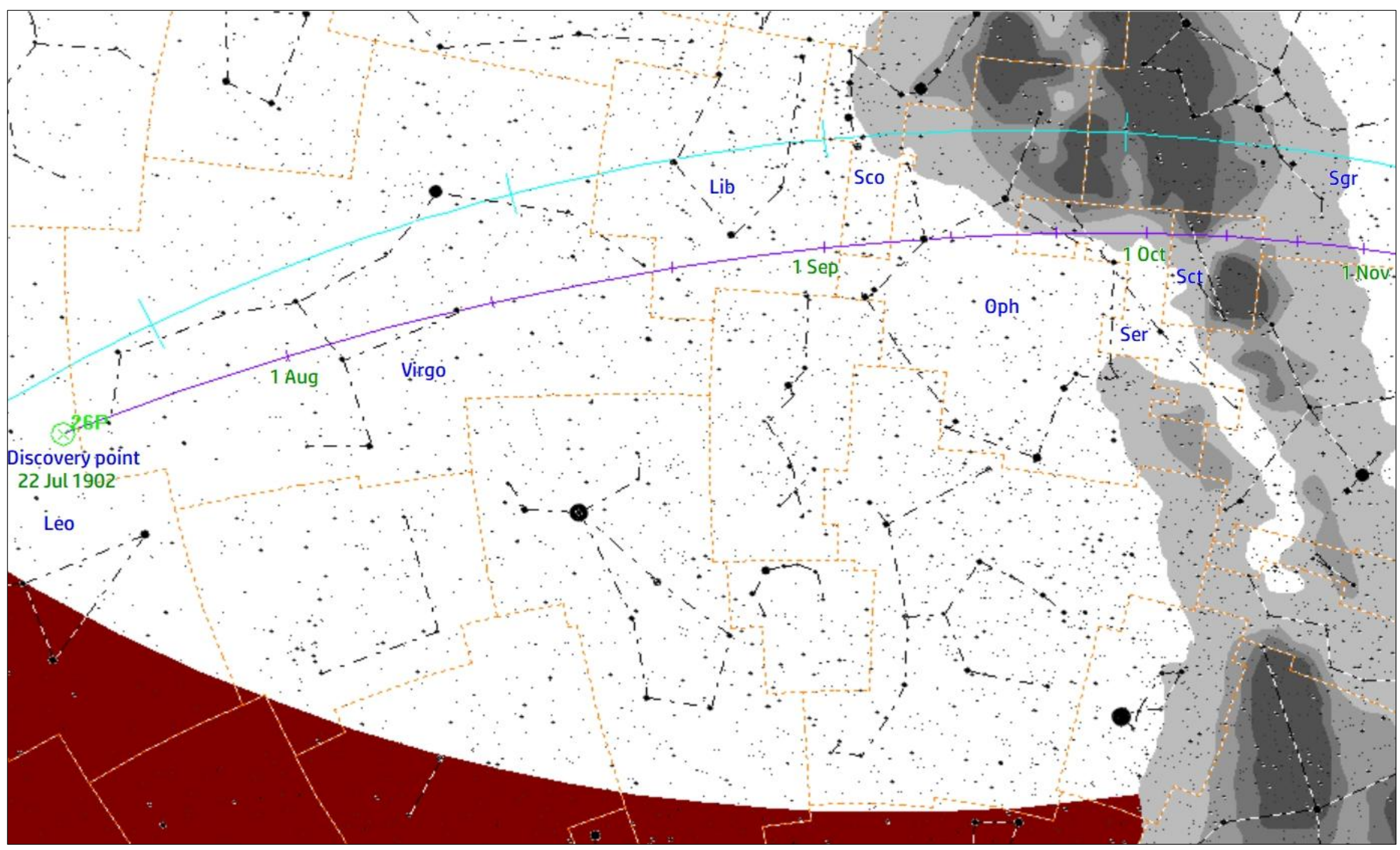

Figure 9: The path of Comet 26P/Grigg-Skjellerup (purple) from discovery on 23 July 1902 (UT) until early November. The constellations it traversed are labelled in blue font. The start of each month is in green. The blue line is the ecliptic. The red region is the northern horizon as seen from NZ. Note how the comet never crossed the ecliptic (got too close to the Sun) since perihelion had already occurred in April 1902, over three months before discovery. Diagram produced by GUIDE software.

Satisfied that it was a newly discovered comet, Grigg wrote to the prominent Australian amateur astronomer Walter Gale (1865–1945; Orchiston, 2017: 402–403) and Pietro Baracchi (1851–1926; Orchiston, 2017: 335–336), the Director of Melbourne Observatory, which was the designated Australasian reporting centre for discoveries (Grigg, 1902c: 389–390; Orchiston, 2001: 17; 2016: 485–486). Baracchi did not receive the letter until 6 August (Baracchi and Grigg, 1902: 213). Grigg also sent a notice to the *New Zealand Press Association* for publication in NZ newspapers. News of the comet discovery was published in 35 newspaper articles across NZ between 28 July and 5 August. However, there were no reports of other NZ astronomers observing it (based on Papers Past searches between 23 July and 17 August). Grigg (1902c: 391–392) confirmed this by stating: "... I have not heard of any other observations than my own …" on 6 August 1902. Even though Grigg was a competent astrophotographer by 1902, it would have been pointless to attempt to photograph his comet as the sensitivity of photographic plates at that time and his modest equipment would not have recorded the faint, diffuse light of the comet.

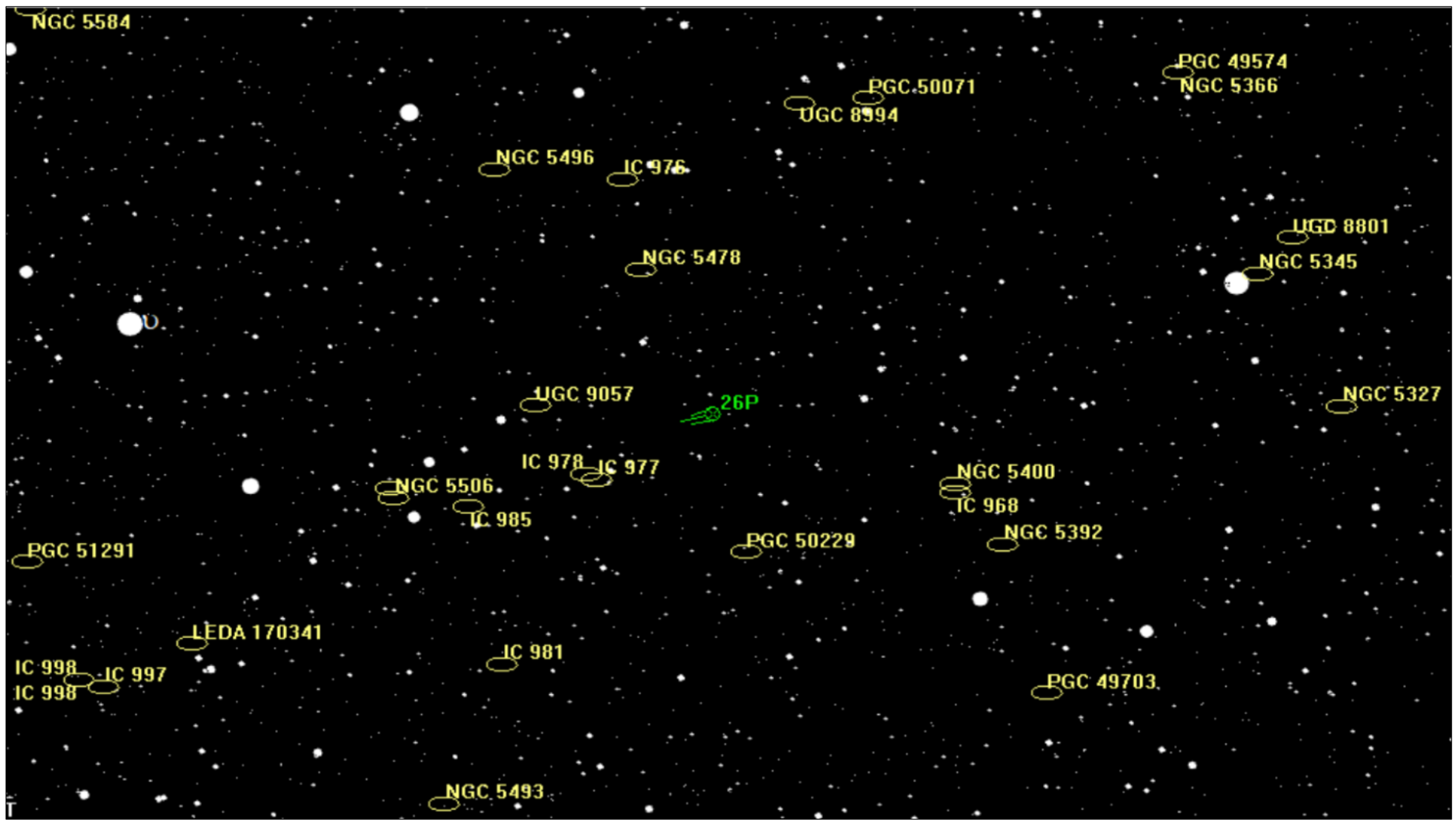

Figure 10: Comet 26P/Grigg-Skjellerup and the crowded field of galaxies it lay in when Baracchi tried to locate it on 14 August 1902 UT. The faintest galaxies shown are magnitude 15. The field of view is 8° × 5°. The field is centred on RA 14h 02m and Dec –02° 06'. From GUIDE software.

Orchiston (1993; 1999a) and Kronk (2007) have highlighted the problems Australian astronomers had in confirming the existence of Grigg's comet. Indeed, they could not. The crude positions that Grigg supplied—obtained simply by reading the RA and Dec positions off the circles, not from micrometric observations—were not accurate enough (Orchiston, 2001: 17; 2016: 485–488). Orchiston (2001:17) writes that Baracchi

> … shirked his responsibility by not immediately cabling Grigg's positions and discovery circumstances to Kiel (the international comet centre) and to Australasian cometary astronomers … so that they could confirm it. However, Baracchi stated that he sent a discovery notice to the *Astronomische Nachrichten* the same day that he received Griggs' letter on 6 August—a full two weeks after initial discovery (Baracchi and Grigg, 1903: 213–214). Baracchi also searched the area for the new comet on 14 August 1902 (the first clear night after notification) between RA 13h and 15h and Dec +05° to –05°, "... but the comet could not be found." (*ibid.*). Barachi noted that it was a moonlit night. GUIDE software confirms this: on 14 August 1902 (UT) the Moon was 11 days old, therefore very bright. To see the 'faint nebula' of the comet, especially if the degree of condensation was low (that is, it had little central condensation) would have been very difficult due to the bright sky overwhelming the low signal to noise (comet to background sky) ratio. It should also be noted that this region in Virgo is rich in faint galaxies that resemble faint comets (see Figure 10).

Barachi also notified John Tebbutt (1834–1916; Orchiston, 2017) of Windsor Observatory, near Sydney, asking him for confirmatory observations. Tebbutt replied that

> ... there were not sufficient data for finding so faint an object, and [as] I was closely engaged in minor planet work, I did not make a search. (Orchiston, 2016: 488).

"Not sufficient data" may possibly refer to not accurate enough ephemerides. Tebbutt also noted that Grigg was the only one to observe this comet and "... roughly …" too (Tebbutt, 1907: 286). He also lamented that "...Grigg failed to communicate the discovery in time to those who were better able to make observations".

Perhaps if Grigg had sent a cable to Baracchi as soon as he confirmed the comet's movement on 23 July 1902, Baracchi would have had a greater chance of confirming the moving target. He tried, unsuccessfully, on 14 August. One wonders if the numerous galaxies in the vicinity put Baracchi off confirming which faint smudge moved. Figure 10 shows the comet's field and the number of galaxies brighter than magnitude 15 surrounding it on the date Baracchi did his search. Confirmation was also attempted from the Harvard College Ob-servatory (USA) after the astronomers there received an Astronomical Telegram from Professor Kreutz of Kiel Observatory on 16 September 1902 stating that John Grigg discovered a comet on 22 July (Pickering, 1902: 1). A prediction of its position on 17 September was given: RA 15h 37m 16s, Dec –07° 58' (refer to Table 3). It stated that the predicted position was a "... rough approximation …" and that the comet was fading. Any observations were asked to be reported. Using GUIDE software (set for a 1900 epoch), which utilised many later observations of 26P/Grigg-Skjellerup as it would later be called (see below) for better orbital elements, the position for the date given was a full two hours in RA and four degrees in Dec off. It is not surprising that there were no additional observations!

Table 3: Grigg's computed ephemerides of Comet 26P/Grigg-Skjellerup compared to GUIDE's suggested positions.

| Date (GMT) | Grigg RA | GUIDE RA | Grigg Dec | GUIDE Dec | Notes |
|---|---|---|---|---|---|
| Jul 23.8 | 11h 40m | 11h 40m 23s | +06° 35′ | +06 39′ 30″ | Grigg's ephemerides |
| Jul 26.8 | 12h 00m | 11h 59m 28s | +05° 30′ | +05 31′ 58″ | |
| Jul 29.8 | 12h 20m | 12h 19m 03s | +04° 20′ | +04 20′ 57″ | |
| 14 Aug | 13–15 h | 14h 02m | +05 to –05° | –02 06′ | Barachi's search |
| 17 Sep | 15h 37m | 17h 18m | –07° 58′ | –12 02′ | Harvard's search |

Based on NASA/JPL and Kronk (2007: 18–19), the comet was discovered at its brightest, when it was around magnitude 9 in July 1902. This stands to reason, as the limiting magnitude for a 3.5-inch telescope is typically around 12.5 for stellar objects. For cometary objects, and other diffuse targets with a lower surface brightness, this limiting magnitude is typically at least two magnitudes brighter than this. As a result, for Grigg's telescope, the limiting magnitude for a comet was around 10–11. In addition, the Moon was very bright at 17.8 days old when discovered, thus, only four days past full (GUIDE). Once the Moon was above the horizon, this would have significantly impacted the visibility of faint objects, such as the comet, due to scattered moonlight/glare. However, it is likely that Grigg used the short dark window between late twilight (6:30pm local) and the rising of the Moon (7:45pm) to find the comet.

Fortunately, on 17 May 1922, the Australian amateur astronomer John Francis (Frank) Skjellerup (1875–1952; Orchiston, 1999; 2003b) discovered a new comet from South Africa. It was then tracked for four months. Orchiston (1993: 71) states that Dr. Gerald Merton (1893–1983) of Oxford University Observatory realised that the orbital elements of the new comet and those of Grigg's comet were almost identical. Sadly, Grigg died two years before Skjellerup's confirming discovery. This comet is now known as 26P/Grigg-Skjellerup (Hughes, 1991; Kronk, 2007: 392; Orchiston and Drummond, 2024) and is one of the best-researched of all short period comets. Indeed, it was visited by the Giotto probe in July 1992, which approached within 200 km of the comet's nucleus (Reinhard, 1987: 525). It is also the parent body for the episodic Pi-Puppids meteor shower (Vaubaillon and Colas, 2005: 1139–1144) which peaks around 23 April of each year. The nucleus is estimated to be approximately 2.5–4.0 km in diameter (depending on the method used) and to be relatively dust-poor (Reinhard, 1987: 525). It currently has an orbital period of 5.26 years (NASA/JPL; Orchiston and Drummond, 2024), though its orbit is frequently modified as a result of gravitational perturbations by the giant planet Jupiter.

Because Grigg was the only person to observe this comet in 1902, there were no additional observations of it from NZ detailed in Papers Past searches for 1902.[3] However, Grigg's next comet proved to be much different.

### 3.2 1903 H1 (Grigg): 1903 III = 1903b

Initial details of this comet are listed below in Table 4, its orbit is illustrated in Figure 11 and its path through the sky is plotted in Figure 12. For further information see Kronk (2007: 38).

Grigg's second official comet discovery, C/1903 H1 (Grigg), was more straightforward than his first in terms of confirmation and international observability. He discovered the comet in the evening sky on 17 April 1903 (NZST) (17.29 April 1903 UT) (Kronk, 2007: 38; Ross, 1903: 77) near the star ζ Eridani (*Ashburton Guardian*, 1903: 2; Kronk, 2007: 38–39; Vsekhsvyatskii, 1964: 357). Initially he was searching for Comet 1903a (C/1903 A1 Giacobini), after receiving updated ephemerides from Kiel which were forwarded to him by Baracchi, however,

> ... there were … some clearer places elsewhere, and I turned my [telescope] tube towards them. Soon I found a nebulous object, roughly noting by circles R.A. 3h 7m, S. Dec. 11° 6′, with two stars in a field of diameter 1° 5′. (Grigg, 1903: 320).

Of note is that Grigg said that he centred "… on a field of 1° 5′ diameter …", thus revealing the possible true field of view of this telescope and eyepiece combination. The comet was described as moving ESE at 1.25° a day (*Evening Star*, 1903: 4). Learning from his first comet dis-covery, Grigg promptly cabled Pietro Baracchi in Melbourne, Australia, on 22 April, rather than sending a letter (*Auckland Star*, 1903: 4). Grigg wrote,

> As soon as I verified its cometary motion [Geddes observed it on Friday 17 and Saturday 18 April], I cabled to Australia and have received information from Messrs. Tebbutt, Gale, and Butterfield that they have located it. (Ross, 1903: 77).

Table 4: An overview of Griggs' second comet. Epoch 25.0 March 1903 (MPC). Based on the Minor Planet Center (MPC). If not listed in MPC, information was used from NASA/Jet Propulsion Laboratory Horizons System, Kronk (2007), or GUIDE.

| | |
|---|---|
| Discovery Date | 17 April 1903, ~7pm NZST (17.29 April 1903 UT) |
| Discovery Magnitude | "… extremely faint …" (Kronk, 2007: 38) |
| Discovery Declination | –11° (Eridanus) |
| Perihelion date | 25.920 March 1903 UT (MPC) |
| Perihelion distance | 0.499 au (MPC) |
| Perigee date | 21 April 1903 UT (Kronk, 2007: 38) |
| Perigee distance (q) | 1.3735 au (Kronk, 2007: 38) |
| Brightest | Not stated |
| Visible from NZ | 17 April (discovery) to late May 1903 (too faint) (GUIDE) |
| Last observed | 28 May 1903 (UT), magnitude: "… very faint …" (Kronk, 2007: 38) |
| Eccentricity of the orbit (e) | 1.0000 (MPC) |
| Semi-Major axis (a) | Not defined by MPC or NASA/JPL |
| Aphelion distance (Q) | Not defined by MPC or NASA/JPL |
| Inclination (i) | 66.484° (MPC) |
| Epoch | 25.0 March 1903 (MPC) |
| Period | Not defined by MPC or NASA/JPL |
| Observations in Kronk (2007) | 19 (all visual), 1 from NZ (discovery) |
| Observations in MPC | None (32 used to determine orbital elements) |
| Observations in COBS | None |
| Papers Past newspaper articles | 67 (17 Apr–30 May 1903) |

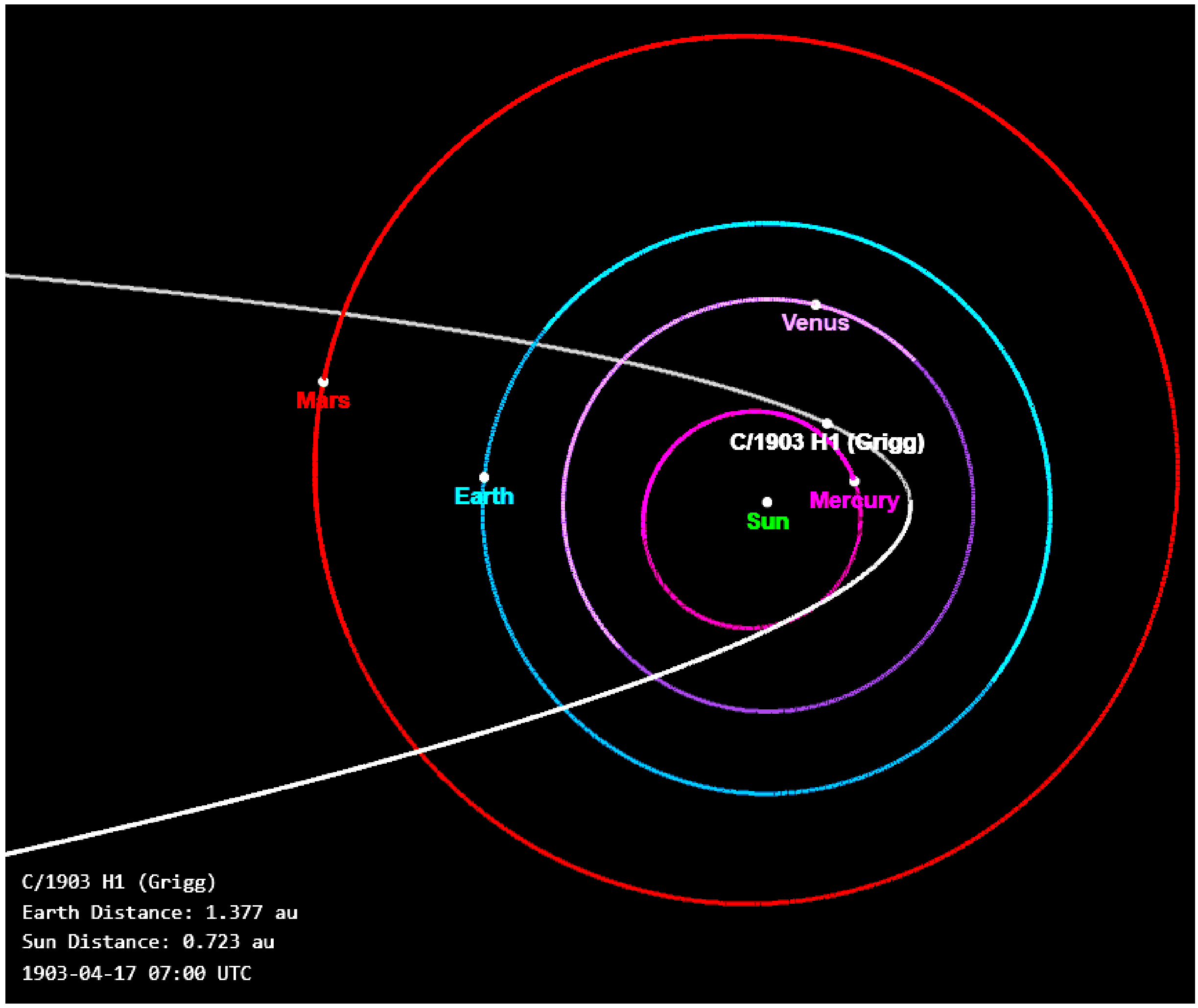

Figure 11: The orbit of Comet 1903 H1 (Grigg) at the time of discovery on 17 April 1903 UT. The comet has an anti-clockwise orbit, so it came up from lower centre, rounded the Sun and then headed off to the left (from NASA/Jet Propulsion Laboratory (JPL) Small-Body Database Lookup (https://ssd.jpl.nasa.gov/)).

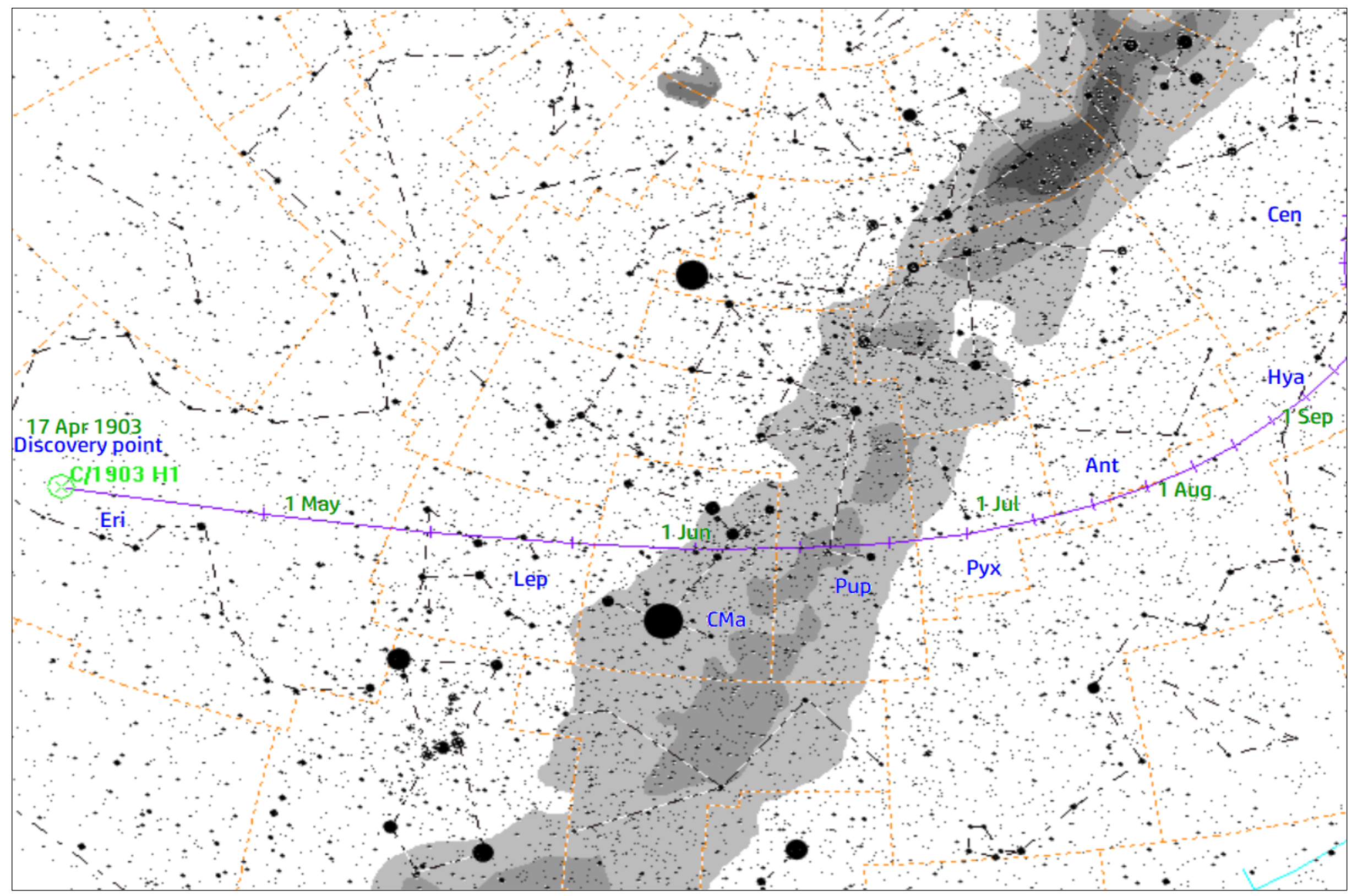

Figure 12: The path of Comet 1903 H1 (Grigg) from discovery on 17 April 1903 (UT) until September 1903 is shown in purple. It was too faint for NZ observation from late May 1903, but was seen by Tebbutt (Australia) and Cox (South Africa) on 28 May 1903 (UT). The constellations C/1903 H1 traversed are labelled in blue. The start of each month is in green. The light blue line (lower right) is the ecliptic. North is down (for the NZ sky), East is right. GUIDE software.

In his letter to the British Astronomical Association, Grigg apologised for not being more precise with his determination of the comet's position, stating,

> I regret that I cannot yet obtain close readings. My chief difficulty is that I have no means of illuminating the wires for viewing a faint object, and no good catalogue or atlas showing the small stars. (*ibid.*).

This deficiency undoubtedly influenced his rather poor astrometry for his first comet discovery (above).

According to GUIDE software, the comet set at 8:20 pm local time on 17 April. Astronomical twilight ended at 7:08 pm. The 19-day old Moon rose at 9:30 pm. Thus, it was an evening comet discovery. GUIDE suggests that the comet was brightest three weeks prior to discovery, around 27 March (NASA/JPL not stating the magnitudes in the ephemerides). At that stage it only had an eastern elongation of 10° from the Sun. Based on GUIDE, and this is just an approximate indication, on the discovery date, it was possibly around magnitude 8–9. By the evening of the last observations by Tebbutt and Cox (see below) on 28 May 1903, it was "... very faint … [and] only just visible." (Kronk, 2007: 38). Interestingly, no observer submitted a magnitude estimate in Kronk, choosing rather to simply say that the comet was 'extremely faint'—or words to that effect. The largest aperture used for observations stated in Kronk was Tebbutt's 8-inch (20-cm) Grubb refractor shown here in Figure 13.

Observations of this southern target were made exclusively from the Southern Hemisphere, namely NZ, Australia and the Cape (later called South Africa in 1910) (Kronk, 2007: 38–39; Vsekhsvyatskii, 1964: 357). John Tebbutt, Walter Gale, and Mr Butterfield confirmed the discovery from Australia (Ross, et al., 1903:

77–78). Tebbutt and Gale were well-known observational astronomers with special interests in comets, but Butterfield's forte was astronomical outreach and he was best known for his planispheres (see Orchiston, 2003a).

C/1903 H1 (Grigg) reached perigee on 21 April 1903 UT at a distance of 1.37 au. Perihelion occurred earlier on 26.01 March 1903 (UT) (Kronk, 2007: 38), ten days before discovery. According to Kronk (2007: 38–39) the following observations were made. In Australia, John Tebbutt confirmed the comet on 25 April from Windsor Observatory, near Sydney (Orchiston, 1993: 71). He also observed it with his 8-inch refractor on a number of occasions between 10 and 28 May. He described it as "... very faint …" On 4 May, Walter Hubert Cox (1864–1932) from the Royal Observatory, Cape of Good Hope, South Africa, used an old 7-inch (18-cm) refractor and described it as "... an extremely faint nebulous mass with no visible nucleus …" Cox's last observation of 1903b (C/1903 H1) was on 28 May 1903 and he described it as being "... only just visible …" with the 7-inch refractor. Tebbutt and Cox were the last to see it.

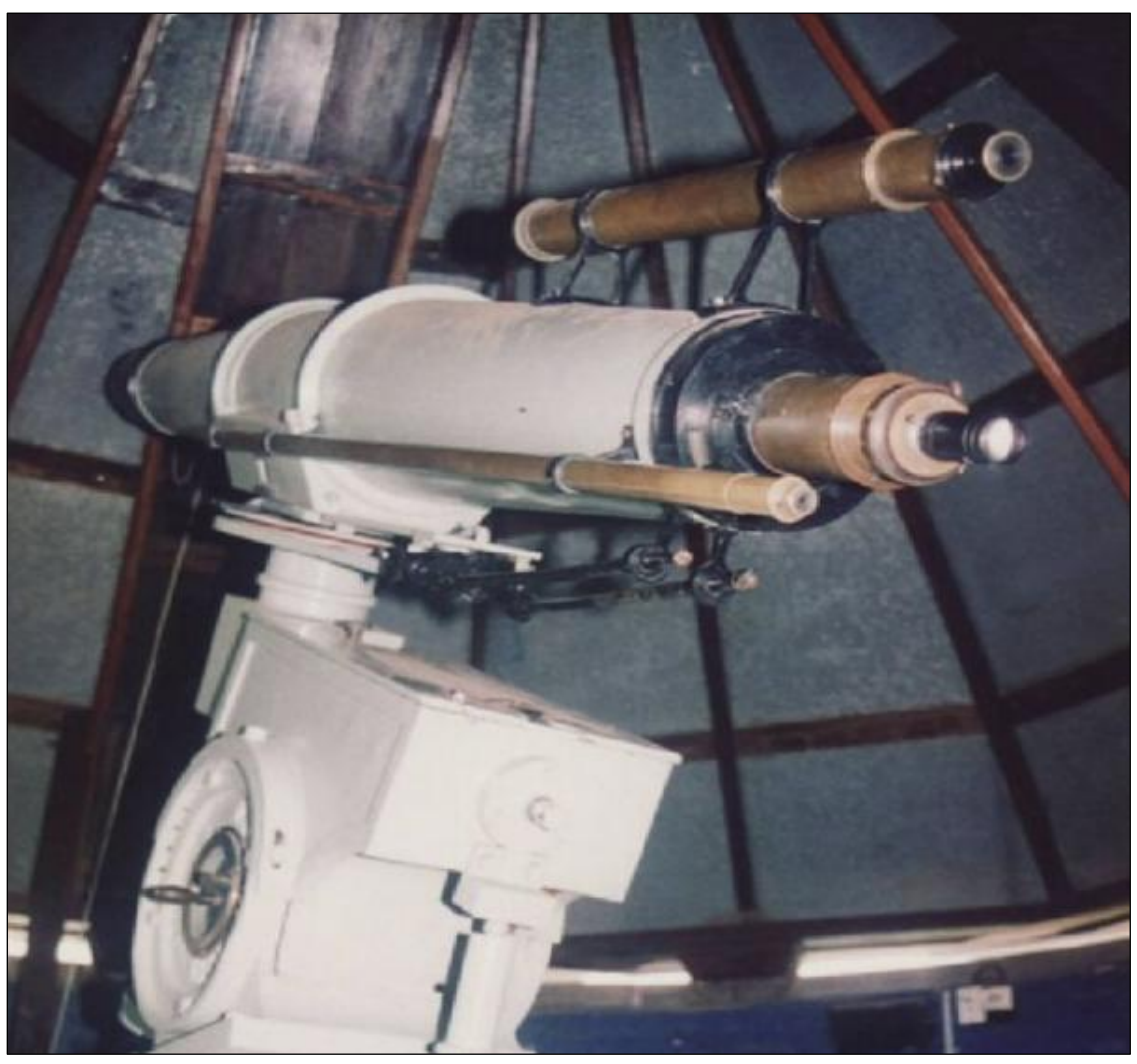

Figure 13: The 8-inch Grubb refractor used by John Tebbutt to observe Comet 1903 H1 (Grigg) in its original dome at the renovated John Tebbutt Observatories in 1996 at Windsor, Sydney, Australia (photograph: Wayne Orchiston).

Orchiston (1993: 71) points out that Grigg observed his comet for a further 29 nights after his discovery up until 26 May. On 24 May he said that it was "... extremely faint though still large …" As stated, the limiting magnitude for comets with his Wray refractor was approximately magnitude 10–11. Of the 67 newspaper articles published in NZ about Comet Grigg (many repeats of the same discovery information), only one reports a positive observation of the comet (besides Grigg's observations). This was made from Wanganui (NZ) by Joseph Thomas Ward (1862–1927), Director of the Wanganui Astronomical Society's Observatory, probably around mid-May 1903. Ward wrote:

> In appearance, in the field of a low power eye-piece, it is a faint object, of its class, and might easily be mistaken for one of the thousands of little nebulous objects visible in the telescope on any clear night. (*Wanganui Herald*, 1903: 6).

Ward made an interesting side-note, writing,

> Mr Grigg, who has spent between twenty and thirty years of his active life many a vigil searching for these objects, may be congratulated on his discovery, making as it does the second in a few months. The two other comets are, I think, the only discoveries of (telescopic) comets made by NZ astronomers.

In fact, we know that Grigg began to systematically search for comets in 1886, eighteen years earlier (see Grigg, 1902b; 1906b).

From discovery (29° elongation east) to last observation (60° elongation east), C/1903 H1 had passed from Eridanus, through Lepus and into Canis Major. Ross et al. (1903: 77) reveal that the orbit was first computed by the prominent Australian astronomer, Charles James Merfield (1866–1931; Orchiston, 2015). Kronk (2007: 38–39) points out that the German astronomers Heinrich Carl Friedrich Kreutz (1854–1907) and C.W.L.M. Ebell also determined the orbit of C/1903 H1 (Grigg) using Tebbutt's observations, which was later refined by H.A. Peck in 1905 (*ibid.*). Grigg's second comet never reached naked-eye visibility, and neither would his third.

### 3.3 C/1907 G1 (Grigg-Mellish), 1907 II = 1907b

Initial details of this comet are listed below in Table 5, its orbit is illustrated in Figure 14 and its path through the sky is plotted in Figure 15. For further information see Kronk (2007:106).

Grigg's third and final comet discovery occurred on 8.33 April 1907 UT (Grigg, 1907: 364; Kronk, 2007: 106; Vsekhsvyatskii, 1964: 371). On 27 April 1907 Grigg described the 8 April comet observation as "... a large faint comet, a diffused mass of light about 15′ in diameter …" and that "… it was moving with great rapidity." He received a confirmation observation, stating "Mr. Ward, of Wanganui, found it on the 10th [April], and afterwards reported it on the 12th …" (Grigg, 1907: 364). Grigg sent a telegram to the Press Association from Thames which then issued the following, "... Mr John Grigg, astrono-mer, states under date April 8: 'There is a comet in Right Ascension 4hr 35min; declination 44deg; moving north-west.'" (*Ashburton Guardian*, 1907: 2). He also cabled Mr. Baracchi in Australia, stating on the 27th April "On the 9th I cabled to Mr. Baracchi, but he has not replied." (Grigg, 1907: 364). At discovery, 1907 II = 1907b (later, C/1907 G1) had an elongation of 70° east and was thus an evening target setting just after 1:00 am local time (GUIDE). The com-et was discovered in Caelum (*Evening Post*, 1907: 8). The *Timaru Herald* (1907: 5) stated that *Whittaker's Almanac* predicted that there were only two known periodical comets due in 1907, the names were not provided, however GUIDE suggests that they were Comet Westphal (20P, D/1913 S1), and Comet Tuttle-Giacobini-Kresak (41P, P/1907 L1). Both of these comets were brightest in October many months after Grigg's discovery, thus, Grigg had probably discovered a new comet (*Timaru Herald*, 1907: 5).

Table 5: An overview of Grigg's third comet, C/1907 G1 (Grigg-Mellish). Epoch: 28.0 March 1907. Based on the Minor Planet Center (MPC). If not listed in MPC, information was used from NASA/Jet Propulsion Laboratory Horizons System, or Kronk (2007).

| | |
|---|---|
| Discovery Date | 8.33 April 1907 UT (Grigg), 14.1 April UT (Mellish) |
| Discovery Magnitude | "Faint", around magnitude 8 (Ward and Allison) |
| Discovery Declination | –46° (Dorado) |
| Perihelion date | 28.05 March 1907 UT (MPC) |
| Perihelion distance (q) | 0.924 au (MPC) |
| Perigee date | 11 April 1907 UT (Kronk, 2007: 106) |
| Perigee distance | 0.2065 AU (Kronk, 2007: 106) |
| Brightest | Described as "faint". Peaked 3 days after discovery (11 April 1907 UT). Magnitude 8 (Ward and Allison) |
| Visible from NZ | 8 April (discovery) to late April 1907 – too far north (NASA/JPL) |
| Last observed | 14 May 1907 (UT), magnitude 12.0 (visual) (Kronk, 2007: 107) |
| Eccentricity of the orbit (e) | 1.0000 (MPC) |
| Semi-Major axis (a) | 62.427 au (NASA/JPL) |
| Aphelion distance (Q) | 123.931 au (NASA/JPL) |
| Inclination (i) | 110.057° (MPC) |
| Epoch | 28.0 March 1907 (MPC) |
| Period | 493.25 years (NASA/JPL) |
| Observations in Kronk (2007) | 23 (visual and photographic) – 3 from NZ (including discovery) |
| Observations in MPC | 5 (20 used to determine orbital elements) |
| Observations in COBS | None |
| Papers Past newspaper articles | 94 (8 Apr–8 May 1907) |

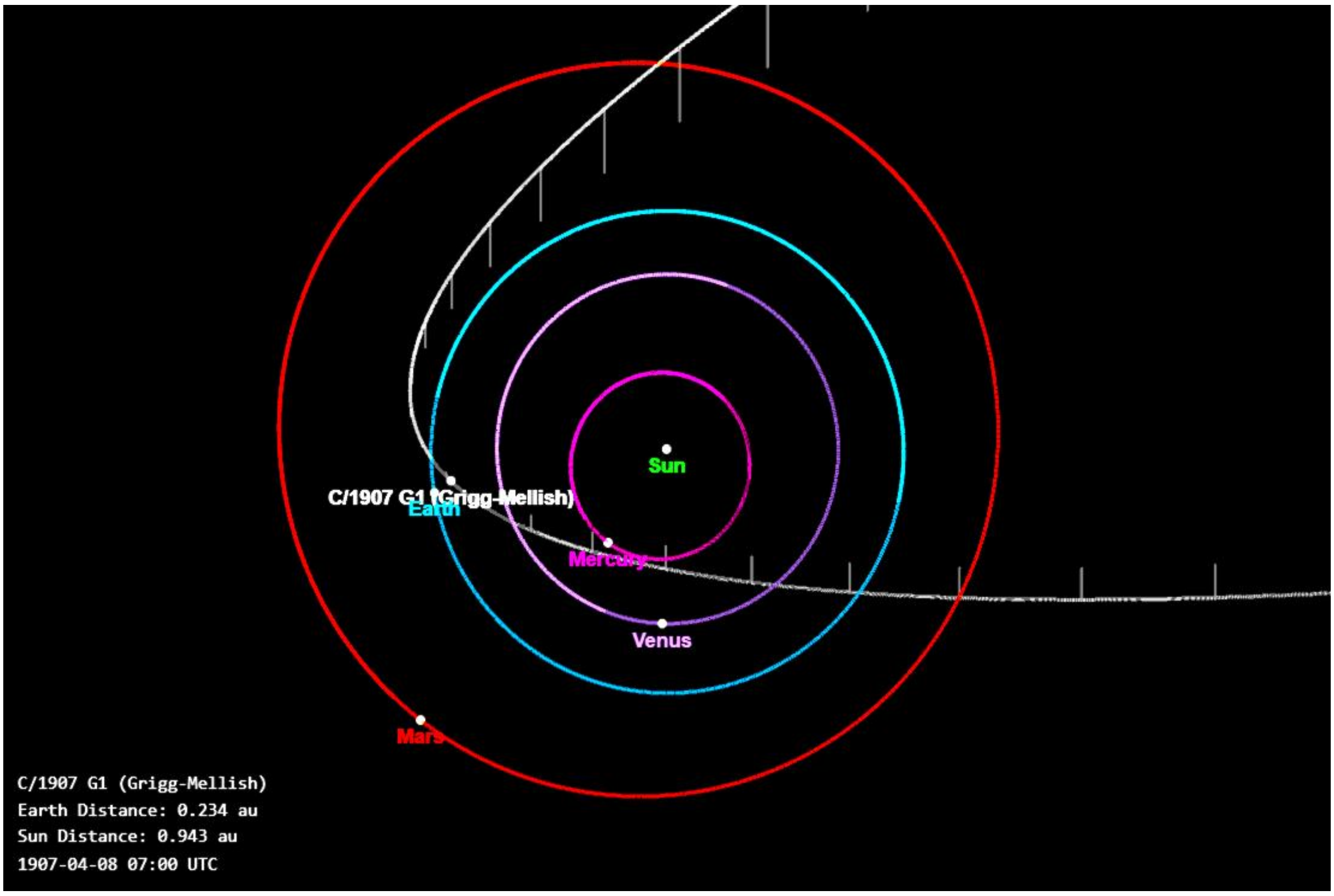

Figure 14: The orbit of Comet C/1907 G1 (Grigg-Mellish) at the time of discovery on 8 April 1907 (UT). The comet followed a retrograde (clockwise) orbit, entering from the lower right and exiting to the upper centre. From the Jet Propulsion Laboratory (JPL) - California Institute of Technology (https://ssd.jpl.nasa.gov/).

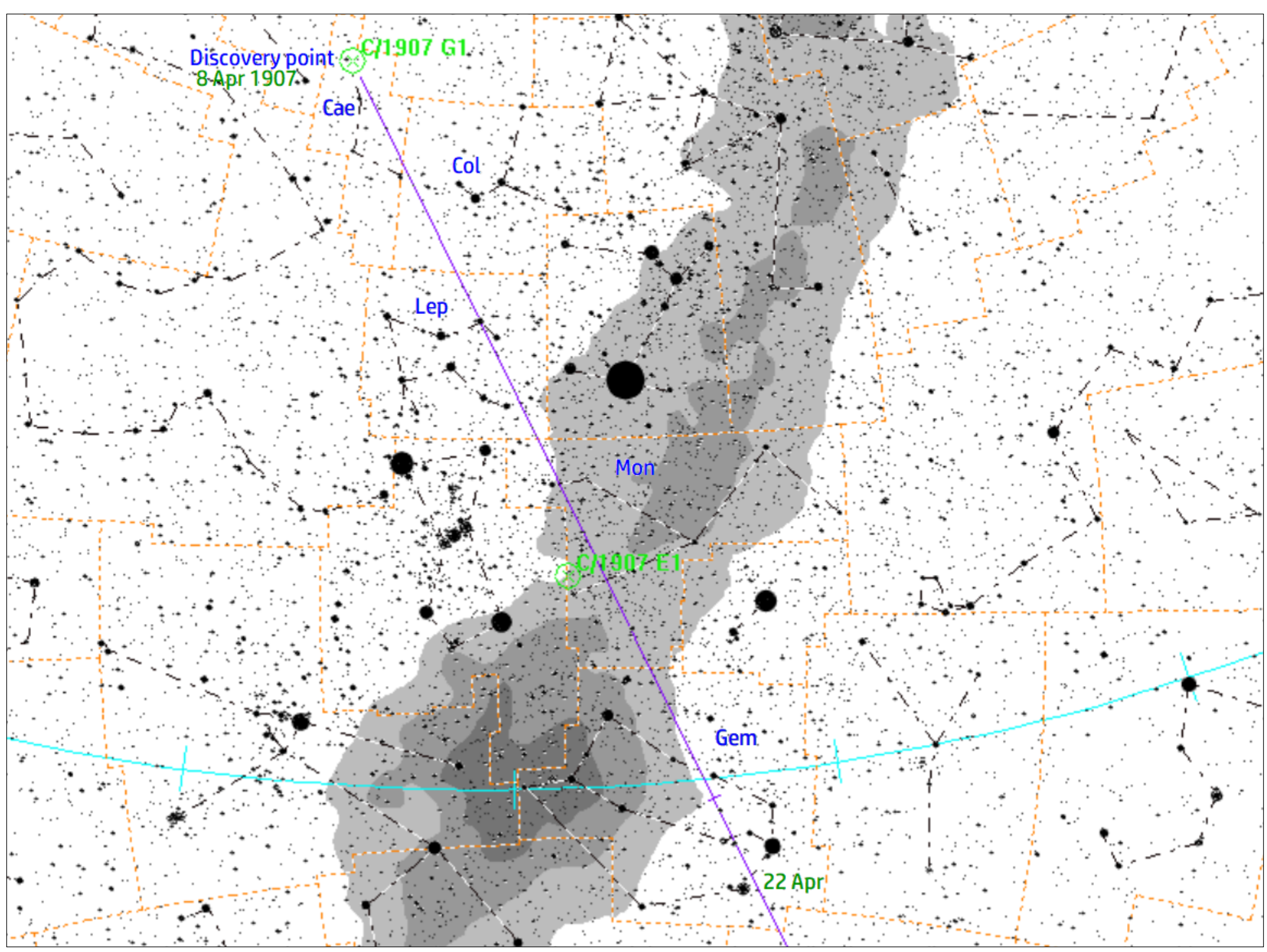

Figure 15: The path of Comet 1907 G1 (Grigg-Mellish) (purple) from discovery on 8 April 1907 (UT) until late April 1907. Note the rapid movement of the comet towards the north in a short time span. The constellations C/1907 G1 traversed are labelled in blue. The dates are in green. The light blue line is the ecliptic. North is down (for the NZ sky), East is right (GUIDE software).

Grigg observed C/1907 G1 again on 9 April stating,

> Its position at nine o'clock was right ascension 4 hours 59 minutes, declination south 37 degrees 22 mins; daily motion in right ascension plus 23 mins, daily motion in declination 6 degrees 40 mins. northwards. (*Auckland Star*, 1907: 4).

Note the very rapid rate that the comet moved towards the northern sky. NASA/JPL Horizons indicates that by the end of April 1907, that is, three weeks after discovery, it was too far north to be seen from NZ.

Based on Grigg's and Ward's observations (Grigg, 1907: 364), the comet had moved north by 32° and towards the east by 2 hours in RA between 8 and 12 April. Grigg was fortunate to discover it before it became a Northern Hemisphere object. Grigg wrote to John Torrens Stevenson (Auckland) that he saw the comet on Monday 8 April and that he was unsure if it was a known comet or not. He described it as faint (*Auckland Star*, 1907: 4) and needed a telescope to be seen (*Thames Star*, 1907: 2). News of the discovery travelled across NZ rapidly. Nelson amateur astronomer Frederick Giles Gibbs (1866–1953) tried to locate it on 9 April with the Atkinson telescope but was thwarted by clouds (*Nelson Evening Mail*, 1907a: 2), however, he did locate it on 10 April, describing it as "... a nebulous head without discernible tail." He concluded that the comet was only visible by means of a telescope (*Nelson Evening Mail*, 1907b: 2). Mr F. Stuart (FRAS) agreed (*NZ Herald*, 1907: 4), as did Ward, suggesting that it "... requires a telescope of from one and a half to two inches aperture to render it easily visible." (*Wanganui Herald*, 1907: 7). Joseph Ward and his assistant, prominent lawyer Thomas Allison (1858–1924), at the Wanganui Observatory estimated the comet to be magnitude 8 on 11 April (*Rangitikei Advocate and Manawatu Argus*, 1907: 3). A newspaper article released from Wellington stated that, "The Thames comet was observed again last night." (*Bay of Plenty Times*, 1907: 2). This indicates that it was seen from Wellington over two nights on 8 and 9 April.

Mr George Vernon Hudson (1867–1946), who had a 4.3-inch (11-cm) telescope at Karori (Western Wellington) (Mackrell, 1985: 78, 80), also confirmed the existence of the comet on 11 April (*Evening Post*, 1907: 6). Neither Kronk (2007: 106–108) nor Vsekhsvyatskii (1964: 371) recorded Hudson's, Ward's or other NZ observations of C/1907 G1—apart from Grigg's discovery and follow-up observations. Hudson stated (*Evening Post*, 1907a: 6):

> I picked up the comet discovered by Mr. Grigg on the 8th inst., at 7:45 last evening. It is now in the constellation Lepus, in a trapezium of stars above and to the south of Orion. Right ascension 5 hours 42 minutes, south declination 21 degrees 45 minutes. The comet is very faint, invisible to the naked eye, diffuse, with no defined tail or nucleus. Its substance is extremely attenuated, the faintest stars shining through the midst of it. Its motion is very rapid, the change in the position relative to the surrounding stars was visible after fifteen minutes observation with a low power. Its apparent movement across the sky is about equal to that of the moon. As the comet is so very faint, and its motion so rapid, it may be reasonably inferred that it is a very small comet, probably in close proximity to the earth.

Prominent US amateur astronomer John Edward Mellish (1886–1970) (Madison, Wisconsin) independently discovered Grigg's comet on 14 April 1907 (Kronk, 2007: 106; Vsekhsvyatskii, 1964: 371), nearly a week after Grigg. Note that Ward, Hudson and other NZers had confirmed Grigg's comet before Mellish's independent discovery. Mellish also noted the rapid movement in declination, stating it was moving north at 7° a day (and 3° east). As a result of two people independently discovering the comet, it officially came to be known as Comet Grigg-Mellish.

Kronk (2007: 106) suggests that C/1907 G1 (Grigg-Mellish) was closest to Earth between Grigg's and Mellish's discovery dates, on 11 April 1907 UT, and perihelion occurred in late May (Kronk, 2007: 107).

Based on the observations of the comet during its apparition, it is believed that it moves on an orbit with a period of approximately 490 years. It is interesting to note that comet Grigg-Melish has been identified as the parent of the Delta Pavonids meteor shower, which is active in late March each year (Jenniskens et al., 2020).

This ends our account of Grigg's three comet discoveries, but in fact he nearly ended with a fourth comet.[4]

## 3.4 GRIGG'S FINAL YEARS

Grigg officially discovered three comets in a span of five years (1902–1907), however his search program was much longer. He stopped hunting for and observing comets possibly prior to the 1910 appearance of Halley's Comet, for there are no published records of observations made by Grigg in international journals of that famous comet. But he did photograph it.

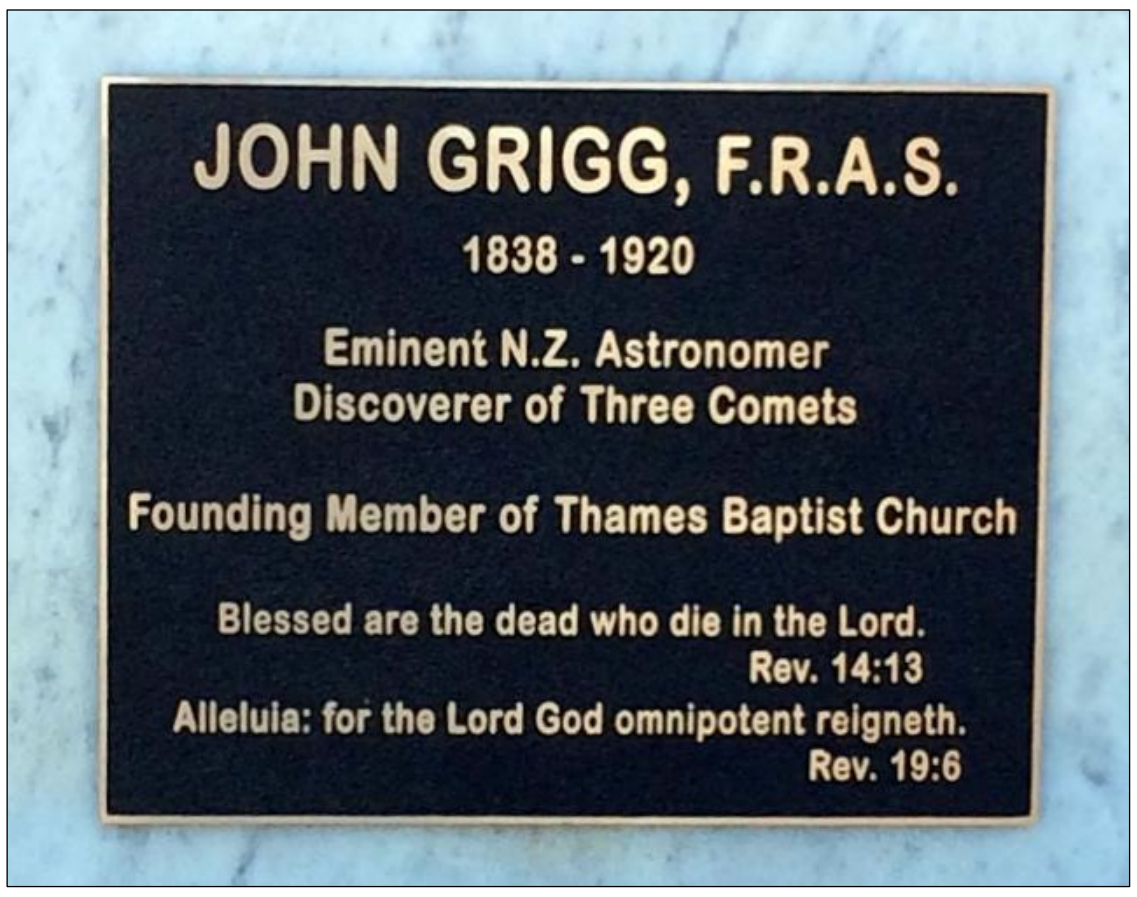

Figure 16: John Grigg's final resting place was the Shortland Cemetery in Thames. Note the recently upgraded headstone; the old one previously was falling into disrepair (photograph: Find a Grave).

In recognition of his discoveries, Grigg was awarded the Donohue Comet Medal (The Thames Comet, 1903: 1) of the Astronomical Society of the Pacific, and made a Fellow of the Royal Astronomical Society of England in 1906 (Mackrell, 1985: 78). Grigg's grandson, R. Grigg, points out (Grigg, 2020) that "In 1970, John Grigg was honoured by the International Astronomical Union by having a crater on the Moon named after him. It is located on the far side of the Moon."

John Grigg passed away on 20 June 1920 aged 82 and is buried in the Shortland Cemetery in Thames (Figure 16).

## 4 MURRAY GEDDES

On the evening of 22 June 1932 (UT), the shortest day of the year for NZ, a humorous note appeared in the Lost and Found column of the *Otago Daily Times* (1932: 1), "Lost, a comet; "Lost, a comet; substantial reward."[5] That same night, Murray Geddes found a comet. It was a visual discovery from Otekura near the coast and towards the southern end of NZ's South Island (see Figure 2). Contrary to Kronk's (2007: 582) claim, Geddes was actively sweeping for comets when he made the discovery and was not "... observing variable stars …" What of Geddes, who was he and what were his astronomical interests?

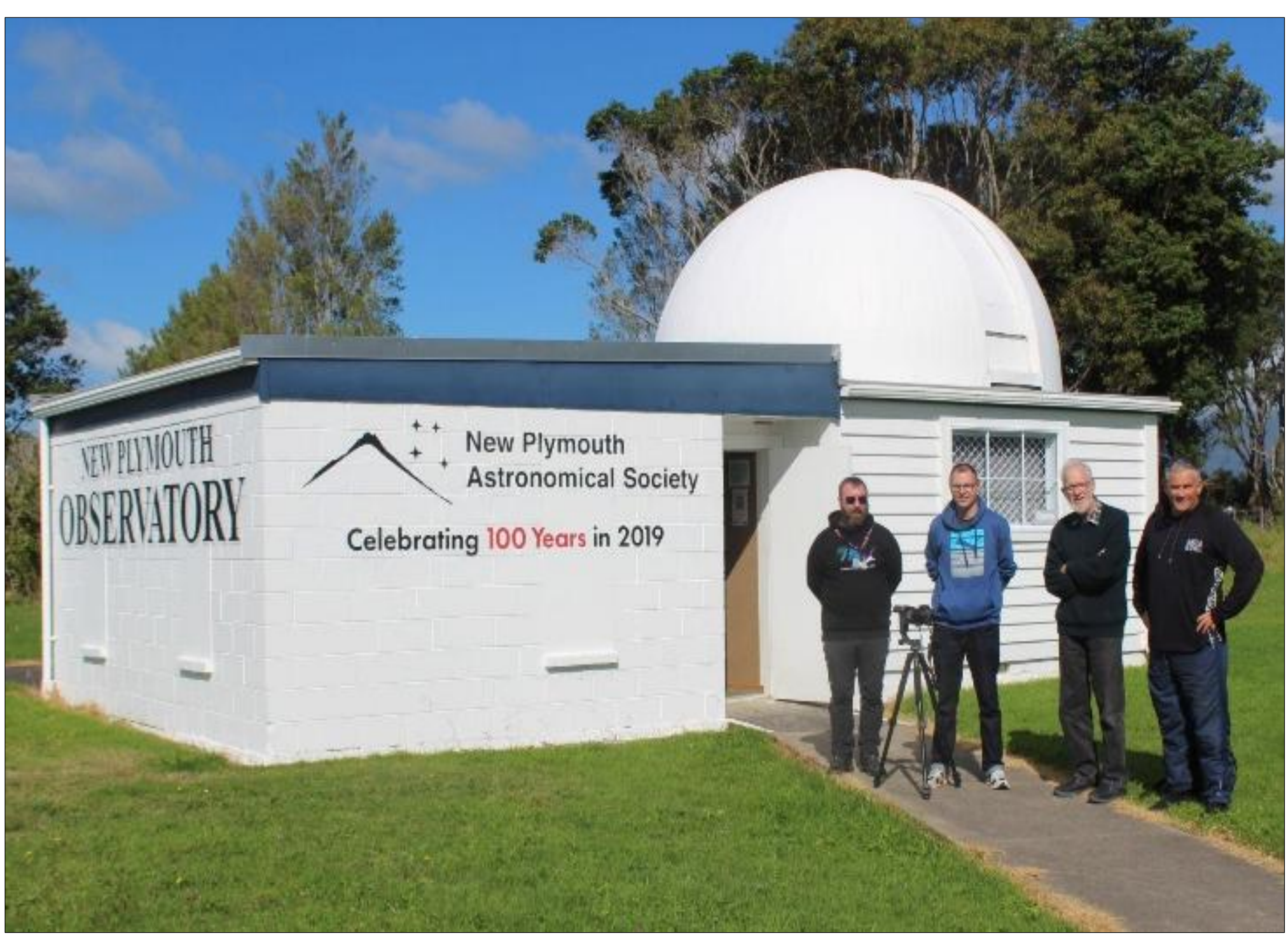

Figure 17: The New Plymouth Astronomical Society as photographed by the lead author of this paper. The people in the foreground are (left to right) Kyle Francis, Brendan Larsen, Rod Austin, and John Drummond (photograph: John Drummond, circa 2010).

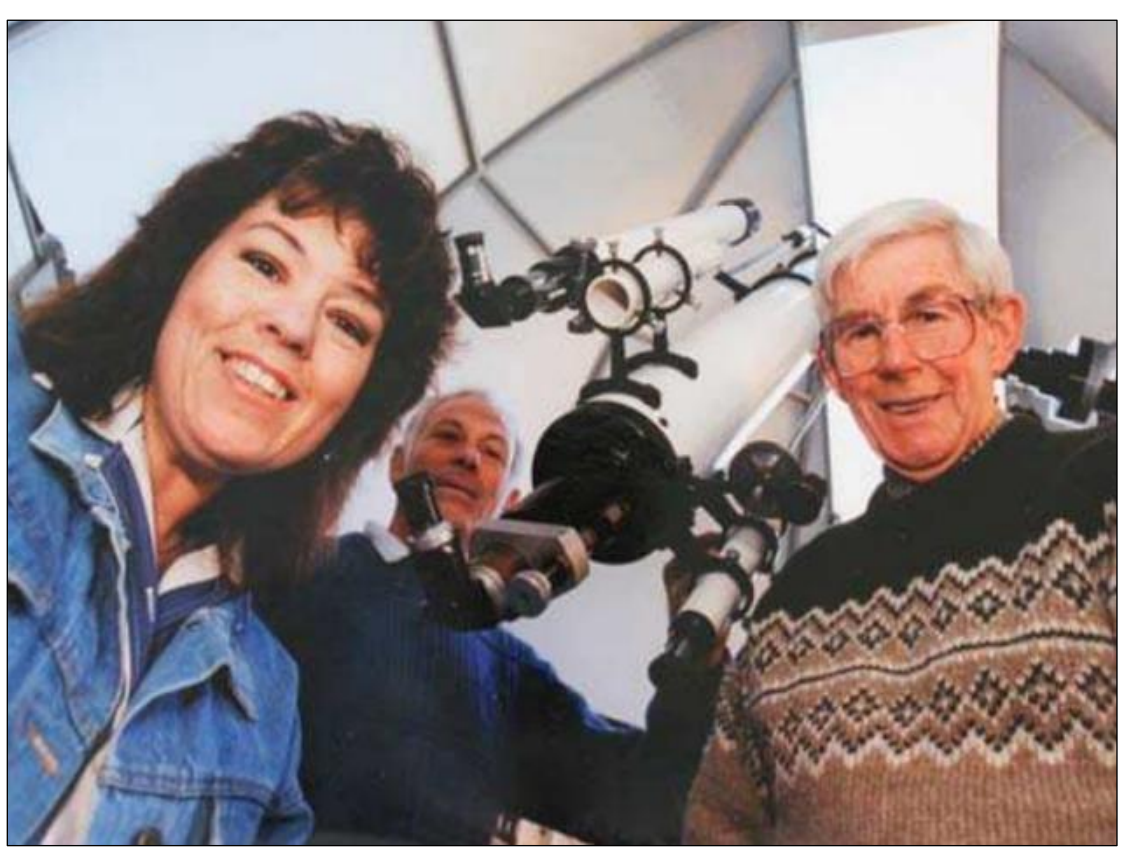

Figure 18: A 1995 New Plymouth Observatory photograph by Ashley Marles and Carolyn Jones showing, left to right, Heather Couper (BAA President 1984–1986), Peter Knowles, the historic 6-inch refractor, and Albert Jones (after Toone, 2016: 90).

Murray Arthur Geddes (1909–1944; Dickie, 2010) was born in Glasgow, Scotland, on 27 February 1909, and later died there, on 23 July 1944. His parents, Joseph and Edith Geddes (Personal, 1932: 4), immigrated to NZ when Murray was a young child. They established their family at Tarata, 27 km East of New Plymouth in Taranaki (North Island). Geddes went to the New Plymouth Boys' High School. During his high school years in the 1920s, he regularly visited the New Plymouth Astronomical Society Observatory (Figure 17). The Society had (and still operates) an historic 6-inch (15-cm) refractor (Orchiston and Austin, 2025) that Geddes would have looked through (Figure 18). He developed interests in observing meteors and had soon made more than 1500 observations (Dickie, 2010: 3–4). Dickie notes that at the time this "... was the second highest number recorded by anyone in New Zealand." (*ibid.*). Geddes also sent a detailed paper to *Popular Astronomy* in 1941 about the 1939 opposition of Mars (Geddes, 1941: 2–12), along with one to *Nature* co-authored by I.L. Thomsen (Geddes and Thomsen, 1939).

Geddes decided to enter the teaching profession and enrolled at Dunedin Teachers Train-ing College in 1928. He also studied English at the University of Otago in Dunedin. His first teaching position was at Westown Primary School in New Plymouth. He taught there from 1930 to 1931 and during this period joined the NZ Astronomical Society (which would later become the Royal Astronomical Society of New Zealand). He then moved to Otekura near the bottom of the South Island in 1932 to take up a teaching position at a small one-classroom school (Figure 19). While teaching there, he also enrolled at the University of Otago to study Mathematics.

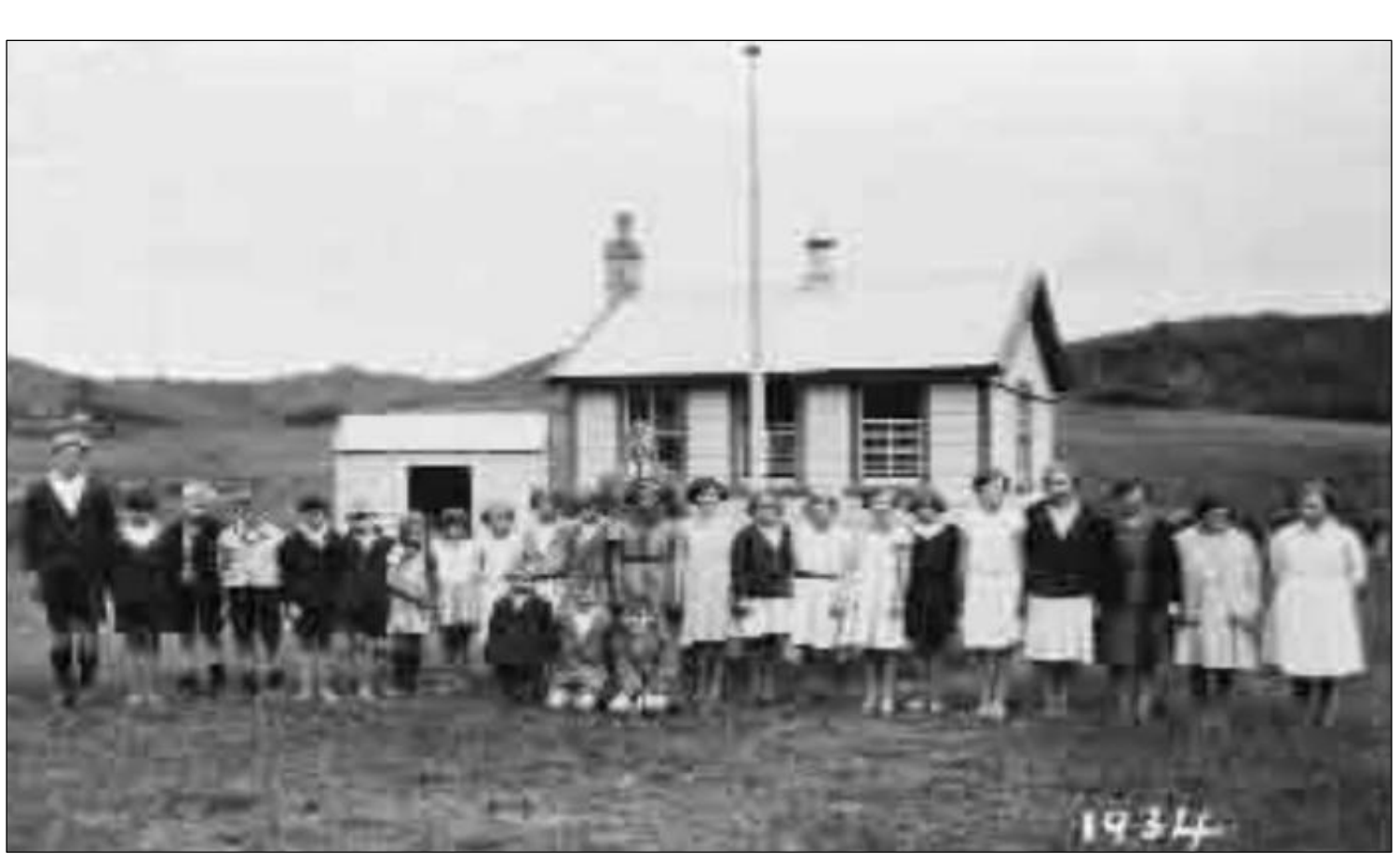

Figure 19: A grainy photograph of Geddes and his 25 students at Otekura School in 1934. Geddes is wearing a black tie in the back row, centre (after Dickie, 2010: 4).

At Otekura, Geddes built an observatory that housed a 5-inch (12.7-cm) Cooke refractor with a 1.5-inch (3.8-cm) finderscope. This telescope was lent to him by the Dominion Observatory in Wellington after it was used in a solar eclipse expedition to Niuafo'ou to view the 21 October 1930 (UT) solar eclipse (Personal, 1932: 4; The New Comet, 1932d: 5). Being in the deep south, 105 kilometers east of Invercargill, one of the world's most southern cities, Geddes developed an interest in aurorae. The closer an observer is to the Earth's magnetic poles, the greater the chance of observing aurorae. Murray Geddes was in an ideal location. He became the Director of the Auroral and Zodiacal Light Section of the New Zealand Astronomical Society in 1932 at the age of 23. Volunteer aurora observers from across NZ submitted observations to him. His collaboration with Professor Carl Størmer (1874–1957) in Norway led to new understandings of these so-called Northern and Southern Lights.

Geddes was elected President of the New Zealand Astronomical Society (NZAS) on 30 November 1939 (RASNZ minutes, 1939), having earlier in the year been appointed the founding Director of the newly built Carter Observatory in Wellington (NZ)(Thomson, 1945: 89). This position was awarded to him for the following achievements: gaining a Masters degree in mathematics from Otago University, discovering a comet, making many meteor observations and detailed observations of Mars, successfully leading the Auroral and Zodiacal Light Section of the NZ Astronomical Society to international acclaim, and serving as the President of the NZAS.

On 24 February 1942, a few months after the official opening of Carter Observatory, Geddes went overseas to aid the war effort, joining the Royal New Zealand Naval Volunteer Reserve Radio Direction Finding Section with the rank of Sub-Lieutenant (Thomson, 1945: 88). He served in Colombo (Ceylon, now Sri Langka), Madagascar, the Mediterranean, the United States of America, and the United Kingdom. His expertise in radar and fighter detection were so valued that he later became a Lieutenant-Commander in the Admiralty Signals Establishment (Fraser, 2022). While serving in Scotland, Geddes died from a brain haemorrhage on 23 July 1944 at a hospital in Glasgow, Scotland. He was 35 years old. His cenotaph record (Auckland Museum) states that "It was believed this had been due to overwork as he understood the importance of his work and worked night and day on the task". He is buried at Cardonald Cemetery, Glasgow, Scotland (Pease, 2024).

Murray Geddes discovered one comet.

## 4.1 C/1932 M2 (Geddes) - 1932 VI = 1932g

Initial details of this comet are listed below in Table 6 its orbit is illustrated in Figure 20 and its path through the sky is plotted in Figure 21. For further information see Kronk (2007: 106).

According to Geddes' personal log for 22 June 1932 (Dickie, 2010: 4; Geddes, 1932), using his 5-inch Cooke refractor (Figure 22) he

> Swept for comets for 1h 15m through Apus, Chameleon, Octans, Mensa, Hydrus. At 19h 20m observed [a] hazy object close to the star z [ζ] Octanis.

Returning to the suspect object after 50 minutes he noted that it appeared to have "... moved slightly …" He then made a confirming observation at 20h 45m and "Decided it was a comet … [at the] limit of vision in the finder …" scope. He then placed the eyepiece from the finder scope into the main 5-inch telescope and determined that this combination gave a field of view of 46′. He wrote that the "Magnitude [was] possibly 10." (Geddes, 1932). He also made a drawing with 60× magnification (see Figure 23). This is compared below to a GUIDE screenshot set for the same date and at the same scale.

Table 6: An overview of Geddes' first and only comet. Epoch: 10.0 September 1932 UT. Based on the Minor Planet Center (MPC). If not listed in MPC, information was used from NASA/Jet Propulsion Laboratory Horizons System, or Kronk (2007).

| | |
|---|---|
| Discovery Date | 22 June 1932 NZST (22.33 June 1932 UT) |
| Discovery Magnitude | 9 (Geddes 1932) |
| Discovery Declination | –84° (Kronk, 2007: 582) |
| Perihelion date | 21.074 September 1932 UT (MPC) |
| Perihelion distance (q) | 2.314 au (MPC) |
| Perigee date | 1 July 1932 (Kronk, 2007: 581) |
| Perigee distance | 1.9688 au (Kronk, 2007: 581) |
| Brightest | Magnitude 8 on 26 June 1932, G. Hudson (Kronk, 2007: 582) |
| Visible from NZ | 22 June (discovery) to late April 1932 (NASA/JPL) |
| Last observed | 19.29 July 1934 (UT), magnitude 17.5 (photographic) (Kronk, 2007: 585) |
| Eccentricity of the orbit (e) | 1.0014 (MPC) |
| Semi-Major axis (a) | 3,280 au (NASA/JPL) |
| Aphelion distance (Q) | 2,445.28 au (MPC) |
| Inclination (i) | 124.989° (MPC) |
| Epoch | 10.0 September 1932 (MPC) |
| Period | 42,810 years (MPC – for the 2025-11-21.0 epoch) |
| Observations in Kronk (2007) | 90 (visual and photographic), 3 from NZ (including discovery) |
| Observations in MPC | 22 (17 used to determine the orbital elements) |
| Observations in COBS | 21 |
| Papers Past newspaper articles | 220 (June 1932–May 1933) |

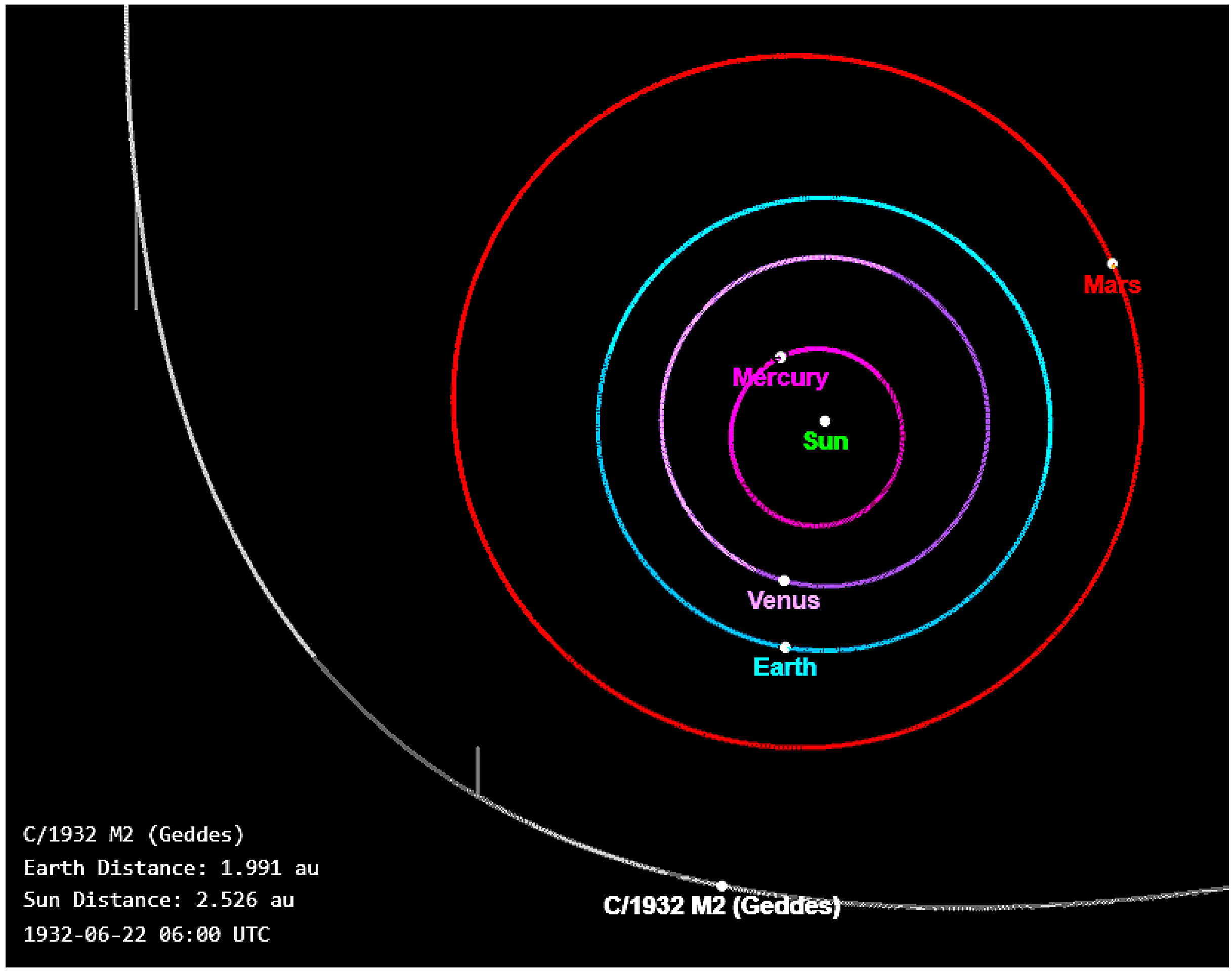

Figure 20: The orbit of Comet C/1932 M2 (Geddes) at the time of discovery on 22 June 1932 (UT). The comet followed a retrograde (clockwise from above) orbit (from the Jet Propulsion Laboratory (JPL), California Institute of Technology; https://ssd.jpl.nasa.gov/).

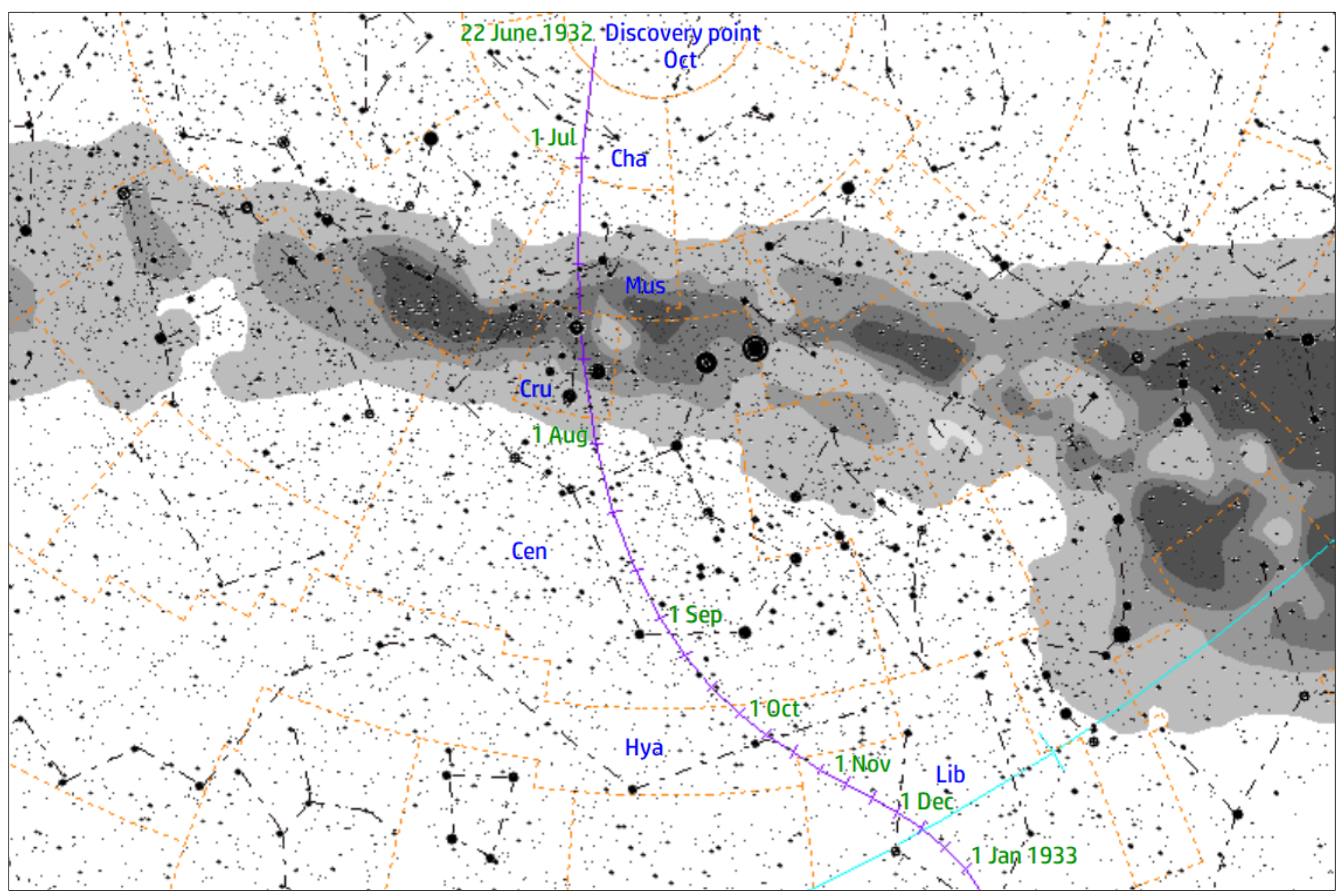

Figure 21: The path of Comet 1932 M1 (Geddes) (in purple) from discovery on 22 June 1932 (UT) until January 1934. Note the gradual movement of the comet towards the north. The constellations that C/1932 M1 traversed are labelled in blue. The dates are in green. The light blue line is the ecliptic. North is down (for the NZ sky), East is right (GUIDE software).

Geddes observed his suspect comet the following night (23 June, NZST) and determined that the comet was, at that stage, moving 55′ a day, being "... slightly more than [the] angular diameter of [the] finder eye-piece." He observed a "... definite nucleus & slight traces of a tail." (Geddes, 1932). This second night, he determined that the magnitude was 9 (rather than 10 as he stated in his log on 22 June). It was after this confirming observation that Geddes "Decided to send following telegram to Dominion Observatory." (*ibid.*).

Dr Charles Edward Adams (1870–1945) was the NZ Government Astronomer when he received the following telegram from Geddes on 23 June 1932: "Daily motion [of] comet about fifty five [arc] minutes … direction zeta Chamelionis. Visible one and quarter inch finder. Magnitude nine. Tail suspected." (A New Comet, 1932: 13). Adams reported the discovery to the newspapers; the first NZ newspaper report of Geddes' comet discovery was published on 23 June 1932 (one day after discovery). The discovery position on 22 June was RA 9h 15m and the Dec –84° 36′ and the estimated magnitude was 10 with a brighter nucleus—so perhaps a stellar-like nucleus surrounded by the coma. It was 5° from the South Celestial Pole (A New Comet, 1932: 13). The discovery was attributed to "... Mr. Geddes, the well-known observer in Otekura, Otago …" (New Comet, 1932: 5).

Two days after discovery, Mr George Vernon Hudson (1867–1946) of Karori, Wellington,

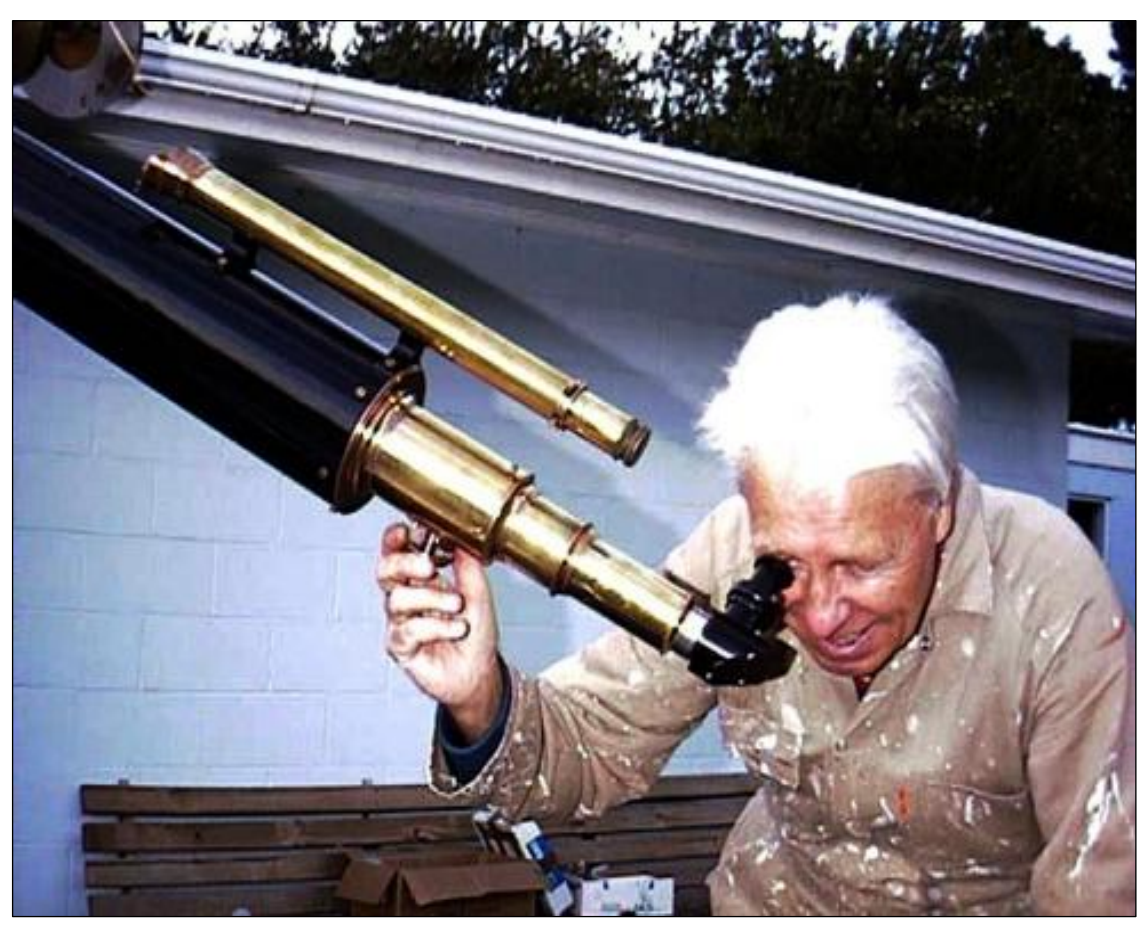

Figure 22: The 5-inch Cooke refractor that Geddes used to discover his comet in 1932. Malcolm Flain and Phil Barker refurbished the telescope in the early 2000s. Malcolm is pictured here looking through the instrument when it was at the Canterbury Astronomical Society's West Melton Observatory near Christchurch (photograph courtesy: Phil Barker, early 2000s).

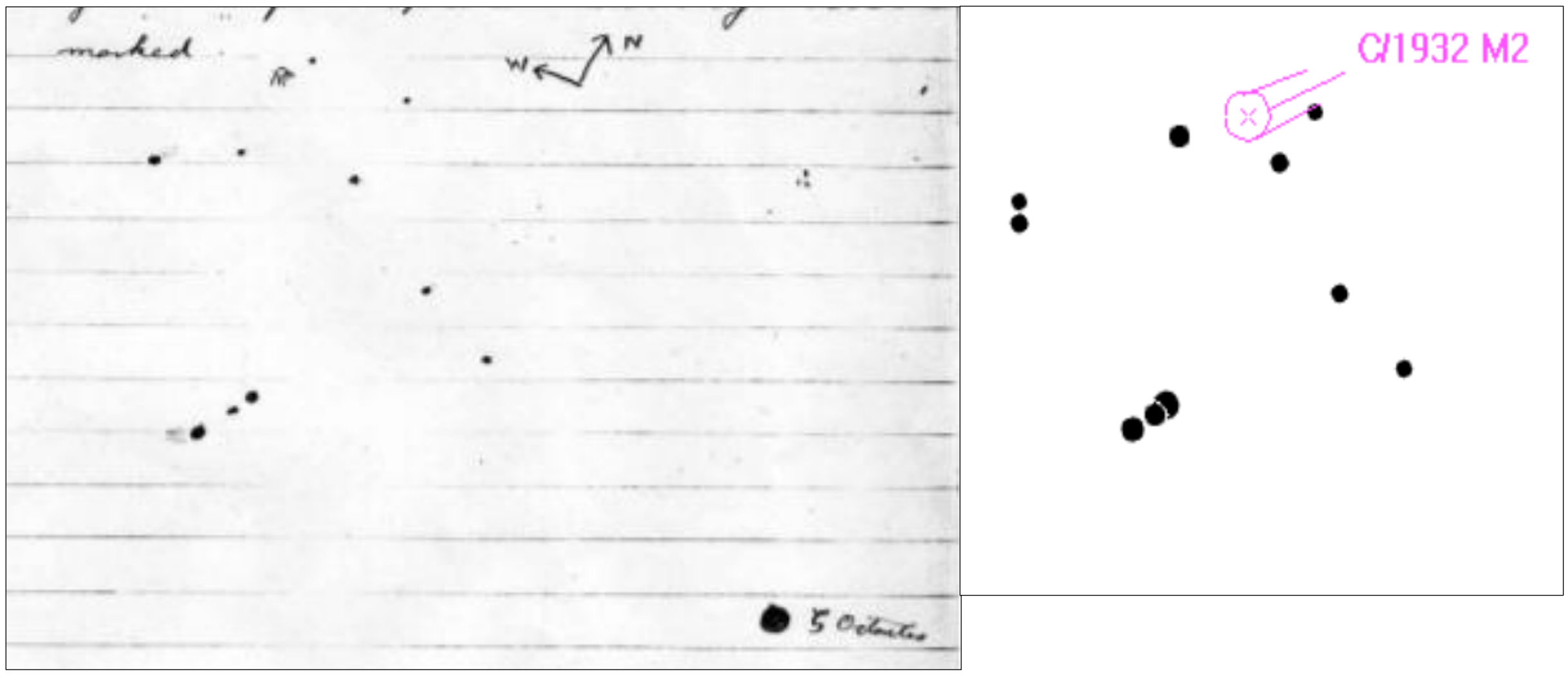

Figure 23 (left): Geddes' discovery sketch made with the 5-inch refractor at 60× magnification (after Geddes, 1932) compared (right) with the same field using GUIDE (sources: Record of astronomical activities from 1 January 1932–12 August 1936, by Murray Geddes, Carter Observatory archives 17 May 1989; and GUIDE).

estimated the comet to be magnitude 9 on 24 June 1932. He possibly used his 4.3-inch refractor for these observations. He also determined the RA and Dec on 24 and 26 June to help Dr Adams at Dominion Observatory (Wellington) determine the orbital elements of the comet (New Comet's Position, 1932: 8). Kronk (2007: 582) points out that the comet attained its maximum solar elongation on 23 June 1932 (110° - GUIDE) and its most Southerly declination (–86° - NASA/JPL Horizons) a few days earlier on 18 June, all fairly close to the 22 June (NZST) discovery date.

The Sun set close to 5pm and astronomical twilight ended at 6:30pm on that eventful night of 22 June 1932 (NZST). According to GUIDE, the 17.9-day old Moon rose at approximately 8:45pm (NZST), that is, 1.5 hours after Geddes first observed the comet that would soon bear his name. At 17.9 days old, the Moon was a waning gibbous; large, and bright. The sky offered a dark window from 6:30pm until 8:45pm, that is, approximately 2 hours and 15 minutes of dark sky until the Moon rose. As stated, it was customary for comet hunters to visually sweep the evening Western sky a few days after full Moon to pick up new comets that may have been hidden by moonlight in the previous week. For more on this, see Mobberley (2011: 148–152). This was undoubtedly the strategy adopted by Geddes.

According to Geddes' (1932) personal 'Record of astronomical activities from 1 January 1932–12 August 1936', Geddes observed his comet on 22, 23, 24, 26, 27, 30 June, 1, 3, 4, 10, 13, 15, 29, 30 July, and 1, 6, 12 August 1932. Only Geddes' discovery observation and that of Hudson on 24 and 26 June have been included in international publications such as Kronk (2007: 582) and Vsekhsvyatskii (1964: 467).

The first newspaper reports about the comet discovery stated that it was decreasing in both RA and Dec (New Comet's Position, 1932: 8). However, NASA/JPL Horizons System reveals a decrease in declination, that is, it was moving approximately North, but that it was increasing in RA. At discovery, it was in the 9 hour arc, by 24 June it was at 10 hours and on 28 June it was in the 11 hour RA sector, thus it was moving in a NE direction. The comet's orbit was such that it would have been South of the Celestial Equator from October 1926 (well before perihelion) until February 1933, after which it moved as far Northward as the constellation Draco, by March 1935. Interestingly, the comet was reportedly photographed on a plate taken on 14 August 1931 with the 10-inch (25-cm) Metcalf triplet from the Southern Station of Harvard Observatory at Bloemfontein, South Africa. It went undetected until Dr Fred Lawrence Whipple (1906–2004) noticed the faint smudge post-discovery. Whipple estimated it to be magnitude 13 and with a position of RA 01h 59.4m and Dec –44° 20′ (Kronk, 2007: 582). This photograph was taken ten months before Geddes visually detected the comet.

Geddes also observed Comet Geddes (for his name was now attached to the comet—The New Comet, 1932b, 1932: 7) on 24 June at 18h 45m with the 5-inch refractor and noted in his observing log that it had a bright nucleus with 120× magnification. He also viewed the comet on 26 June and found that the comet had moved 4° in declination since discovery, that is, towards Crux, roughly Northwards. He also added in his log on 26 June that, "There is still a suspicion of a tail … but it is quite possible that this is only an optical illusion." (Gedde's, 1932). Bernhard Hildebrandt Dawson (1890–1960; La Plata Observatory, Argentina) wrote that the nucleus was displaced from the centre to PA 315° on 25 June (Kronk, 2007: 582). Perhaps this nucleic offset gave the impression of a tail as seen by Geddes.

On 28 June a description and basic star chart (see Figure 24) showed newspaper readers where the comet was located (New Comet's Position, 1932: 8). Fortunately, it was in the deep Southern sky not far from the Southern Cross (Crux) and The Pointers, asterisms that most readers probably knew. Members of the public were warned that it was not a naked eye object but there was hope that it may be in the future.

Mr Artha, of Auckland, possibly used the newspaper star chart published on 28 June 1932 (Figure 24) to observe the comet on 28 June with a 3.5-inch (8.9-cm) telescope. The *Northern Advocate* stated on 29 June that two earlier newspaper reports unfortunately told readers to search in the wrong celestial location (Local and General, 1932a: 4). Allan Bryce (1885–1964), founder of the Hamilton Astronomical Society (McIntosh, 1973: 19), also observed C/1932 M2 (Geddes) on 28 June from Hamilton. He gave the following positions, RA 11h 14m, Dec –79° 56′ and described it as faint, approximately as bright as Neptune (magnitude 8) and that it resembled a star with a "... delicate glow surrounding it, and extending out to a diameter of probably two or three minutes of arc." He noticed considerable movement over a 20-minute period (The New Comet, 1932b: 6). Two days earlier, on 26 June, B.H. Dawson (La Plata Observatory, Argentina) also described the coma surrounding the nucleus, stating the comet had a 1′ coma and star-like 3″ nucleus of magnitude 10.5 displaced from the centre in PA 315° (Kronk, 2007: 582). Kronk writes that international observers were submitting observations of C/1932 M2 (Geddes) from Argentina and South Africa in June 1932. They were estimating the comet's magnitude to be 9. Observations from Australia during this period are not reported in Kronk, Vsekhsvyatskii or in NZ newspapers.

In early July, at the New Plymouth Observatory, where Geddes spent much of his youth, his parents Joseph and Edith Geddes observed the comet that their son had discovered (Local and General, 1932b: 6). Mr F.J. Moorshead, the Observatory Director, used the Society's historic 6-inch (15-cm) refractor to show it to them. Since the comet's discovery on 22 June, Society members had observed it over seven nights. Joseph and Edith were not the only ones proud of their son and his discovery for by early July the comet was so well known by the NZ public that drawings of it were used in decorations at public shows (Attractive Displays, 1932: 12). Geddes was also coming into public view with his photograph appearing in newspapers (e.g., see Figure 25).

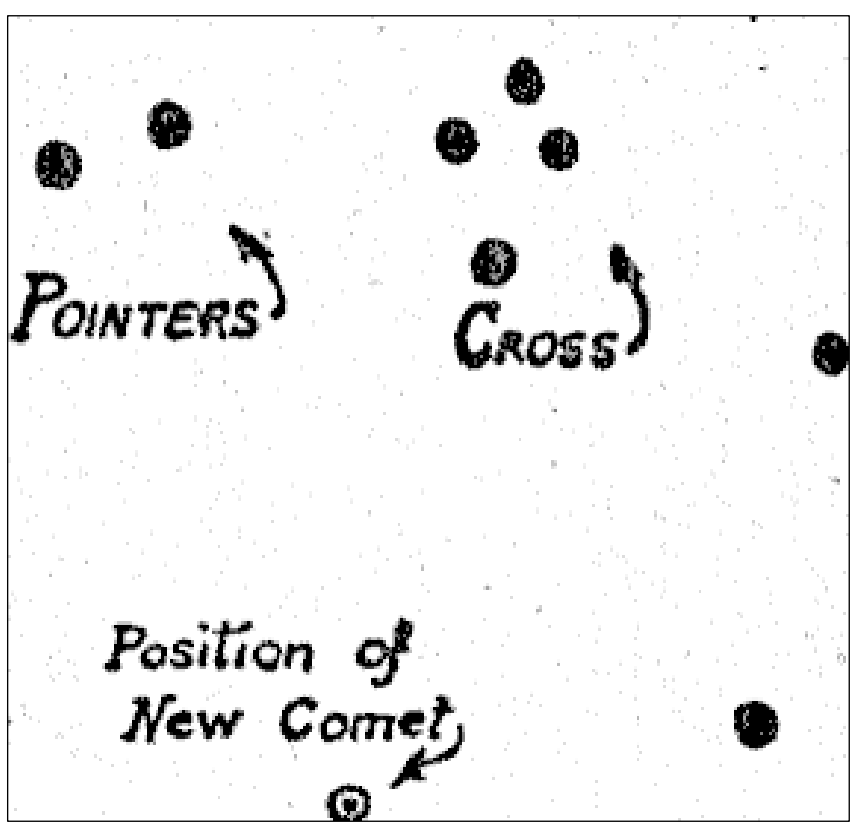

Figure 24: The position of Comet Geddes in relation to the Southern Cross and Pointers, published on 28 June 1932 (Sketch map, 1932: 8).

Ronald Alexander McIntosh (1904–1977; Orchiston, 2016: 523–563) observed the comet from Auckland over several nights around 4 July. The instrument used was not stated, but it may have been with his 14-inch (35.6 cm) equatorially mounted reflector (Orchiston, 2016: 540–542). He described the nucleus as magnitude 7 and that the coma was still approximately 2′ in diameter (Local and General, 1932c: 4). He also published astrometric positions for the nights of 29 June, 2 July and 3 July UT (The New Comet, 1932d: 8). These confirmed that C/1932 M2 (Geddes) was travelling in a NE direction. Orchiston (2016: 545) writes that McIntosh had

> ... very great pleasure … observing … C/1932 M2, which was discovered by his friend and fellow meteor-observer, Murray Geddes. McIntosh observed this comet on thirteen different nights between 30 June and 15 July, on each occasion

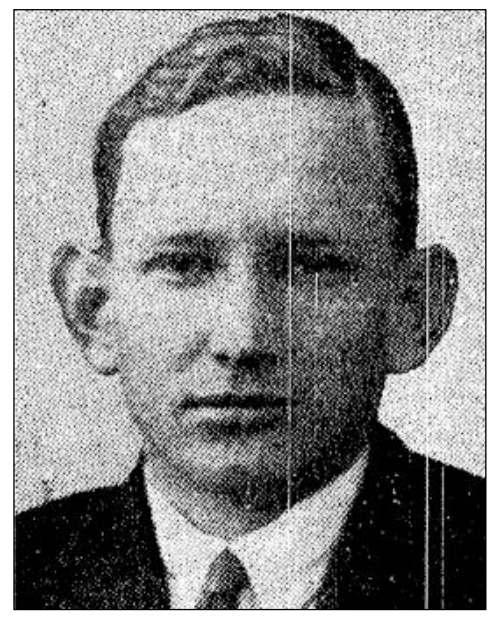

Figure 25: A newspaper photograph of Murray Geddes. The original caption read "Mr. M. Geddes, an amateur astronomer of Otekura, Otago, who recently discovered the new Geddes comet." (photograph: *NZ Herald*, 1932: 6).

Table 7: NZ observers of Comet C/1932 M2 (Geddes) and their observation numbers for the three weeks after initial discovery by Geddes on 22 June 1932 in order of total observations (after Comet Geddes, 1932: 8). WAS = the Wellington Astronomical Society 9-inch Cooke photovisual refractor.

| Observer | Location | Instrument | Observations |
|---|---|---|---|
| Mr. I.L. Thomson | Wellington | WAS 9-inch (23-cm) refractor | 14 |
| Mr. G.V. Hudson | Wellington | 4.3-inch (11-cm) reflector | 13 |
| Mr. R.A. McIntosh | Auckland | 14-inch (36-cm) reflector | 7 |
| Mr. R.C. Hayes | Wellington | WAS 9-inch (23-cm) refractor | 6 |
| Mr. F.M. Bateson | Wellington | WAS 9-inch (23-cm) refractor | 2 |
| Mr. M.S. Butterton | Wellington | WAS 9-inch (23-cm) refractor | 2 |
| Mr. A. W. Burrell | Stratford | Not stated | 1 |
| Mr. F.J. Morshead | New Plymouth | 6-inch (15.2-cm) refractor | 1 |
| Mr. A. Bryce | Hamilton | 8-inch (20-cm f/8) reflector | Not stated |

noting its appearance and magnitude, and its position relative to nearby stars.

By mid-July, reports were being published that astronomers from around the country were following the comet. In Hamilton, *The Waikato Times* (Geddes Comet, 1932: 8) stated that Mr Bryce had observed C/1932 M2 (Geddes) on nine nights since the start of the month and that, "It is sensibly brighter than at discovery, but is still quite invisible to the naked eye." On 17 July, the comet would be within one degree of Alpha Crucis and by the following weekend it would be very close to Gamma Crucis. Thus, it was travelling approximately 5° towards the NE in one week. Mr W.H. Ward of Wanganui Observatory also observed the comet in mid-July.

Members of the Wellington Astronomical Society were using the historic 9-inch (23-cm) Cooke photovisual refractor at the City Observatory (see Orchiston, 2016: 352–355) to observe C/1932 M2 (Geddes) during its weekly meetings in July. It was said to be a "... round hazy blob with a nucleus resembling a faint star." (The Observatory, 1932: 2). 47 observations of the comet were made on most nights from then and up to the date of the newspaper article on 22 July (Comet Geddes, 1932: 8). In this same *Evening Post* article, the observers and observation numbers were stated (Table 7).

The *New Zealand Herald* (The New Comet, 1932e: 12) stated on 23 July that the comet was slowly getting brighter and was now visible in binoculars and was "... quite prominent …" in small telescopes. It was a "... faint, rather hazy oval of light, with a brighter condensation at the northernmost focus of the oval …" and moving nearly a degree a day. GUIDE software indicates that any tail present would be pointing away from the Sun which lay to the NW, so perhaps the elongation to the South from the Northern condensation was an extremely short tail. It also showed that the comet was circumpolar but by month's end would dip below the horizon from 3am.

On 27 July, Mr Jim Rolston of Levin was able to observe the comet with a pair of marine binoculars. Rolston also described the oval appearance of the comet (*The Levin Daily Chronicle*, 1932: 4). William Herschel Ward (1900–1973) of Wanganui stated it was magnitude 8 (The Comet, 1907e: 3). Of note however is an article from the Wellington Astronomical Society stating that the comet was too faint to be seen with binoculars (Wellington City Observatory, 1932: 9). Perhaps they were basing this on observations with binoculars from within a (then) partially light-polluted city that probably had "... smoke and haze …" (Comets and Their Messages, 1932: 5).

Kronk (2007: 582) writes that "Although numerous positions were obtained around the world during July, no physical descriptions were published." ADS search results for July 1932 revealed only one descriptive set of observations of C/1932 M2 (Geddes). L. Cap (Sao Paulo) wrote that the comet ranged in magnitude from 8.5 to 9.5 between 1–6 July 1932 (Cap, 1932: 116–117). Given its unique longitudinal and latitudinal position, observations during this period from NZ (when C/1932 M2 was circumpolar) would have contributed significantly to the international astronomical community.

At the start of August 1932, 'Crux Australis' published an elaborate article in the *New Zealand Herald* (Glory of the Stars, 1932a: 6) about comets, using Comet C/1932 M2 (Geddes) as an example. They stated that "Both coma and nucleus are visible in Comet Geddes." We are unsure who the astronomer (amateur or professional) using this pseudonym was but it would seem that they were both an observational and theoretical astronomer.

One day after Crux Australis' article, on 2 August, W. Gardner also wrote a small treatise on comets in the *Otago Daily Times* (Comets and their Messages, 1932: 5). He reported that he observed Comet Geddes between $\mu$ and $\gamma$ Crucis on 30 June 1932 from the Beverley-Begg Observatory on Māori Hill in Dunedin. Gardner described it as magnitude 7 or 8 and predicted that it may get brighter as it approached the Sun. He hoped it would, as the last bright naked eye comet was Halley's Comet which put on an "... awe-inspiring display …" in 1910 (a future article about this apparition over NZ will be published by the authors of this paper). The Great January Comet of 1910 several months before 1P/Halley was also a brilliant comet (see Drummond et al., 2025). Gardner's article laid out an elaborate description of comets and the astrophysics involved, using bright historical comets to illustrate his points.

Geddes also spoke about the astrophysical nature of comets at a meeting of the Astronomical Branch of the Otago Institute on 23 August (Otago Institute, 1932a: 2). He said that when he discovered his comet, it was close to perigee at approximately 160 million miles (257 million kilometers). By the end of 1932, astronomers in the Northern Hemisphere would see the comet. Geddes stated that it would return in 9.6 years; this no doubt being explained to him by Dr Adams of Wellington's Dominion Observatory (see The New Comet, 1932f: 10). Future observations and orbital recalculations showed that C/1932 M2 (Geddes) was in fact a long-period comet rather than a short-period one. Geddes was congratulated by the Institute on his discovery (Otago Institute, 1932b: 5).

In early October 1932, Allan Bryce gave a talk to the Waikato Society of Model and Experimental Engineers about an 8-inch (20-cm) f/8.4 reflecting telescope that he made for £15 (approximately NZ$1,000 in today's currency) (Big Telescope, 1932: 11). Bryce observed and noted the RA and Dec of Comet Geddes on over 30 occasions and sent them to Dr Adams to help determine an orbit (with other NZ observations).

Another telescope that was used for observations of C/1932 M2 (Geddes) was the 6-inch refractor at the New Plymouth Observatory. At the end-of-year report at a meeting of that Society on 29 November 1932 (Astronomical Society, 1932: 3), the Director, Mr F.J. Morshead, emphasised that the number one goal of their Society was 'To promote the study of astronomy among interested ladies and gentlemen'. He stated that visitor numbers to the Observatory were stimulated by the discovery of the comet by Murray Geddes, a member of the Society who was residing in the South Island. In 1932, 156 visitors attended public viewing nights on 42 evenings (3.7 people per night). Since records began thirteen years prior, the total number of visitors was 6,906 (531 visitors a year). Observers used the telescope 328 times in the year. He mentioned that

> The equatorial mounting still remains unsatisfactory, and is a continuous source of annoyance to the observers. It has been impossible to make reliable position observations of comet Geddes owing to the unreliability of the equatorial …

Mr Geddes was held in extremely high regard by the Society and mentioned five times in the report. During the year he submitted 63 solar drawings (plus 41 enlargement sketches of selected sunspots), many observations of aurora (of which Geddes was the Director of the NZ Astronomical Society Aurora Section), and an unstated number of meteor observations.

In a newspaper article titled 'Glory of the Stars' (1932b: 9) 1932 was seen as a highly productive year in the discovery of comets internationally. 13 were found, 6 being periodic returns and 7 previously unknown comets, C/1932 M2 (Geddes) being one of them. In another annual report, this time by the Dominion Observatory in Wellington (Kelburn Observatory, 1933: 16), it was stated that

> Astronomical observations made in New Zealand are considered by astronomers in other parts of the world as of special importance, for well-equipped observatories in the Southern Hemisphere are none too numerous and New Zealand is very favourably situated for observation purposes. A regular interchange of the results of research work with all the leading observatories in the world takes place.

After perihelion on 21 September 1932 UT (NASA/JPL) Comet Geddes moved away from its small solar elongation of 7° from the Sun on 7 November 1932 UT (Kronk, 2007: 582). The comet-to-Earth distance decreased faster than the comet-to-Sun distance increased, so the comet's brightness was expected to remain unaltered for some time (Comet Reappears, 1933: 10). As the new year rolled over, Geddes observed his comet on 9 and 10 January 1933 with the 6-inch refractor at New Plymouth. He was probably visiting his parents during the NZ summer school holidays. He declared to the astronomers at Dominion Observatory in Wellington, on his way home, that the comet had not changed in appearance much (Comet Geddes, 1933: 9). While there, he showed the staff the Donohoe Medal for comet discoveries that he received from the Astronomical Society of the Pacific. The article stated that "This medal is awarded to comet discoverers all over the world, and is specially coveted by amateur astronomers". His was the 144th presentation of the medal. Other accolades that Geddes would receive were a congratulatory letter from the NZ Astronomical Society in March 1933 (Astronomical Society, 1933: 15), becoming a member of the American Meteor Society, being an active member of the American Association of Variable Star Observers, and also being awarded the Donovan Medal by the Donovan Trust in Sydney, Australia (Astronomical Activity, 1933: 4).

Table 8: Newspaper articles about 'Comet' and 'Comet C/1932 M2 (Geddes)' in Papers Past. Note the high article numbers in June and July 1932, around the time of discovery (22 June).

| Month/Year | Total for 'Comet' | Total re 'Comet Geddes' | Notes |
|---|---|---|---|
| June 1932 | 109 | 67 | Discovery month |
| July 1932 | 207 | 72 | |
| August 1932 | 134 | 26 | |
| September 1932 | 118 | 13 | |
| October 1932 | 97 | 1 | Solar conjunction |
| November 1932 | 181 | 2 | " |
| December 1932 | 267 | 9 | " |
| January 1933 | 165 | 15 | Comet reappears in morning |
| February 1933 | 207 | 8 | |
| March 1933 | 387 | 3 | |
| April 1933 | 302 | 0 | |
| May 1932 | 275 | 5 | Comet too far north |

On 26 January 1933, Comet Geddes was seen from Auckland in the morning sky by an unnamed observer. It was 3° north of β Librae, with a declination of 6° South. The magnitude of the nucleus was 10 and the surrounding 1.5′ coma was one magnitude fainter (Comet Reappears, 1933: 10). By May 1933 Comet C/1932 (Geddes) was too far North to be seen from NZ (The Comet Geddes, 1933a: 6).

A newspaper article in the *Otago Daily Times* (The Comet Geddes, 1933a: 6) provided a useful summary of the comet as seen from NZ, stating it was discovered on 22 June 1932 by Murray Geddes at Otekura. Initially it was believed to be a short-period comet with an orbital period of 9 years, however, after additional astrometry, it was found to be a long-period comet with a hyperbolic orbit. It never reached naked-eye visibility and was probably brightest a few weeks after discovery. A tail was never observed visually. The appearance was that of a "... diffuse point of light surrounded by a faint coma, looking in the telescope very similar to a star seen through a fairly thick haze". After discovery, it was followed by numerous NZ telescopes until it approached the Sun at the beginning of October. It remained within the Sun's realm for three months until the beginning of January when it was visible once more in moderately sized telescopes from January and into February until it sank too low in the Northern sky for any further observations from NZ. The article stated that

> Since its reappearance in January it has been within the range of powerful northern hemisphere telescopes, and it is certain that these will succeed in obtaining observations of sufficient accuracy to fix definitely its path in space.

This was confirmed by a Letter to the Editor of the *Otago Daily Times* (Thomson, 1933: 19) by Ivan Leslie Thomson (1910–1969; Eiby, 1970; Dominion Observatory) on 18 May 1933 that stated:

> ... Professor G. Van Biesbroeck, of the Yerkes Observatory, in Popular Astronomy for February, 1933, said that 'On a plate taken here on December 26, 1932, the magnitude [of C/1932 M2] comes out about 10; the coma has a diameter of about three minutes, and it is strongly condensed in the centre. Besides, there is a tail recorded over the length of 15 minutes in position angle 160 degrees.'

Further observations from the Northern Hemisphere can be found in Kronk (2007: 584–585). I.L. Thomson (The Comet Geddes, 1932b: 19) concluded by stating that the comet had been observed for "... at least ten or eleven months …" and was well observed from NZ (the lead author found 103 observations in Papers Past), Australia, and South Africa.

Of the 2,449 newspaper articles that contained the word 'comet' between June 1932 and the end of May 1933, 221 newspaper articles related to Comet Geddes. See Table 8 for a more detailed analysis.

Geddes died while on overseas duty in WWII. He was buried in Scotland (Figure 26). To honour Murray Geddes, the Royal Astronomical Society of New Zealand established the Murray Geddes Award in 1945. Their website (RASNZ - Murray Geddes Memorial Prize) states that "The Murray Geddes Memorial Prize is awarded by the Royal Astronomical Society of New Zealand to a person or persons for contributions to astronomy in New Zealand". A photograph of the obverse of the medal is seen in Figure 27. Interestingly, the first New Zealander to receive it was the next person to discover a comet from NZ, Albert Jones.

## 5 ALBERT FRANCIS ARTHUR LOFLEY JONES

Albert Jones (1920–2013; Toone, 2016; Figure 28), one of the world's most prolific variable star observers, discovered two comets (C/1946 P1 and C/2000 W1), with a 54-year gap between discoveries! This is a world record (Orchiston, 2016: 499). In addition, Jones discovered his second comet when he was 80 years old, the oldest person known to discover a comet, beating American Lewis Swift by a year. Swift was 79 when he discovered his last comet in 1899 (Toone, 2016: 91).

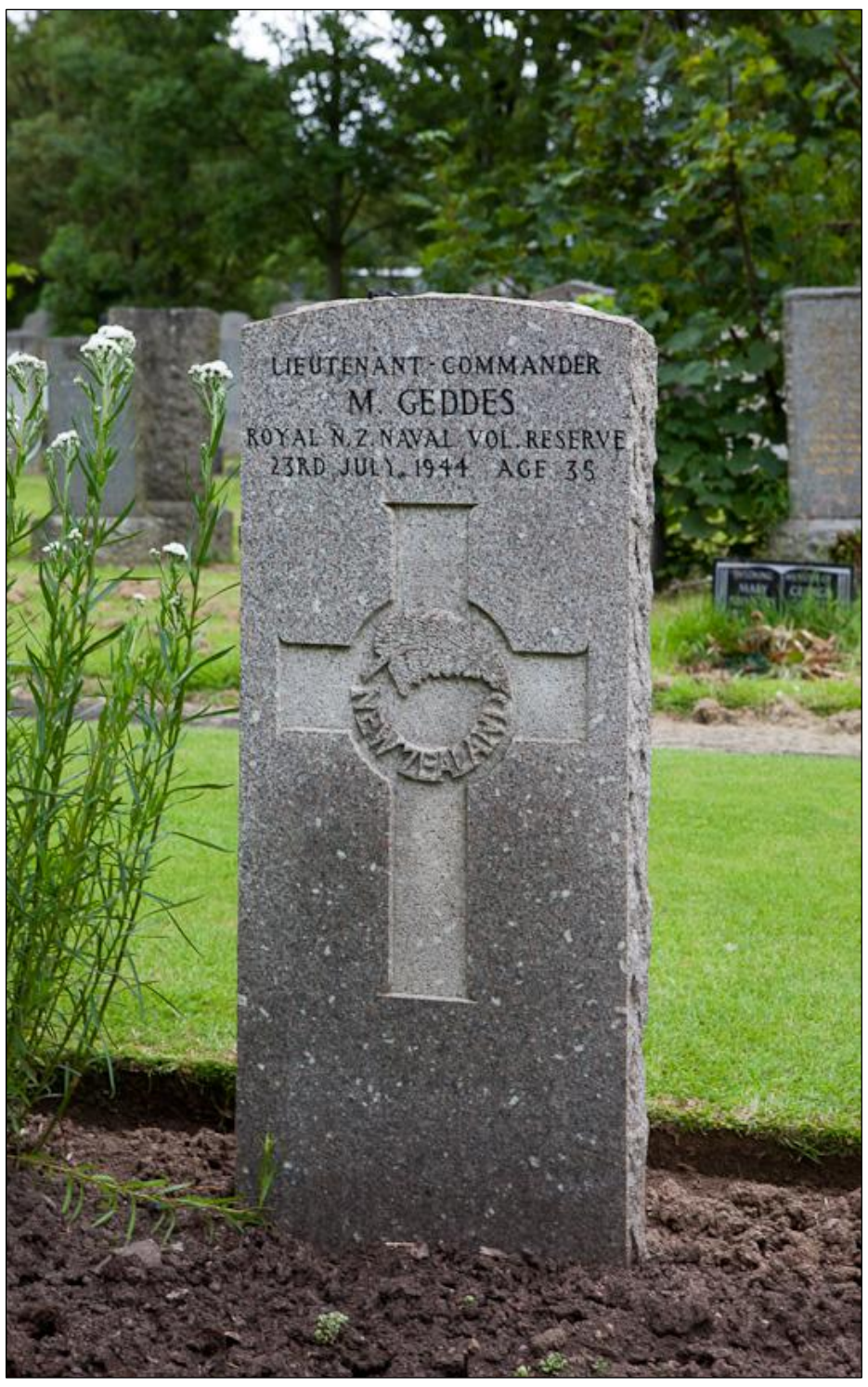

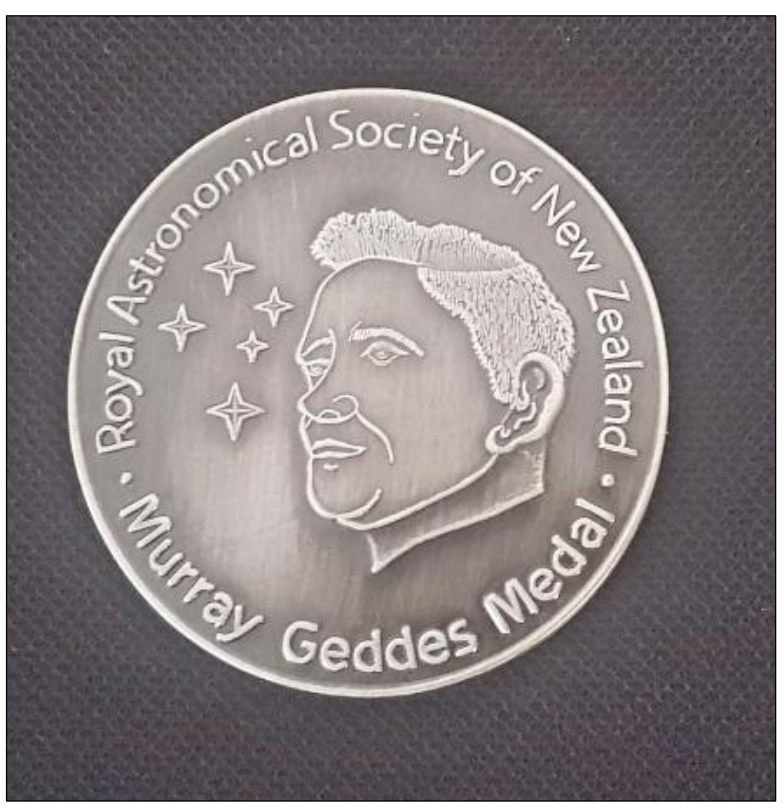

Figure 26 (left): Murray Geddes' final resting place in Scotland (photograph: NZ War Graves).

Figure 27 (above): The Murray Geddes Award medal, presented to 'a person or persons for contributions to astronomy in New Zealand' (photograph: Antony Gomez, 2020).

Much of the following is based on a *Journal of the British Astronomical Association* biographical paper about Albert Jones by John Toone (2016), an autobiographical paper (Jones, 2011); an interview published in *Astronomy Now* (Orchiston, 1990), a *Southern Stars* paper by Rod Austin (1994), Wayne Orchiston's (2016) book and from personal friendships with Jones that the first two authors enjoyed for many years.

Albert Jones was born at 263 Worcester Street, Linwood, Christchurch, NZ, on 9 August 1920 to Edward and Clara Jones (Austin, 1994: 36) who had married in 1910 (the year that Halley's Comet returned). They had two boys (one being a step-son) and one daughter. In October 1926 the family moved to Timaru, also in the South Island, settling at 40 Trafalgar Street. Jones witnessed a particularly active aurora as a child and this roused an interest in astronomy. He attended Timaru Boys' High School between 1933 and 1936 where he passed the matriculation and school certificate exams; Toone (2016: 84) points out that his favourite subject was chemistry. Jones started reading books on a range of topics, but it was those on astronomy that piqued his interest. He often went on fishing trips with his father to Rakaia River, a dark-sky site, that helped foster his interest in the stars. After submitting a description of an aurora to Carter Observatory in Wellington, he received a friendly and encouraging reply from Murray Geddes (see above), the Director of the New Zealand Astronomical Society's Auroral Section, welcoming Jones and inviting him to submit more observations. After two years, Geddes nominated Jones to become a member of the Society (now the RASNZ).

For a career, Jones opted out of university and began working at the Timaru Milling Company which his father managed between 1937 and 1963 (Toone, 2016: 84). With the advent of World War II in 1939 Albert joined the Home Guard in 1940 and then the 2nd Battalion Canterbury Regiment. However Private Jones was deemed unfit for overseas duty in 1942 (Austin, 1994, 37) due to "... having sustained foot injuries during long marches". In addition, it was thought that he could contribute more to the war effort as a miller that produced food for the army and general population (Toone, 2016: 84). In 1963, Jones' father retired (aged 86) and the role of head miller was offered to Albert, however, Jones chose rather to move to Nelson in 1964 and open a successful grocer's shop with his brother Eric and sister Goldie. In light of his numerous astronomical activities, Jones was also offered a job at Carter Observatory in Wellington (reputedly, including the Directorship) and a job by Frank Maine Bateson (1909–2007; Christie, 2014) doing site-testing for a new university observatory (which would later become Mount John Observatory at Tekapo). He turned both offers down, being content with his grocery store business.

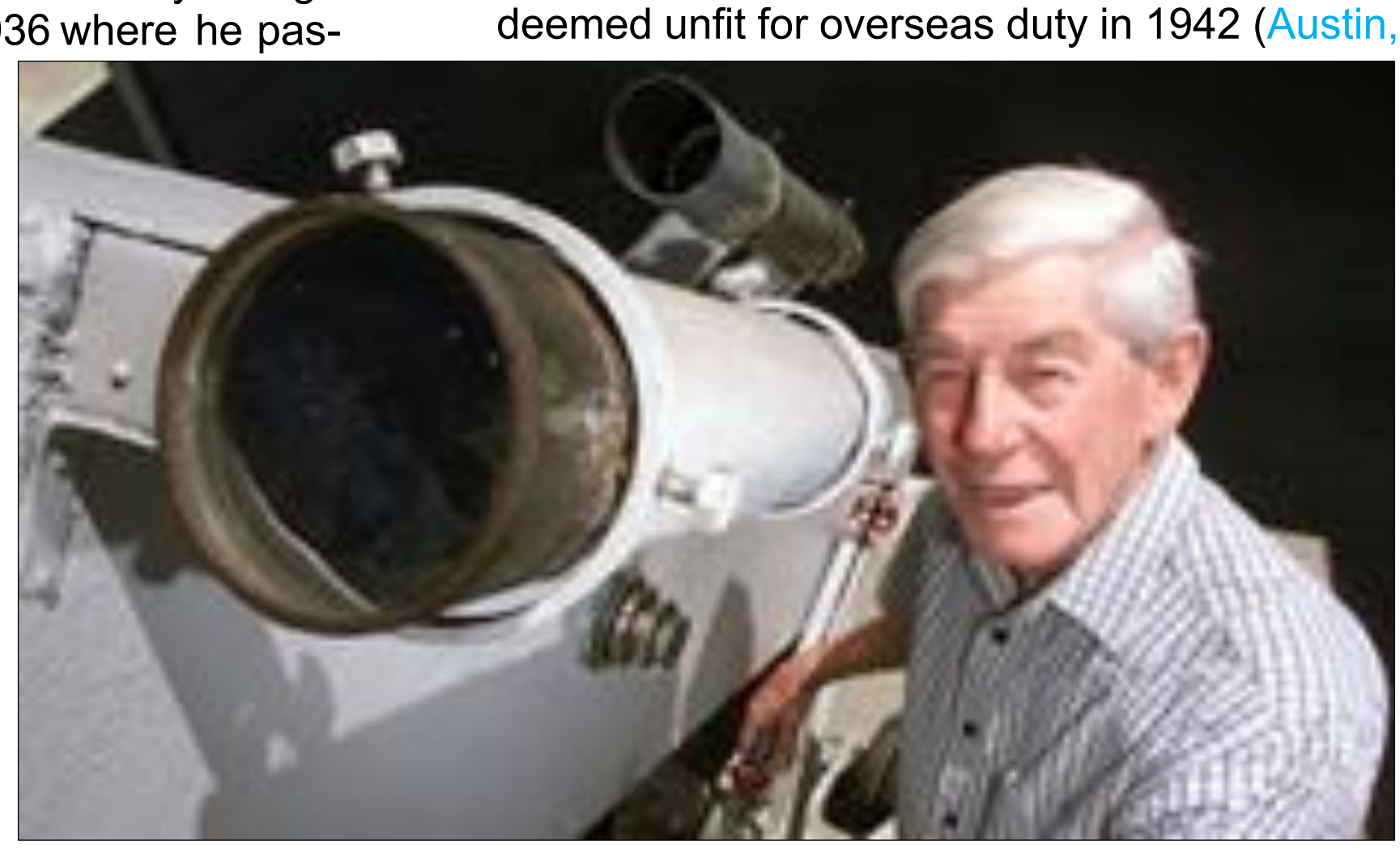

Figure 28: Albert Jones, the world's leading visual variable star observer and discoverer of two comets at his 3-inch (7.5-cm) finder scope, which is bolted on to his trusty 'Lesbet' reflecting telescope (after Toone, 2016: 86)

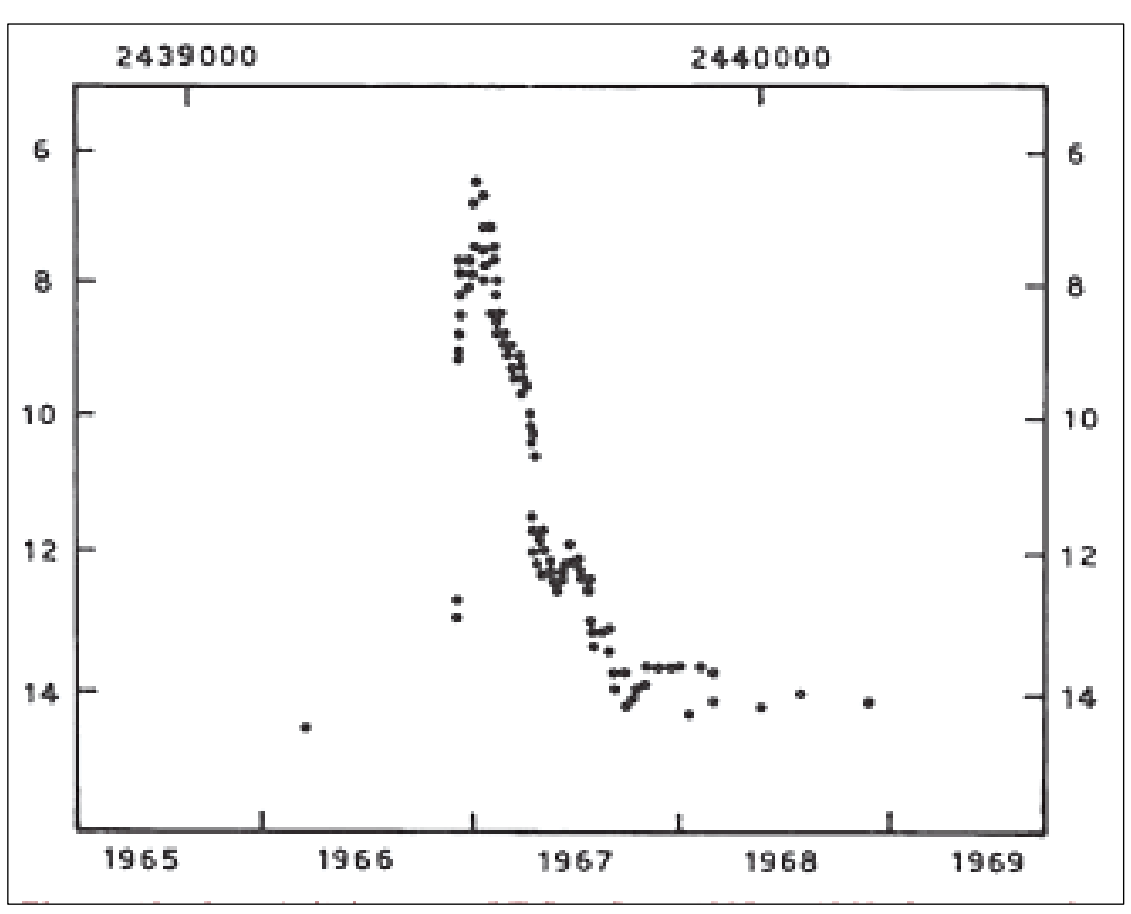

Figure 29: An example of a variable star's (T Pyxis) light curve based on Jones' observations. The years are on the X-axis and the variable star's magnitude on the Y-axis (after Toone, 2016: 87).

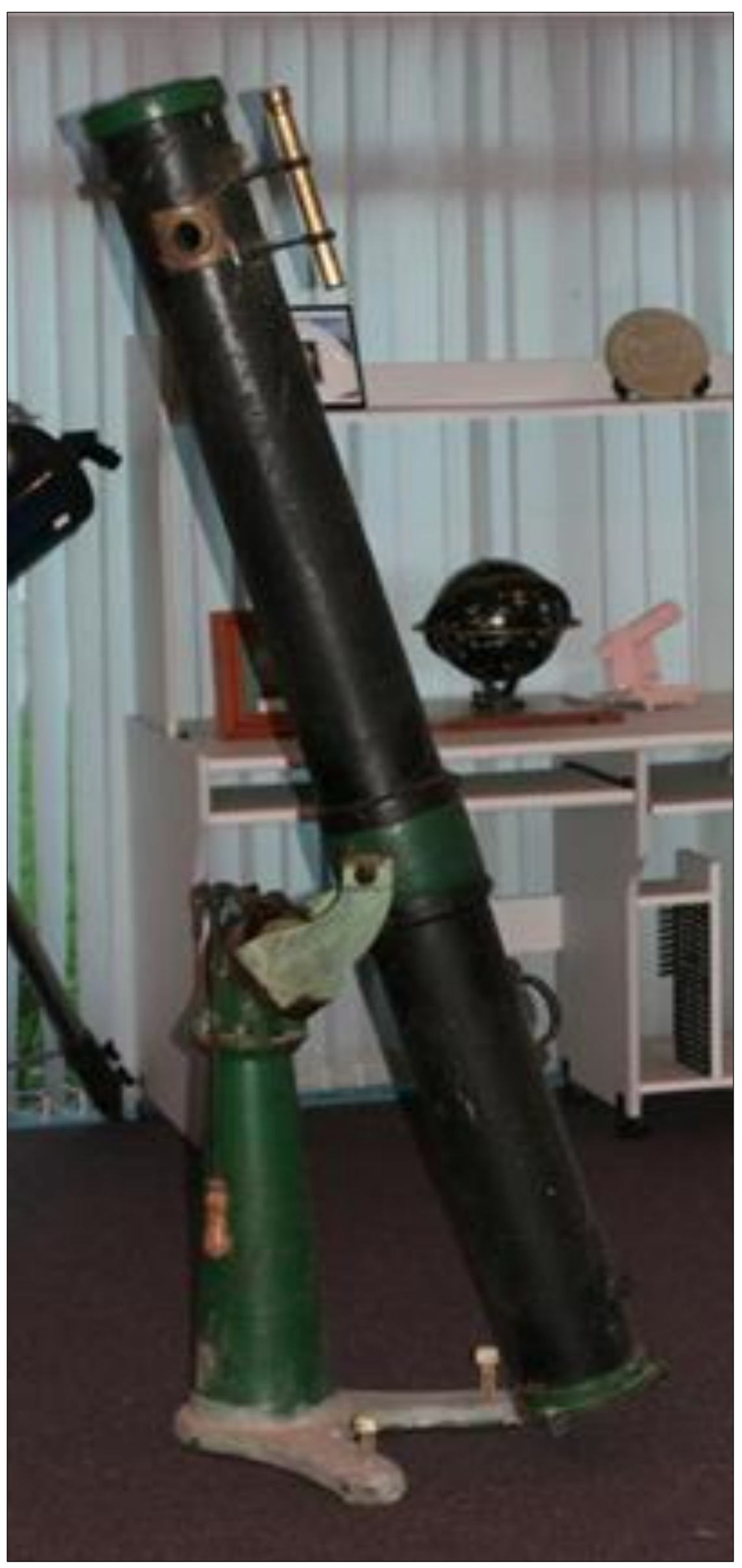

Figure 30: Albert Jones' 5-inch (12.7-cm) f/15 Calver reflector telescope (photograph: Wikicommons).

In late 1942, Nova CP Puppis erupted (Pettit, 1942: 259). It attained a maximum visual magnitude of 1 (Stoy, 1942: 3–4), the brightest in many decades until supernova SN1987A in 1987. Jones submitted magnitude change observations to Alec G.C. Crust, who published them and also gave instructions on variable star observing in *Southern Stars*. This led to an interest in variable star observing under the tutelage of Frank Bateson who founded the NZ Astronomical Society Variable Star Section in 1927 (Toone, 2005: 6). Austin (1994: 37) recalls Jones telling him that he was overwhelmed with how the variable star charts did not seem to match the actual telescopic views and the large volume of charts that Bateson sent him. With perseverance he submitted his first observations to the Variable Star Section in January 1943, aged 22. This was the start of a practice that lasted ~68 years. An example of a variable star's light curve derived from Jones' observations is seen in Figure 29.

Jones' first telescope was self constructed, made from cardboard tubes with lenses at each end (Austin, 1994, 36). Toone (2016: 85) points out that "The lens alone cost the equivalent of five loaves of bread …", the views it afforded were feeble but it helped trigger a desire to see farther into the Universe. In 1941, aged 21, Jones purchased a 5-inch (12.7-cm) f/15 Calver reflector with a frustratingly small finder scope (Figure 30). This was followed in 1945 by a 13-cm (5.1-inch) Cooke refractor with a larger 1.2-inch (3-cm) finder scope with which he made his first comet discovery in 1946 (Figure 31). The greater diameter finder permitted an easier location of targets. Austin (1994: 37) and Toone (2016: 85) state that he also purchased an 8-inch (20-cm) telescope built by J.T. Ward.

Desiring to see even fainter variable stars, in 1947 Jones turned to NZ born (Austin, 1994: 40) Dr Leslie Comrie (1893–1950; Figure 32; Massey 1952) in the United Kingdom (known as the 'father of astronomical computing', the Director of the BAA Computing Section from 1919 to 1922). Jones asked Comrie to help him secure a mirror for a new telescope (Toone, 2016: 85) so Comrie sourced a 12.5-inch (31.7-cm) f/5 mirror ground by Frederick James Hargreaves (1891–1970; Figure 33; Steavenson, 1971), one of Britain's foremost telescope makers. Hargreaves later jointly formed Cox, Hargreaves and Thompson, precision telescope makers. Comrie brought the mirror with him to NZ in 1948 when "... he made a complete tour of NZ, revisiting old friends and places." (McLintock, 1966). Jones' new telescope had a square tube and included a 1.8-inch (4.5-cm) finder and 3.1-inch (7.9-cm) guide scope (Figure 34). Jones named it 'Lesbet' out of appreciation for the encouragement that Les and Betty Comrie gave him in his early years of astronomy (Austin, 1994: 40). In operation, Jones would not point the telescope's equatorial axis at the South Celestial Pole when observing, as is the traditional method, but rather rotated the base in azimuth on its rollers to permit comfortable viewing of a star no matter its azimuth. Liller (1992: 128) in his book *The Cambridge Guide to Astronomical Discovery* calls it "The most novel mount of all those described in the [his] previous chapter is, without question, Albert Jones' …" Jones decided against constructing an observatory, opting rather to store Lesbet in his garage or laundry and wheel it out as needed.

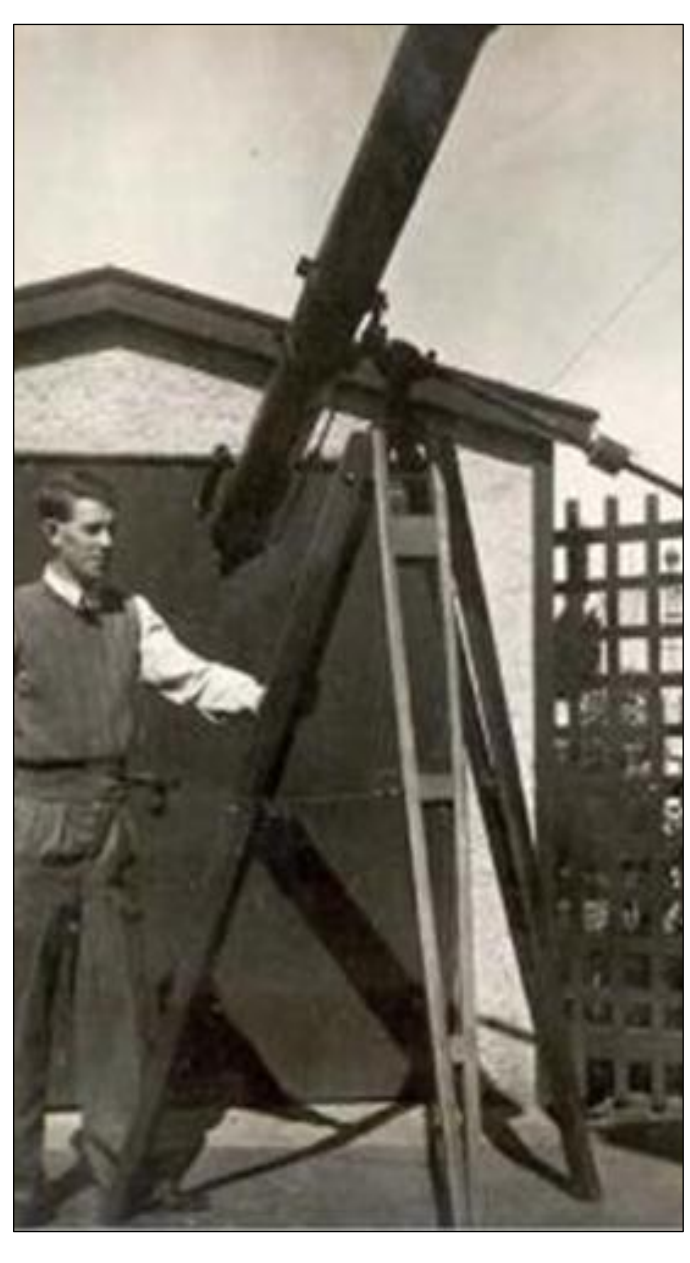

Figure 31: Jones' 5.3-inch (13.5-cm) refractor and Albert as a young man (after Toone, 2016: 85; from Alan Gilmore's collection).

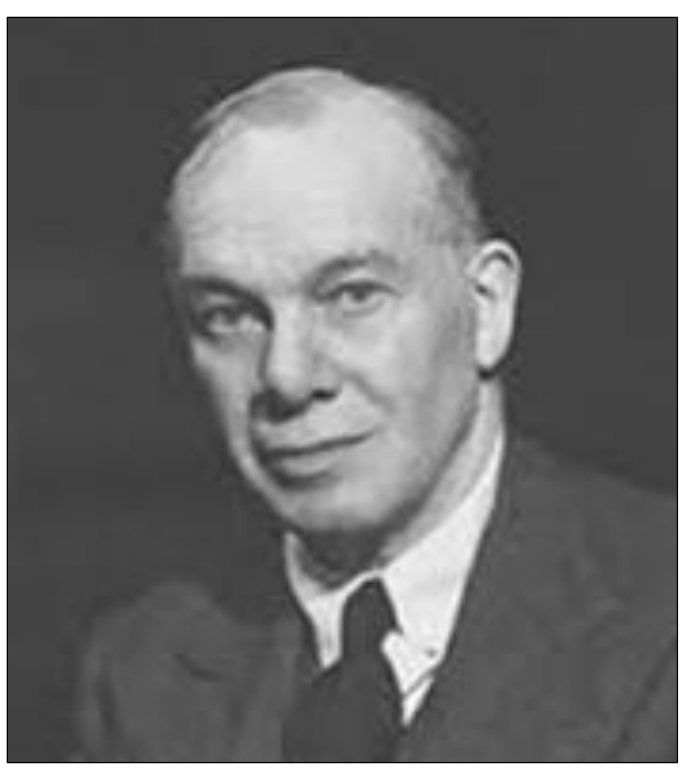

Figure 32: Lesley John Comrie (1893–1950). Director of the BAA Computing Section 1919–1922. He arranged for Hargreaves (Figure 33) to produce the 12.5-inch 'Lesbet' mirror (photograph: Alchetron).

With these telescopes, and the ultimate culmination of his most used telescope, 'Lesbet', Jones conducted a extensive visual variable star observing program that lasted 68 years—from his first observation of Nova CP Puppis on 18 January 1943 (aged 22) to his last observation of V766 Centauri on 31 August 2011 (aged 91) (Toone, 2016: 89, 93). Toone (2016: 89) highlights the record numbers of observations that Jones achieved, emphasising that Jones never kept a tally of his observation numbers as he deemed such a task unimportant. Toone points out that Jones "... had become only the second observer to reach

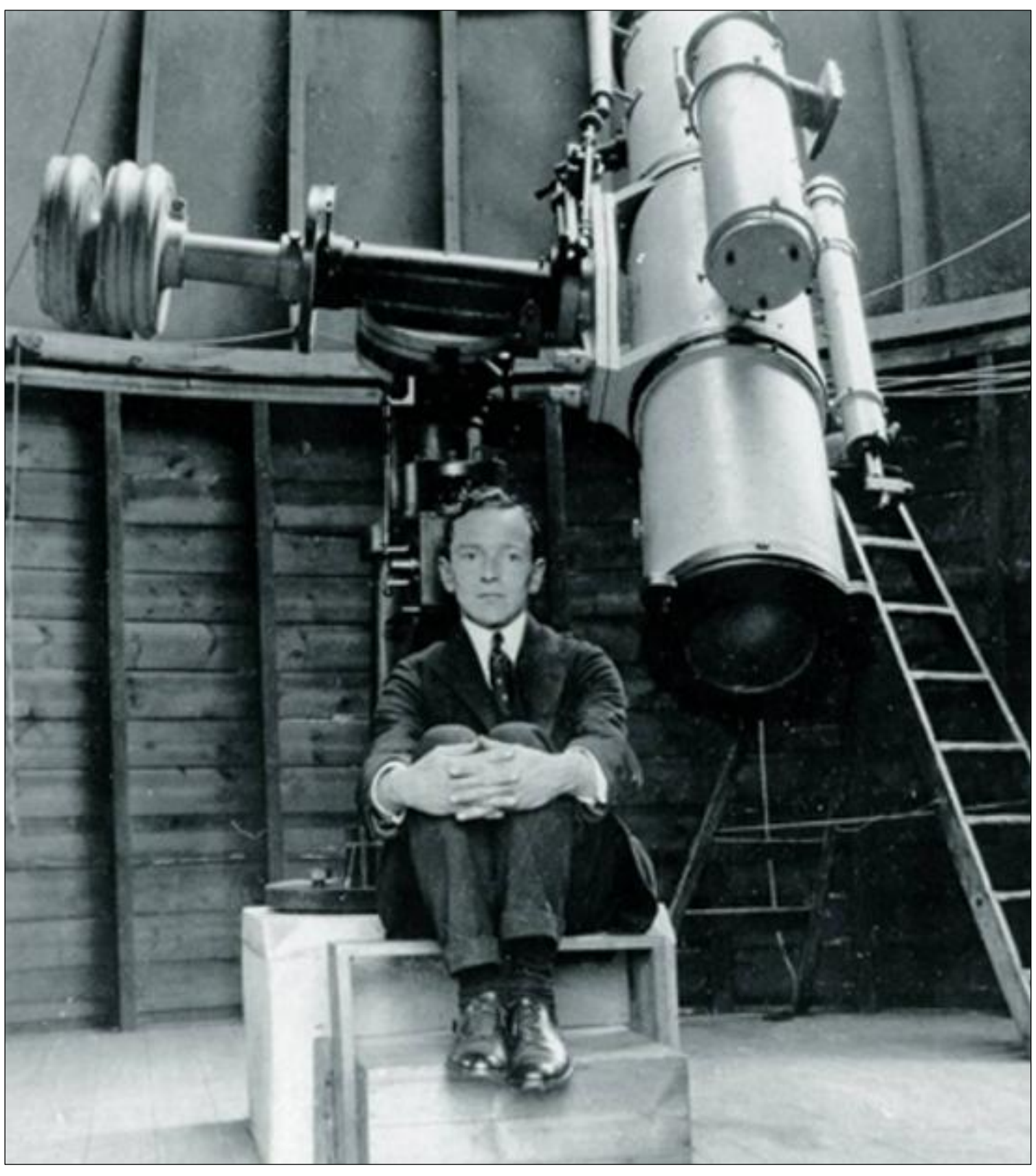

Figure 33: Frederick James Hargreaves (1891–1970). BAA President 1942–1944. Hargreaves ground Jones' 12.5-inch 'Lesbet' primary mirror (after Toone, 2016: 85; photograph: Bob Marriott).

100,000 observations following BAA VSS observer Charles Butterworth in 1939." (*ibid*.). Jones achieved this in just 15 years, the next shortest period was 24 years done by Cyrus Fernald of the USA. Jones would go on to make approximately 515,000 observations in his life, a world record that still holds today.[6] Jones stated that "Being a mere amateur has the advantage that I can observe what I want to observe and whenever I wish." (Toone, 2016: 92). This freedom certainly paid off. During his ~7,500 observations a year (515,000 / 68), he determined the periods of numerous variable stars, including R Doradus, AR Pavonis and VW Hydri; detected outbursts of recurrent novae such as T Pyxis in 1966 (Figure 29), V1017 Sagittarii in 1973, and V3890 Sagattari in 1990, and did pioneering work on dwarf Novae.

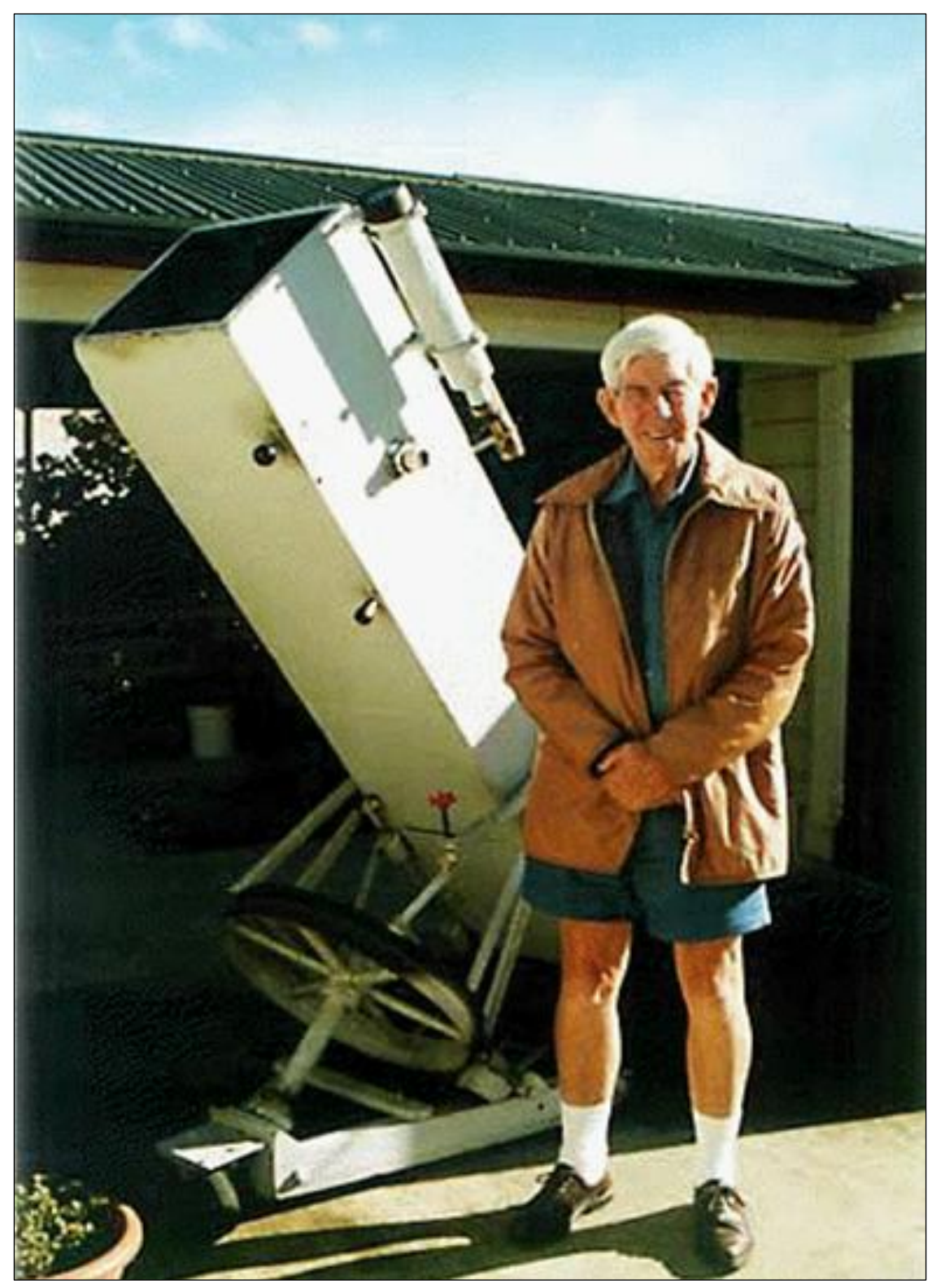

Figure 34: Albert Jones and his moveable 12.5-inch (31.7-cm) f/5 equatorially mounted Newtonian reflector 'Lesbet' at his Nelson home in the mid-1990s (photograph: Rod Austin, 1990s).

In addition, Jones was the first to visually detect the outburst of supernova SN1987A. He estimated the progenitor star reached magnitude 5.1 on 24 February 1987. It rose to magnitude 2.9 after three months. Jones, who observed a "... bright bluish thing …" beside the Tarantula Nebula in the Large Magellanic Cloud (Liller, 1992: 89) phoned Bateson about this new 'star' after he performed a magnitude estimation. Bateson then contacted Siding Spring Observatory (Australia). Toone (2016: 90) wrote that "... everyone [at Siding Spring] stopped what they were doing and focused their attention on what appeared to be an incredibly bright supernova on the rise." SN1987A is the most well studied supernova to date. Carolyn would later state, "... he [Albert] only found one supernova, but it was a GOOD one." (Austin, 1994: 41).

Despite his great achievements with variable star observing, Jones often said to the lead author and Wayne Orchiston that the greatest 'star' that he ever found was his second wife, Carolyn. They got married in July 1984 (Austin: 1994: 41). Jones told Drummond that, if she woke up when Albert was asleep, Carolyn would check the sky conditions and if it was clear, she would whisper to Albert as she returned to bed, "… it's starry outside". Jones would then wheel his trusty telescope out of the garage and contribute more variable star observations to the international astronomical community.

When the second author of this paper became Executive Director of Carter Observatory in the 1990s, Albert Jones already had an international reputation as a visual observer of variable stars. At this time, about a dozen leading NZ amateur astronomers were appointed Honorary Research Associates of the Observatory, and Albert was one of these. In his case, the primary objective was to network effectively with professional astronomers, who could make effective use of the long datasets Albert had on many different variable stars, and also compare his concurrent visual observations with those they were making photoelectrically. This proved very successful, as the following report from the Observatory's 1997–1998 Annual Report indicates:

> Mr Jones … continued to make visual observations of selected variable stars using his own 32cm telescope, and to forward specific observations to professional colleagues in Australia, Austria, Belgium, Bulgaria, Germany, Mexico, Netherlands, New Zealand, Poland, Russia, Slovakia, Switzerland, the UK, and the USA. (Sears and Orchiston, 1998:12).

A second objective in appointing Albert a Carter Observatory Honorary Research Associate was to use his renown to encourage Australian and NZ amateur astronomers to take up visual variable star observing. He did this through conference presentations and posters, and publications (e.g. see Jones, 1995a; 1995b).

It was while observing variable stars, just prior to dawn on 7 August 1946, that Jones was moving his telescope to the southern variable star U Puppis when he saw an unknown fuzzy blob in the finder scope. This would be his first comet discovery. It was provisionally designated Comet 1946h ('h' being the 8th letter of the English alphabet and this being the 8th comet discovery in 1946). After an orbit and perihelion date was derived, it received the permanent title of Comet 1946 VI since it was the 6th comet to reach perihelion in 1946. Today, after the cometary naming convention change in 1994, it is known as C/1946 P1 (Jones). The following highlights the discovery circumstances and observations made from New Zealand.

Table 9: An overview of Jones' first comet. Epoch: 27.0 October 1946 UT. Based on the Minor Planet Center (MPC). If not listed in MPC, information was used from NASA/Jet Propulsion Laboratory Horizons System, or Kronk (2009).

| | |
|---|---|
| Discovery Date | 7 August 1946, 7pm NZST (6.76 August 1946 UT) |
| Discovery Magnitude | 9 |
| Discovery Declination | –13° (Puppis) |
| Perihelion date | 26.779 October 1946 UT (MPC) |
| Perihelion distance (q) | 1.136 au (MPC) |
| Perigee date | 4 October 1946 (Kronk, 2009: 236) |
| Perigee distance | 1.9944 au (Kronk, 2009: 236) |
| Brightest | Magnitude 7 on 3 October 1946 UT (Kronk, 2009: 237) |
| Visible from NZ | 6 August 1946 (discovery) to late April 1947 (GUIDE, NASA/JPL) |
| Last observed | 23 November 1948 (UT), magnitude 19.3 (photographic) (Kronk, 2009: 239) |
| Eccentricity of the orbit (e) | 1.0008 (MPC) |
| Semi-Major axis (a) | Not stated |
| Aphelion distance (Q) | Not defined |
| Inclination (i) | 56.965° (MPC) |
| Epoch | 18 March 1947 (MPC) |
| Period | Not defined by MPC or NASA/JPL. Kronk (2009: 239) suggests elliptical orbits of 3.4 million year (original) and 15.6 million years (future) |
| Observations in Kronk (2007) | 72 (visual and photographic) – 18 from NZ (including discovery) |
| Observations in MPC | 12 (9 used to determine the orbital elements) |
| Observations in COBS | 30 |
| Papers Past newspaper articles | 64 (August 1946–May 1947) |

## 5.1 C/1946 P1 (Jones)

Initial details of this comet are listed below in Table 9, its orbit is illustrated in Figure 35 and its path through the sky is plotted in Figure 36.

The first newspaper reports of Jones' 6 August 1946 comet discovery from Timaru were published on 9 August (NZST) (e.g. Evening Star, 1946a: 4). Jones stated that he first saw the comet at dawn on Wednesday 6 August "... while studying variable stars …" He was using the 5.3-inch (13.5-cm) refractor (before owning Lesbit) at 42× magnification (Kronk, 2009: 236; Toone, 2016: 91). To confirm the cometary nature, as opposed to a suspected star cluster, Jones observed it again the following morning. It had moved. He then sent a full report to Ivan Thomsen (1910–1969; Eiby, 1970), Director of Carter Observatory in Wellington (*Evening Star*, 1946a: 4; Local and General 1946: 2). Thomsen then cabled the discovery news and details to the Harvard College Observatory in the USA and The Canberra Solar Observatory in Australia (The New Comet 1946: 6)—note that the *Marlborough Express* calls it the Solar Observatory in Sydney (Timaru Man's Observations, 1946: 4). The details about the discovery, based on Austin (1994), Toone (2016: 83–93) and personal conversation with Jones (around 2009) are as follows.

Jones was doing his usual visual variable star observations and then decided to sweep for comets South of 40° declination. At approximately 6 am he ceased this endeavour in order to observe some variable stars that were rising low in the East just prior to astronomical twilight. At 6:15 am he decided to observe U Puppis. In Jones' own words:

> ... as dawn had arrived, I did not waste time getting the box to stand on to reach the finder but sighted roughly along the telescope. But my aim was out and realizing that it was too high I swept down wards, but did not get as far as U Puppis for I came across an object that I felt sure was a stranger. Quickly I made a field drawing and plotted the position on H. B. Webb's Atlas. (Toone, 2016: 90).

He rechecked the suspicious object after 20 minutes, but was unsure as to whether it had moved in the brightening dawn sky. The next morning, attempting to confirm the comet's existence, he was initially thwarted by clouds. However, as Jones recalls,

> Shortly after 06.00 hours I had a chance to look up the previous day's position but the comet was not to be seen. A quick sweep soon located it not far away, and another field drawing was made, and the field plotted on Webb's Atlas.

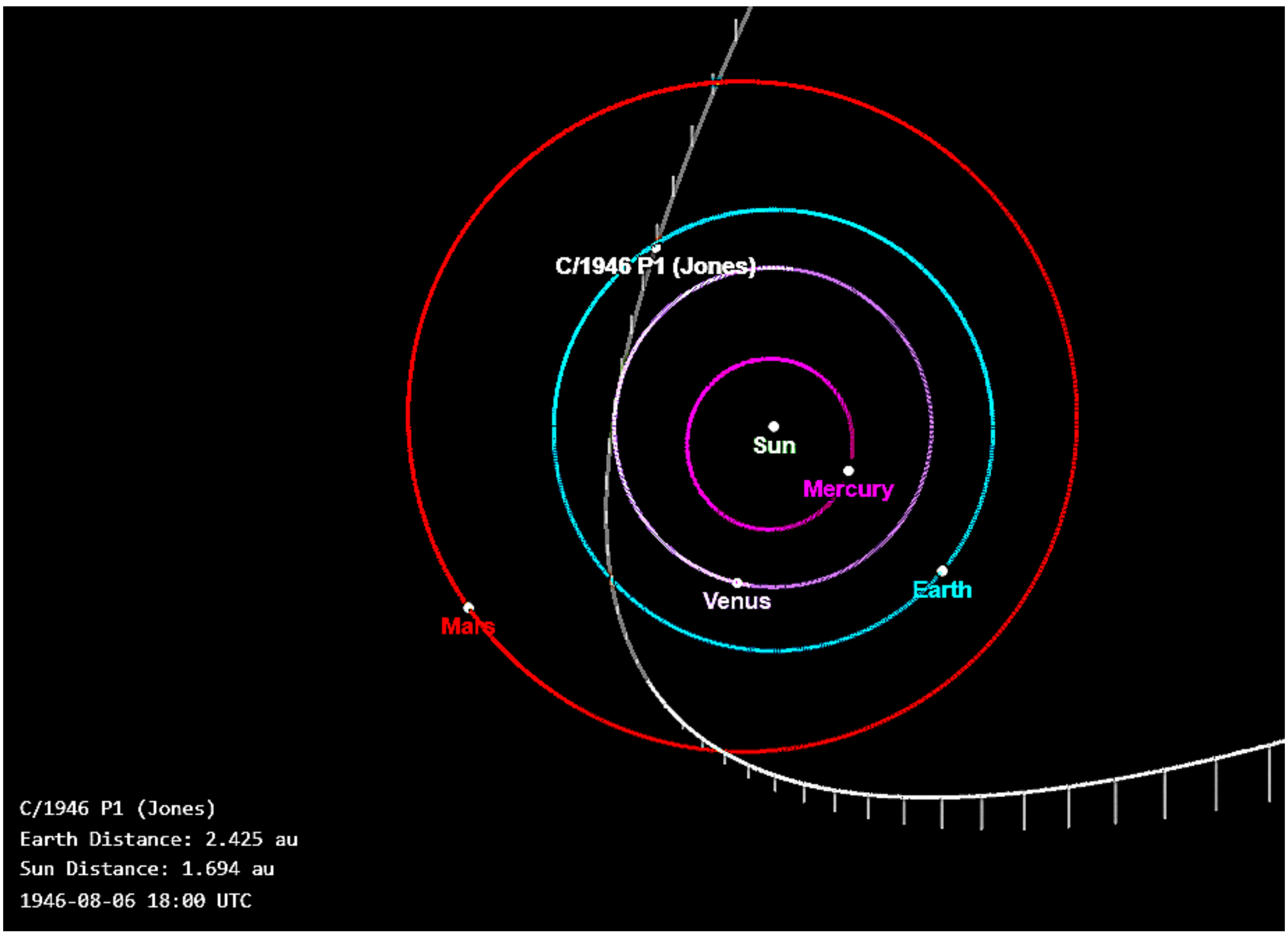

Figure 35: The orbital path of Comet C/1946 P1 (Jones). The comet came from the top of the diagram (South of the ecliptic, as evidenced by the vertical strips) and existed bottom right (North of the ecliptic). Note the close proximity of the comet to the Sun as seen from Earth—truly a dawn discovery (NASA/JPL Horizons)!

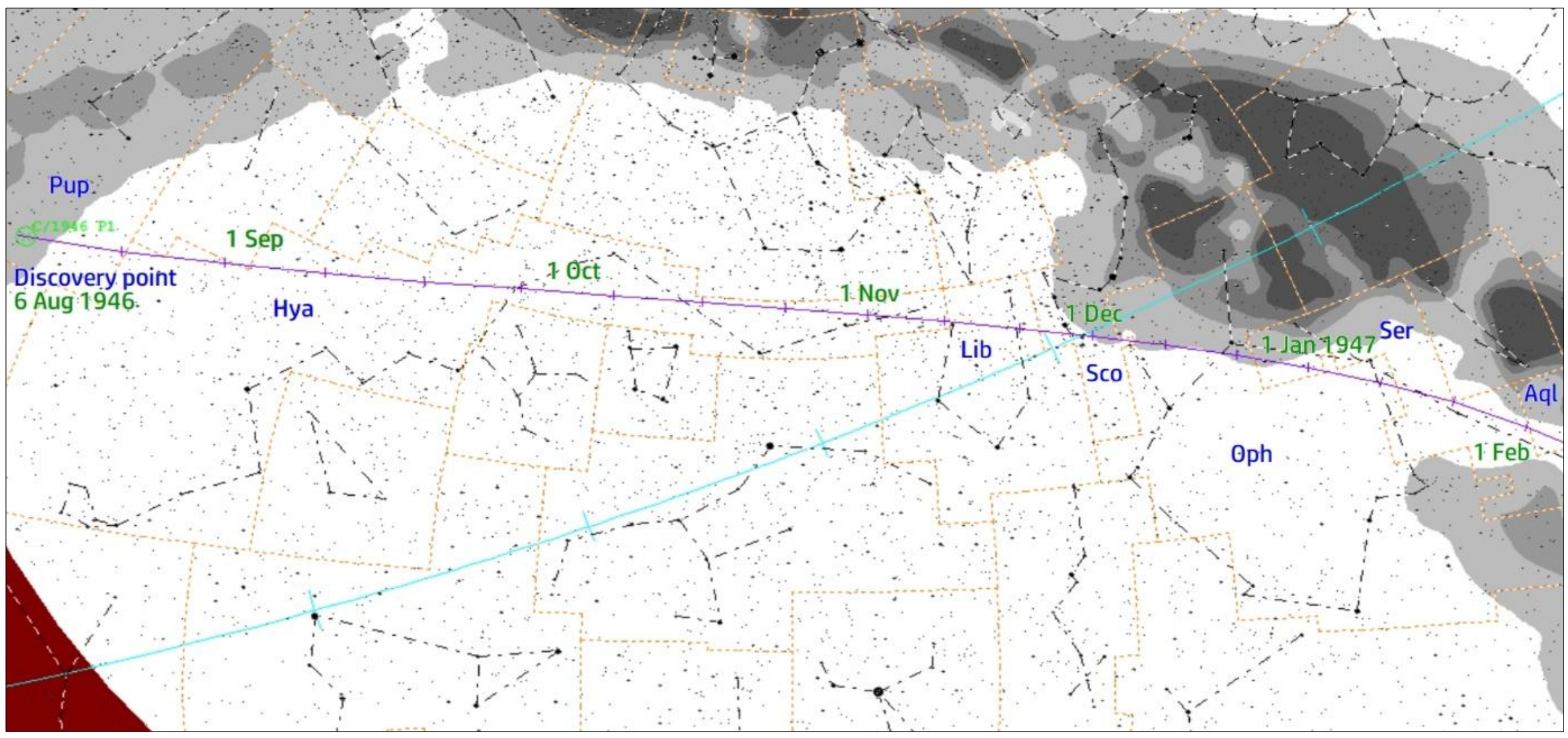

Figure 36: The path of Comet 1946 P1 (Jones) (in purple) from discovery. The constellations that C/1946 P1 traversed are labelled in blue. The dates are in green. The light blue line is the ecliptic. North is down (for the NZ sky), East is right (GUIDE software).

He estimated the coma diameter to be 1′ and of the ninth magnitude. The position was RA 7 hrs 56 mins.; Dec –13° 30′ (Timaru Man's Observations, 1946: 4), approximately 15° below Sirius. It was moving in the RA at the rate of three minutes a day toward the East, in Dec it was moving 15′ a day South. To determine the magnitude, Jones used a method described by British amateur astronomer Dr. William Herbert Steavenson (1942: 189–191) in the July 1942 issue of the *Journal of the British Astronomical Association*. Steavenson (1894–1975; Dewhirst, 1977) admitted the difficulties associated with determining a comet's magnitude due to the non-stellar (spread out) nature of the head. Since a comet's head cannot be reduced to a stellar point and then compared to stars of known magnitudes, the next best option was to unfocus the stars to the same apparent diameter as the comet's head and find a comparison star or stars of the same brightness and use these to determine a magnitude. Jones used comparison stars from his U Puppis variable star chart. Today, we know this method as the 'Sidgwick Method'. John Bortle (2006) describes this as "... the most widely used procedure, popularized by John Benson Sidgwick (1916–1958) within the British Astronomical Association in the 1950s." Of note is that Steavenson described this method approximately ten years before Sidgwick.

In regards to his discovery, Jones summarised (*ibid.*):

> I have been lucky to find it, but it surprises me that someone with a bigger telescope has not picked it up. It might have been visible for several weeks, and in all probability has been picked up already by astronomers in other countries south of the Equator. Last year, when I discovered the comet Kopff, an American had picked it up a month beforehand. (Timaru Man's Observations, 1946: 4).

This article (*ibid.*) went on to explain that in December 1945 Jones was awarded the Murray Geddes Memorial Prize by the New Zealand Astronomical Society for nearly being the first to pick up Comet Kopff and for his work on Society affairs.

On 9 August 1946, Mr G.G. Couling, of the Beverley-Begg Observatory (Dunedin), advised readers that the comet "... should be low in the south-eastern sky in the region to the right of Canis Major, some 18 degrees away from Sirius." (*Evening Star*, 1946b: 10). If viewable, Douglas Charles Berry (1918–2002; Kronk and Meyer, 2024: 159, 186; Kronk et al., 2026; Orchiston, 2016: 613–618) would attempt to photograph Jones' comet. News of the discovery was published in newspapers across NZ in Ashburton, Auckland, Dunedin, Gisborne, Greymouth, Levin, Marlborough, Northland, Te Awamutu, the Hawkes Bay, and Wanganui over the following weeks. The *Marlborough Express* (Timaru Man's Observations, 1946: 4) stated that "All active amateur astronomers in New Zealand have been informed by telegram."

The comet was claimed to be confirmed photographically on 16 August by D.C. Berry at the Berry–Thomas Observatory in Wakari, Dunedin (New Comet, Discovery Confirmed, 1946: 6). Berry was the Director of the Comet Section of the NZ Astronomical Society (soon to become the Royal Astronomical Society of New Zealand). Berry stated that

> Since Mr Jones first detected the comet 10 days ago, there has been no claim made from overseas identifying it, so that it can be safely assumed that he is the discoverer.

However, confusingly, the *Evening Star* (The New Comet, 1946: 6), stated on 10 August (six days prior to the 16 August confirmation) that

> Mr D. O. Berry observed the comet … and reports that it is fairly faint. A telescope of an aperture of at least one inch is necessary to observe it.

Did Berry observe it on 10 and/or 16 August? Kronk (2009: 236–237) states that Ernest Leonard Johnson (1891–1977) from the Union Observatory, Johannesburg, South Africa, photographed the comet on 13 August. Both he and Jones submitted observations in the ensuing weeks, stating the comet was magnitude 8–9. The first photographic observation by Berry mentioned in Kronk (2009: 237) was on 3 September 1946, nearly a month after initial discovery. One wonders if Berry's 10 or 16 August observation(s) were ever submitted internationally.

After Thomsen's notice of Jones' discovery was released to the national and international astronomical community, many astronomers, both amateur and professional, observed Comet Jones either visually or photographically. Vsekhsvyatskii (1964: 520–521) and Kronk (2009: 236–240) describe observations from (in approximate chronological order) NZ, South Africa, USA, Algeria (Alger then), Germany, Russia, and England. Of the 72 visual and photographic observations in Kronk (2009: 236–239), 12 (17%) were attributed to Jones (including the discovery observations). These were between August and October 1946. In October, Jones was the only astronomer to officially observe his comet, this being primarily due to the comet reaching its greatest Southern declination of –29° on 16 October 1946. He said the magnitude was 7 on the 3 October (UT) and 8 on 7 October (UT) and that a very short tail (less than 1 arc minute) extended to the WNW (Kronk, 2009: 237). The comet passed within 6° of the Sun on 25 November 1946 (UT).

C/1946 P1 (Jones) was too close to the Sun in November and December for observations. George van Biesbroeck (1880–1974) pointed out in *Popular Astronomy* that it would emerge from the Sun's vicinity in late January and then be better placed for Northern observers (van Biesbroeck, 1947a: 53). He predicted it would be approximately magnitude 9 and a morning object.

On 3 January 1947, Henry L. Giclas (1910–2007) of Lowell Observatory (Arizona, USA), observed it and estimated it to be magnitude 10.2. The position was RA 17h 33m 49.17s and Decl. +10° 51′ 15.3″. This was the first observation of the comet North of the celestial equator (van Biesbroeck, 1947b: 110). Based on updated orbital calculations, the ephemerides of Leland Erskin Cunningham (1904–1989) indicated Comet Jones would be too far North for observations from NZ from early May 1947 when it would be magnitude 13.0. Jones' 7 October 1946 magnitude 8 observation was the last official observation of C/1946 P1 (Jones) from NZ (Kronk, 2009: 237). Kronk (2009: 237–239) describes numerous visual and photographic observations from Northern Hemisphere observatories. The comet slowly faded from magnitude 10 in January 1947 to 19.3 in November 1947 when the last observation of the comet was made.

Table 10: Newspaper articles about 'Comet' and 'Comet C/1946 P1 (Jones)' in Papers Past. Note that August was the discovery month.

| Month and Year | Total for 'Comet' | Total for 'Comet Jones' |
|---|---|---|
| Aug 1946 | 137 | 38 |
| Sept 1946 | 66 | 13 |
| Oct 1946 | 88 | 0 |
| Nov 1946 | 13 | 0 |
| Dec 1946 | 62 | 0 |
| Jan 1947 | 29 | 0 |
| Feb 1947 | 19 | 0 |
| Mar 1947 | 89 | 0 |
| Apr 1947 | 0 | 0 |
| May 1947 | 40 | 0 |
| Totals: | 543 | 51 |

In the December issue of *Popular Astronomy*, van Biesbroeck (van Biesbroeck, 1947c: 560) highlighted that the comet had been observed for 15 months since its August 1946 discovery. Photographs with larger telescopes, e.g. the 36-inch (91-cm) Crossley Reflector at Lick Observatory, USA, revealed a short tail of no more than 2′ (Kronk, 2009: 238). Kronk (2009: 236) points out that the comet was discovered two months before perigee (at 1.99 au) and three months before perihelion.

L.E. Cunningham and Miss Martha Stahr (Berkeley) used Jones' and other observations to determine a "... preliminary parabolic orbit giving 4.873 October 1946 as the time of perihelion passage." (van Biesbroeck, 1946: 420). They estimated that Northern Hemisphere observers would see it towards the end of 1946. The constellations that C/1946 P1 (Jones) transversed are shown in Figure 36. The Northerly motion is evident. It was in Cygnus when last photographed on 23 November 1948 with a 36-inch (91-cm) reflector (Kronk, 2009: 239).

In relation to Comet Jones and Jones the man, Papers Past revealed that of the 543 NZ articles found using the search-word 'comet' from August 1946 to May 1947 (when the comet was observable from NZ), 51 related to the comet and Albert Jones. These all occurred in August (the discovery month) and September 1946. Refer to Table 10. Not only was Jones suddenly thrust onto the national stage, but he also gained significant international respect through his comet discovery, variable star estimates, and meteor and comet observations (Orchiston et al., 2021; Taibi et al., 2026; Toone, 2016). He also appeared as co-author on numerous scientific papers. His awards and recognitions are listed below.

Fifty-four years after his first comet discovery, in late 2000, Jones was continuing his almost nightly practice of visually observing variable stars when he accidentally discovered his second comet. This was a co-discovery with Syogo Utsunomiya in Japan. The comet was officially called C/2000 W1 (Utsunomiya-Jones) and Section 5.2, below, describes it.

## 5.2 C/2000 W1 (Utsunomiya-Jones)

Initial details of this comet are listed below in Table 11 its orbit is illustrated in Figure 37 and its path through the sky is plotted in Figure 38.

The discovery circumstances were similar to Jones' first comet discovery in 1946. Jones was observing variable stars on 25 November 2000 (UT) when he came across a fuzzy blob that he knew should not be there. Gilmore and Jones (2001: 9–11) relay the story of how Jones was swinging his 12.5-inch reflector to the variable star T Apodis around dawn:

> While 'star-hoping' to the field, viewing through the 7.8 cm 30× finder, he saw a hazy spot just 50′ (arc-minutes) north-west of T Aps which he instantly recognised as a comet.

Determining a magnitude of 8 and an approximate position, he phoned Mount John Observatory (Tekapo, NZ). Alan Gilmore and his wife, Pam Kilmartin, notified the IAU Central Bureau for Astronomical Telegrams (CBAT) with the discovery information. Gilmore confirmed the existence of the comet both with a CCD and visually with a 6-inch (15-cm) finder scope within a day. Gilmore (pers. comm., 2025) writes:

> Brian Marsden immediately suspected that Jones' comet was the same object that Utsunomiya had reported – three visual observations – and fitted a plausible parabolic orbit to the sightings. From that Brian was able to produce an ephemeris. On the following evening I used the 15-cm finder on Mt John's OC [Optical Craftsman] telescope to sweep around the predicted position and found the comet. Knowing its position I then got Glen Bayne, who was doing CCD imaging on the 1-metre telescope, to take pictures of the comet. These I measured and reported the positions to Brian at the IAU Bureau. Later in the night we got another set of images and more positions. These confirmed that Utsunomiya and Jones had seen the same comet. See IAUC 7526, 2000 November 26.

Table 11: An overview of Jones' second comet. Epoch: 26.0 December 2000 UT. Based on the Minor Planet Center (MPC). If not listed in MPC, information was used from NASA/Jet Propulsion Laboratory Horizons System.

| | |
|---|---|
| Discovery Date | 18 and 25 November 2000 UT |
| Discovery Magnitude | 8 |
| Discovery Declination | –76° (Chameleon) |
| Perihelion date | 26.559 December 2000 UT (MPC) |
| Perihelion distance (q) | 0.32118 au (MPC) |
| Perigee date | 25 November 2000 UT (Kammerer) |
| Perigee distance | 0.28 AU (NASA/JPL) |
| Brightest | Magnitude 8.0 on 25 December 2000 UT (NASA/JPL) |
| Visible from NZ | 25 November 2000 (discovery) to late April 2001 (NASA/JPL) |
| Last observed | 23.8 Jan 2001 UT (MPC) |
| Eccentricity of the orbit (e) | 1.00000 (MPC) |
| Semi-Major axis (a) | At perihelion, the eccentricity stated in the NASA/JPL database is greater than 1, and as such, the orbit at that point is hyperbolic and unbound. |
| Aphelion distance (Q) | Not stated |
| Inclination (i) | 160.165° (MPC) |
| Epoch | 26.0 December 2000 UT (MPC) |
| Period | Outbound orbit is unbound; the comet will likely never return |
| Observations in Kronk | The last year covered by Kronk is 1993, thus before C/2000 W1 |
| MPC Observations | 455 (147 observations used to determine orbital elements) |
| COBS Observations | 122 |
| Papers Past newspaper articles | 0 (November 2000–June 2001) |

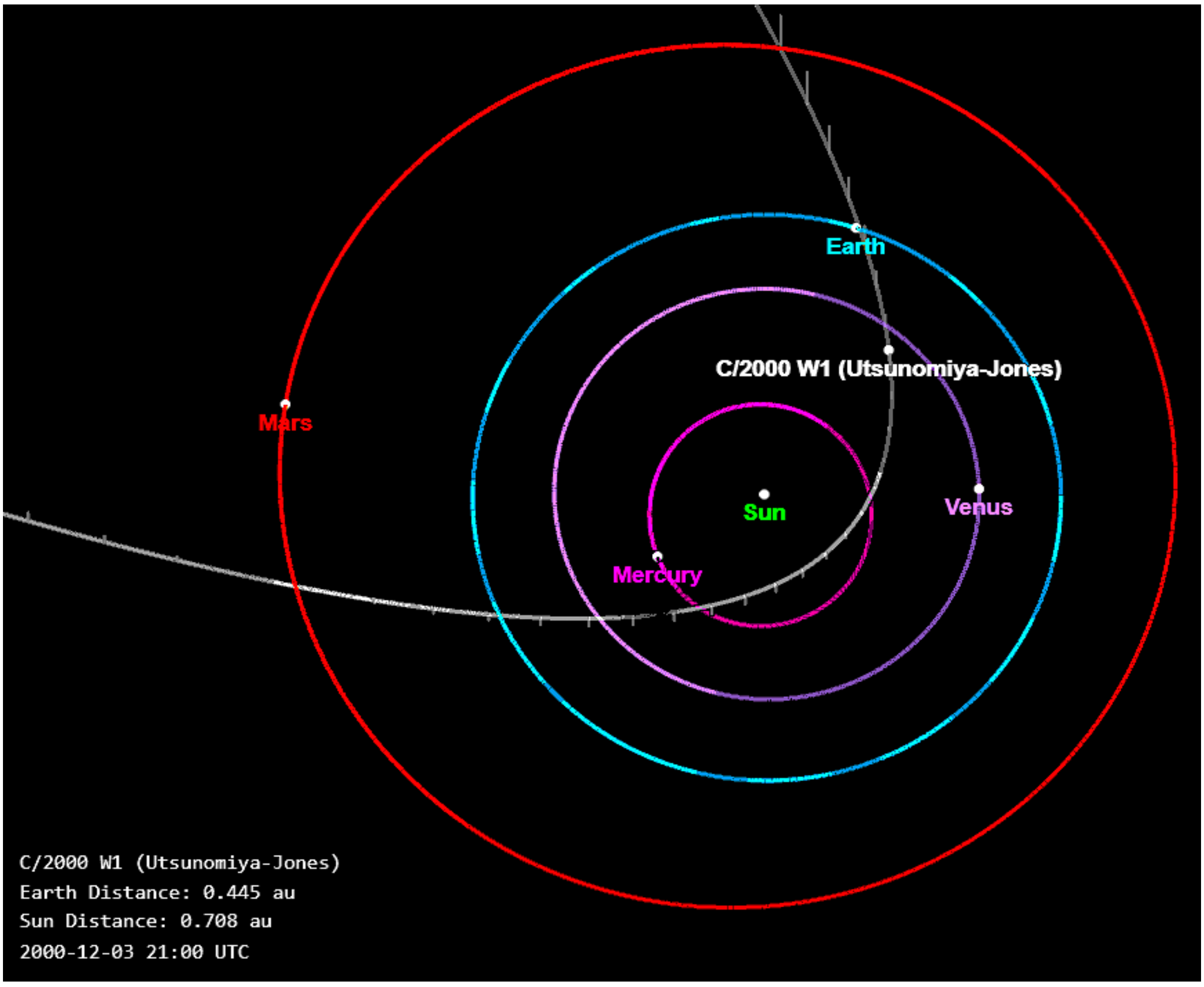

Figure 37: The orbital path of Comet C/2000 W1 (Utsunomiya-Jones). The comet came from the top of the diagram (South of the ecliptic - as evidenced by the vertical strips) and exited to the right (NASA/JPL Horizons).

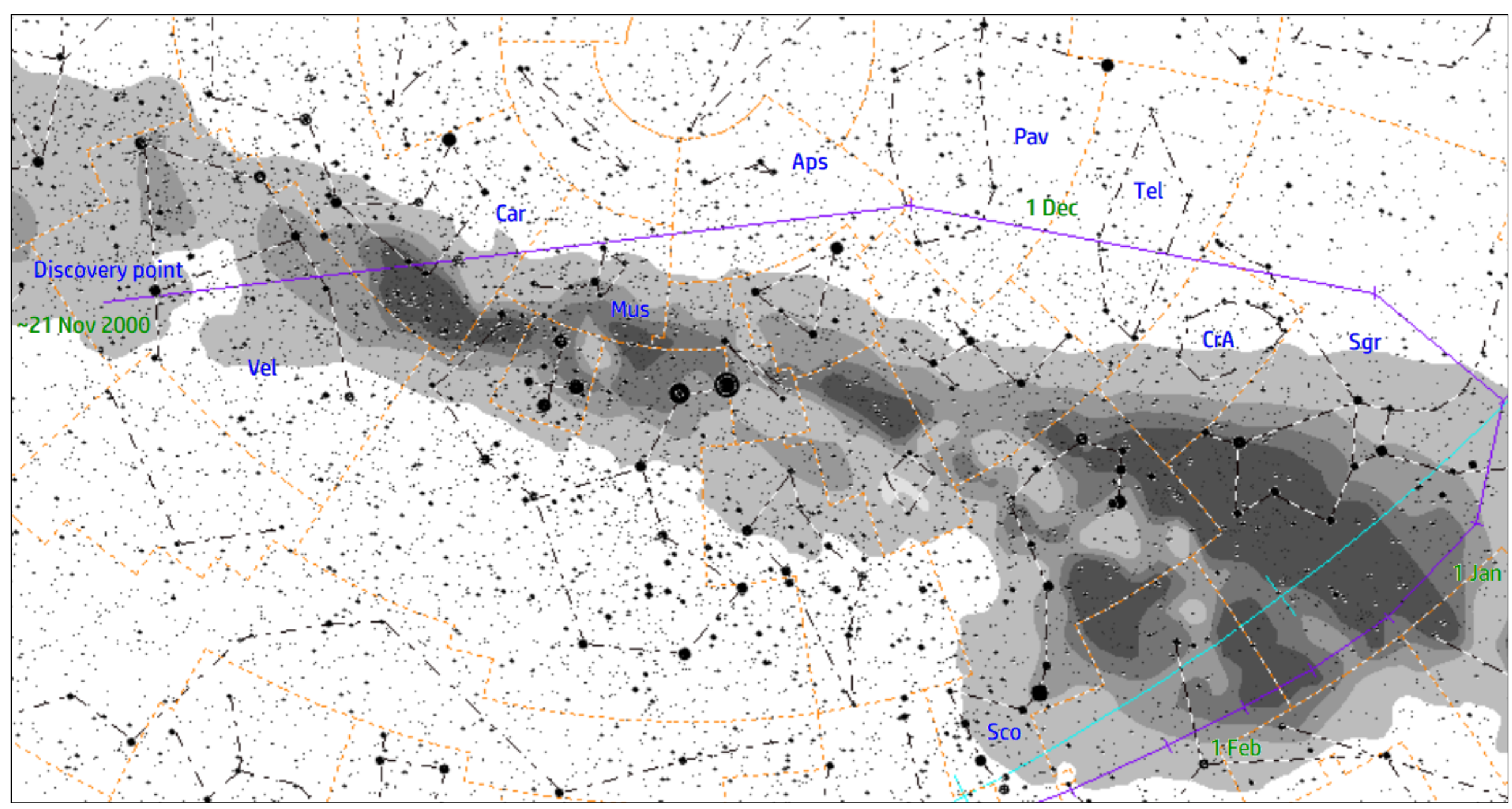

Figure 38: The path of Comet C/2000 W1 (Utsunomiya-Jones) (in purple) from discovery in late November 2000 until May 2001. Note the gradual movement of the comet towards the North. The constellations that C/1932 M1 traversed are labelled in blue. The dates are in green. The light blue line is the ecliptic. North is down (for the NZ sky), East is right (GUIDE software).

However, as Gilmore alludes to above, Jones was not the first to observe the comet. Nakano et al. (2000) reported that on 18.8 November 2000 (UT), seven days before Jones sighted it, Japanese comet hunter Syogo Utsunomiya saw a suspect comet in Vela from Japan as he was searching with a pair of 25 × 150 mm binoculars. It was low in his Southern sky two degrees North of λ Velorum. He noted that it moved 10′ to the SE during 40 minutes. A report was sent to the IAU Central Bureau for Astronomical Telegrams (CBAT) (*ibid.*). Numerous observers tried in vain to confirm the discovery. It was then that Jones accidentally discovered it and Gilmore confirmed Jones' co-discovery. Because Utsunomiya and Jones were the first two observers to see the comet, it was named Comet C/2000 W1 (Utsunomiya-Jones) after them. Unfortunately, neither Papers Past nor Kronk (and Vsekhsvyatskii) cover the year 2000, when Jones made his second comet discovery. However, Kammerer states that

> Observations on Feb. 12, 2001 with the 1.5m Catalina Reflector showed that this comet had undergone a rapid fading with the 1.7′ coma showing a R-band brightness of only 16.5m while observers reported visual magnitudes near 11.5m at the end of January. No condensation brighter than 21m could be detected (IAUC 7586). Observations by A.C. Gilmore with the 1m-Telescope of the Mount John Observatory on March 3 showed only a diffuse parabolic glow at the expected comet's position. The glow was brighter and 1′ wide at the 'head' end; the 'tail' was more than 10′ long (PA = 80°) and 2′ wide at the end. No condensation brighter than R = 20m was found (IAUC 7594).

It would seem that C/2000 W1 (Utsunomiya-Jones) disintegrated possibly in Scorpius. For the mostly Southern celestial path it took during this time, refer to Figure 38.

Albert Jones was 80 years old when he made his second comet discovery. The oldest human in history to make an official comet discovery (IAUC 7594). In addition, due to the 54-year gap between his first (1946) and second (2000) comet finds, he holds the world record for the longest time span between comet discoveries. Sadly, he passed away in 2013 and his wife, Carolyn, passed on 18 June 2018.

In light of all his astronomical achievements, Jones received the following awards (not exhaustive), based mostly on Austin (1994: 41–42) and Toone (2016: 91–92):

- 1945 – The Murray Geddes Award from the Royal Astronomical Society of New Zealand (RASNZ)

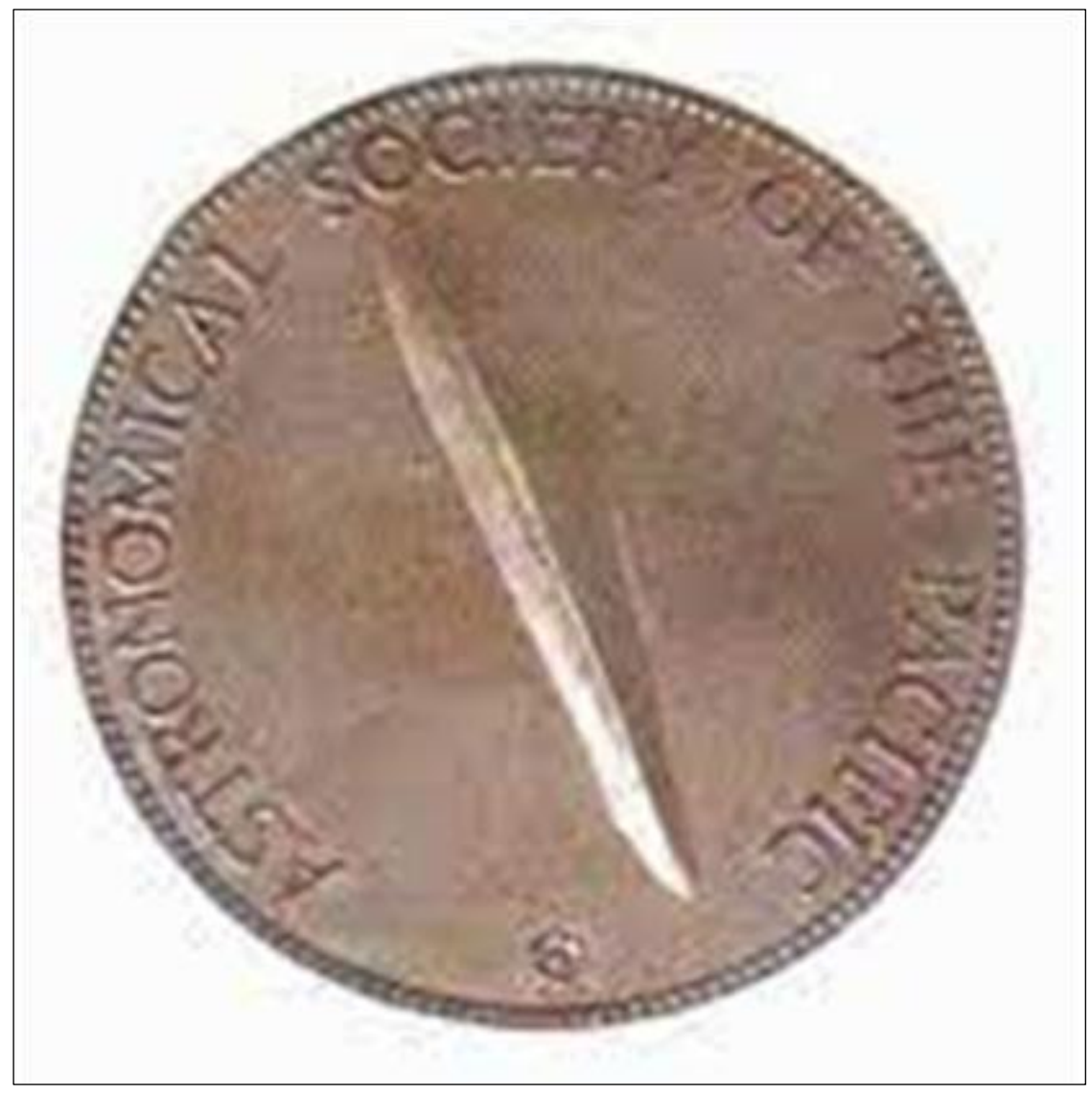

Figure 39: The obverse side of the Donohue Comet Medal from the Astronomical Society of the Pacific. Jones received it for his comet discovery in 1946 (photograph: Wikicommons).

- 1947 – The Donohue Comet Medal from the Astronomical Society of the Pacific (ASP). See Figure 39.
- 1949 – The Donovan Medal and Prize from the Donovan Astronomical Trust (Sydney)
- 1956 – The Michaelis Gold Medal & Prize from the University of Otago
- 1960 – The Jackson Gwilt Medal & Gift (jointly with Frank Bateson) from the Royal Astronomical Society (RAS)
- 1963 – Made a Fellow of the RASNZ
- 1968 – The Merlin Medal & Gift from the BAA
- 1973 – The Bronze Comet Medal from the ASP
- 1987 – Made a [NZ] Officer of the British Empire (OBE) for services to astronomy
- 1987 – Received the Nelson City (his home town) Council's 'Certificate of Achievement'
- 1988 – Asteroid 3152 (Jones) was named after him by the discoverers (Alan Gilmore and Pam Kilmartin). See Figure 40.
- 1997 – The Director's Award from the AAVSO
- 1998 – The Edward A. Halbach Amateur Achievement Award from the ASP
- 1998 – The Steavenson Memorial Award from the BAA
- 2001 – The Edgar Wilson Award from the Smithsonian Astrophysical Observatory (SAO)

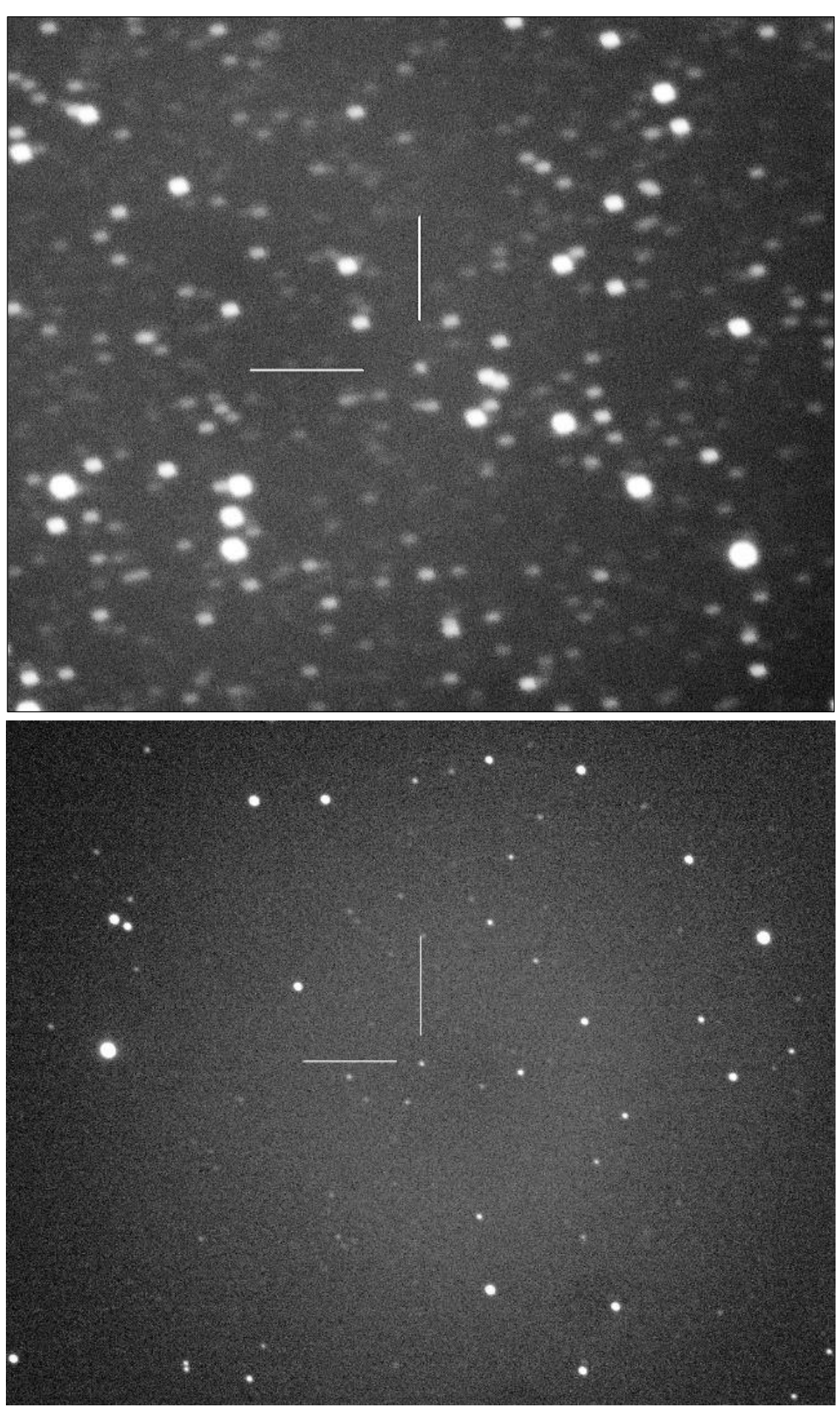

Figure 40: Main-Belt Asteroids 3152 (Jones), top, and 9171 (Carolyndiane), bottom, as imaged by the lead author of this paper on 26 January 2026. North is up, East is left. Field of view is 10′ × 8′. 3152 was 2.8 au and 9171 was 2.1 au from Earth when imaged (photographs: John Drummond, 2026).

- 2004 – An honorary DSc from Victoria University in Wellington, at the age of 83. See Figure 41, and Gilmore and Kilmartin (2004), and Sullivan (2004).
- 2005 – The Murray Geddes Award (jointly with Carolyn Jones) from the RASNZ
- 2008 – The Merit Award from the AASVO
- 2011 – Made an Honorary Life Member of the AASVO

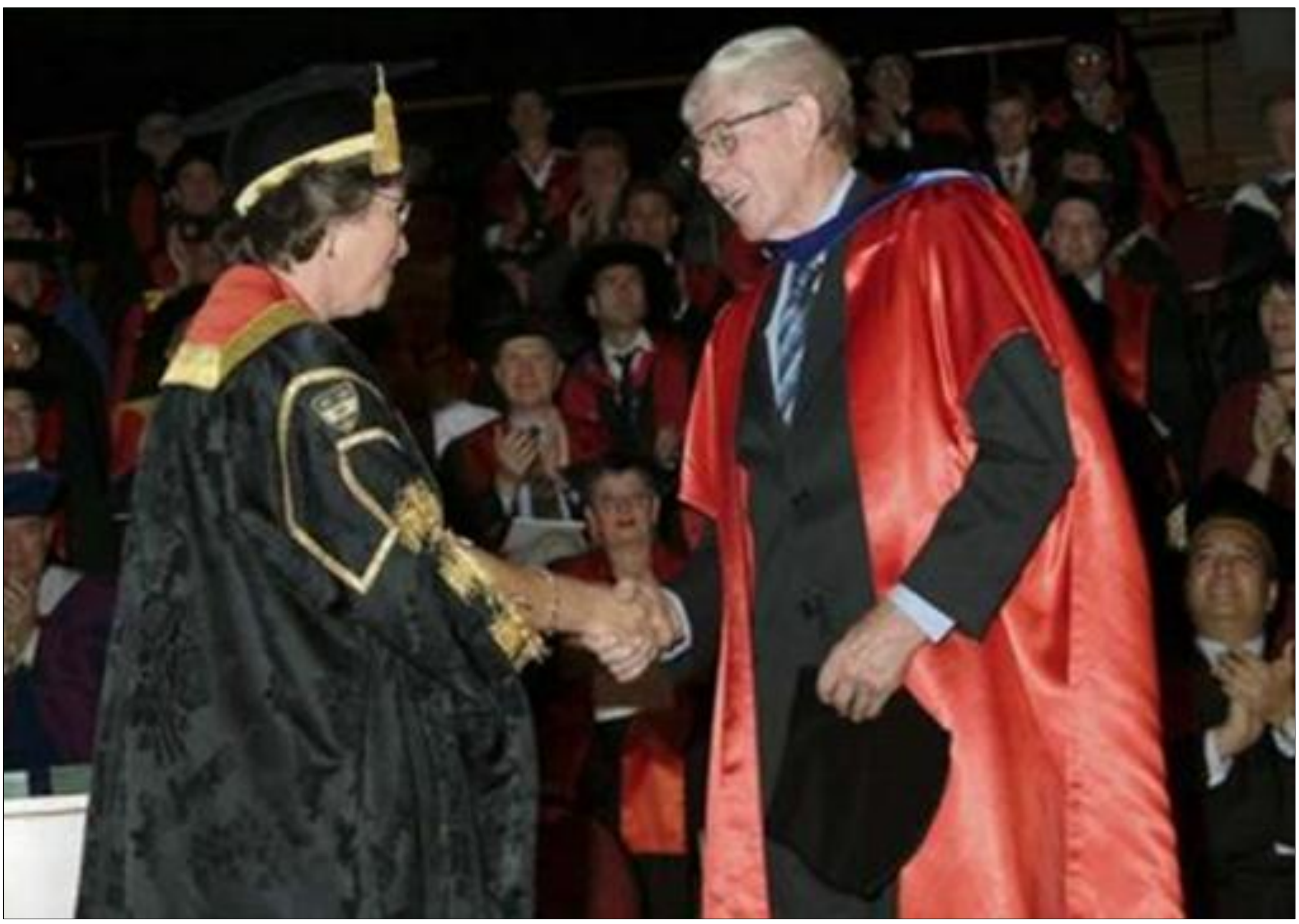

Figure 41: Jones receiving his Honorary Doctorate of Science from the Chancellor (Dr Rosemary Barrington) of Victoria University of Wellington on 7 May 2004 (after Toone, 2016: 92; photograph: Bruce Leadley).

## 6 MICHAEL CLARK

British-born Michael (Mike) Clark (1942–2019; Figure 42) worked as an observer-technician at the University of Canterbury's Mount John Observatory (Lake Tekapo, South Island) from 1971 and was Superintendent from 1980 until March 1996. He discovered a comet when examining photographic plates that he had taken (Hearnshaw, 2015: 38). According to Gilmore (2023) and Austin (2025), Clark previously narrowly missed detecting an 8th-magnitude supernova in the galaxy NGC 5253 in Centaurus because he was not in the habit of examining his plates, but would send them to Germany to be scrutinised (Orchiston, 2016: 501). Austin (2025) recalls that it was "… the brightest supernova in an external galaxy for nearly a century." After this missed opportunity, Clark began inspecting the plates himself, before forwarding them, looking for galactic supernovae. On 9 June 1973, while Clark was studying a pair of 60-minute exposures (Kronk, 2010: 379) taken on 1 June, Gilmore (2023) bemusedly noted that one of the galaxies had moved between the exposure dates. In fact, Clark had discovered a comet.

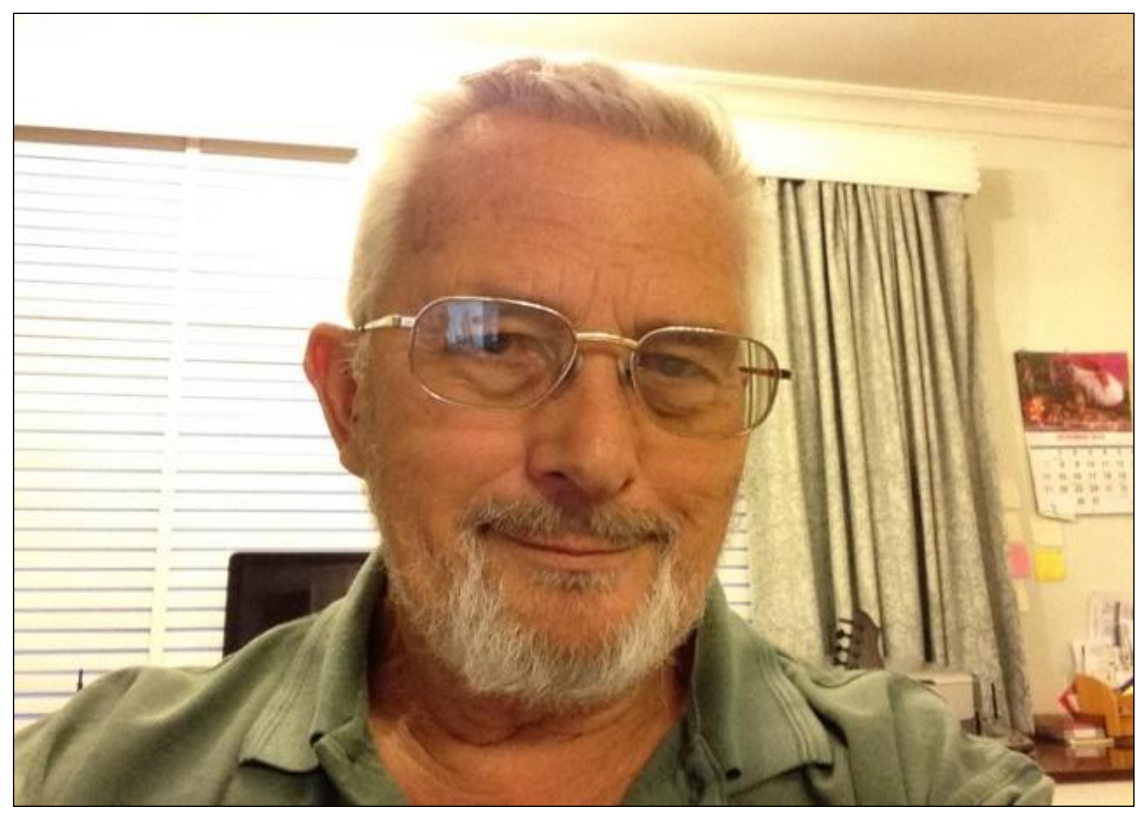

Figure 42: Mount John Observatory observer-technician and then superintendent, Michael Clark, discovered a comet in 1973 (photograph: Michael Clark, 2014; after Orchiston 2016: 501).

News of the discovery was sent to the media which highlighted that this was "The first comet discovery from New Zealand for more than a quarter of a century …" (Mount John Find, 1973: 17), the last one being discovered by Albert Jones of Nelson in 1946. Surprisingly, only this and one other NZ newspaper wrote about Clark's discovery (namely, Bright Planets in July, 1973: 9)—see Table 12. Details of the discovery were published in the International Astronomical Union's (IAU) Central Bureau for Astronomical Telegrams (CBAT), e.g. IAUC 2550 1973i (Marsden, 1973).

Clark, a carpenter by trade, who previously worked for Kodak Ltd in Christchurch (Hearnshaw, 2015: 63), was employed at Mount John Observatory (Tekapo) from February 1971 and ran the already initiated Bamberg variable star survey which searched the Southern sky for new variable stars and to help classify known ones (Hearnshaw, 2015: 41). It ran for ten years until 1977. A similar survey ran at the Boyden Observatory in South Africa from 1962 until 1977, as well as the San Miguel Observatory in Argentina (FAU). The programme was run by the Remeis Observatory of Bamberg, Germany, the University of Florida, and the University of Canterbury. The Dr. Karl Remeis-Sternwarte Astronomical Institute (FAU) states that:

> All southern stations were equipped with identical Aero Ektar (aperture: 10cm, focal length: 61cm, plate scale: 388″/mm) cameras. The major southern station operated by Bamberg staff was set up at Boyden observatory in South Africa, which was equipped with 10 cameras and operated from 1963 to 1972 [see Figure 43], while 4 cameras were operated from 1967 to 1976 at Mount John station and 6 at San Miguel from 1969-1972. A total of 22000 plates of the southern sky were obtained.

Table 12: Newspaper articles about ‘Comets’ and ‘Comet 71P/Clark’ in Papers Past. Note that June was the discovery month.

| Month/Year | Total for ‘Comet’ | Total for ‘Comet Clark’ |
|---|---|---|
| June 1973 | 11 | 2 |
| July 1973 | 14 | 0 |
| August 1973 | 22 | 0 |
| September 1973 | 2 | 0 |
| October 1973 | 14 | 0 |
| November 1973 | 19 | 0 |
| Totals: | 82 | 2 |

These cameras and lenses were arranged on a mount which pointed to the same RA but covered a range in Dec as a strip (Gilmore, 2023). The survey used recently produced Kodak 103a-e emulsion film which was sensitive to the red-end of the spectrum to capture red variables. Clark was running the program at Mount John from 1971, after Ian Patterson retired (Hearnshaw, 2015: 40–41). The survey was highly productive, an example being in the first month of operation in June 1967, 131 good quality plates were obtained from Mount John (*ibid.*). Clark retired from Mount John in 1996 (Orchiston, 2016: 499).

Comet Clark was the first comet discovered using photography from NZ by a NZer (however, note that a US satellite-tracking observer based at Mount John, Sergeant Len Edwards, photographed a comet in 1971, although it was not confirmed until 2023 by Maik Meyer and Gary Kronk when they were checking old photographs from other photographic surveys).[7] The following Section highlights the observations of Comet Clark.

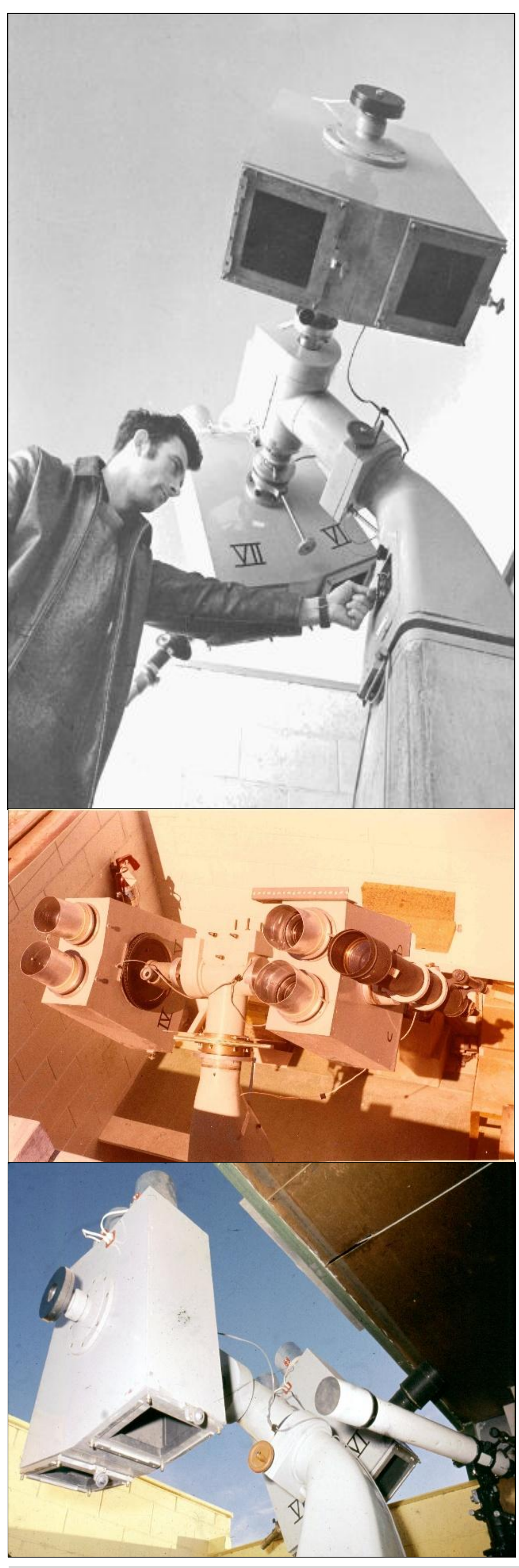

Figure 43 (Top): Mike Clark with the Bamberg variable star survey cameras and lenses (after Hearnshaw and Gilmore, 2015: 64). Centre and Bottom: The Aero Ektar cameras with 3.9-inch (10-cm) aperture f/6 lenses (courtesy: John Hearnshaw).

Table 13: An overview of Clark's comet. Epoch: 7.0 June 1973 UT. The MPC has four epochs for this comet, the most recent being 13 September 2023, so the 7 June 1973 epoch from NASA/JPL was primarily used instead, with Kronk and Meyer (2010: 379) being used for additional information.

| Discovery Date | 10 June 1973, 4am NZST (9.66 June 1973 UT). Prediscovery 1.70 June 1973 (Kronk and Meyer, 2010: 379) |
|---|---|
| Discovery Magnitude | 13 |
| Discovery Declination | –31° (Microscopium) |
| Perihelion date | 24.88 May 1973 UT (NASA/JPL) |
| Perihelion distance (q) | 1.560 au (NASA/JPL) |
| Perigee date | 7 July 1973 UT (Kronk and Meyer, 2010: 379) |
| Perigee distance | 0.6587 au (Kronk and Meyer, 2010: 379) |
| Brightest | Magnitude 12.0 in July 1973 (Kronk and Meyer, 2010: 379) |
| Visible from NZ | 10 June (discovery) to September 1973 (GUIDE, NASA/JPL) |
| Last observed | 21 November 1973 (UT) (photographic) (Kronk and Meyer, 2010: 380) |
| Eccentricity of the orbit (e) | 0.5003 (NASA/JPL) |
| Semi-Major axis (a) | 3.139 au (2019: NASA/JPL) |
| Aphelion distance (Q) | 4.691 au (2019: NASA/JPL) |
| Inclination (i) | 9.504° (NASA/JPL) |
| Epoch | 07 June 1973 (NASA/JPL) |
| Period | 5.561 au (2019 epoch) |
| Observations in Kronk (2010) | 24 (visual and photographic) – 8 from NZ (including discovery) |
| Observations in MPC | 58 (77 used to determine orbital elements) |
| Observations in COBS | None |
| Papers Past newspaper articles | 2 (June–November 1973) |

## 6.1 C/1973 L (71P/Clark) - 1973 V = 1973i

Initial details of this comet are listed below in Table 13, its orbit is illustrated in Figure 44 and its path through the sky is plotted in Figure 45.

Following Clark's comet discovery and the ensuing discovery notices, the then head of the University of Canterbury's Department of Physics, Professor A.G. McLellan (1919–2012) said in a newspaper article, "... in accordance with astronomical practice, the comet would be named after the discoverer." (i.e. Clark). He continued:

> Later observations taken at the Carter Observatory [Wellington] will assist in determining the comet's orbit. At present, it is a faint object barely visible in the 24-inch telescope at Mount John; it is as yet too early to predict whether it will become visible without telescopic aid … Its motion is being studied continuously with various instruments at the Mount John observatory. (Mount John Find, 1973: 17).

A few nights later, on 11 June 1973, Alan Gilmore and Pam Kilmartin imaged the comet from Carter Observatory in Wellington. Carter Observatory technician Russell Millington and Pam Kilmartin then got astrometric positions using the plate-measurer at the Department of Scientific and Industrial Research's Physics and Engineering Department at Gracefield (Wellington). These were sent to British-born astronomer Dr. Brian Geoffrey Marsden (1937–2010; Hurst, 2011) at the International Astronomical Union's Central Bureau. Clark also sent his discovery plates to Gilmore and Kilmartin for astrometric positions to be established. From these measurements, Marsden (1973) was able to determine a 5.5-year orbital period, so Clark had found a short-period comet. Based on their photographs from Carter Observatory, Gilmore, Kilmartin and Millington described the coma as being 10″ in diameter and strongly condensed. There was a hint of a fan-shaped wispy tail 1′ long in a p.a. of 260° (Kronk and Meyer, 2010: 379). According to Marsden's ephemerides, it would fade to magnitude 15.5 by early October 1973.

Clark determined 71P/Clark to be magnitude 13 when the discovery photographs were taken (Bright planets in July, 1973: 9; Kronk, 2010: 379). The RA was 20h 52.5m, Dec –31° 28′. GUIDE software indicates that it rose just after 8pm local time and was in the constellation Microscopium with a Western elongation of 130° and an apparent movement of 1.04′/hour to the ESE. It was barely brighter at its brightest, reaching magnitude 12 on 21 June 1973 UT. When discovered on 9.66 June (UT) it was 0.71 au from Earth and 1.57 au from the Sun (Kronk, 2020: 379). It was closest to Earth on 7 July 1973 UT at 0.66 au after passing perihelion on

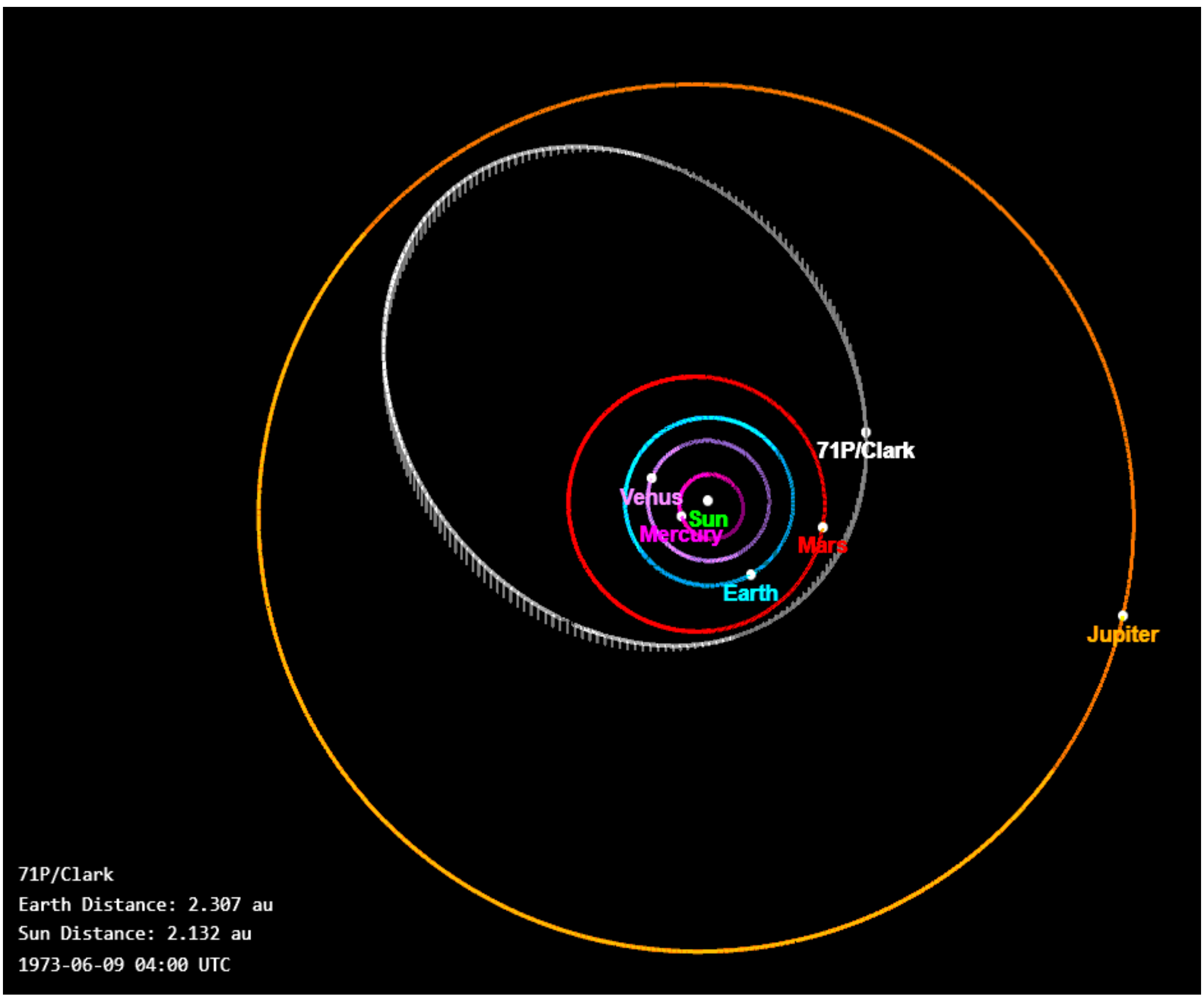

Figure 44: The orbital path of Comet 71P/Clark at the time of discovery. Note that the aphelion almost reaches the orbit of Jupiter. The comet's orbit is prograde (anti-clockwise from above) (source: NASA/JPL Horizons).

24 May UT (Kronk, 2010: 379). During its periodic returns, 71P never attains naked eye status. According to NASA/JPL Horizons, Comet Clark will next be at its brightest in August 2028 when it may reach magnitude 15 in Virgo. Figure 46 displays the small, faint nature of the comet when photographed near the Antares/ Rho Ophiuchi nebulosity by the lead author on 24 May 2017 NZST (see Figure 46). It was just weeks away from its brightest magnitude for that apparition (magnitude 13.4 in mid-June 2017). Orchiston (2016: 501) writes and quotes Clark, stating,

> Periodic comet 71P/Clark has ... a period of almost exactly 5.5 years. It is normally quite faint, so every 11 years it comes around at about 11th mag.

The 0.5 year (six months) in an 11.5 year orbit means the Earth is on the other side of the Sun at every second orbit.

Table 14 provides an overview of observations of Comet 71P/Clark obtained from around the world during its 1973 apparition. Based on Kronk and Meyer (2010: 379−381), observations of the comet were made from Argentina, Australia, Japan, Kazakhstan, NZ, South Africa, and the United States of America. Being published in 1964, Vsekhsvyatskii's *Physical Characteristics of Comets* was written too early to include Comet 71P/Clark.

For a complete list of submitted observations of Comet 71P/Clark from 1973 until 2024 see https://minorplanetcenter.net/db_search/show_object?object_id=71p. Lamy et al (2009) derived a nucleus diameter of 1.36 km, based on Hubble Space Telescope observations on 12 March 2000. This is quite small when compared to the brighter and more famous comets discussed above, which explains the comet's relatively faint nature.

Hearnshaw (2015: 64) states that

> [John] Baker and Clark were to become two of the [Mount John] observatory's longest-standing employees over the next three decades.

Clark retired from Mount John in March 1996.

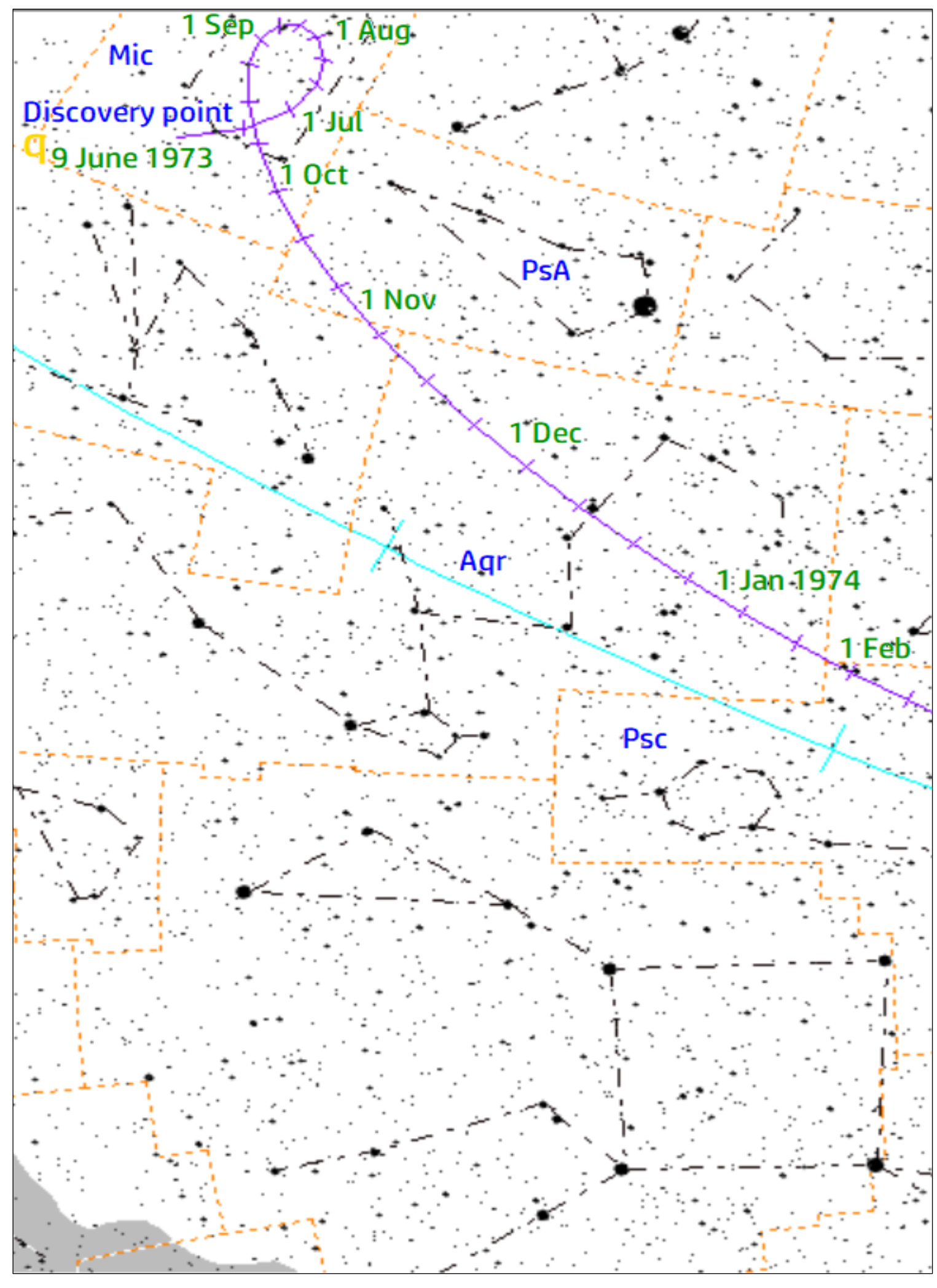

Figure 45: The path of Comet 71P/Clark from discovery in early June 1973 (UT) until February 1974 (at magnitude 20) is shown in purple. Note the gradual movement of the comet towards the North. The constellations that 71P/Clark traversed are labelled in blue. The dates are in green. The light blue line is the ecliptic. North is down (for the NZ sky), East is right (source: GUIDE software).

Austin (2019) recalls that Clark briefly joined a Christchurch software company and then rejoined the University of Canterbury's Physics and Astronomy Department, first as a computer technician, then as the purchasing officer. Austin thinks that he was there until at least December 2005 (pers. comm., 2025).

In the mid-2010s, Mike, and his wife June, moved to Melbourne (Australia) to be with their son Terry. Then towards the end of 2019 they decided to return to NZ to temporarily live with their daughter Mel, but sadly Mike passed away from a stroke on 27 November 2019. Austin (2019), who spent many years observing (and fishing) with Clark recalls what an incredibly talented man he was and how the music from the electric guitar that Clark built in the Mount John Observatory workshop would reverberate around the domes as he was waiting for the next Bamberg survey exposure to end (Figure 47).

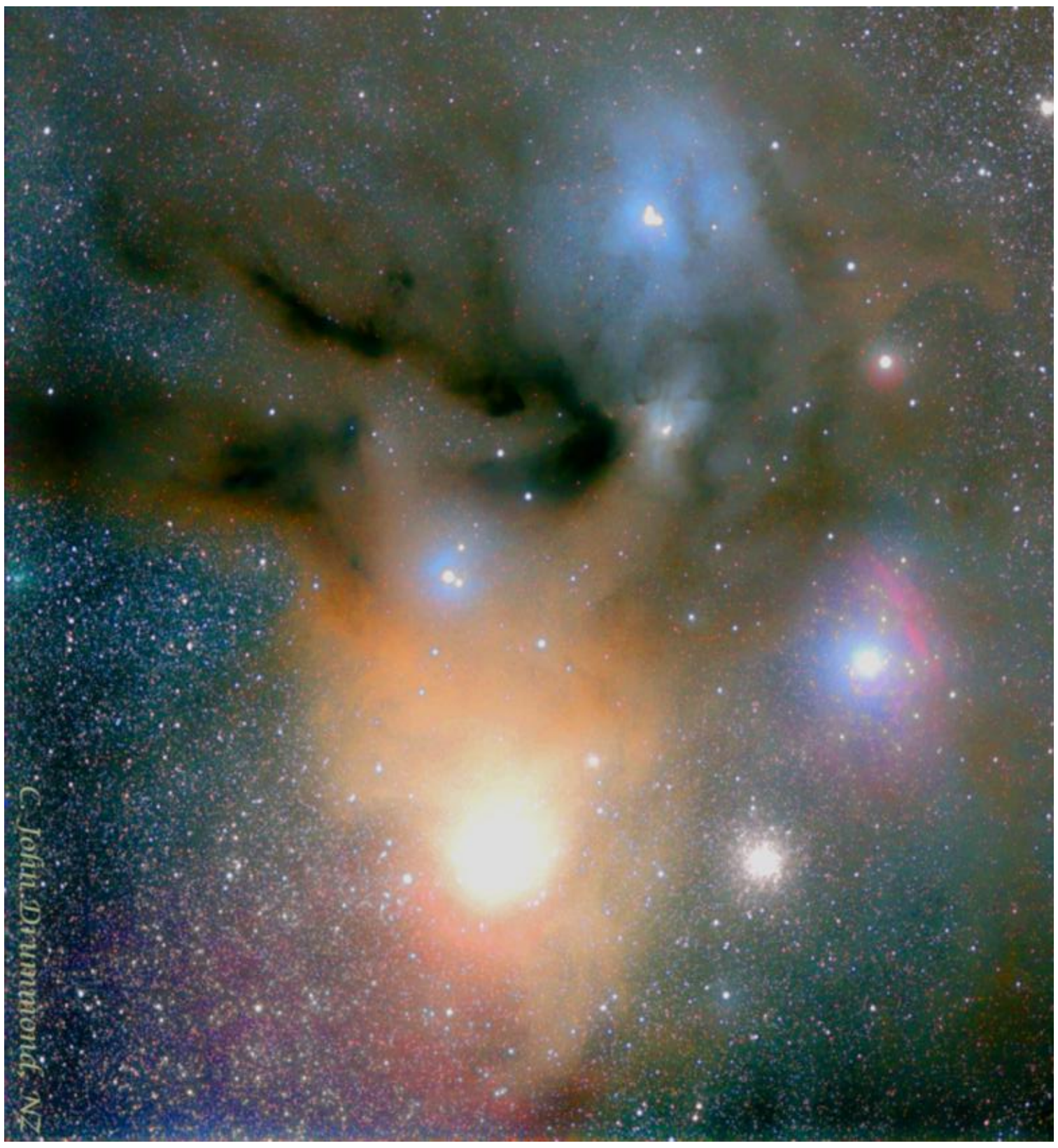

Figure 46: Comet 71/P Clark and the Rho Ophiuchi and Antares region photographed by the lead author with a Canon 550D DSLR camera and a 200mm Sigma lens at f/2, 800 ISO, on 24 May 2017. 100 × 3 minute exposures were stacked. Comet 71P/Clark is the small green smudge on the extreme left edge, halfway between the top and bottom. A comparison photograph taken around the same time can be seen at the *Astronomy Picture of the Day* (APOD) at https://apod.nasa.gov/apod/ap170527.html (2017 May 27).

Table 14: Observational overview of Comet 71P/Clark. 'Mag' is the apparent visual magnitude of the comet.

| Date 1973(UT) | Observer | Observatory | Country | Telescope | Mag | Notes |
|---|---|---|---|---|---|---|
| 9.66 Jun | Mike Clark | Mount John Obs | NZ | 3.9-inch (10-cm) Bamberg | 13.0 | Discovery images (first taken June 1.70) |
| 10.66 Jun | Mike Clark | Mount John Obs | NZ | 3.9-inch (10-cm) Bamberg | 13.0 | Confirmation image; 1′ tail |
| 11 Jun | Gilmore et al. | Carter Obs | NZ | 16-inch (41-cm) | 13.0 | Astrometry. 30″ tail |
| 12 Jun | Kilmartin | Carter Obs | NZ | 16-inch (41-cm) | 13.0 | Photography |
| 17 Jun | Gilmore et al. | Carter Obs | NZ | 16-inch (41-cm) | 13.0 | Diffuse coma with condensation. |
| 20 Jun | Gilmore | Carter Obs | NZ | | - | 10″ coma |
| 23 Jun | Kojima | | Japan | | 13.5 | |
| " | Seki | Kochi Obs | Japan | | 13.5 | |
| 1 Jul | Bruwer | SA Astron Obs | South Africa | 10-inch (25-cm) camera | 12.0 | |
| 2 Jul | Roemer | Steward Obs | USA | 90-inch (229-cm) | [18.8] | Nuclear magnitude |
| 3 Jul | Gilmore | Carter Obs | NZ | 16-inch (41-cm) | [14] | Nuclear magnitude. Condensed nucleus. |
| 4 Jul | Bruwer | SA Astron Obs | South Africa | | 12.0 | |
| 7 Jul | — | | | | | Perigee (0.6587 au) |
| 15 Jul | Gorodetskiy | | Kazakhstan | 20-inch (50-cm) | 12.8 | Diffuse, central condensation |
| 25 Jul | Bruwer | SA Astron Obs | South Africa | | 12.5 | |
| 27/28 Jul | Sim | Siding Spring | Australia | 48-inch (122-cm) Schmidt | 13.0 | |
| 29 Jul | Gorodetskiy | | Kazakhstan | 20-inch (50-cm) | 13.2 | |
| 31 Jul | - | | | | | Maximum elongation (158°) |
| 1, 21, 29 Aug | Bruwer | SA Astron Obs | South Africa | | 13.0 | |
| 6 Aug | - | | | | | Comet at most Southerly declination (–39°) |
| 22 Sep | Roemer | Steward Obs | USA | 90-inch (229-cm) | [18.8] | Nuclear magnitude |
| 26 Sep | Gibson | Yale-Col Uni | Argentina | | | |
| 21 Nov | Roemer | Steward Obs | USA | 90-inch (229-cm) | | Last time imaged for 1973 apparition. |

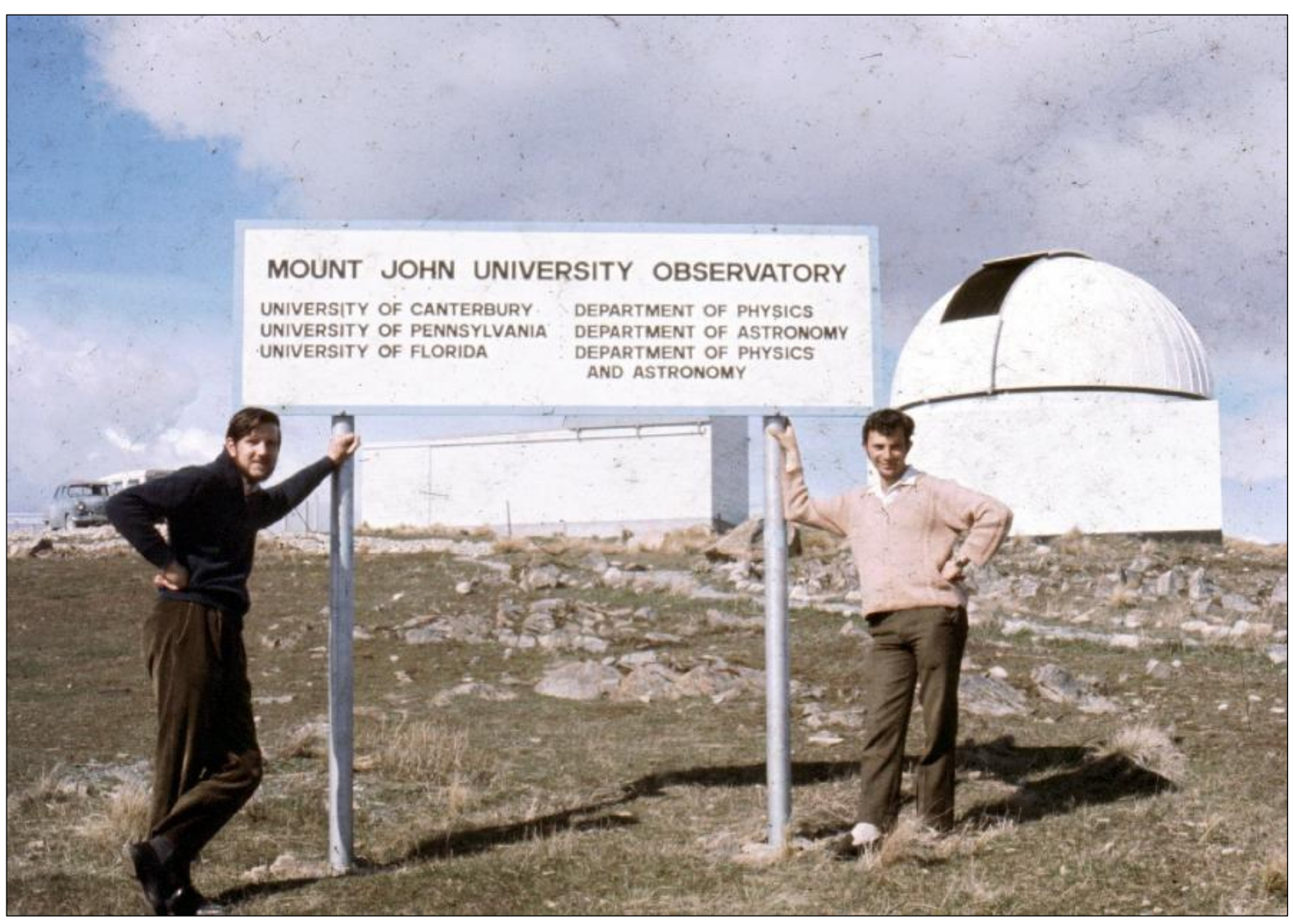

Figure 47: One-third of all NZ–comet discoverers as photographed in 1973, with Mike Clark on the right and the next comet discoverer to be-discussed here, Rod Austin, on the left (photograph: John Baker; with Rod Austin's camera).

## 7 RODNEY RICHARD DACRE AUSTIN

Rodney (Rod) Austin (b. 1945) was employed at Mount John Observatory at the same time as Mike Clark, but for a shorter duration. He also was a comet discoverer. Austin visually discovered three comets in the 1980s. He stands with John Grigg as the most prolific discoverer of comets from NZ. All of his discoveries were made from the Taranaki region and his name was the only name associated with them, that is, there were no co-discoverers. His first discovery was C/1982 M1 (Austin), his second C/1984 N1 (Austin), and his third, C/1989 X1 (Austin), as the X in X1 indicates, was found just weeks before the start of 1990.

Who is Rod Austin? Much of the following is based on an interview conducted by the first author with Austin on 28 August 2025 (but also see Drummond and Orchiston, 2026).

Rodney Richard Dacre Austin was born on 4 February 1945 in Christchurch, NZ; then the family moved to Oamaru in 1946. His father was a school teacher at Waitaki Boys' High School. Rod recalls that the first comet that he saw was early in 1948, when he was about three. It did not really trigger an interest in astronomy for Rod at such a young age. Rod's mother Una, and grandmother Selena Dacre, were interested in astronomy. His grandmother used to take her children outside at night to look at bright comets. Rod's mother's first comet memory was the brilliant Comet C/1917 F1 (Mellish) in 1917. It was visible to the naked eye in April and May. Observers stated it reached magnitude 1 (Kronk, 2007: 337–338).

The family moved to New Plymouth in 1952. Rod watched a partial solar eclipse in 1956 (where, according to GUIDE, 70% of the Sun was covered from New Plymouth). This rare phenomenon activated his interest in astronomy. Solar eclipses were and still are one of his main interests in astronomy. Austin jokes that comets "… were his side-line …", something to observe between solar eclipses. He recalls that he saw several brightish comets in the early 1960s.

Austin's first job out of school was as a trainee technician with broadcasting in Wanganui. After eight months he decided to join the Royal New Zealand Air Force as a radar mechanic. He then switched over to photography with the Air Force, based at Wigram in Christchurch. This career lasted eight years until the end of 1970. He then went to the University of Canterbury's Mount John Observatory in Tekapo for employment. While there, his primary role was working on the photometry of eclipsing binary stars. Austin worked at Mount John for 8.5 years and then returned to New Plymouth in the late 1970s. In New Plymouth he did computer programming for income, then got a job at the *Taranaki Daily News* as a photolithographer for four years. He worked briefly at the U.S. Naval Observatory's Southern Station at Black Birch Observatory (South Island) writing computer programs to control the atomic clocks (Figure 48) and then returned to the *Taranaki Daily News* to continue as a photolithographer until 2006, when he was made redundant. Austin next delivered mail for the Post Office and then worked for the Hertz Rental Car Company cleaning cars. Of interest is that his own cars (he's had six) have all had the personalised number plate 'FMERIS' (as in 'ephemeris')—which his father suggested—to Rod after his third comet discovery.

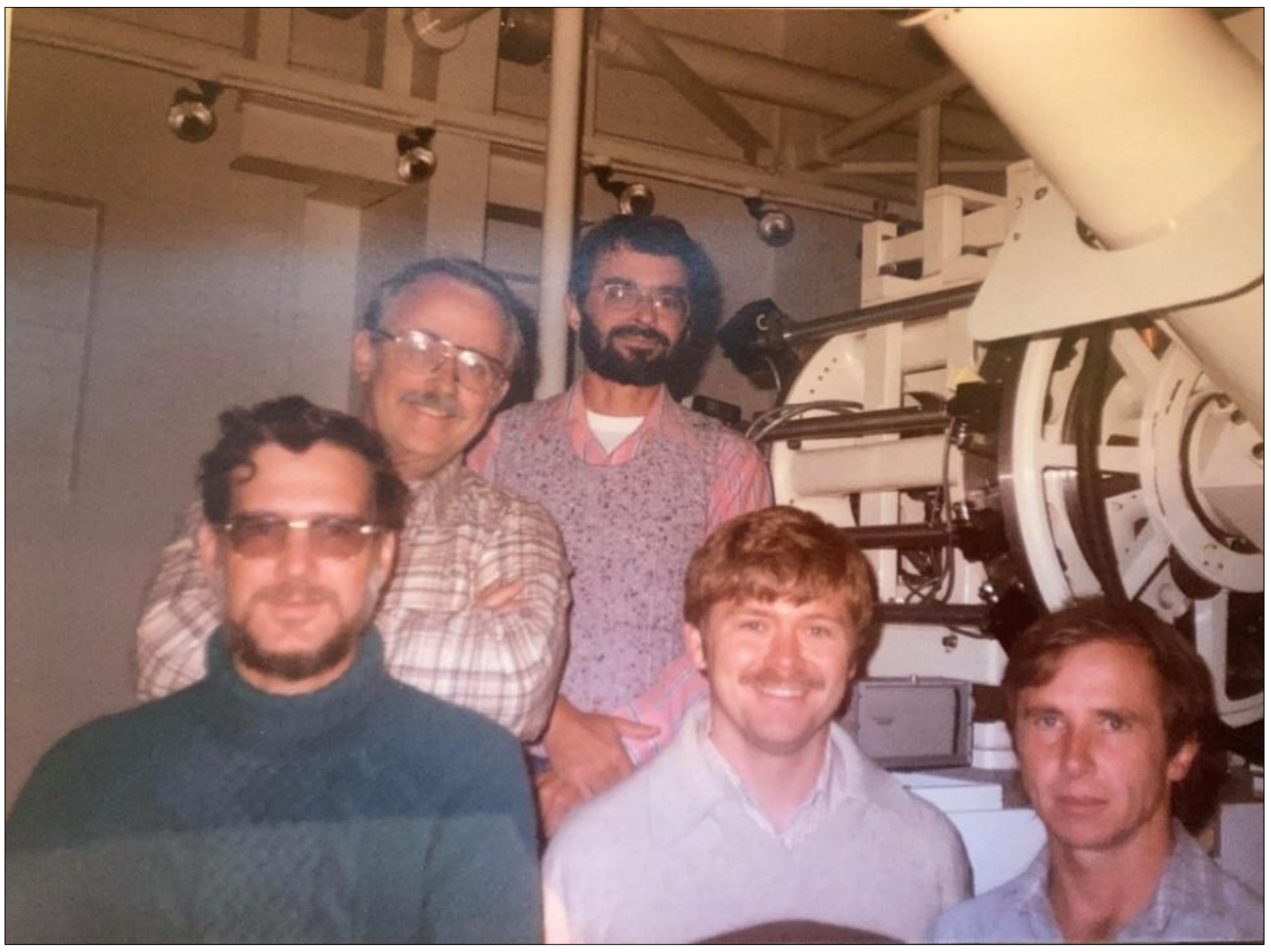

Figure 48: Rod Austin (front row, left) with U.S.N.O. colleagues and the 7-inch transit circle at Black Birch. Others of interest in this historic photograph are ex-Carter Observatory's Russell Millington (front row, right), who is mentioned elsewhere in this paper, while between Rod and Russell is Dr. Steven J. Dick, a close friend and research collaborator of the second author of this paper; Steve is also one of the Associate Editors of the *Journal of Astronomical History and Heritage* (photograph: Steve Dick).

Living back in New Plymouth and being a member of the New Plymouth Astronomical Society, Austin realised that they had limited observing equipment to use, so he took up comet hunting with his own telescope. Bortle (1982b: 504) points out that Austin began casual comet hunting when he accidently discovered Comet 1968e (C/1968 Q2/Honda) three months after Minoru Honda (1913–1990) found it from Japan (see Kronk 2010: 215–218). Austin initially used 7 × 50 binoculars and a 60-mm spotting scope for searches, accumulating 74 hours without success. Then in 1976 he built a 6-inch (15-cm) f/8 refracting telescope with a Jaegers lens (Figure 49). Austin said it was "… awkward to use …", even though he discovered his first comet, C/1982 M1 (Austin), with it. He then rebuilt the telescope into a 'cross-four configuration' with two mirrors so that the telescope rotated in altitude around the eyepiece (Figure 50). Austin found his second comet with this altered telescope, but it was a "… massive great thing." Kronk (2010: 98) says it weighed 235 pounds (107 kg) and caused Austin a back injury one night when his foot slipped while dismantling it after an observing run. Austin recalls that he was out of action for quite a while. The realisation that his refurbished telescope was too unwieldy prompted Austin to get a lighter telescope, but with more aperture: an 8-inch (20-cm) f/4 Meade Schmidt–Newtonian. A friend built a specialised mount for the telescope whereby the eyepiece was always at eye-level when Austin was standing (Figure 51). He found his third comet (C/1989 X1) with this telescope.

Working for the *Taranaki Daily News* as a photo lithographer in the 1980s, Austin used to finish work at 2 am each work day and then drive out into the hills surrounding New Plymouth to hunt for comets. Prior to this job, he had searched unsuccessfully for 125 hours up to July 1981 (Bortle, 1982b: 504). He would search on any clear night when there was not much Moon interference. Due to work constraints, Austin did the vast majority of searching in the morning sky before dawn. He would sweep along the Southern and Eastern horizons, hoping to find an unknown rising comet. With his first comet hunting telescope, the 6-inch f/8 refractor, due to the telescope mount being a challenge to move horizontally, he would search in a vertical up-down pattern. He could not see anything above approximately 40° in altitude due to the eyepiece placement. However, with the later 6-inch (15-cm) configuration and then the 8-inch (20-cm) telescope, Austin utilised the more traditional method of sweeping parallel to the horizon, then moving down half a field and sweeping back the opposite way and so on. There was less time to search for comets in summer due to the early onset of morning twilight at approximately 4am. By the time of his third comet discovery in 1989, Austin had done "… 183 actual hours of hunting over ten years …" (Austin, pers. comm., 2025), about 18 hours a year. New Plymouth can be fairly cloudy due to the prevailing Westerly winds in NZ being onshore there and producing orographic cloud. When it was clear, the bright Moon could also thwart hunting. Averaging 183 hours between three comets results in 61 hours of hunting per comet. The following Sections discuss Austin's three comets.

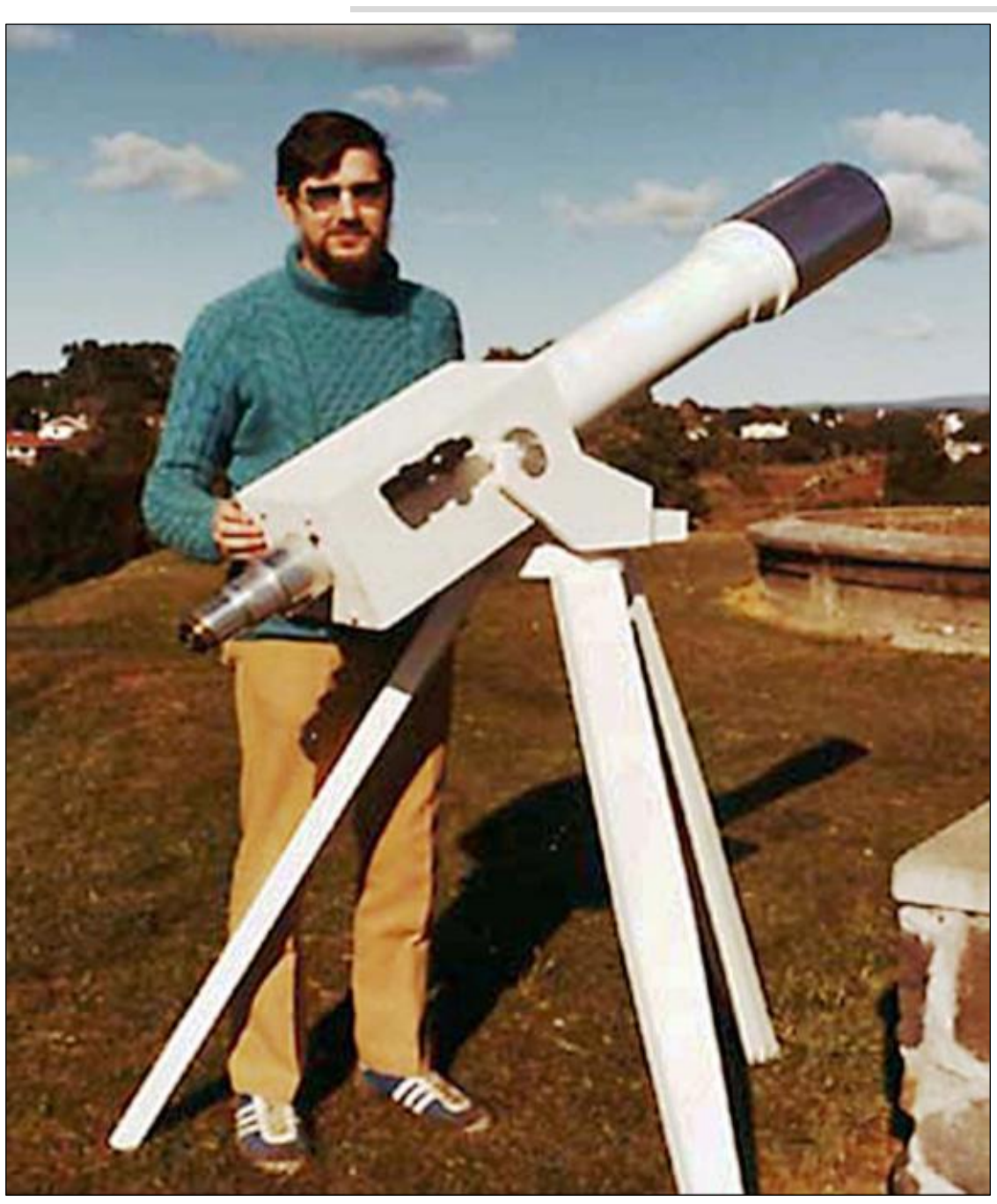

Figure 49: Austin standing beside the telescope, a 6-inch (15-cm) f/8 refractor, with which he discovered his first comet (photograph: Rod Austin, 1980s).

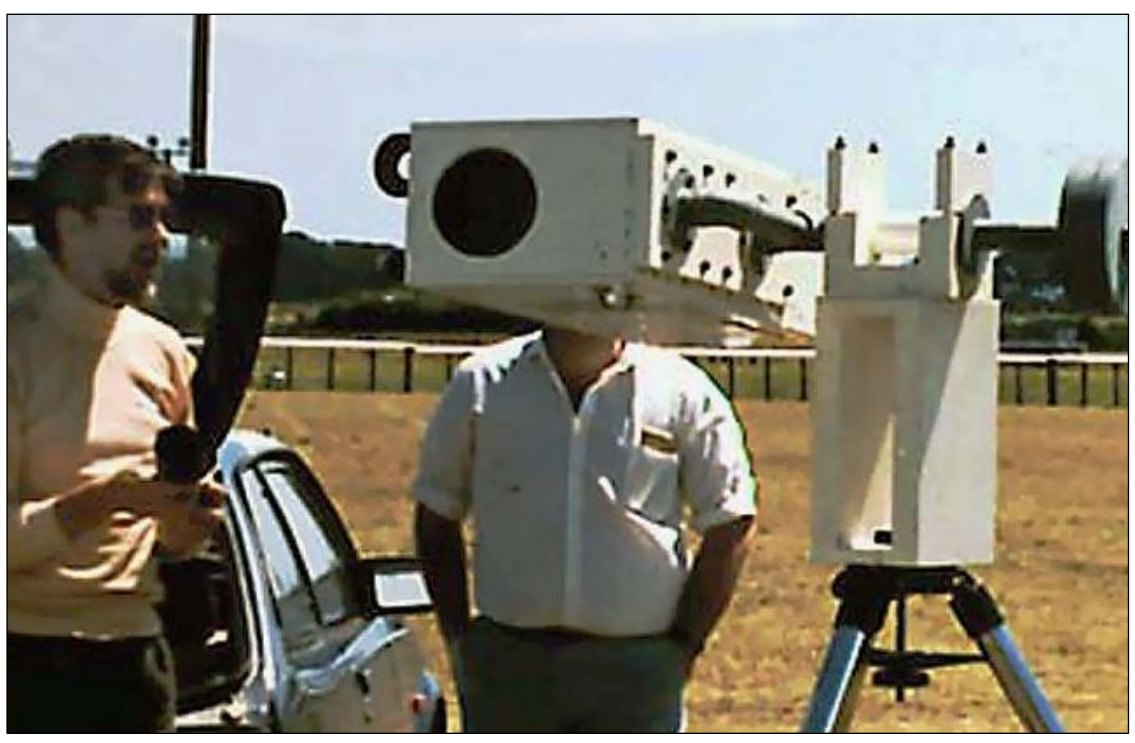

Figure 50: Austin at a star party in 1988, standing beside the modified 6-inch (15-cm) f/8 'cross-four configuration' refractor with which he discovered his second comet (photograph: Ian Cooper, 1988).

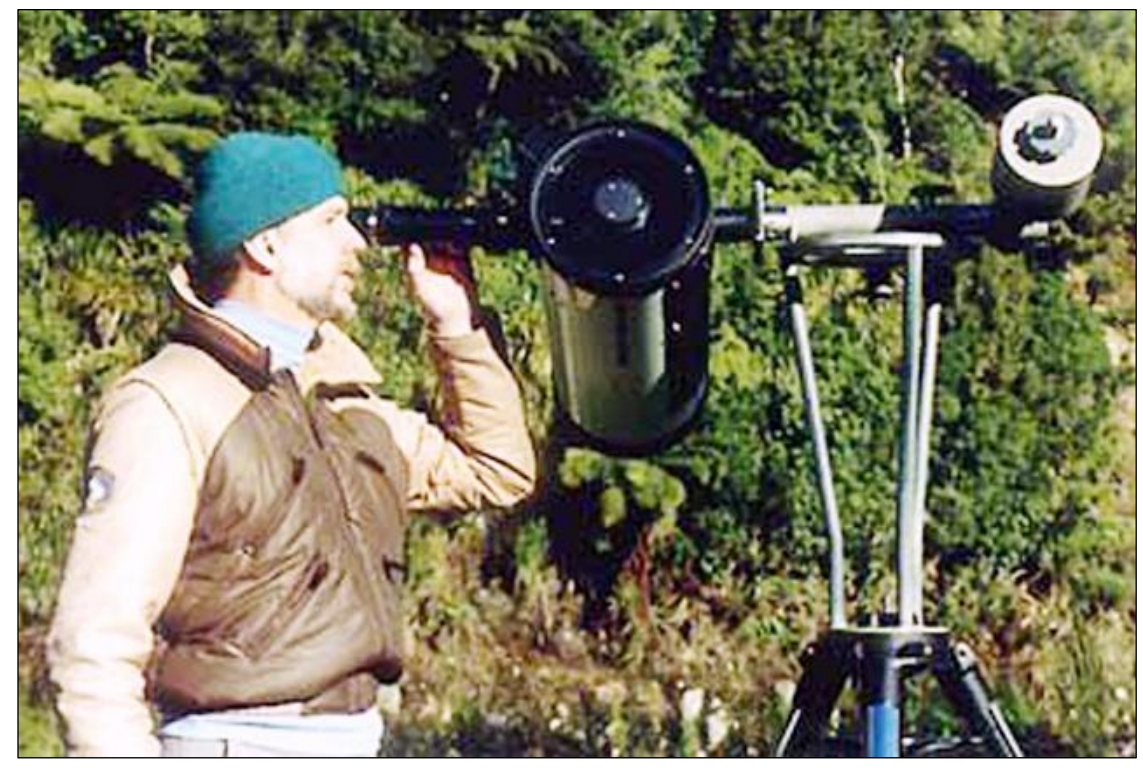

Figure 51: Austin standing beside the 8-inch (20-cm) f/4 Meade Schmidt–Newtonian telescope with which he discovered his third comet. Note how the eyepiece always remains at Austin's eye height, no matter where it is pointed in the sky (photograph: Rod Austin, late 1980s).

Table 15: An overview of Austin's first comet, C/1982 M1 (Austin). Epoch: 19.0 August 1982. Based on the Minor Planet Center (MPC). If not listed in MPC, information was used from NASA/Jet Propulsion Laboratory Horizons System, or Kronk and Meyer (2010).

| | |
|---|---|
| Discovery Date | 19 June 1982, 4am NZST (18.67 June 1982 UT) |
| Discovery Magnitude | 10 |
| Discovery Declination | −40° (Horologium) |
| Perihelion date | 24.73 August 1982 UT (MPC) |
| Perihelion distance (q) | 0.648 au (MPC) |
| Perigee date | 10 August 1982 UT (Kronk and Meyer, 2010: 758) |
| Perigee distance | 0.3245 au (Kronk and Meyer, 2010: 758) |
| Brightest | Magnitude 4 on 17 August 1982 UT (Kronk and Meyer, 2010: 760) |
| Visible from NZ | July to mid-August 1982 (GUIDE) |
| Last observed | 3 April 1983 (UT), magnitude 20 (photographic) (Kronk and Meyer, 2010: 761). |
| Eccentricity of the orbit (e) | 0.9994 (MPC) |
| Semi-Major axis (a) | 1,070 au (NASA/JPL) |
| Aphelion distance (Q) | 2,163.217 au (MPC) |
| Inclination (i) | 84.495° (MPC) |
| Epoch | 19.0 August 1982 UT (MPC) |
| Period | 35,588 years (MPC) |
| Observations in Kronk (2010) | ~114 (visual and photographic). 26 (~23%) from NZ (including discovery) |
| Observations in MPC | 312 (256 used to determine orbital elements) |
| Observations in COBS | 1,183 |
| Papers Past newspaper articles | 2 (June 1982–October 1982). |

## 7.1 C/1982 M1 (Austin) - 1982 VI = 1982g

Initial details of this comet are listed below in Table 15, its orbit is illustrated in Figure 52 and its path through the sky is plotted in Figure 53.

Thirty-seven year old Rodney Austin discovered his first comet on 18 June 1982 (UT) in the local morning hours with his 6-inch telescope. Up to this point he had searched for 151 hours spread over 15 years (Success in a Comet Search, 1982: 25). At around 6am he spotted the unknown comet. Astronomical twilight started just minutes after 6 am (Time and Date), however, the comet had risen many hours before, at 1:45 am NZST, and was in the far-Southern constellation of Horologium (GUIDE). Austin (pers. comm., 25 August 2025) recalls:

> Comet 1982g (C/1982 M1) was discovered with a 6" f/8 refractor on an alt-az mounting, from a site up Carrington Road about 15 km south of New Plymouth [on the] morning of 19 June 1982 [18 June UT]. An unstable night with the wind gusting and increasing. Got to the point of threatening to blow the telescope over, so I decided 'just one more scan' up from the horizon. At an altitude of about 40 degrees and close to the telescope's limit, something faint and fuzzy passed through the field …

Austin had to rush home and find his star atlases that were still in boxes after his return from Mount John. At the telescope he noted the approximate position and made a field drawing of the object's location. Careful analysis showed that this was indeed an unknown comet. Austin says that this was the greatest moment in his comet-hunting career, as "… nothing beats the first discovery. You can only discover a first comet once." (*ibid.*).

The suspect comet was at RA 04h 04.5m, Dec −40° 05'. Austin described it as "... diffuse, with a condensation." (Kronk and Meyer, 2010: 758). Austin confirmed motion of the comet the following morning (19.63 June UT). The discovery was reported to Mount John Observatory at Tekapo by phone (Success in a Comet Search, 1982: 25), Austin's place of former employment. Alan Gilmore of Mount John Observatory observed the comet on 19.74 June UT (Bortle, 1982a: 198; Kronk and Meyer, 2010: 758). He also estimated the magnitude to be 10. Gilmore's wife, Pamela Kilmartin, then notified the International Astronomical Union Central Bureau for Astronomical Telegrams of Austin's dis-covery. Brian Marsden (1982b) sent out IAUC Circular No. 3705 on 21 June 1982 (UT) informing the international comet community of the discovery and observational astrometry three days and six hours after the initial sighting (Bortle, 1982b: 504).

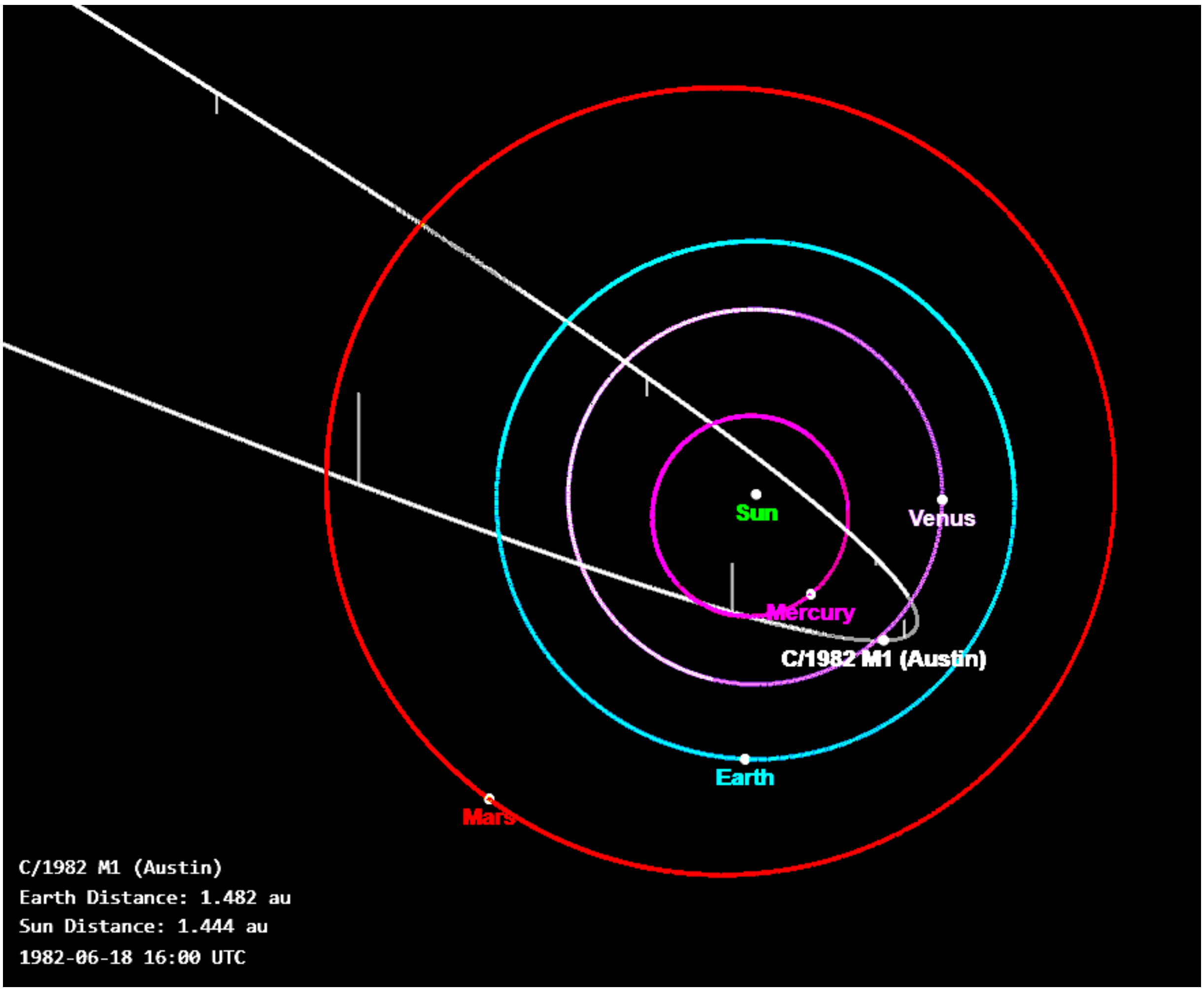

Figure 52: The orbital path of Comet C/1982 M1 (Austin). The comet's orbit is prograde (anti-clockwise from above). At discovery it was in the Southern sky (source: NASA/JPL Horizons).

As Figure 53 reveals, the comet was rapidly moving in a Northeastward direction (Bortle, 1982a: 198) and was too far North for NZ observations from mid-August 1982 (Volcanic theory of Venus supported, 1982: 14). When it was visible from NZ, the most prolific observer was Albert Jones of Nelson. Jones observed C/1982 M1 (Austin) 19 times between 1-27 July 1982 (Kronk: 2020: 759). Bortle (1984b: 504) states that D. Goodman (probably Dennis Goodman of New Zealand and not Australia, as stated) also sent in "... particularly notable … reports …" of the comet. Of note is that Papers Past only mentioned Comet Austin three times between June and October 1982. No newspapers reported Jones' or anyone else's observations. One article stated that Comet Austin could be the "... brightest [comet] in six years." (Comet 'the brightest', 1982: 29).

Comet C/1982 M1 (Austin) steadily brightened during June and July and peaked at magnitude 3.1 on 14 August 1982 (Kronk and Meyer, 2010: 760), three weeks before the photograph in Figure 54 was taken. At that stage GUIDE indicates that it had a Western elongation of 15° from the Sun. The Declination was +24°. From NZ, on 14 August, it rose at approximately 8 am and set at 4 pm, so it was impossible to view. It reached perigee on 10 August 1982 UT (Kronk and Meyer, 2010: 758) and perihelion on 24 August 1982 UT (Kronk and Meyer, 2010: 762). At its best it was a northern hemisphere target. Bortle (1982b: 504) wrote that the comet "... became the most widely observed comet in the past year and a half."

Kronk and Meyer (2010: 760) state that August and September were also "… big month[s] for observations of this comet." No Southern Hemisphere observations were mentioned after 9 August 1982, for it was a Northern Hemisphere

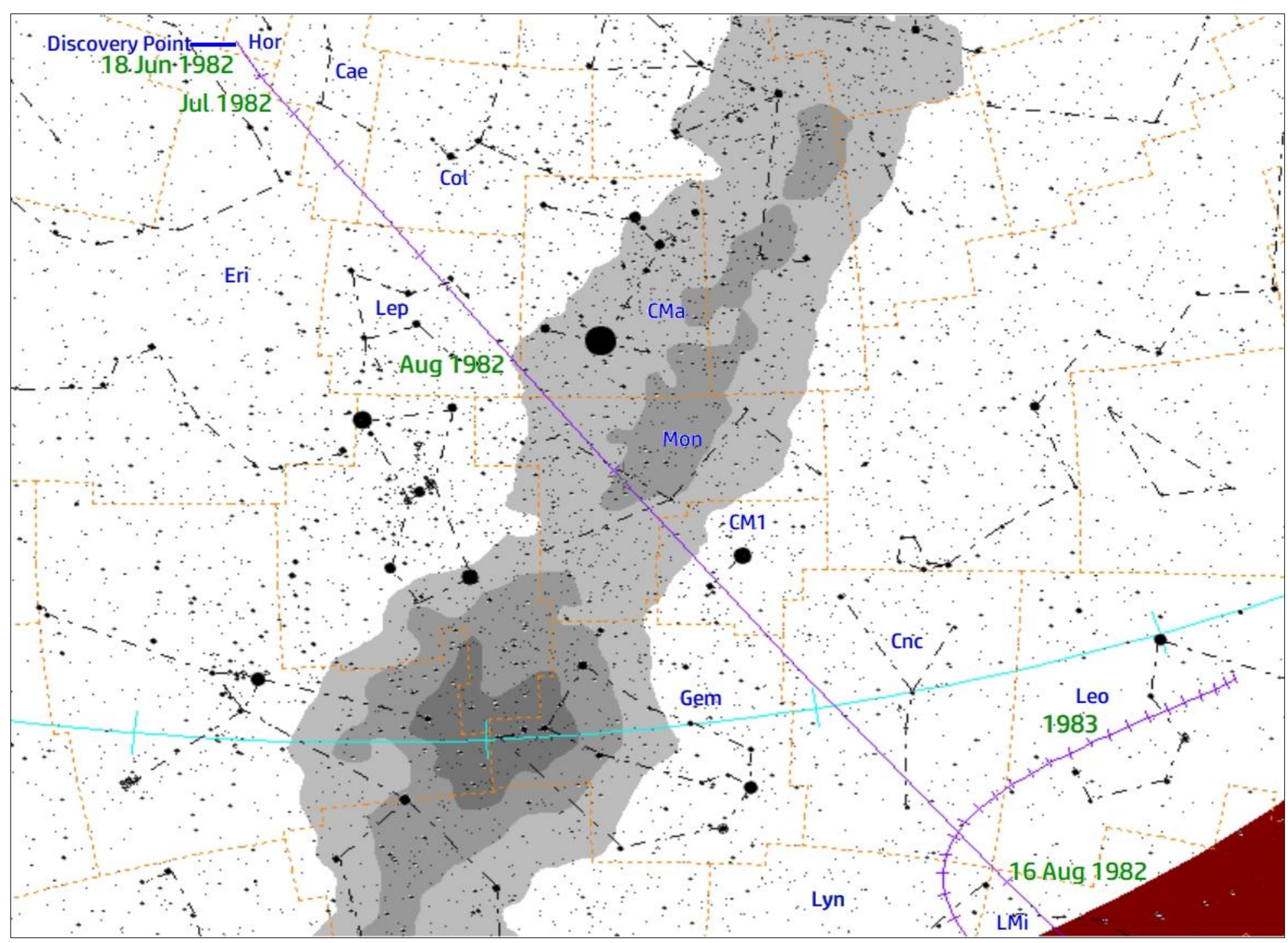

Figure 53: The path of Comet C/1982 M1 (Austin) in purple. The constellations that the comet passed through are in blue font. The light blue line is the ecliptic. North is down, East is right - this being the view from NZ so that the Northern horizon is at the bottom (source: GUIDE).

Hemisphere target by then. British amateur astronomer Dr. Jonathan Shanklin (Cambridge, England) reported that the coma reached an apparent angular size of 14′ on 25 August. That is nearly half the apparent size of the Full Moon. The small-angle formula of

$$D = ad/206265 \quad (1)$$

can be used to calculate the size of the coma, where D is the linear size (in km), a is the angular size (in arc-seconds) and d is the distance (in km). Using Shanklin's coma size of 840″ and GUIDE's Earth–comet distance of 96 million km, we found the coma diameter to be 391,000 km.

Kronk and Meyer (2010: 760) state that the tail of Comet C/1982 M1 (Austin) was observed to gradually grow to an apparent length of 5° visually in late August by Charles Morris (Massachusetts, USA) and 12° photographically on 22 August (by Harold B. Ridley, West Chinnock, England). The comet faded in the following months and was last detected photographically on 3 April 1983 (UT) by J.B. Gibson with the 1.2-metre Schmidt Telescope at Mount Palomar, USA. He estimated it to be magnitude 20. Based on the large number of observations (Kronk and

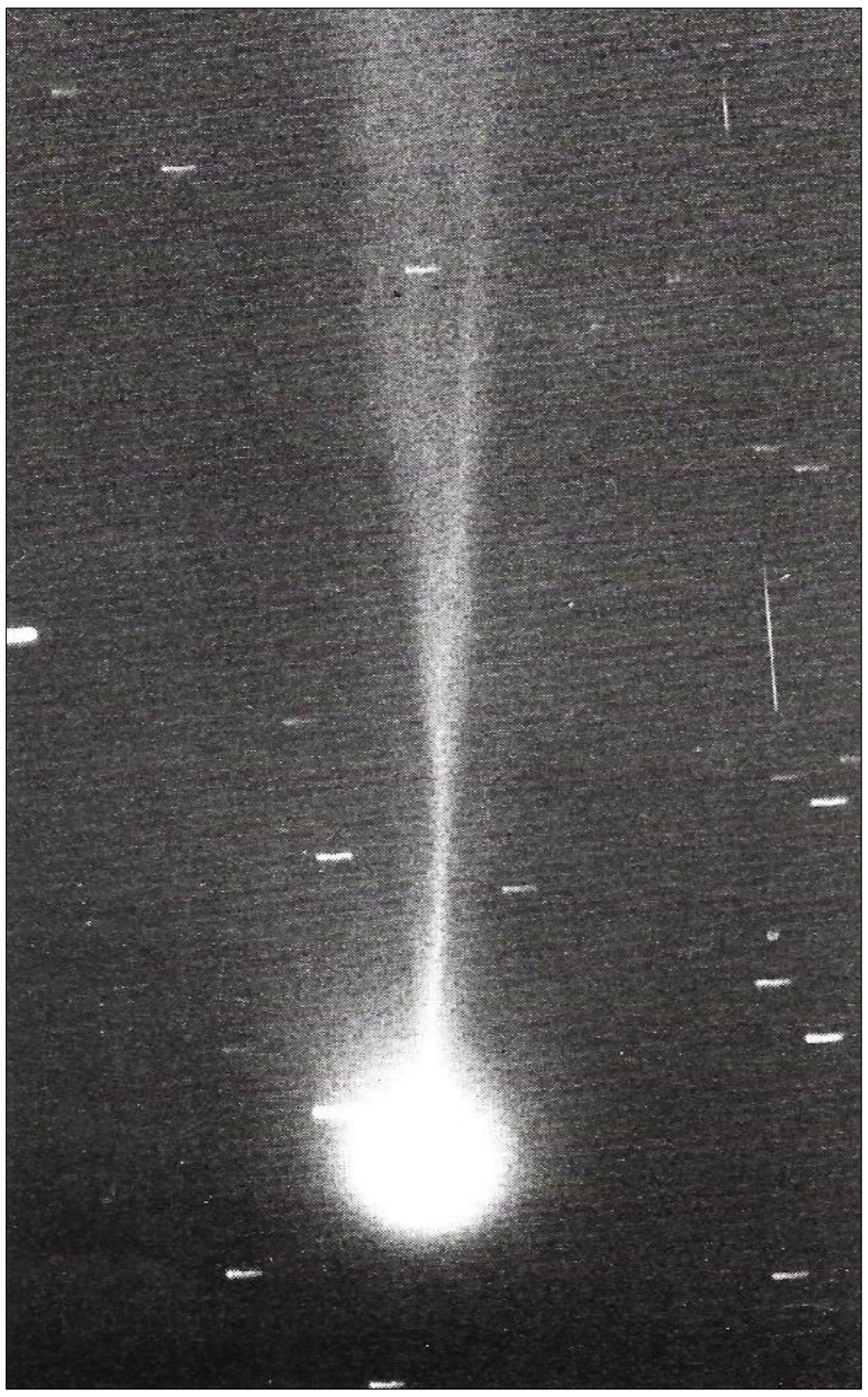

Figure 54: Comet Austin (C/1982 M1 Austin) on 6.13 September 1982 UT., photographed by Edgar Everhart and Tom Dadisman with a 16-inch (41-cm) f/5.5 reflector from Colorado, USA (after Bortle, 1982b: 506).

Meyer record 114 of them), Brian Marsden determined a period of 36 thousand years (Kronk and Meyer, 2010: 762).

Papers Past searches for the word 'comet' between 18 June and 31 October 1982 produced 125 results; however, only 3 of these related to Rodney Austin or his first comet (Table 16). Sadly, Austin's second comet discovery resulted in even fewer newspaper articles.

Table 16: Newspaper articles about 'Comet' and 'Comet C/1982 M1 (Austin)' in Papers Past. Note that June was the discovery month.

| Month and Year | Total for 'Comet' | Total for 'Comet Austin' |
|---|---|---|
| June 1982 | 18 | 1 |
| July 1982 | 23 | 1 |
| August 1982 | 23 | 1 |
| September 1982 | 25 | 0 |
| October 1982 | 36 | 0 |
| Totals: | 125 | 3 |

## 7.2 C/1984 N1 (Austin) - 1984 XIII = 1984i

Initial details of this comet are listed below in Table 17, its orbit is illustrated in Figure 55 and its path through the sky is plotted in Figure 56.

Austin found his second comet on 9 July 1984 NZST (8.73 July 1984 UT) using his modified 6-inch refractor for the third time from Carrington Road, 16 km South of New Plymouth. Austin (pers. comm, 2006) bemusedly recalls,

> 1984i (C/1984 N1) was found on the morning of July 9th 1984. The morning of the 9th was very humid and cold but clear. I started by observing Periodic Comet Clark (the one Mike Clark found from Mt John, on a night I was comet-hunting only about 100 metres away). Because of my observations of his comet on July 9th, I missed discovering what became Periodic Comet Takamizawa [discovered three days later], but two hours later nearly dropped with shock when my number 2 [comet] entered the field. Mag 5.8 almost filling the field and moving over its own diameter in just 20 minutes …

Comet C/1984 N1 (Austin) was discovered in the Southern constellation of Eridanus at a Declination of –39°. It was rising at 1:00 am local time (GUIDE). Austin then contacted Mount John Observatory and the comet's existence was confirmed from two photographs of the area taken on the same morning by Mike Clark (Bortle, 1984: 284; Marsden, 1984a). Refer to Clark's photographic patrol above. Alan Gilmore from Mount John Observatory then notified the IAU Central Bureau for Astronomical Telegrams in Massachusetts by telex. Brian Marsden sent out a discovery notice (IAU CBAT Circular 3957; Marsden, 1984a) on 9 July 1984 (US time), "... only hours after Austin's sighting …" (Bortle, 1984: 284). The magnitude given was 8 and the coma described as diffuse but with slight condensation and without a tail. However, a follow-up circular, IAU CBAT Circular 3958 on 11 July 1984 (Marsden, 1984b) had a revised magnitude estimate made by Austin of 6.5 with a coma diameter of 12′. In the same circular, Marsden (*ibid.*), using six astrometric observa-tions done by P. Birch (Perth) and Gilmore and Kilmartin (Mount John Observatory) stated that perihelion would occur at 12.13 August 1984 Eastern Time. It also contained ephemerides for 9–25 July 1984. The magnitude was estimated to rise from 6.8 (9 July) to 6.6 (11–14 July) and then drop to 6.8 again from 15 July. Bortle (1984: 284) produced ephemerides indicating that the comet would be magnitude 11.1 by 7 October 1984. However, as Kronk (2010: 101) points out below, observations revealed that C/1984 N1 would be brighter than first anticipated and did not fade to the 11th magnitude until late November 1984.

Table 17: An overview of Austin's second comet, C/1984 N1 (Austin). Epoch: 8.0 August 1984 UT. Based on the Minor Planet Center (MPC). If not listed in MPC, information was used from NASA/Jet Propulsion Laboratory Horizons System, or Kronk et al. (2017).

| | |
|---|---|
| Discovery Date | 9 July 1984 (8.73 July 1984 UT) |
| Discovery Magnitude | 5.8 (faint to naked eye) |
| Discovery Declination | –39° (Eridanus) |
| Perihelion date | 12.14 August 1984 UT (MPC) |
| Perihelion distance (q) | 0.291 au (MPC) |
| Perigee date | 10 July 1984 UT (Kronk et al, 2017: 98) |
| Perigee distance | 0.2554 au (Kronk et al, 2017: 98) |
| Brightest | Magnitude 4.8 on 7 August 19842 UT (Kronk et al, 2017: 99) |
| Visible from NZ | July–September 1984 (after which it was too far north) |
| Last observed | 23 December 1984 (UT), magnitude 19.8 (photographic) (Kronk et al, 2017: 101) |
| Eccentricity of the orbit (e) | 0.9998 (MPC) |
| Semi-Major axis (a) | 1,920 au (NASA/JPL) |
| Aphelion distance (Q) | 3,788.2 au (MPC) |
| Inclination (i) | 164.153° (MPC) |
| Epoch | 8.0 August 1984 (MPC) |
| Period | 82,443 years (MPC) |
| Observations in Kronk (2017) | ~95 (visual and photographic). 5 (5.3%) from NZ (including discovery) |
| Observations in MPC | 113 (85 used to determine orbital elements) |
| Observations in COBS | 457 |
| Papers Past newspaper articles | 1 (July 1984–October 1984) |

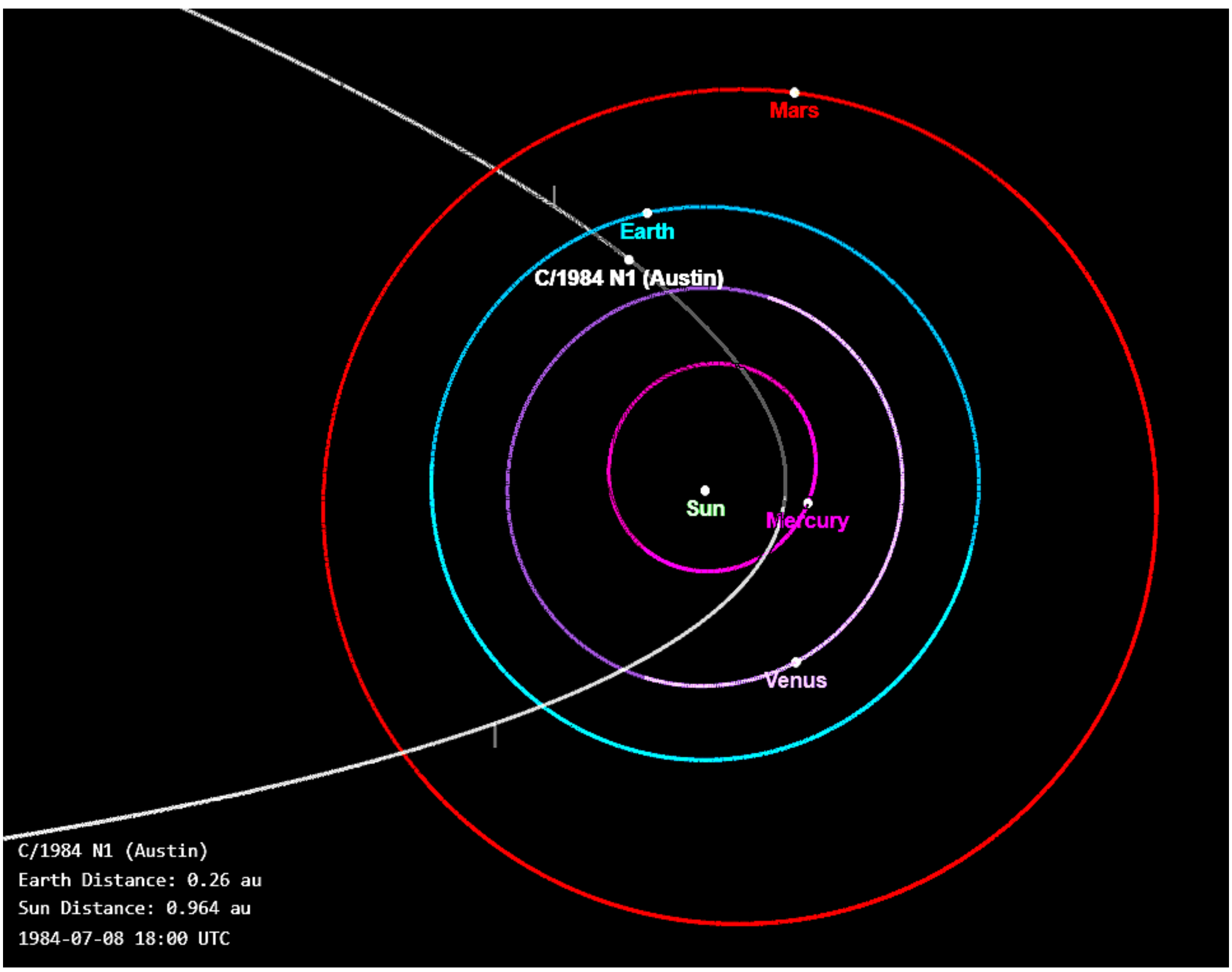

Figure 55: The orbital path of Comet C/1984 N1 (Austin). The comet came from the upper left and disappeared towards the lower left. At discovery it was in the Southern sky (source: NASA/JPL Horizons).

Unfortunately, as opposed to NZ comet discoveries from 1902 (Grigg) until 1946 (Jones) where numerous informative newspaper articles and observational reports by both professsional and amateur astronomers were published across NZ (on average, 96 per comet), no comet observations (apart from discovery notices) were published in New Zealand newspapers relating to 'NZ comets' after 1946 (e.g. see Table 18 in relation to Austin's latest discovery). In 2025 a request sent out by the lead author of this paper to NZ astronomical societies for observations, photographs or sketches of Austin's comets only resulted in one photograph of note, of C/1984 N1 (Austin) taken by Noel Munford of Palmerston North (see Figure 57). It was supplied by Ian Cooper, who wrote (pers. comm., 2025) that, "... it was taken on the Astrograph Mounting that Barrie Ward used to photograph both Ikeya-Seki and Bennett back in the 1960s." Barrie Ward was a prominent NZ astro-photographer back in the late 1960s and first half of the 1970s (Orchiston, 2016: 618–619; c.f. Ward, 2022).

Kronk et al. (2017: 98–101) reveals that prior to perihelion on 12 August 1984, all observations of C/1984 N1 (Austin) came from the Southern Hemisphere. After perihelion, when the comet became a Northern object, observations were made from the Northern Hemisphere.

These observations revealed that C/1984 N1 (Austin) increased in brightness from the discovery magnitude of ~6 in early July to magnitude 4.8 on 7 August 1984 (UT). At this stage it was sinking lower into the evening sky as it approached perihelion on 12 August (UT). In early August, Robert McNaught (Australia) stated that the comet had a tail of up to 3.1° in length (Kronk, 2017: 99). On 16 August the comet was closest to the Sun at 1.8°. It then became a morning target and was mostly observed by Northern Hemisphere observers. The magnitude was estimated to be 5–6 in late August and the tail appeared shorter at approximately 1.0–1.25°. It continued to fade in September from magnitude 6 to magnitude 8 (Kronk, 2017: 100). By the end of October John Bortle determined the magnitude to be 8.9. Around this time J. Bouma (Netherlands) observed an anti-tail of up to 24′ with 6-inch (15-cm) and 8-inch (25-cm) reflectors (*ibid.*). The anti-tail was first observed in early September.

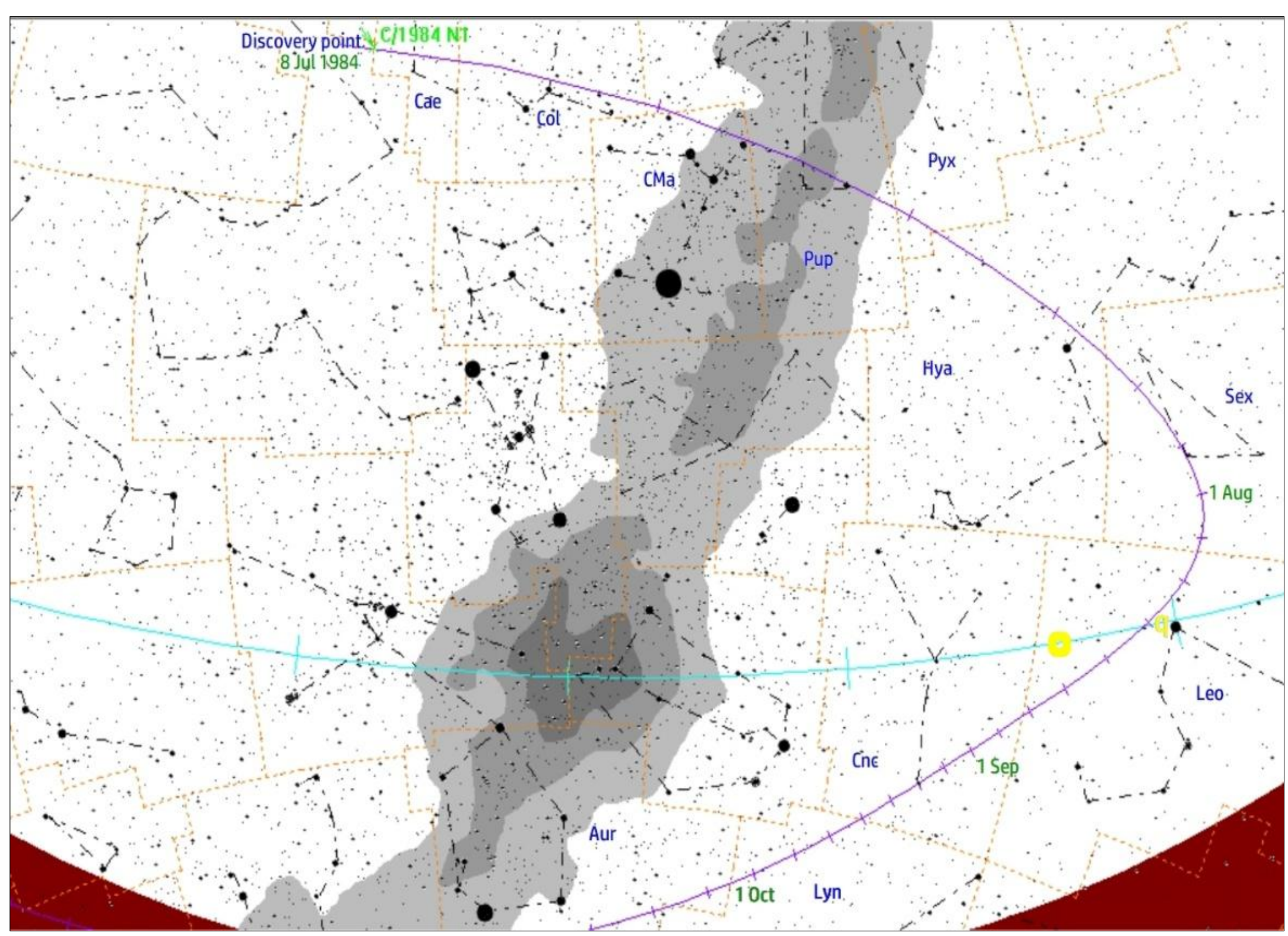

Figure 56: The path of Comet C/1984 N1 (Austin) in purple. The constellations that the comet passed through are in blue font and the dates in green. The light blue line is the ecliptic. During the dates in question, the Sun passed through Gemini (until 20 July), Cancer (21 July–10 August), Leo (11 August–16 September), perihelion being 12 August (note how the ecliptic and comet path lines intersect), and Virgo (17 September–31 October). The tick-points are set to 10-day intervals. The position of the Sun and comet are given on the date of perihelion (12 August 1984 UT). The Sun appears as a yellow circle and the comet as a yellow 'q' - just to the left (West) of α Leonis (Regulus). North is down, East is right - this being the view from NZ so that the Northern horizon is at the bottom (source: GUIDE).

Table 18: Newspaper articles about 'Comet' and 'Comet C/1984 N1 (Austin)' in Papers Past. Note that July was the discovery month.

| Month and Year | Total for 'Comet' | Total for 'Comet Austin' |
|---|---|---|
| July 1984 | 23 | 1 |
| August 1984 | 22 | 0 |
| September 1984 | 25 | 0 |
| October 1984 | 36 | 0 |
| Totals: | 106 | 1 |

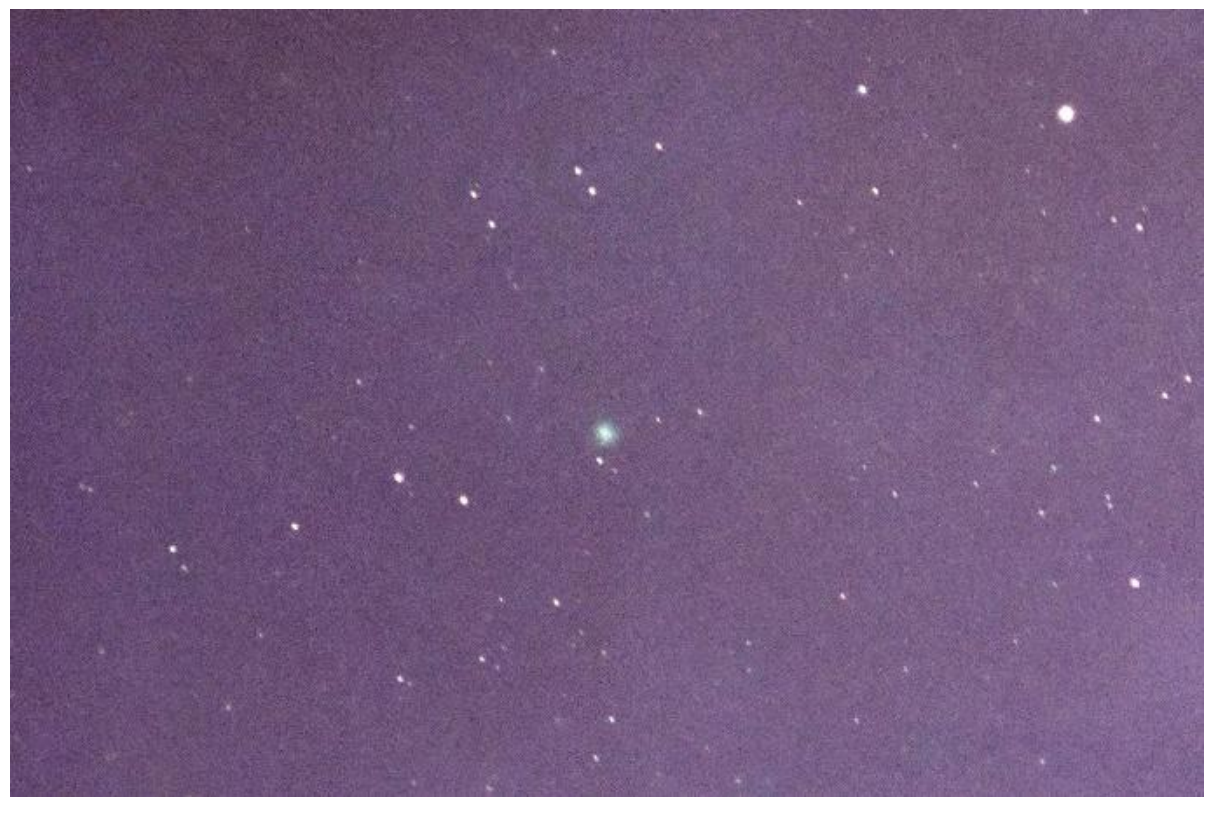

Figure 57 (right): Comet C/1984 N1 (Austin)—the nebulous object in the centre—photographed by Noel Munford (Palmerston North, NZ) on 20 July 1984 at around 8:10pm (NZST). RA 09h 28.5m, Dec –09° 45' (in Hydra) with a 400-mm lens at f/6.3. The field of view is approximately 3.5° × 2.5° (courtesy: Ian Cooper).

On 19 October it had a Northerly declination of +50°. Kronk et al (2017: 101) state that the last visual observation was on 28 November when the comet was magnitude 11.6. A 3.5′ tail was still evident.

E.P. Ney (O'Brien Observatory, Minnesota, USA) took infrared images of C/1984 N1 (Austin) and B.H. Foing, M. Festou, and S. Char (La Silla Observatory, Chile) took spectroscopic images. A significant scientific study of the comet was made via this spectral and infrared analysis (Kronk, 2017: 101). Foing et al. (*ibid.*) "... detected diatomic carbon, the amidyl radical, sodium, ionized water, and oxygen." Sekanina also determined the particle size in the anti-tail as several tens of microns (Green, 1984). Brian Marsden ascertained the orbital period to be 82 thousand years (Kronk, 2017: 102).

Austin's next comet would attract a great deal of international media attention.

### 7.3 C/1989 X1 (Austin) - 1990 V = 1989c1

Initial details of this comet are listed below in Table 19; its orbit is illustrated in Figure 58 and its path through the sky is plotted in Figure 59.

Austin's third and final comet discovery was made on 6 December 1989 (UT). Because Carrington Road (17 km south of New Plymouth), his usual observing site, was starting to be developed residentially, Austin drove to the back of Inglewood to conduct his comet search under that location's dark sky with his 8-inch Schmidt–Newtonian telescope. He used a 50 mm diameter 2 times Barlow and 2-inch diameter eyepiece. This provided a field of view of 0.75°, with a limiting magnitude of 13. He ingeniously designed a mount whereby the telescope's eyepiece remained at the same height no matter what azimuth or altitude it was pointing at (Figure 51).

Austin (pers. comm., 2006) recalls:

> 1989c1 (1989X1) was discovered on the morning of December 7th 1989 [NZDT]. I started by observing [Comet] Okazaki-Levy Rudenko, and with about an hour to twilight I decided to use the time hunting. A quick decision was made as to the area. I remember thinking 'over there will do'. Half an hour later it certainly did! The confirmation process was fairly swift but full of hilarious errors. All hell broke loose after that one …

Being the late 1980s, cell phones were expensive and large. Not owning one, Austin drove home to New Plymouth to check where known periodic comets were and to scrutinize star charts, after drawing a location chart at the telescope. He stated that there was no comet "… within a bull's roar …" (*ibid.*) of where his suspect object lay. Austin phoned Peter Birch at the Perth Observatory to ask him to confirm if it was an unknown comet, dawn having arrived at Mount John Observatory. Austin knew Birch as a result of them both attending the conference 'Asteroids and Planets X' in Tucson, Arizona, in 1979. Austin recalls that he initially told Birch a position that was around 0.5° East of its actual position. Realising his error, he rephoned Birch to tell him to search 0.5° West of the initial pos-position. Birch did and located the comet. Birch then sent a discovery notice and astrometric positions to the Minor Planet Center. These were published on 6 December 1989 by Daniel Green (1989a) in IAU Circular No. 4919.

Table 19: An overview of Austin's third comet, C/1989 X1 (Austin). Epoch: 19.0 April 1990. Based on the Minor Planet Center (MPC). If not listed in MPC, information was used from NASA/Jet Propulsion Laboratory Horizons System, or Kronk et al. (2017).

| | |
|---|---|
| Discovery Date | 7 December 1989 NZST (6.6 December 1989 UT) |
| Discovery Magnitude | 11 |
| Discovery Declination | –62° (Tucana) |
| Perihelion date | 9.97 April 1990 UT (MPC) |
| Perihelion distance (q) | 0.350 au (MPC) |
| Perigee date | 25 May 1990 UT (Kronk et al, 2017: 466) |
| Perigee distance | 0.2368 au (Kronk et al, 2017: 466) |
| Brightest | Magnitude 4.3 (25 April 1990) |
| Visible from NZ | 7 Dec 1989 (discovery) until April 1990 (too north afterwards) (GUIDE) |
| Last observed | 27.6 July 1990 UT (MPC) |
| Eccentricity of the orbit (e) | 1.0002 (MPC) |
| Semi-Major axis (a) | Outbound orbit is unbound; comet will never return |
| Aphelion distance (Q) | N/A |
| Inclination (i) | 58.956° (MPC) |
| Epoch | 19.0 April 1990 UT (MPC) |
| Period | Comet will never return |
| Observations in Kronk et al (2017) | ~1,600 (visual and photographic) – 2 from NZ (including discovery) |
| Observations in MPC | 279 (144 used to determine orbital elements: 6 Dec 1989–17 Jun 1990) |
| Observations in COBS | 2,748 |
| Papers Past newspaper articles | 1 (December 1989–April 1990) |

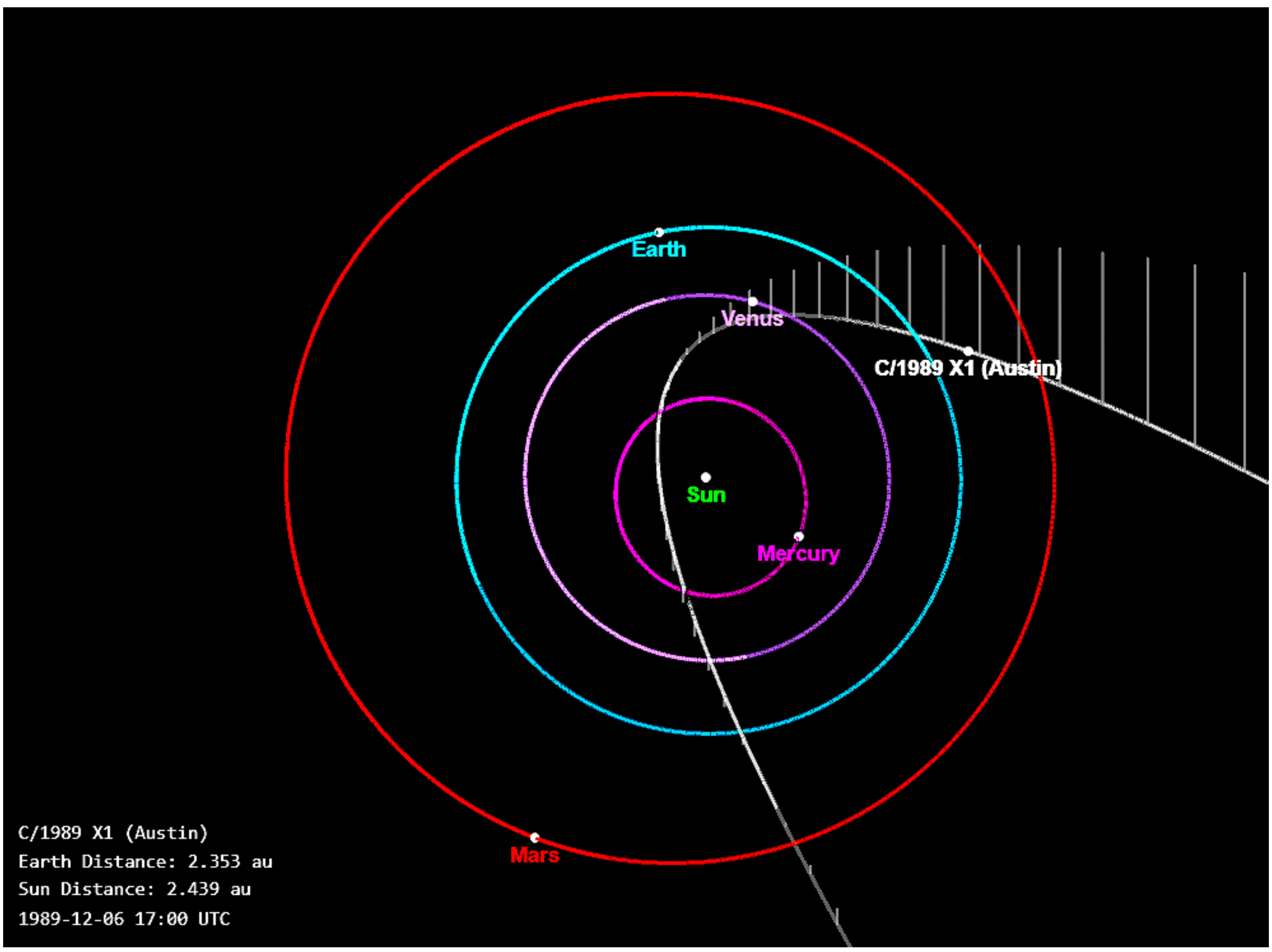

Figure 58: The orbital path of Comet C/1989 X1 (Austin). The comet approached from the upper right and disappeared to the bottom in this chart (source: NASA/JPL Horizons).

After confirmation by Birch, it was designated Comet Austin (1989c1). Austin (pers. comm., 2006) recalls that post confirmation, "All hell broke loose!" On 11 April, Green (1989b) wrote in IAUC Circular 4921 that "The … preliminary parabolic orbital elements from 10 observations, Dec. 6-11, suggest that this may become a moderately bright object in 1990 March and April." By 20 December, circulars were predicting a bright magnitude of 2.7 in April 1990 (Green, 1989c). This was also reported in *Sky and Telescope* (Feb, Mar, Apr, May 1990) magazine. Unfortunately, as the comet got closer, the optimistic predictions diminished. Austin humorously recalls that the front cover of the February 1990 issue of *Sky & Telescope* said 'Monster Comet Coming!', in March the front cover stated 'See Comet Austin!', in April it said 'Comet Austin Observing Guide', finally in May 1990 it stated 'How to View Comet Austin' (Figure 60).

Numerous publications predicted a bright comet. David Levy (Austin's friend) stated in the *Journal of the Royal Astronomical Society of Canada* Newsletter (Levy, 1990: 26) that, "In early February the comet is brightening steadily and looks very promising as it prepares to greet the earth [sic] in April." Levy highlighted that observations revealed the comet may be making its first journey to the inner Solar System, hence the 'bright' magnitude at such a large distance from the Sun. He warned that it may not live up to the hype, similar to Comet Kohoutek in 1974, stating

> ... its [Kohoutek] early promising indications of brightness were not fulfilled as the burn off of materials stopped as the comet approached near the Sun. (*ibid.*).

Levy then pointed out that two years after Comet Kohoutek, Comet West "... produced a fabulous morning show." He concluded that it would be fun to watch and see if Comet Austin would put on a performance like Comet Kohoutek or Comet West. Levy also pointed out what a productive year for comets 1989 was, having 34 discovered or recovered.

Comet Austin also featured several times on television, the most popular outlet no doubt being Patrick Moore's *The Sky at Night* (BBC1) where the 19-minute episode was titled 'Visitor from Space' (referring to Comet Austin). Moore (1923–2012) stated that it could be "... a really good one …", "... the best for many years …" and even "... the comet of the century …" but warned that comets are unpredictable. A photograph taken with the European Southern Observatory Schmidt telescope at La Silla in the Atacama Desert (Chile) on 25 February 1990 (UT) was displayed, revealing a "... long tail …" (Moore) (Figure 61). Viewers were told by Harold Ridley (BAA Comet Section), Moore's studio guest, to look West after sunset in the first ten days of April 1990, then from mid-April Comet Austin would be a nice morning sky target for a few weeks. Ridley predicted that it may reach magnitude 1 with a 10–20° tail (refer to the magnitude estimate graph in Figure 61). The orbit was described as slightly hyperbolic (open curved rather than close curved) and that it may leave the Solar System altogether.

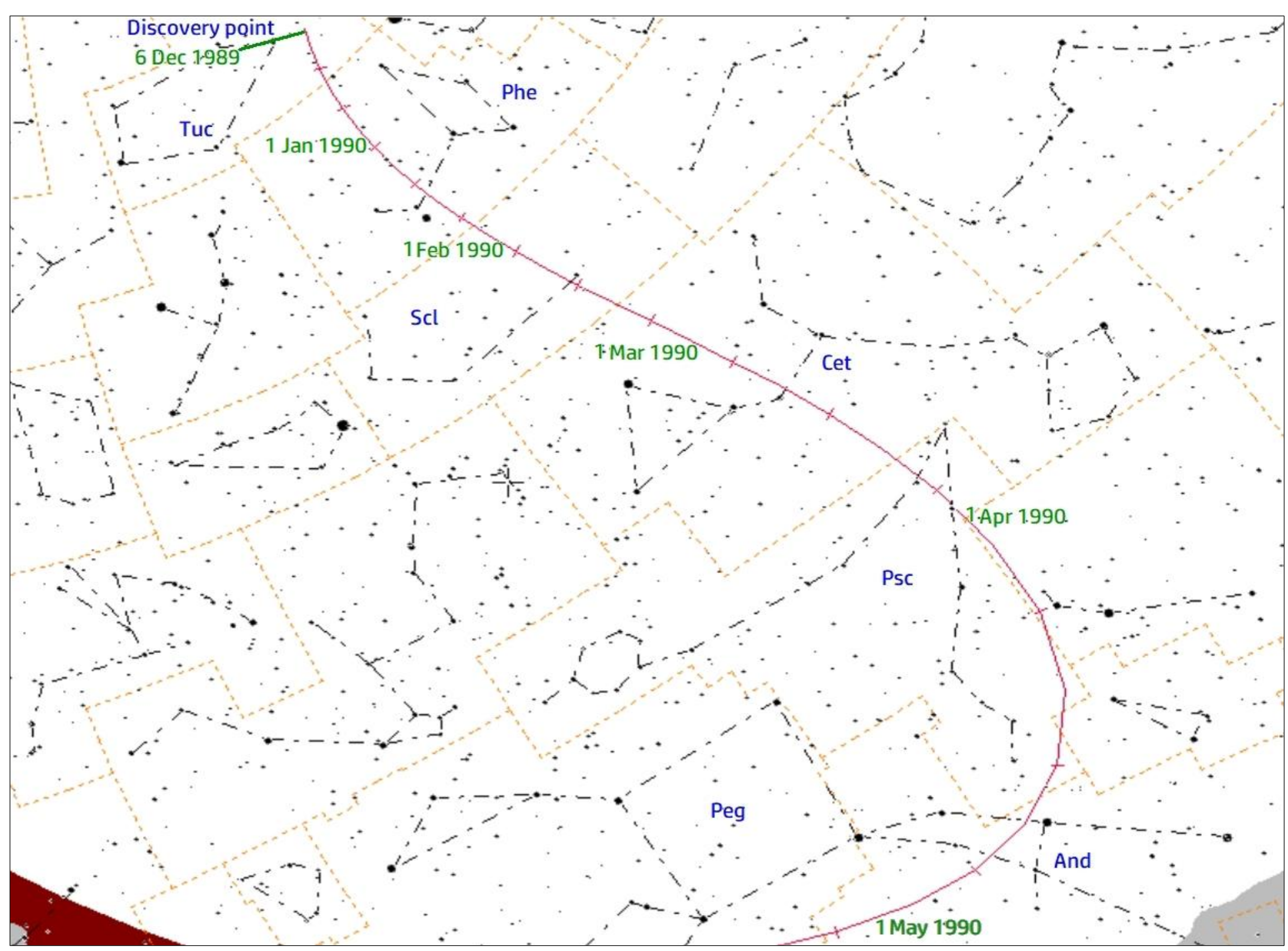

Figure 59: The path of Comet C/1989 X1 (Austin) in purple. The constellations that the comet passed through are in blue font and the dates in green. North is down, East is right, this being the view from NZ so that the Northern horizon is at the bottom (source: GUIDE).

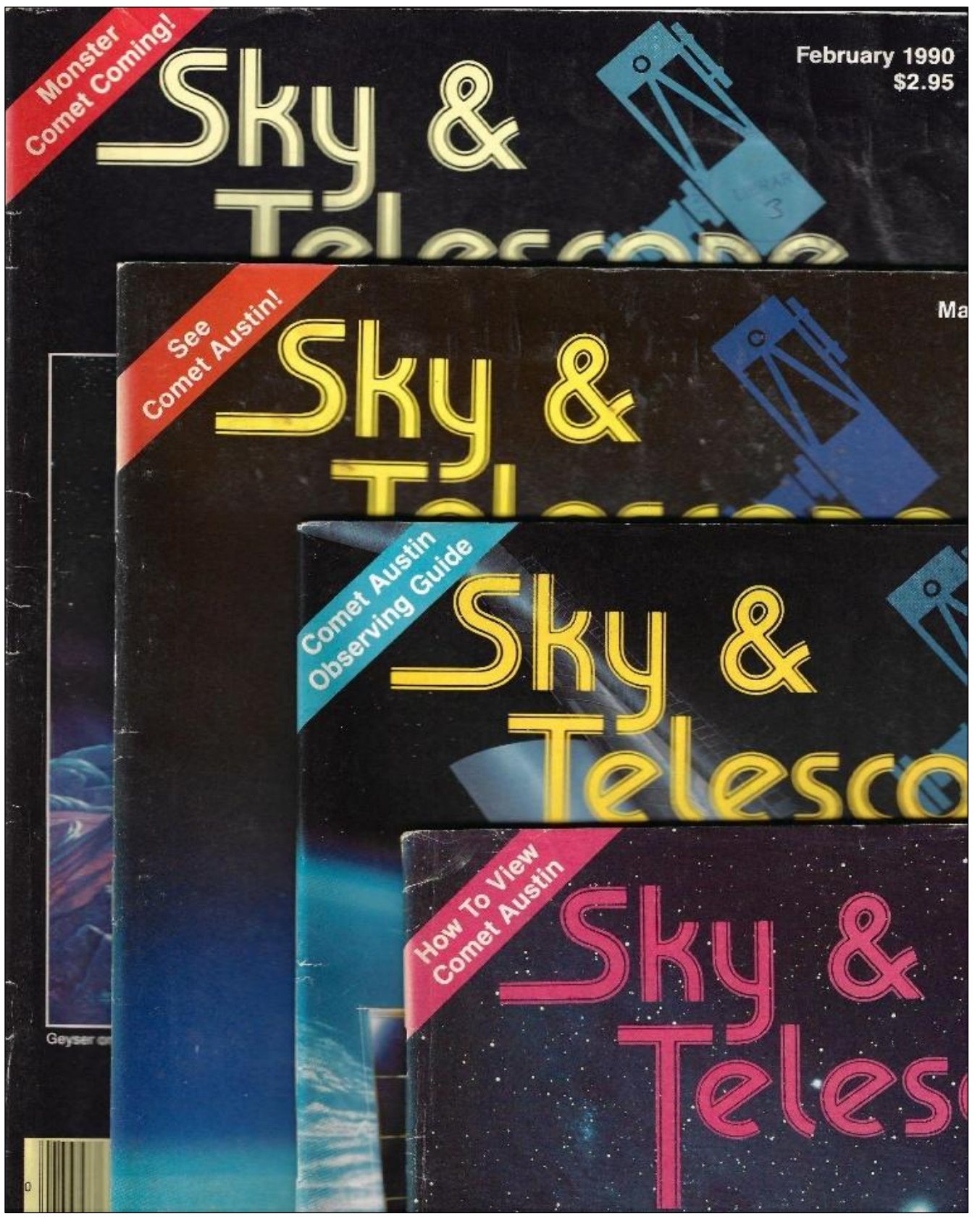

Figure 60: The front covers of *Sky & Telescope* magazine from February to May 1990 showing the decreasing optimism for a bright Comet Austin.

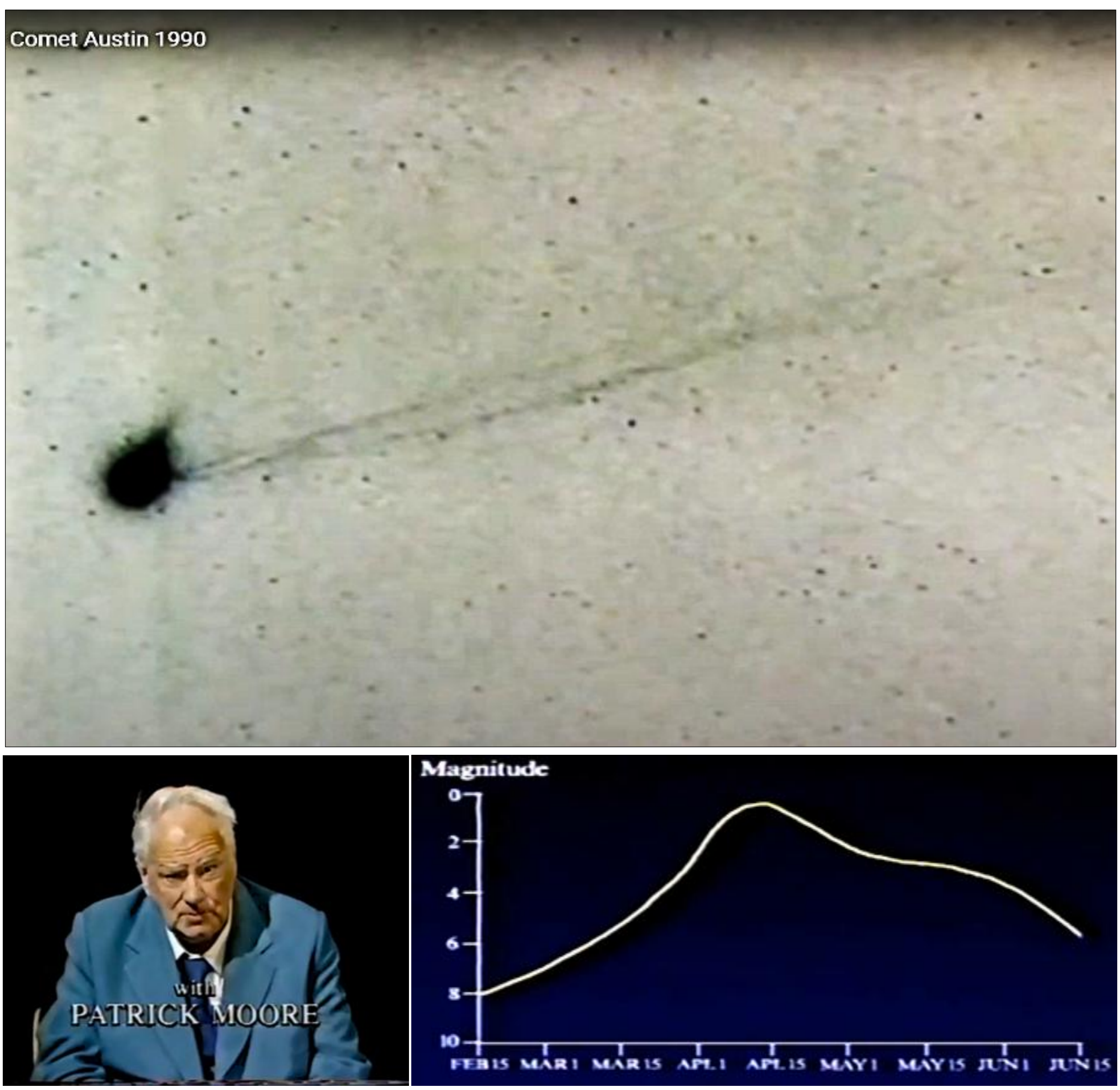

Figure 61: Screenshots of Patrick Moore's *The Sky at Night* (BBC1) episode about Comet Austin. Patrick Moore is pictured bottom left, and above him is a photograph of Comet Austin taken with the European Southern Observatory's Schmidt telescope at La Silla in the Atacama Desert (Chile) on 25 February 1990 (UT) (see https://www.eso.org/public/images/eso9004a/). The predicted magnitude of the comet is shown to the right of Patrick Moore—note that it was predicted to almost reach magnitude 0.

Kronk et al. reveal with hindsight that the optimistic outlook for C/1989 X1 (Austin) was overly confident. R.W. Panther (England) estimated the comet's magnitude to be 3.9 on 3 April 1990 (Kronk et al., 2017: 468–469). Several nights later, other noted observers such as John Bortle (USA) and Alan Hale (USA) were describing it as a magnitude 5 object (*ibid.*). Interestingly, the first naked eye observations were on 21 April, possibly when the comet had risen out of the atmospheric murk. R.A. Keen (Colorado, USA), R. Haver (Italy), M. Moller (Germany), J.V. Scotti (Kitt Peak Observatory, USA), and Charles Morris (USA) all described it as a magnitude 4–5 object in the later half of April (c.f. Figure 62). Austin (pers. comm, 2006) recalls that he saw his comet from the Arizona desert (USA) in April/May 1990 with his naked eye and that it had a 7–8° tail. This agrees with articles in the August (Eicher, 1990a: 82–84) and September (Eicher, 1990b: 71–73) 1990 issues of *Astronomy* magazine where observers were reporting "... a faint tail that nearly stretched 10°." It remained at this brightness into May 1990. Kronk et al. (2017: 469) highlight the large number of observations as it rose higher in the morning sky, stating that

> Checking the various issues of the International Comet Quarterly revealed that nearly 200 observers reported over 1300 observations in May … [and] 19 observers reported a total of 47 naked-eye ob-servations throughout the month.

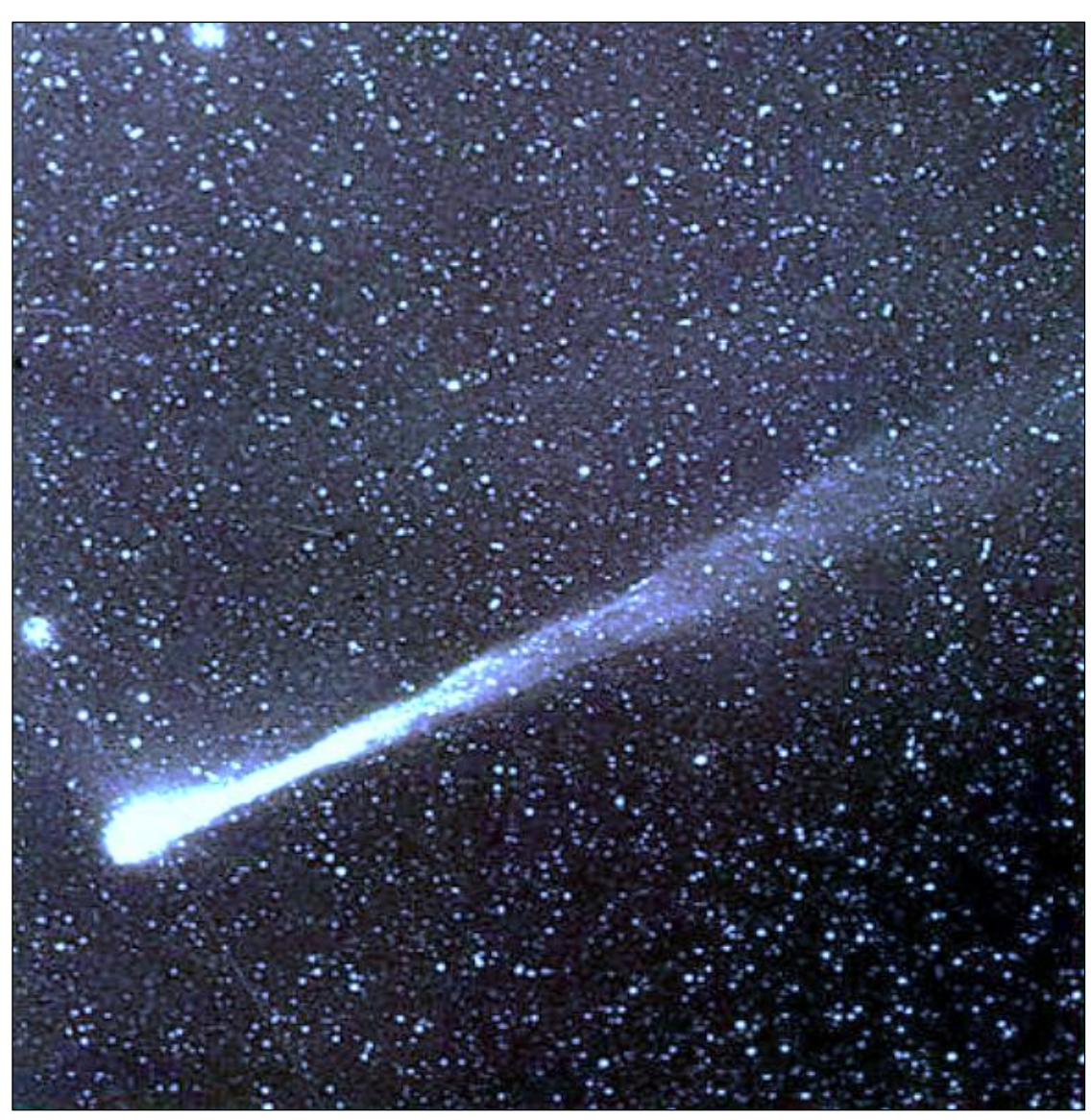

Figure 62: Comet Austin (C/1989 X1) as photographed on 29 April 1990 by Bob Yen (Eicher, 1990b: 72). The colour has been slightly tweaked by the first author. The comet never attained the brightness that was first predicted.

It did not reach magnitude 1, as many hoped it would, but around magnitude 4, which was still easily visible to the naked eye (Austin). However, it was not 'the comet of the century'.

The comet reached perigee on 25 May 1990 (UT). Some observers, such as Bortle, reported a 60′ coma (twice the apparent diameter of the full Moon) with the naked eye on 27 May (Kronk et al.: 2017: 469–470). Morris reported seeing several tails in May, including a 2° antitail on the 20th. The comet faded in June and was last detected on CCD images taken on 22 July 1990 by K.J. Meech (Mauna Kea, Hawaii). An attempt to image it was made in May 1991, but no comet was detected down to magnitude 24 (*ibid.*).

Of note is that C/1989 X1 (Austin) was going to be one of the first targets for the newly launched (24 April 1990 UT) Hubble Space Telescope. However, poor optics prevented this from happening. After the famous HST optical retrofit, the first comet to be imaged was discovered by Austin's friend, David Levy.

Table 20: Newspaper articles about 'Comet' and 'Comet C/1989 X1 (Austin)' in Papers Past. Note that December was the discovery month.

| Month and Year | Total for 'Comet' | Total for 'Comet Austin' |
|---|---|---|
| Dec 1989 | 48 | 1 |
| Jan 1990 | 0 | 0 |
| Feb 1990 | 1 | 0 |
| Mar 1990 | 0 | 0 |
| Apr 1990 | 0 | 0 |
| May 1990 | 0 | 0 |
| Jun 1990 | 0 | 0 |
| Jul 1990 | 0 | 0 |
| Aug 1990 | 0 | 0 |
| Totals: | 49 | 1 |

Sekanina used West's and Gilmore's observations of C/1989 X1 (Austin)'s tail for analysis and determined that the tail formed farther out than most comets, approximately 900–500 days prior to perihelion. He concluded that "Early tail formation of this kind is characteristic of new, Oort-cloud comets." (Kronk, 2017: 471). Thus, in Levy's two possible comet scenarios, a West or Kohoutek, Comet Austin leaned towards a Comet Kohoutek.

Interestingly, despite all the media 'hype' around Comet C/1989 X1 (Austin), the authors could only find one related article published in NZ newspapers using Papers Past (see Table 20). One wonders why.

Austin recalls his three comet discoveries in the 1980s with fond memories. They opened up numerous doors for him, including sponsored trips to the United States and visits to venerable observatories and meeting famous astronomers such as Clyde Tombaugh, the discoverer of Pluto in 1930. Austin jokes that he and Tombaugh share the same birthday (but not year), 4 February. On one trip he met Alan Hale who had been hunting comets for many years. Hale asked Austin, "What's it like to discover a comet?" He would soon find out as he and Thomas Bopp later discovered the extremely popular Comet Hale-Bopp in July 1995.

Austin no longer hunts for comets, but he particularly enjoys solar eclipses. He has been to six eclipses and seen four. He has also been to two annular eclipses and seen one. He still lives in New Plymouth with his wife of over thirty years, Lorraine. His astronomical activities of late have been determining the light curves of variable stars with the New Plymouth Astronomical Society's 35-cm (14-inch) Schmidt-Cassegrain and a DSLR camera. As with determining eclipse predictions, he loves the precision involved in this work.

## 8 ALAN GILMORE

Alan Gilmore (b. 1944; Figure 63) serendipitously discovered the most recent comet from New Zealand while doing astrometry of a newly discovered asteroid in 2007. At the time, Alan, and his wife Pam Kilmartin, were working for the Department of Physics and Astronomy at the University of Canterbury's Mount John Observatory at Tekapo in NZ's South Island. Before delving deeper into the discovery, we shall first look at Alan and Pam who are two of the most productive astronomical observers in NZ. Much of the following is from *in situ* interviews that the lead author conducted with them in May 2023.

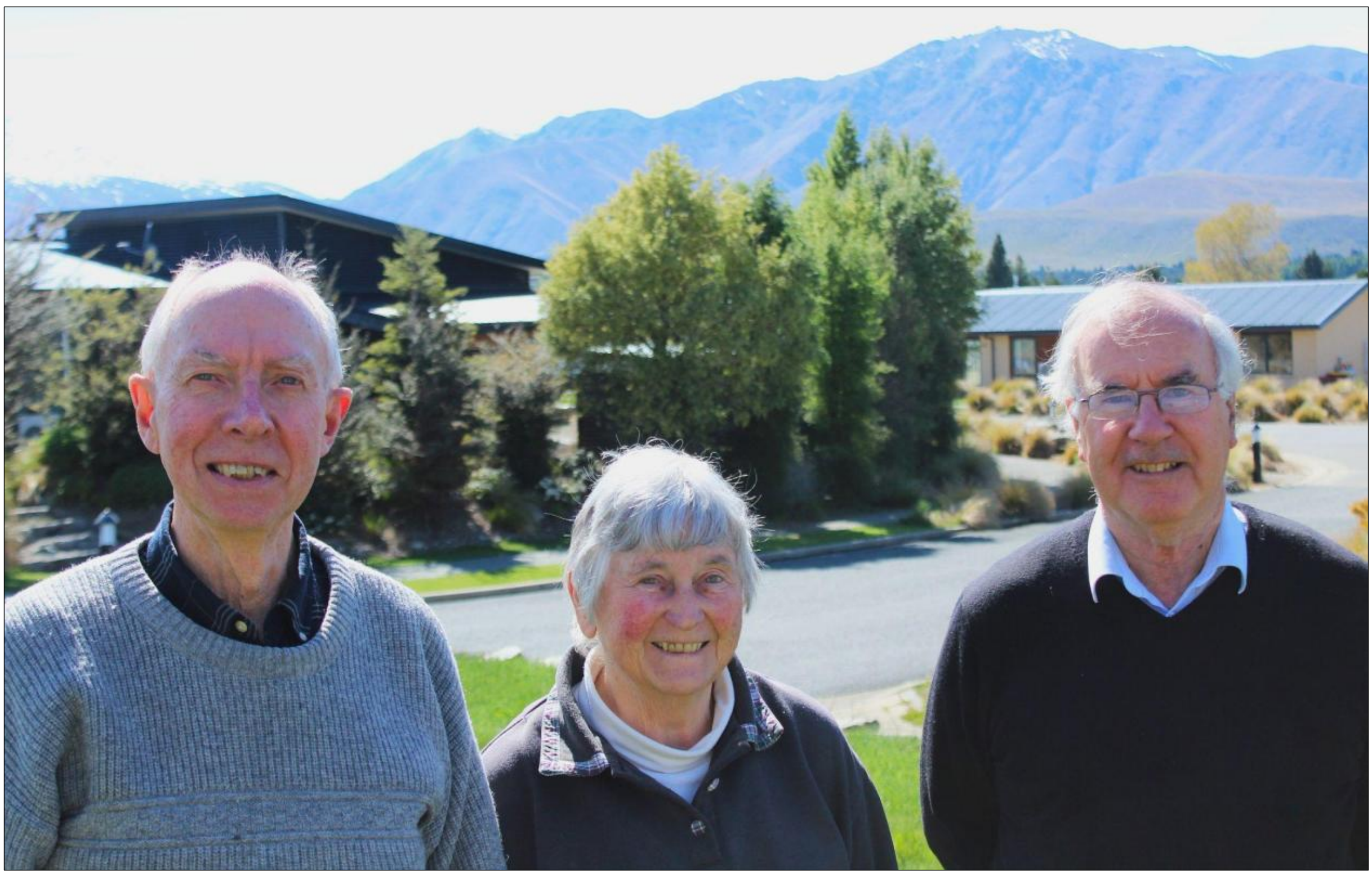

Figure 63: Three NZ 'pillars of astronomy'. L-R: Alan Gilmore, his wife Pamela Kilmartin, and Professor John Hearnshaw (photograph: John Drummond, 2021).

Alan Gilmore was born in Greymouth in 1944. His parents, Frederick Harry Gilmore and Catherine Mary Johnston, were married in Greymouth in 1942. They had two children. The family moved to the elevated dark-sky site of Otoko Valley, 40 km north-west of Gisborne (North Island) in 1947 and lived there until 1950. The sight of a bright meteor intrigued the young Gilmore who had not quite reached five. His father, a keen reader of scientific literature, explained the phenomena to Gilmore.

Two years later, the family visited Carter Observatory in Wellington and looked at the Moon and various other targets through the 23-cm (9-inch) Cooke refractor (see Andrew and Budding, 1992). Mr Gilmore also purchased a bundle of newspaper reprints (Gifford, 1931) by local astronomy expert and Cambridge graduate Algernon Charles Gifford (Eiby, 1972; Gifford, 2005) for Alan to read, which he did. His first telescope was lent to him by his neighbours when he was at intermediate school (about 11), but it had no eyepiece. The inexperienced Gilmore did not realise that one was needed and gave up using the telescope because, naturally, he could not see anything. In 1956 he read the book *Observing the Heavens* by Peter Hood (Oxford University Press, London, 1951) in the Hutt Intermediate School library. This explained how to operate a telescope, so Gilmore placed a microscope eyepiece in the telescope and observed the moons of Jupiter, craters on the Moon, etc. After this, Alan's parents bought a small telescope from General Science Supplies. He took delight in showing people the Moon, projected sunspots, stars and more.

Gilmore got an after-school job at a local pharmacy delivering prescriptions, which happened when the influenza epidemic hit NZ in 1957. It was the second major world-wide influenza pandemic to occur in the twentieth century. He used the money to save for a larger telescope. With help from his father, he built a 15-cm (6-inch) reflector, using his savings to purchase the f/8 mirror from General Science Supplies. Gilmore first started using his new telescope in 1958. After exploring the heavens with this larger aperture, he wrote to Carter Observatory asking what he could do with it. Ivan Thomsen, the then Director (Orchiston, 2016: 265), replied encouraging Gilmore to start variable star observing. He was also instructed to write to Albert Jones in Timaru to get guidance in this new venture. He did. Albert promptly replied, sending Frank Bateson's book *The Observation of Variable Stars* (1958) and a set of variable star charts. Gilmore began observing variables and submitted his first observations to Bateson's Variable Star Section of the Royal Astronomical Society of New Zealand in July 1959. In 1961, Jones suggested that Gilmore also try comet observations and comet magni-tude estimates. Gilmore recalls that "… there were quite a few bright telescopic comets back then …", in the early 1960s.

In 1960 Frank Bateson (1909–2007; Christie, 2014) was commissioned by the University of Pennsylvania to manage a site survey of New Zealand with the goal of establishing a Southern Station. NZ partly filled a longitude gap between Australia and South America. With an obvious enthusiasm for astronomy and some observing experience, Gilmore was invited by Bateson to join the site testing in his school holidays. The work involved weather observations, estimating seeing (air turbulence) and air transparency. Gilmore worked with Bateson on Black Birch Range in Marlborough in August 1961, his 6th Form year, and returned to Black Birch in May and August 1962. In the summer of 1962–1963, before starting University, Gilmore worked at Mount John in Central Otago. Hearnshaw and Gilmore (2015) go into much detail about the process, progress and adventures in their book *Mount John: The First 50 Years*, as does Frank Bateson (1964) in his *Final Report on the Site Selection Survey of New Zealand*. After the selection of Mount John as the preferred site in 1963, Pennsylvania formed a partnership with the University of Canterbury to run it. Mount John is above Lake Tekapo at an altitude of 1,030 metres, 3.5 hours drive from Christchurch (Hearnshaw and Gilmore, 2015). In 1962, Gilmore's work on the site testing and his variable star and comet observing were recognised with the award of the Royal Astronomical Society of New Zealand's 'Murray Geddes Memorial Prize'.

After leaving secondary school, Gilmore went to Victoria University and graduated with a BSc in Physics in 1968. After graduation, he worked for the Insulation and Acoustics Division of Fletcher Construction for one year, then, in 1970, a position opened at Carter Observatory (Wellington). At Carter, a 16-inch (41-cm) Boller and Chivens telescope lay mostly dormant, or, as Gilmore put it, "… had done almost nothing …" and the Carter Observatory Board was keen for it to be utilised, especially in the production of scientific results. Gilmore then began taking test photographs with it to determine tracking, limiting magnitude and so on. He also used it to do photographic photometry.

Working from a description in Elizabeth Roemer's chapter on comets in *The Moon Meteorites and Comets* (Middlehurst and Kuiper, 1963) Gilmore built a device to move a photographic plate at the predicted speed and direction of the image of a target comet or asteroid. At this stage, Garry Nankivell (1929–2001) told Gilmore about a machine, called a Trioptic, at the New Zealand Government Department of Scientific and Industrial Research Physics and Engineering Laboratory. It could measure astrometric positions from photographic plates on a light table in the X–Y axes with a microscope. It was at Gracefield in Lower Hutt, 30 minutes' drive from Carter Observatory. A bequest to Carter Observatory enabled the purchase of an early programmable calculator, a Sony Sobax, that Gilmore programmed to convert x–y coordinates into RA and Dec.

Gilmore (2023) recalls that astrometry of comets in the Southern Hemisphere was lacking back then. The noted British astronomer, Dr. Brian Marsden, Director of the IAU's Minor Planet Center at Cambridge, Massachusetts, was extremely welcoming and supportive of Gilmore's astrometry. Perth Observatory and a US station in Argentina were the only other observatories regularly producing cometary astrometry South of the Equator. Interestingly, one of Gilmore's first astrometric targets was Bill Bradfield's first comet discovery in mid-1972, C/1972 E1 = 1972 III = 1972f (see IAUC 2392; Marsden, 1972a).

Gilmore's astrometric observing was intermittent through 1972 due to daytime work demands at Carter Observatory. Observing increased when Pamela Kilmartin joined the Observatory as Librarian and Information Officer at the beginning of 1973. Kilmartin possessed an interest in astronomy since her rural childhood but had few opportunities to pursue it. This changed when, while a student at the University of Auckland (NZ), she attended an astronomy course at Auckland Observatory. This led to working with a ladies' group doing photoelectric photometry at the Observatory. She was also lent a 6-inch (15-cm) refracting telescope by Graham Loftus (1924–2015) to do visual observations of variable stars.

After graduation with an MA in French and English, Kilmartin enrolled at the Library School in Wellington in 1972. She and Gilmore had already met at a RASNZ Conference in Christchurch the previous year and started going out together. When the position of Librarian–Information Officer at Carter was advertised, at the end of 1972, Kilmartin was the best-qualified applicant. The Carter Board knew of the friendship with Gilmore and hoped that they would continue to get along well together.

Once on the Carter staff, Kilmartin became adept at using the Trioptic machine and reducing the measurements on the Sobax. A third member of the team was Russell Millington (see Figure 48), the Observatory's technician. The Carter astrometry came to international attention after Michael Clark at Mount John Observatory discovered a comet in 1973 (see above).

Measurement of Clark's plates and others obtained at Carter Observatory enabled Brian Marsden to derive the short-period orbit of the comet. A month later, Kilmartin and Gilmore followed up the discovery of near-Earth Asteroid 1973 NA, the first found in a NEO search programme begun by Eleanor Helin (1932–2009) at Mt Palomar (USA).

The astrometric programme was greatly assisted by the fortuitous acquisition of an old plate-measuring machine from Yale University Observatory. This came about when Donald Kimball and his wife visited Carter while on a world tour. Seeing the need for a measuring machine, Kimball mentioned that there was an old one at Yale University that was not being used and suggested that someone at Carter Observatory contact Yale and ask for it. A kindly reply was received from Dorrit Hoffleit (1907–2007) of Yale. After further negotiation at the IAU General Assembly in Sydney in August 1973—with Marsden endorsing the Carter work—the machine arrived in March 1974, the month that Gilmore and Kilmartin married.

At the end of 1975, Gilmore and Kilmartin decided to leave Carter Observatory. Nonetheless, the Board agreed that they be given such telescope time as they requested to continue the astrometric programme. They did this on the weekends. They also built an 8-inch (20-cm) astrograph at their home in Happy Valley, near Wellington city. At the Royal Astronomical Society of New Zealand conference in Dunedin at the end of 1975, they were invited by Dr Noel Doughty (1939–2001) from the University of Canterbury to apply for telescope time at Mount John Observatory. This they did, visiting Mount John for a week each winter from 1976 to 1980 to use the 24-inch (61-cm) Boller and Chivens telescope. This required some juggling of family arrangements after the birth of their first son in December 1977.

In May of 1980, Professor John Hearnshaw (b. 1946), Director of Mount John Observatory from 1976 to 2008, asked Gilmore and Kilmartin if they would like to move to Lake Tekapo and work at Mount John. A position had opened with the resignation of Rod Austin. They would be em-ployed as technician–observers and their astro-metric programme would be supported by the University of Canterbury. In November 1980 they moved to a University house in Lake Tekapo village. In early 1996 Alan became the Superintendent of Mount John Observatory. This position (which continued until 2014) increased his day-time activities and night-time commitments, and as a result of the increased workload the number of nights dedicated to astrometry dropped sharply to 3–4 a month.

Since P/2007 Q2 (Gilmore) was serendipitously discovered when Gilmore was doing astrometry, we shall discuss astrometry in a little detail. Astrometry is an essential science in determining the position of an object on the celestial sphere. Akin to determining a position on Earth using latitude and longitude, the position of known stars is used to locate an object's celestial location. For Solar System objects, their transitory nature causes movement against the background stars. By knowing the exact date, time, RA and Dec from a number of observations, a moving target's orbit can be determined, much like taking three photographs of a tennis ball in flight; the ball's path and speed can be calculated based on careful time-keeping and photographic comparison to the background.

As stated, in the late 1990s due to other commitments, Gilmore and Kilmartin did little astrometry as by this time Robert McNaught (b. 1956) and his colleagues at Siding Spring Observatory (Australia) were producing astrometry more efficiently using Charge-Coupled Devices (CCDs).

In early 2000 the University of Canterbury commissioned a 1k × 1k × 24-micron CCD for both spectroscopy and direct imaging for photometry. Dr. Ian Griffin (b. 1966), who was the then Director of Auckland's Star Dome, wanted to do some astrometry of Near-Earth Objects (NEOs) with a large telescope from a dark sky location, and he was granted time on the 1-metre McLennan Telescope, with the new CCD. During his stay, in mid-2000, he showed Gilmore and Kilmartin how to do astrometry with the Mount John setup, and he introduced them to the astrometric program 'Astrometrica', developed by Herbert Raab (b. 1969). The introduction of various star catalogues at this time facilitated the use of Astrometrica to get accurate positions of comets, asteroids, and Near-Earth Objects. The field of view of the CCD on the 1- metre telescope was not large by today's standards, covering only 12′ × 12′ (about a quarter of the Moon's disk), but it was sufficient to do meaningful astrometry. The Southern Hemisphere thus gained a valuable astrometry station, especially considering that there were not many South of the Equator in the early 2000s. This was so valuable, in fact, that the head of the Minor Planet Center, Dr. Brian Marsden, named Asteroid 2537 'Gilmore' after them. It was discovered on 4 September 1951 by Karl Wilhelm Reinmuth (1892–1979) at Heidelberg Observatory. The citation reads:

> Named in honor of Alan C. and Pamela M. (Kilmartin) Gilmore, whose program of astrometric observations of comets and

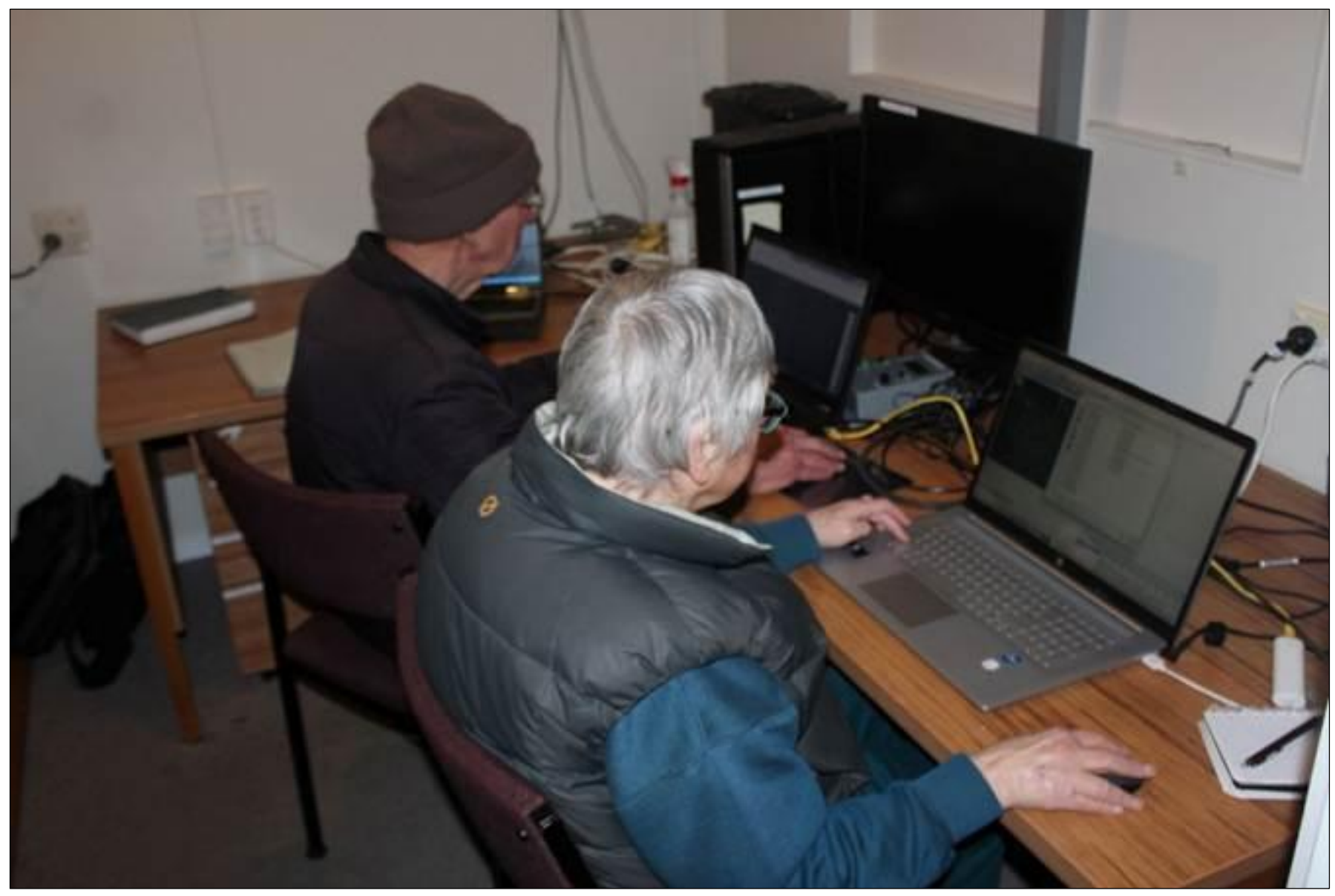

Figure 64: Pam Kilmartin and Alan Gilmore doing an astrometry run at Mount John Observatory in 2023. They are the most prolific astrometrists in New Zealand (photograph: John Drummond, 2023).

> minor planets has for more than a decade been one of the most productive and rapidly responsive such efforts ever to be undertaken in the southern hemisphere. At the Mount John University Observatory since 1980, they were formerly on the staff of the Carter Observatory … Name proposed by C. M. Bardwell and B. G. Marsden, identifiers for this planet. (Schmadel, 2007: 207).

Astronomical friends objected, noting that Kilmartin was known in astronomy by her maiden name. So, Marsden generously named a second asteroid 3907 Kilmartin. It was discovered on 14 August 1904 by Maximilian Franz Joseph Cornelius (Max) Wolf (1863–1932), also from Heidelberg Observatory. Marsden's citation (Schmadel, 2007: 332) reads:

> Named in honor of Pamela Margaret Kilmartin, co-director with her husband, Alan C. Gilmore {see planet (2537)}, of the Comets and Minor Planets section of the Royal Astronomical Society of New Zealand. Originally employed as librarian of the Carter Observatory in Wellington, she quickly became an astronomer in her own right and has been solely responsible for the measurement and reduction of the plates taken in the course of the astrometric programs in Wellington and more recently at the Mount John University Observatory. Name proposed by B. G. Marsden, who found the identifications involving this minor planet, and endorsed by A. C. Gilmore and F. M. Bateson …

After learning how to take images with the CCD (which had to be taken one at a time and saved individually), do image calibration, how to use Astrometrica and submit observations to the Minor Planet Center in Massachusetts, Gilmore and Kilmartin embarked on a journey that would last decades. Gilmore recalls that at that stage, as they learned the processes, their '… early pictures looked awful …', but they were getting results. A major improvement was made in 2011 when they bought their own CCD, a 1k × 1k × 24-micron Apogee U6. It is run by MaxIm DL and downloads and saves images in 0.6 seconds.

Fast-forward to the early 2020s and they are amongst the most prolific astrometrists on the planet (see Figure 64). In 2022 they were the 8$^{th}$ most productive team on Earth (Figure 65). Table 21 is a summary of their astrometric observations over the last ten years based on Royal Astronomical Society of New Zealand Annual Reports. The most recent observations are at the top; each astrometry submission to the MPC counts as one observation.

In addition to the astrometry that Kilmartin and Gilmore carry out at Mount John, they also have an observatory at their home (now on the outskirts of Tekapo township) from which they do astrometry when the Mount John telescopes are time-allocated to other observers and students. This home observatory is called Aorangi Iti Observatory and has the Minor Planet Center code R57. It houses a 35-cm (14-inch) Celestron SCT (see Figure 66). In the last five years (2018–2023) they have made a total of 753 observations with this setup.

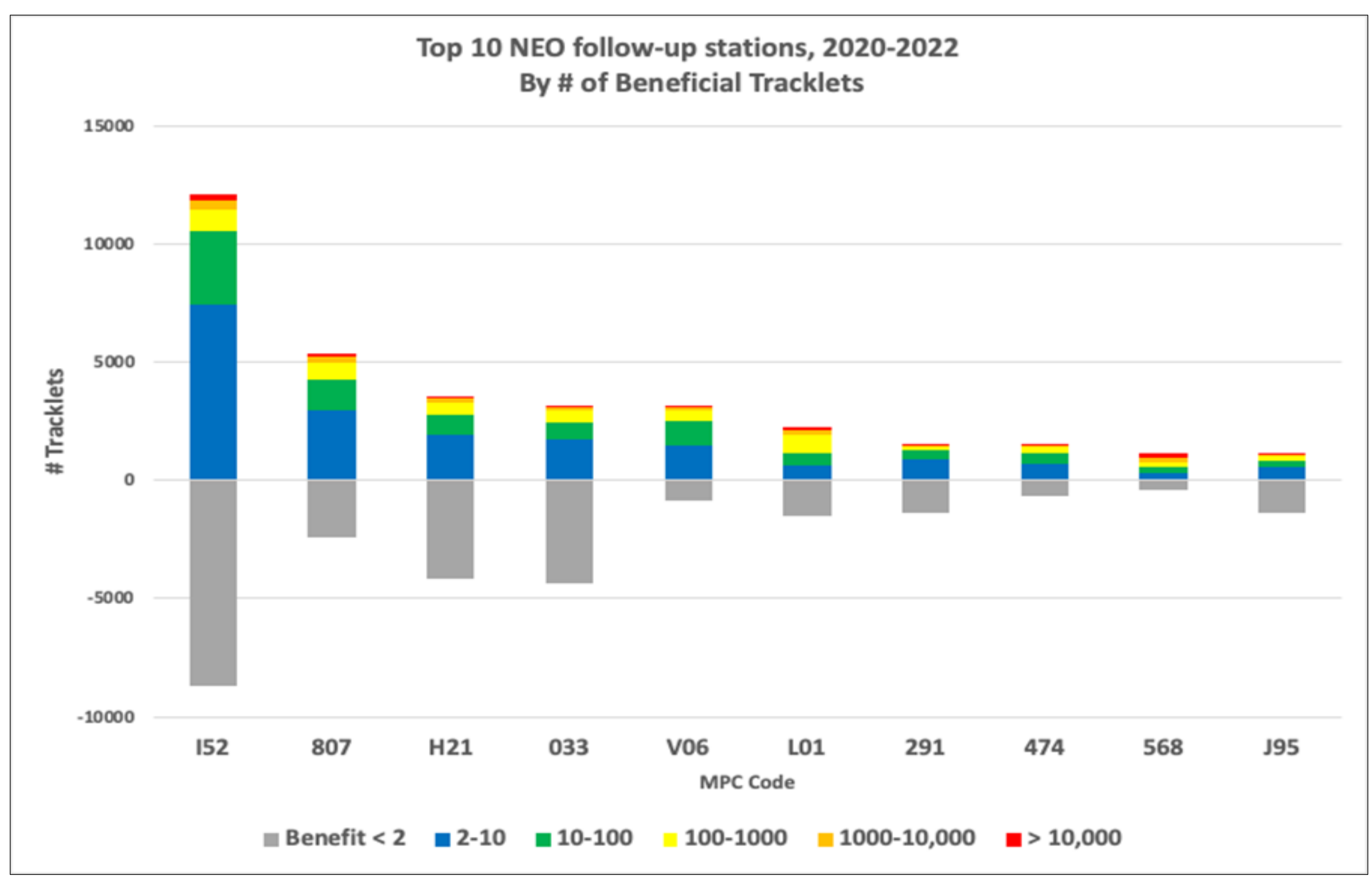

Figure 65: The productivity of Gilmore and Kilmartin (MPC 474) compared to the other top nine Near Earth Object (NEO) stations around the world. Graph compiled by Eric J. Christensen (Director, Catalina Sky Survey, Lunar and Planetary Laboratory, University of Arizona,) who describes it as showing "… everyone's contributions on an absolute rather than relative scale." 474 is the 8th most productive. The other stations, in order of descending rank, are I52 (Mount Lemmon Observatory, USA), 807 (Cerro Tololo Inter-American Observatory, Chile), H21 (Astronomical Research Observatory, USA), 033 (Karl Schwarzschild Observatory, Germany), V06 (Catalina Sky Survey-Kuiper, USA), L01 (Tičan Observatory, the new Višnjan Observatory, Croatia), 291 (LPL/Spacewatch II (1.8-meter telescope), USA), 474 (Gilmore and Kilmartin, Mount John, New Zealand), 568 (Mauna Kea Observatory, USA), and J95 (Great Shefford Observatory, England). Observatory codes from the Minor Planet Center can be found at https://www.minorplanetcenter.net/iau/lists/ObsCodesF.html.

Table 21: Astrometric observations of comets, asteroids, Near-Earth Objects and discovery confirmations made by the Gilmore–Kilmartin team, code 474, at Mount John Observatory using the 24-inch (60-cm) and 40-inch (100-cm) telescopes (source: *Southern Stars* Annual Reports).

| Year | Comets | Asteroids | NEOs | Confirmations | Totals |
|---|---|---|---|---|---|
| 2025 | 16 | 0 | 263 | 1,021 | 1,300 |
| 2024 | 8 | 0 | 1,191 | 901 | 2,100 |
| 2023 | 0 | 0 | 1,502 | 1,278 | 2,780 |
| 2022* | 7 | 0 | 769 | 677 | 1,453* |
| 2021 | 128 | 432 | 3,416 | Many! | 3,976 |
| 2020 | 171 | ---- | 3336 | 518 | 4,025 |
| 2019 | 164 | 1,479 | | 1122 | 2,765 |
| 2018 | | | | | 2,897 |
| 2017 | | | | | 3,378 |
| 2016 | | | | | ? |
| 2015 | | | | | ? |
| 2014 | 179 | 2,082 | | | 2,261 |
| 2013 | | | | | 2,016 |

* The 1-metre was out of action due to a dome malfunction from 23 April – early October 2022, so Mount John's 1-metre totals are not comparable to other years.

Accolades for Gilmore and Kilmartin include:

- Being responsible for the discovery of 41 asteroids.
- Being Fellows of the Royal Astronomical Society of New Zealand.
- Both receiving the RASNZ Murray Geddes Award (Gilmore in 1962 and Kilmartin in 1982).
- Having asteroids named after them: (2537) Gilmore, and (3907) Kilmartin.

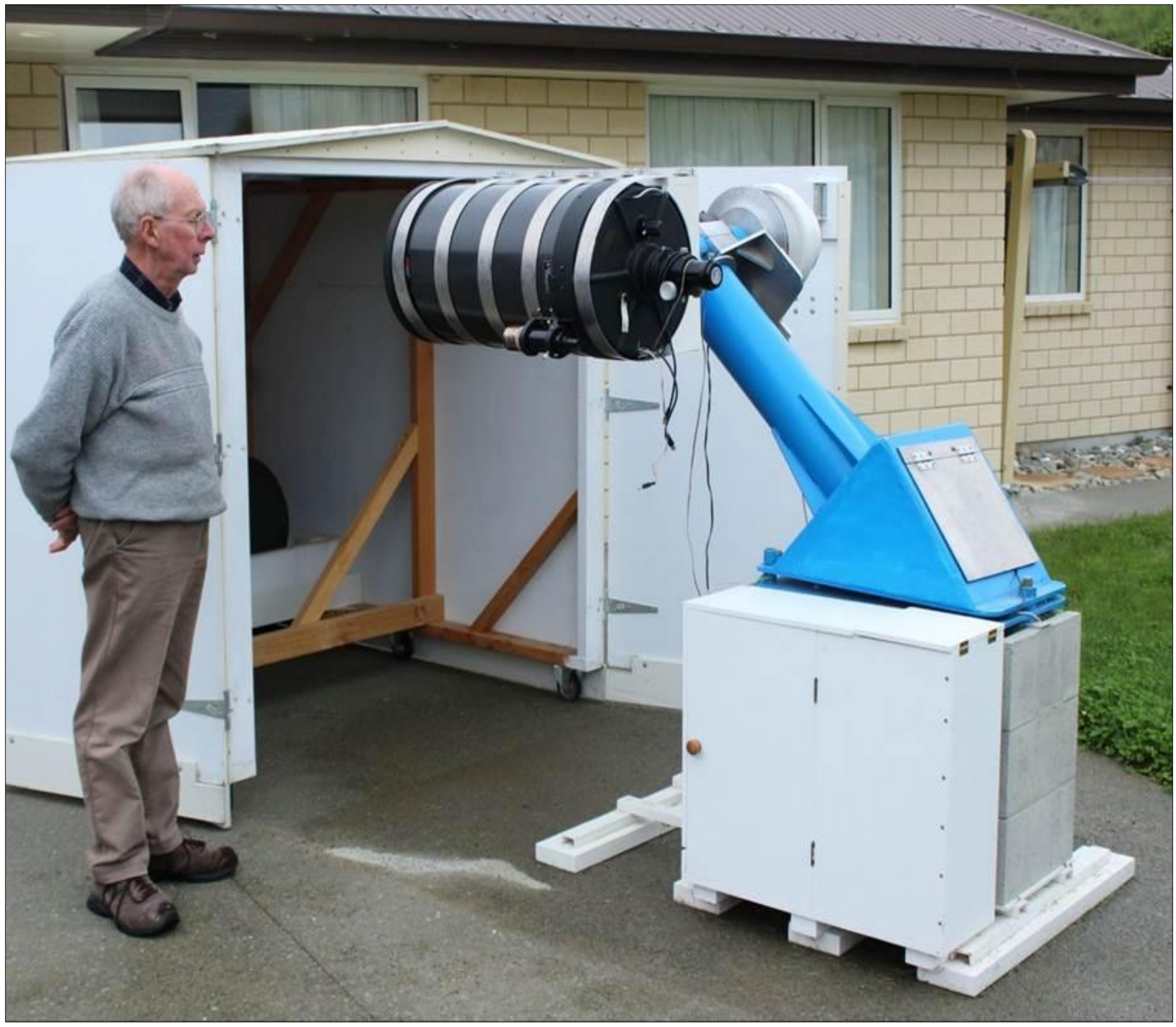

Figure 66: Alan Gilmore standing beside the 35-cm (14-inch) Schmidt–Cassegrain telescope at their Aorangi Iti Observatory (R57) in Tekapo (photograph: John Drummond, 2021).

- Kilmartin is one of eleven voting members of the International Astronomical Union's Working Group on Small Bodies Nomenclature, which is responsible for naming asteroids and comets (WGSBN).
- In May 2019, they appeared on a stamp issued by New Zealand Post in its 'New Zealand Space Pioneers' series.
- In the 2025 King's Birthday Honours, they were appointed Members of the New Zealand Order of Merit (MNZM), for services to astronomy (DPMC, 2025).

## 8.1 P/2007 Q2 (Gilmore)

On 22 August 2007, Gilmore (2007: 5–6) was imaging and doing astrometry of a known asteroid with the 1-metre telescope when an unknown magnitude 19 target moved through the blinked CCD fields. According to Gilmore it looked asteroidal in appearance, with no coma or tail evident. Gilmore imaged it again two nights later on 24 August. After sending notice of the discovery and astrometric positions to Brian Marsden at the Minor Planet Center, Marsden replied that the orbit looked more cometary than asteroidal. He then posted the target on the NEO Confirmation Page for other observers to follow up.

Observations on 27 August with the 1.02-metre (39-inch) f/8 reflector at Lulin Observatory in China revealed a weak coma of 1.9″ diameter and a short 3″ tail in PA 340° (Gilmore, 2007: 5). The cometary appearance was confirmed from Siding Spring Observatory (Australia) on 30 August with a 1-metre reflector. They also noted a diffuse coma and tail of 6″–8″ at p.a. 320°. With these additional astrometric positions Marsden determined that it was a short period comet with an orbital period of 13.5 years. Due to its short period it was named P/2007 Q2 (Gilmore). The discovery notice was announced by Daniel Green (2007). The perihelion was found to be beyond Mars at a distance of 1.843 au, with the aphelion extending out to the orbit of Saturn. The orbit is tilted at 10° to the plane of the ecliptic. Being faint, the comet was not widely followed, nor was it found in any of Kronk's volumes since their last volume covered comets up until 1993 (Kronk et al, 2017).

This was the second comet discovered from Mount John Observatory, after Mike Clark's 1973 discovery (Gilmore, 2007: 6). Initial details are listed below in Table 22, while Figure 67 shows the orbit of Comet P/2007 Q2 (Gilmore) and Figure 68 its path across the sky. Figure 69 shows how faint the comet was at discovery.

Table 22: An overview of Gilmore's comet, P/2007 Q2 (Gilmore). Epoch: 8.0 August 2007 UT. Based on the Minor Planet Center (MPC). If not listed in MPC, information was used from NASA/Jet Propulsion Laboratory Horizons System. Note that in modern times, once the observation of a newly discovered comet is announced, researchers look for pre-discovery observations of the object. Such observations, when available, can significantly improve the speed with which the comet's orbit can be accurately determined, greatly improving the forward prediction of its position in the sky, and thereby facilitating more accurate follow-up. In the case of P/2007 Q2 (Gilmore), the LINEAR survey[8] obtained four pre-discovery observations of the comet on 22 August 2007, which were made just hours prior to Gilmore's discovery of the comet—as detailed in MPEC 2007-Q43 (MPEC 2007).

| | |
|---|---|
| Discovery Date | 23 August 2007 NZST (22.6 August 2007 UT) |
| Discovery Magnitude | 19.1 (Green, 2007) |
| Discovery Declination | –01° (Aquarius) |
| Perihelion date | 23.9 August 2007 UT (MPC) |
| Perihelion distance (q) | 1.84 au (MPC) |
| Perigee date | 1 September 2007 (UT) (Kronk et al., 2017: 98) |
| Perigee distance | 0.834 AU (Kronk et al., 2017: 98) |
| Brightest | Magnitude 18.9 (early September 2007) |
| Visible from NZ | 23 August 2007 (discovery) – too faint for most NZ telescopes |
| Last observed | 15.8 October 2007 UT (MPC) |
| Eccentricity of the orbit (e) | 0.671 (MPC) |
| Semi-Major axis (a) | 5.5937 au (MPC) |
| Aphelion distance (Q) | 9.348 au (MPC) |
| Inclination (i) | 10.238° (MPC) |
| Epoch | 8.0 August 2007 UT (MPC) |
| Period | 13.23 years (MPC) |
| Observations in Kronk et al. | Kronk's et al.'s *Cometography* series does not cover 2007 |
| Observations in MPC | 455 (145 used to determine orbit between 22 Aug 2007 and 15 Oct 2007). |
| Observations in COBS | None |
| Papers Past newspaper articles | Not covered |

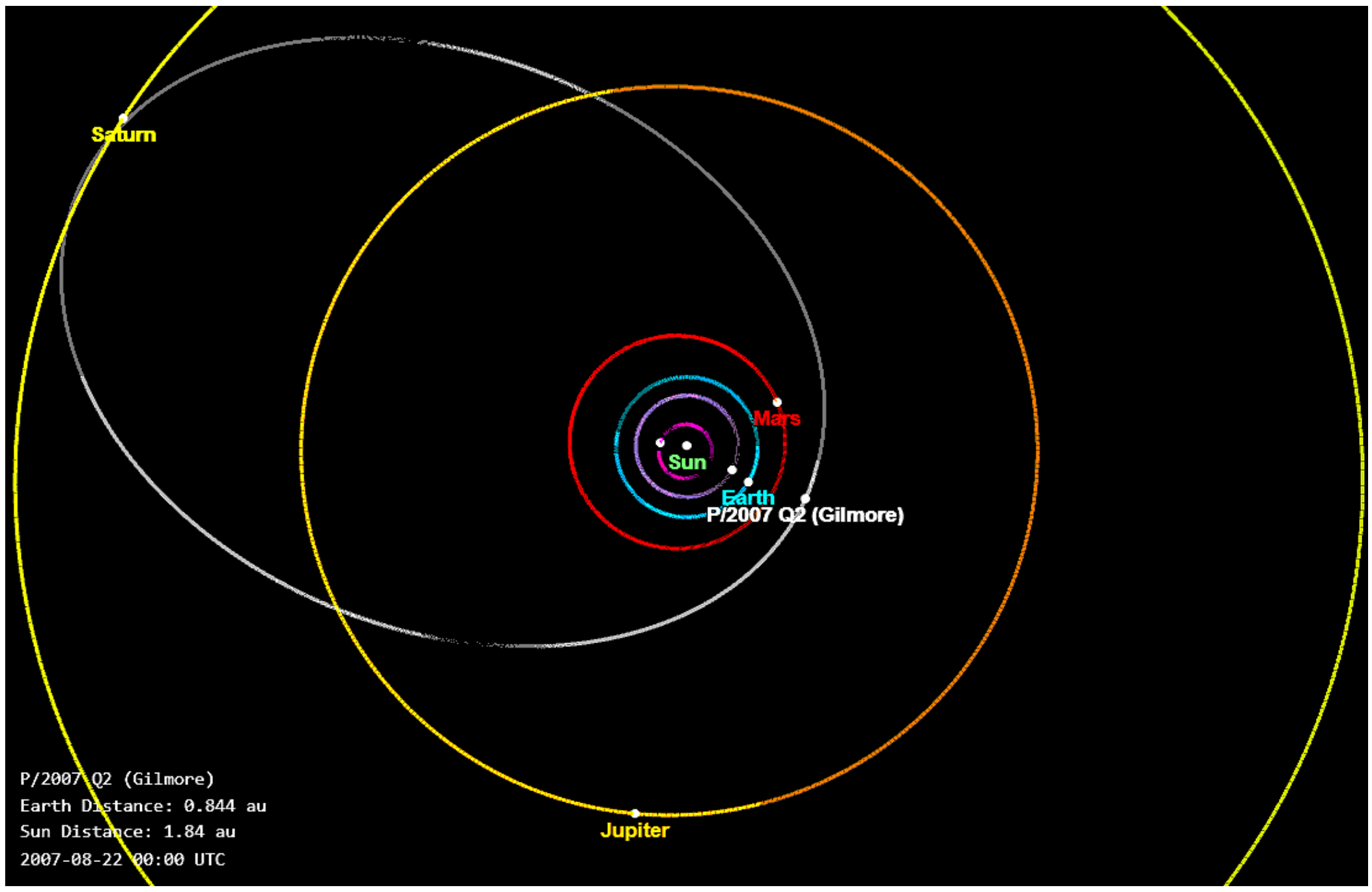

Figure 67: The orbital path of Comet P/2007 Q2 (Gilmore). Note how the aphelion reaches the orbit of Saturn, by analogy to the Jupiter-family comets, of which 71P is a member, P/2007 Q2 (Gilmore) is arguably a Saturn-family comet. On this diagram, P/2007 Q2 (Gilmore) moves in an anti-clockwise direction (source: NASA/JPL).

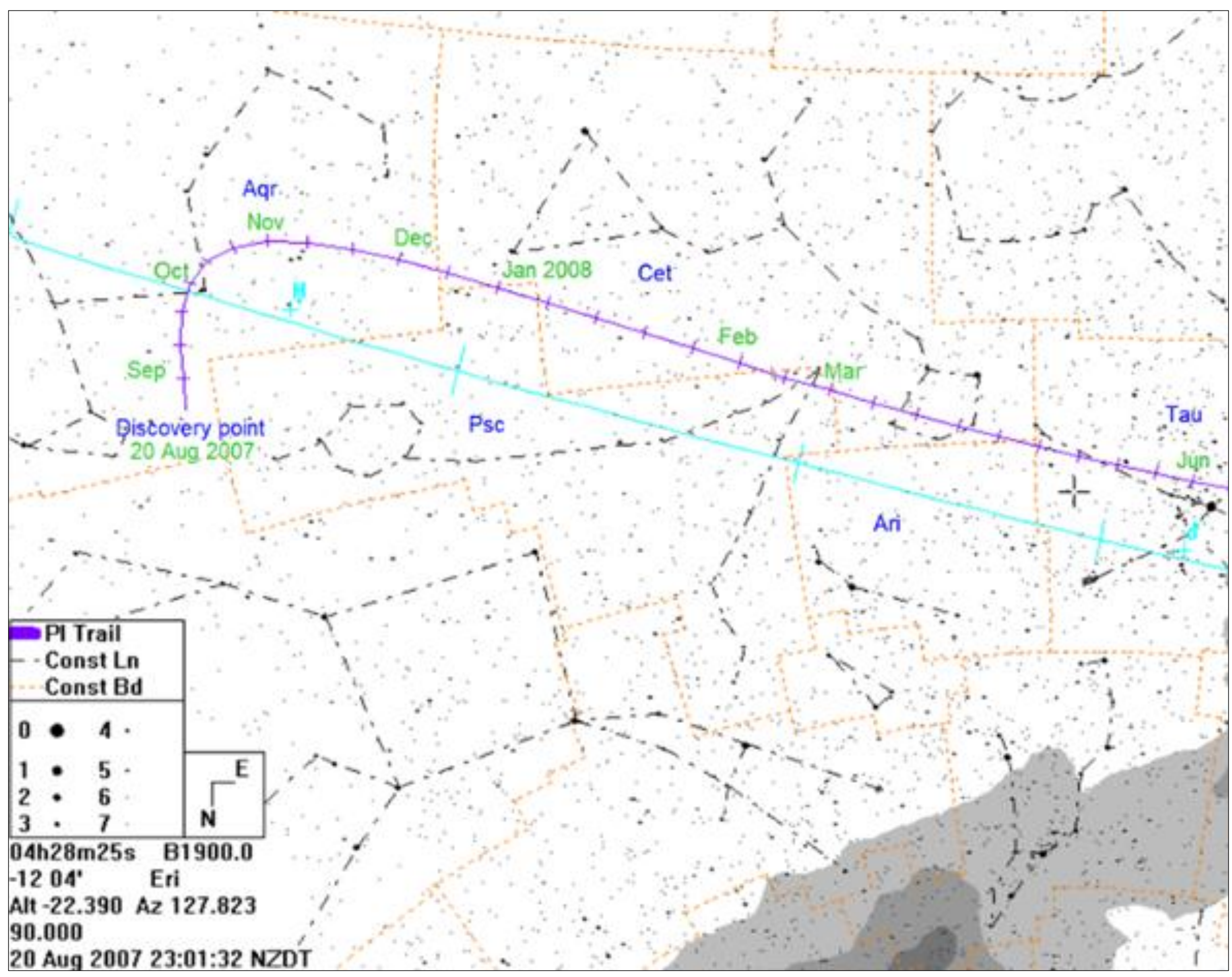

Figure 68: The path of Comet P/2007 Q2 (Gilmore) in purple. The constellations that the comet passed through are in blue font and the dates in green. North is down, East is right - this being the view from NZ so that the Northern horizon is at the bottom (source: GUIDE).

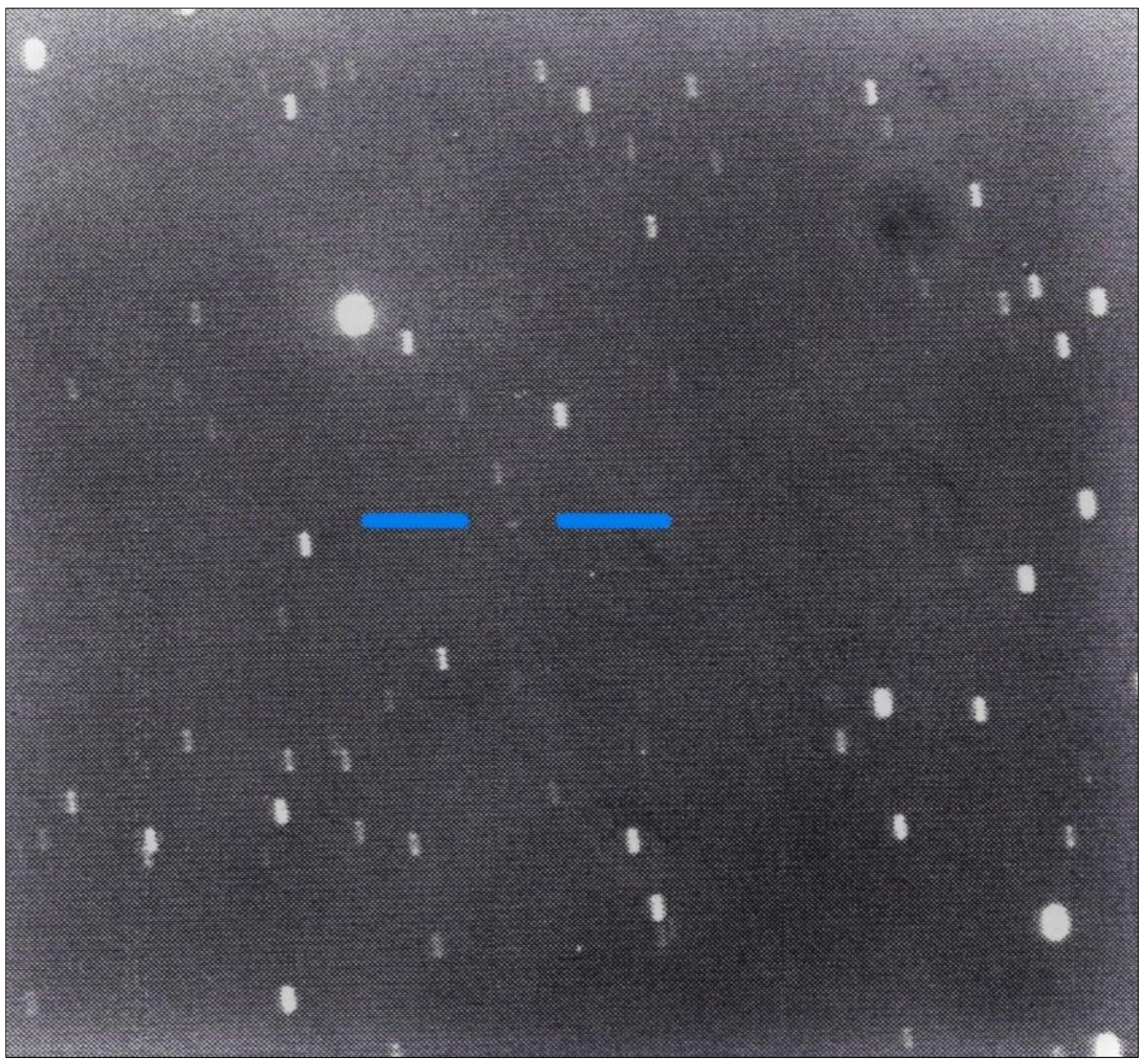

Figure 69: Comet Gilmore P/2007 Q2, as imaged with the 1-metre McLellan Telescope at Mount John Observatory. The original caption in *Southern Stars* states that this was a combination of three 80-second exposures on 24 August 2007. The field of view is about 11′ × 11′ (after Gilmore, 2007: 6).

Gilmore and Kilmartin still live at Tekapo and regularly use the telescopes at Mount John Observatory to do comet and asteroid astrometry. Comet Gilmore P/2007 Q2 was the last official comet found by a New Zealander from New Zealand.

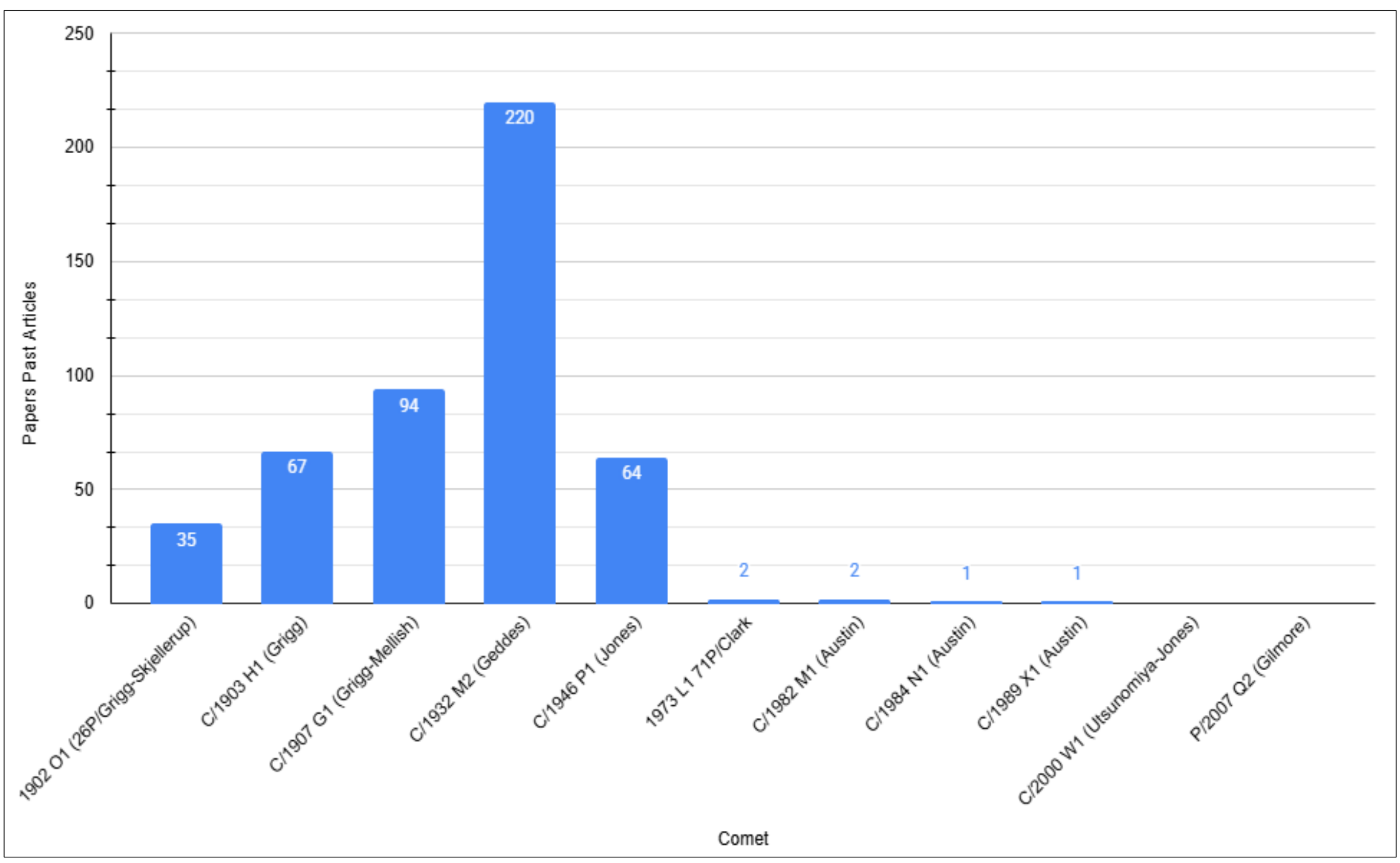

Figure 70: Newspaper article numbers published in New Zealand newspapers regarding nine of the eleven comet discoveries covered above. Based on the New Zealand newspaper article search engine Papers Past. Note that Papers Past does not cover the dates for the 2000 (Utsunomiya-Jones) and 2007 (Gilmore) discoveries (source: the authors).

## 9 DISCUSSION

### 9.1 Newspaper Interest in New Comet Discoveries

One interesting trend that the authors found was the sudden drop-off in newspaper articles across the land in relation to NZ comet discoveries and observations made after 1973 (see Figure 70). This was unexpected because, anecdotally, NZ was supposed to have been blessed with far more than its fair share of people with an interest in astronomy. According to Ivan Thomsen, the long-serving Director of Carter Observatory, who was best placed to make such assessments:

> For the size of its population, New Zealand must be one of the most interested countries in the world on the subject of astronomy. It is not easy to obtain exact figures, but from rough comparisons, there must be more members of astronomical societies per 1,000 of the population than anywhere else. Furthermore, from the numbers of people visiting observatories that are opened to them, there seems to be a fairly high general appreciation of the subject amongst the population as a whole. This has always impressed me as being surprising, in view of the fact that we are completely lacking in any formal instruction on the subject at our schools and colleges … (Thomsen, 1954: 79–80).

Between 1902 (Comet Geddes—the first NZ discovery) and 1946 (Comet Jones), newspapers were laced with discovery notices and observations made by NZers (and occasionally from international observers)—presumably reflecting this high level of popular interest in astronomy among the general population—but from 1973 the number of newspaper articles plummeted. Since there is no evidence for a sudden national disinterest in astronomy, we wonder if this was because people were seeing their news on television from the 1970s and the internet from the 1990s, and they became more attuned to news about rockets, satellites and space missions, and later the beautiful astronomical images provided by the growing assemblage or Earth-based and space-based telescopes, including, of course, the Hubble Space Telescope. So, for most newspaper journalists comet discoveries became rather *passé*—even those made from NZ. We should note that this tendency for the public and the media to focus on specific high-profile areas of astronomy is not new. We saw it in the second half of the nineteenth century in relation to comets, solar eclipses and transits of Venus (e.g. Cottam and Orchiston, 2015; Orchiston, 1998; 2017: 139–171), followed soon after by the possibility of what the aforementioned Australian comet-dis-coverer Walter Gale (1921) later referred to as 'sentient beings' on Mars, a media frenzy whipped up largely by Percival Lowell and his 'canals' (for a recent Australian perspective on this see de Grijs, 2026a; 2026b).

We can also speculate that from the 1970s perhaps people were not as interested in celestial happenings depicted in newspapers (even when coloured print was introduced), their attention being diverted to the beautiful astronomical images seen on the screen. Whatever the actual reason(s), the decline in newspaper articles about comets was dramatic.

### 9.2 Other New Zealand Comet Enthusiasts

The official NZ comet discoveries made from NZ are covered above, however, there were a number of other NZers who had close encounters in that they nearly discovered comets—as well as those who put much time and effort into observing and photographing known comets. In the interests of space, these people will not be dealt with here, but we plan to discuss some of them in future research papers.

## 10 CONCLUDING REMARKS

In this paper we have discussed New Zealanders who discovered comets from NZ. Their comets, discovery circumstances and observations made by other NZ observers of these comets have been examined and often compared to international observations. We looked at the lives of these discoverers, namely John Grigg, Murray Geddes, Albert Jones, Michael Clark, Rodney Austin and Alan Gilmore and their astronomical interests. The eleven comets that these six men discovered were investigated. We found that most were faint, not even achieving naked-eye status. The brightest was discovered by Rod Austin from New Plymouth. Austin's last comet (C/1989 X1) aroused much media speculation due to its high activity (and brightness) at a large distance from the Sun. However, like Comet Kohoutek in 1973–1974, the early awakening of its nucleus was thought to have been because this was the comet's first encounter with the Sun.

The exploits of these astronomers bade well for them. Their achievements were recognised by the international astronomical community, often in the form of medals, prizes and sponsored trips. The strategic astronomical importance of NZ was highlighted, since it is located in the Southern Hemisphere and occasionally witnesses astronomical events that Northern Hemisphere observers cannot see. It also sits in a longitudinal gap between South America and Australia.

Utilising the Papers Past newspaper search engine, we found a dramatic decrease in published newspaper articles about NZ comet discoveries and their related observations from the 1970s. In Section 9 we offered some suggestions as to why this may be. We also noticed that Vsekhsvyatskii (1964), Kronk (2007; 2009), Kronk and Meyer (2010) and Kronk et al. 2017) contain relatively few NZ observations of these comets. This probably was because the observations were often kept in personal journals or published in local newspapers, or shared with 'Comet Sections' of selected overseas astronomical societies, and were not written up and submitted for publication in international journals. One of the goals of this paper was to bring some of these 'lost' observations to an international audience.

In future work, we hope to compile companion papers that focus on some of those New Zealand observers who made important cometary observations without being the discoverers of the comets they studied, and observers or theoreticians such as Bill Bradfield and Professor (later Sir) Ian Axford, respectively, who were born in New Zealand but carried out their pivotal research, or made their comet discoveries, whilst living overseas.

## 11 NOTES

1. In other related studies by the authors they have reviewed NZ cometary astronomy over the past 750 years (Drummond and Orchiston, 2021) and examined comets and Māori cometary astronomy (Orchiston, 2016: 55–63; Orchiston and Drummond, 2019), plus NZ observations of various Great Comets of the nineteenth and early twentieth centuries (Drummond, 2023; Drummond and Orchiston, 2025; Drummond et al., 2025; Orchiston, 2016: 565–583; Orchiston and Drummond, 2022a; 2022b; Orchiston and Orchiston, 2024; Orchiston et al., 2020a), other NZ cometary observations (Kronk et al., 2026; Orchiston, 1983; 2016: 509–522; Orchiston et al., 2020b), NZ paintings of comets (Orchiston et al., 2024a), and NZ meteors (Baggaley and Orchiston, 2024; Orchiston et al., 2021; Taibi et al., 2026) and meteorites (Evans and Orchiston, 2023; Orchiston, 1997; Orchiston and Drummond, 2022c; Orchiston et al., 2024b; 2024c; 2026a; 2026b; 2026c; Warnes et al., 1998). They have also prepared a biographical study of Rod Austin (Drummond et al., 2026).
2. The following is from the 2025 'Easy 2C Calendar' (easy2C, www.easy2c.co.nz). Prior to 1868, each place in NZ kept its own time and this varied across the land by as much as 34 minutes. In 1868 the NZ Government adopted the time corresponding to longitude 172° 30′ east of Greenwich (11.5 hours in advance of Greenwich) as New Zealand Mean Time. This continued until the Standard Time Act in 1945 when NZ became 12 hours ahead of Greenwich Mean Time (GMT) on 1 January 1946 (NZ Standard Time). NZ adopted Daylight Saving time in 1974, where NZ Standard Time jumps ahead one hour).

3. It is now also known that Grigg was not the first to observe Comet 26P/Grigg-Skjellerup, as the French astronomer Jean-Louis Pons (1761–1831), observing from Marseilles, discovered it "… on 1808 February 6. It was again observed by Pons on February 9, but by nobody else." (Kresak, 1987: 65). After this the comet's magnitude declined rapidly as it had already passed what was a very close perigee (<0.12 AU), and with the Moon bright no further observations were made. As yet, there have been no calls from French (or other) astronomers to rename this object Comet 26P/Pons-Grigg-Skjellerup (Orchiston and Drummond, 2025).
4. Had luck gone his way, Grigg could very easily have ended up with four accredited comet discoveries, not three. On 19 March 1906 he discovered a new comet in Cetus and recorded its position that evening and the following one. He then sent the positions to John Tebbutt, noting in his letter that "Possibly this is the comet reported from Australia as having been found in 'Sculptor' but of which no positions were given." (Grigg, 1906a). This actually turned out to be the case, with the comet discovered by Melbourne's leading cometary astronomer of the time, David Ross (1850–1930; Orchiston and Brewer, 1990). But just like Grigg, Ross persisted in supplying Baracchi with imprecise positions for his discoveries (Orchiston, 1999a), and this almost gave Grigg the comet by default. Sydney amateur astronomer Hugh Wright (1868–1957) explains how Ross "… discovered the comet on Feb. 14, but … Baracchi would not accept rough positions, nor would he search. It was only after much delay & worrying that B. said he would give Ross 3 minutes with the telescope to pick up the comet. He failed, the following night was cloudy, & on the following night Ross showed Baracchi the object. A position was taken, & later announced, but none too soon as Grigg of NZ independently detected in the following night." (Wright, 1906).
5. The 'lost comet' in the 'Lost and Found' column was merely a precursor advertisement for an operatic production called the 'Lost Comet' that played a few weeks later in Dunedin. (The Lost Comet, 1932: 16).
6. Rod Stubbings of Australia is approaching Jones' world record of 515,000 visual variable star observations. On 20 November 2025 Stubbings made his 450,000th visual variable star observation after 32.5 years of observing (https://rodstubbingsobservatory.wordpress.com/about-me/).
7. Kronk and Meyer (2024: 260–265) describe how Staff Sergeant Leonard R. Edwards photographically detected a comet from Mount John Observatory while working for the United States Government doing satellite tracking (Gilmore, 2023). He detected it on "… three films (exposures 45 seconds, with intervals of 15 seconds between exposures) taken with the [50-cm] Baker-Nunn camera on 1971 June 19." (Marsden, 1972a: IAUC 2432, dated 1972 August 15). Marsden (1972a) described the comet as magnitude 10 and with a tail 20′ long at 175°. A call for confirmation photographs had already been sent to selected Southern observatories, but none reported photographing the field of interest at the time indicated. Note that the heading of that circular section was 'Possible Comet Edwards'. Gilmore (2023) recalls that the comet was not initially noticed for some weeks due to a long gap between exposures and inspection. This stands to reason since Edwards' photographs were taken on 19 June 1971 and the IAUC 2432 was published on 15 August 1972, 14 months later. The short arc of the initial photographs on 19 June was insufficient to produce reliable ephemerides. Because there was no confirmation, the comet was not officially named, nor is it included in Kronk and Meyer's *Cometography Volume 5: 1960–1982* (2010). Amazingly, inspection of scanned photographic plates taken at Boyden Observatory, Bloemfontein (MPC 074) and Johannesburg-Hartbeespoort (MPC 076) by Maik Meyer and Gary Kronk in 2023 revealed a 5′ tail with a PA of 150°. They determined orbital elements and found the perihelion occurred on 7 December 1971. Based on this later analysis, M.P.E.C. 2023-F148 was issued on 27 March 2023 stating "The MPC is changing the prefix of this comet's designation from X/ to C/, and the WGSBN has assigned it the name Edwards." (MPEC 2023-F148, 2023). Credit must be given to Meyer and Kronk for their outstanding detective work.
8. The Lincoln Near-Earth Asteroid Research (LINEAR) project was a collaboration of the

United States Air Force, NASA, and the Massachusetts Institute of Technology's Lincoln Laboratory for the systematic detection and tracking of near-Earth objects. LINEAR was responsible for most asteroid discoveries from 1998 until it was overtaken by the Catalina Sky Survey in 2005 (Wikipedia).

## 12 ACKNOWLEDGEMENTS

First and foremost, we show gratitude to Rodney Austin, Alan Gilmore and Pam Kilmartin for allowing us to interview them. The insights that they provided have been truly illuminating. We are also grateful to the vast efforts made by Gary Kronk (plus Maik Meyer and David Seargent) and S.K. Vsekhsvyatskii in the production of their volumes on comet discoveries, observations and astrophysical notes. We would like to thank Ian Cooper for his advice and supporting photographs. Thanks are also due to Professor John Hearnshaw for supplying photographs relating to Mount John Observatory. Eric J. Christensen is also thanked for his graph showing Gilmore and Kilmartin's astrometrical productivity. Dennis Goodman is also acknowledged for his helpful emails, and we appreciate the helpful comments supplied by the referees.

Finally, this paper forms part of the first author's PhD project on historical aspects of New Zealand cometary astronomy, and he is grateful to the University of Southern Queensland for providing travel funding and other research support.

## 13 REFERENCES

A New Comet. *Thames Star,* Volume 11399, Issue 11399, 28 July 1902, Page 3 (https://paperspast.natlib.govt.nz/newspapers/THS19020728.2.22?end_date=31-07-1902&items_per_page=100&query=grigg&snippet=true&sort_by=byDA&start_date=01-07-1902: accessed 6 September 2025).

A New Comet. *Evening Post*, Volume CXIII, Issue 147, 23 June 1932, Page 13 (https://paperspast.natlib.govt.nz/newspapers/EP19320623.2.122?end_date=24-06-1932&items_per_page=100&phrase=2&query=comet&snippet=true&sort_by=byDA&start_date=22-06-1932; accessed 16 October 2025).

Alchetron (https://alchetron.com/Leslie-Comrie#google_vignette; accessed 1 February 2026).

Alexander Turnbull Library Collections, National Library of New Zealand

Andrews, F., and Budding, E., 1992. Carter Observatory's 9 inch refractor: the Crossley connection. *Southern Stars*, 34, 358–366.

APOD. *Astronomy Picture of the Day* (https://apod.nasa.gov/apod/ap170527.html)

*Ashburton Guardian*, Volume XXI, Issue 5942, 20 April 1903, Page 2 (https://paperspast.natlib.govt.nz/newspapers/AG19030420.2.6?end_date=20-04-1903&items_per_page=50&query=comet&snippet=true&sort_by=byDA&start_date=20-04-1903; accessed 5 October 2025).

*Ashburton Guardian*, Volume XXVIII, Issue 7146, 9 April 1907, Page 2 (https://paperspast.natlib.govt.nz/newspapers/AG19070409.2.11?end_date=30-04-1907&items_per_page=10&query=Grigg&snippet=true&sort_by=byDA&start_date=08-04-1907#image-tab; accessed 10 October 2025).

Astronomical Activity, 1933. *Taranaki Daily News*, 6 May 1933, Page 4 (https://paperspast.natlib.govt.nz/newspapers/TDN19330506.2.20?end_date=06-05-1933&items_per_page=100&query=comet&snippet=true&sort_by=byDA&start_date=06-05-1933#image-tab; accessed 27 October 2025).

Astronomical Society, 1932. *Taranaki Daily News*, 3 December 1932, Page 3 (https://paperspast.natlib.govt.nz/newspapers/TDN19321203.2.17?end_date=03-12-1932&items_per_page=100&query=comet&snippet=true&sort_by=byDA&start_date=03-12-1932#image-tab; accessed 25 October 2025).

Astronomical Society, 1933. *Evening Post*, Volume CXV, Issue 65, 18 March 1933, Page 15 (https://paperspast.natlib.govt.nz/newspapers/EP19330318.2.129?end_date=18-03-1933&items_per_page=100&query=comet&snippet=true&sort_by=byDA&start_date=18-03-1933; accessed 27 October 2025).

Attractive Displays, 1932. *Taranaki Daily News*, 7 July 1932, Page 12 (https://paperspast.natlib.govt.nz/newspapers/TDN19320707.2.120?end_date=07-07-1932&items_per_page=100&query=comet&snippet=true&sort_by=byDA&start_date=07-07-1932; accessed 20 October 2025).

Auckland Museum Cenotaph Record (https://www.aucklandmuseum.com/war-memorial/online-cenotaph/record/C24272; accessed 17 October 2025).

*Auckland Star*, 1903, Volume XXXIV, Issue 96, 23 April, Page 4 (https://paperspast.natlib.govt.nz/newspapers/AS19030423.2.39?end_date=23-04-1903&items_per_page=50&query=comet&snippet=true&sort_by=byDA&start_date=23-04-1903; accessed 5 October 2025).

*Auckland Star, 1907a,* Volume XXXVIII, Issue 85, 10 April, Page 4

(https://paperspast.natlib.govt.nz/newspapers/AS19070410.2.39?end_date=11-04-1907&items_per_page=10&query=comet&snippet=true&start_date=08-04-1907&title=AS; accessed 12 October 2025).
*Auckland Star*, 1907b, Volume XXXVIII, Issue 87, 12 April, Page 4 (https://paperspast.natlib.govt.nz/newspapers/AS19070412.2.51?end_date=30-04-1907&items_per_page=10&page=7&query=Grigg&snippet=true&sort_by=byDA&start_date=08-04-1907; accessed 10 October 2025).
Austin, R.D., 1994. Albert Jones - the quiet achiever. *Southern Stars*, 36(1-2), 36–42 (http://www.aavso.org/aavso/membership/jones.pdf; accessed 30 October 2025).
Austin, R.D., 2019. Obituary Notices of the *Royal Astronomical Society of New Zealand* Newsletter, December.
Baggaley, J., and Orchiston, W., 2024. New Zealand's contribution to international science: the role of the University of Canterbury's Rolleston Research Station. *Journal of Astronomical History and Heritage*, 27(3), 579–594.
Baracchi, P. and Grigg, J., 1903. Comet 1902 c. *Astronomische Nachrichten*, 160(12), 213–214 (https://articles.adsabs.harvard.edu/full/1903AN....160..213B; accessed 1 September 2025).
Bateson, F., 1958. *The Observation of Variable Stars*. Rarotonga, Royal Astronomical Society of New Zealand.
*Bay of Plenty Times*, 1907. Volume XXXV, Issue 5037, 10 April, Page 2 (https://paperspast.natlib.govt.nz/newspapers/BOPT19070410.2.20?end_date=10-04-1907&items_per_page=50&query=comet&snippet=true&sort_by=byDA&start_date=10-04-1907#image-tab: accessed 13 October 2025).
Big Telescope, 1932. *Waikato Times*, Volume 112, Issue 18763, 11 October 1932, Page 11 (https://paperspast.natlib.govt.nz/newspapers/WT19321011.2.125?end_date=11-10-1932&items_per_page=100&query=comet&snippet=true&sort_by=byDA&start_date=11-10-1932; accessed 26 October 2025).
Bortle, J.E., 1982a. Comet Digest. *Sky and Telescope*, August, 198.
Bortle, J.E., 1982b. Comet Digest. *Sky and Telescope*, November, 504.
Bortle, J.E., 1984. Comet Digest. *Sky and Telescope*, September, 284.
Bortle, J.E., 2006. How to Estimate a Comet's Brightness (https://skyandtelescope.org/observing/celestial-objects-to-watch/how-to-estimate-a-comets-brightness/; accessed 1 November 2025).
Bright Planets in July, 1973. *Press*, Volume CXIII, Issue 33265, 30 June 1973, Page 9 (https://paperspast.natlib.govt.nz/newspapers/CHP19730630.2.76?end_date=31-08-1973&items_per_page=10&query=Comet+Clark&snippet=true&start_date=09-06-1973; accessed 13 January 2026).
Cap, L., 1932. *Gazette Astronomique*, 19, 116–117. September (https://ui.adsabs.harvard.edu/abs/1932GazA...19..116C/abstract; accessed 22 October 2025).
Christenson, E.J., Director, Catalina Sky Survey, The University of Arizona, Lunar and Planetary Laboratory.
Christie, G., 2014. Bateson, Frank Maine. In Hockey, T., et al., (eds.), *Biographical Encyclopedia of Astronomers. 2nd Edition*. New York, Springer. Pp. 171–172.
Comet Geddes, 1932. *Evening Post*, Volume CXIV, Issue 19, 22 July 1932, Page 8 (https://paperspast.natlib.govt.nz/newspapers/EP19320722.2.57?end_date=22-07-1932&items_per_page=100&query=comet&snippet=true&sort_by=byDA&start_date=22-07-1932; accessed 21 October 2025).
Comet Geddes, 1933. *Evening Post*, Volume CXV, Issue 26, 1 February 1933, Page 9 (https://paperspast.natlib.govt.nz/newspapers/EP19330201.2.110?end_date=01-02-1933&items_per_page=100&query=comet&snippet=true&sort_by=byDA&start_date=01-02-1933; accessed 27 October 2025).
Comet Reappears, 1933. *NZ Herald*, Volume LXX, Issue 21401, 27 January 1933, Page 10 (https://paperspast.natlib.govt.nz/newspapers/NZH19330127.2.106?end_date=27-01-1933&items_per_page=100&query=comet&snippet=true&sort_by=byDA&start_date=27-01-1933; accessed 28 October 2025).
Comet 'the brightest'. *Press*, 28 July 1982, Page 29 (https://paperspast.natlib.govt.nz/newspapers/CHP19820728.2.111.7?end_date=30-09-1982&items_per_page=100&query=Comet&snippet=true&sort_by=byDA&start_date=01-06-1982; accessed 14 January 2026).
Comets and their Messages, 1932. *Otago Daily Times*, Issue 21712, 2 August 1932, Page 5 (https://paperspast.natlib.govt.nz/newspapers/ODT19320802.2.38?end_date=02-08-1932&items_per_page=100&query=comet&snippet=true&sort_by=byDA&start_date=02-08-1932; accessed 22 October 2025).
Cottam, S., and Orchiston, W., 2015. *Eclipses, Transits and Comets of the Nineteenth Century: How America's Perception of the Sky Changed*. Cham (Switzerland), Springer.
de Grijs, R., 2026a. Mars in the Australian press, 1875–1899. 1: Interpretation, authority and planetary science. *Journal of Astronomical History and Heritage*, 29(1), 160–189.
de Grijs, R., 2026b. Mars in the Australian press, 1875–1899. 2: Circulation and attribution. *Journal of Astronomical History and Heritage*, 29(1), 190–205.
Dewhirst, D.W., 1977. William Herbert Steavenson. *Quarterly Journal of the Royal Astronomical Society*, 18(1), 147–154.
Dickie, N., 2010. Murray Geddes. *Southern Stars*, 49(2), 3–6.
DPMC. Department of the Prime Minister and Cabinet. *King's Birthday Honours List 2025*

(https://www.dpmc.govt.nz/publications/kings-birthday-honours-list-2025; accessed 26 January 2026).
Drummond, J.K., 2007. Comet McNaught 2006 P1 - the comet that blew our socks off. *Southern Stars*, 46(1), 9–14.
Drummond, J., and Orchiston, W., 2021. Seven hundred and fifty years of cometary astronomy in Aotearoa/New Zealand. Poster displayed at the Centennial Conference of the Royal Astronomical Society of New Zealand, Wellington, 9–11 July 2021.
Drummond, J., 2023. New Zealand observations of the Great Comet of 1881, C/1881 K1 (Tebbutt). In Gullberg, S., and Robertson, P. (eds.), *Essays on Astronomical History and Heritage: A Tribute to Wayne Orchiston on His 80th Birthday*. Cham (Switzerland), Springer. Pp. 367–392.
Drummond, J., and Orchiston, W., 2025. New Zealand observations of the Great Comet of 1881. Poster displayed at the Centennial Conference of the Royal Astronomical Society of New Zealand, Whakatane, 9–11 May 2025.
Drummond, J.K., Orchiston, W., Brown, C., and Horner, J. 2025. The Great January Comet of 1910 (C/1910 A1): a key opportunity missed by New Zealand astronomers. *Journal of Astronomical History and Heritage,* 28(3), 689–710 (https://www.sciengine.com/JAHH/doi/10.3724/SP.J.1440-2807.2025.03.07; accessed 13 January 2026).
Drummond, J., Orchiston, W., Brown, C., and Horner, J., 2026. Rodney Austin: New Zealand's most prolific comet discoverer (along with John Grigg). Poster displayed at the Annual Conference of the Royal Astronomical Society of New Zealand, Blenheim, 15–17 May 2026.
Eiby, G.A., 1970. Obituaries: Ivan Leslie Thomsen. *Southern Stars*, 23, 113–116.
Eiby, G.A., 1972. A visit to Uncle Charlie. A.C. Gifford MA FRAS (1861-1948). *Southern Stars*, 24,109–113.
Eicher, D.J., 1990a. Curtainrise on Comet Austin. *Astronomy*, August.
Eicher, D.J., 1990b. The changing fortunes of Comet Austin. *Astronomy*, September.
European Southern Observatory (ESO), 1992. (https://www.eso.org/public/images/eso9209a/; accessed 21 November 2025).
European Southern Observatory (ESO), Comet Austin develops an ion tail. (https://www.eso.org/public/images/eso9004a/; accessed 5 February 2026).
Evans, R.W., and Orchiston, W., 2023. Accredited meteorites of New Zealand. 1: The Makarewa Meteorite. *Southern Stars*, 62(3), 7–15.
*Evening Post*, 1907a. Volume LXXIII, Issue 86, 12 April, Page 6 (https://paperspast.natlib.govt.nz/newspapers/EP19070412.2.72?end_date=30-04-1907&items_per_page=10&page=7&query=Grigg&snippet=true&sort_by=byDA&start_date=08-04-1907; accessed 15 October 2025).
*Evening Post*, 1907b. Volume LXXIII, Issue 83, 9 April, Page 8 (https://paperspast.natlib.govt.nz/newspapers/EP19070409.2.98?end_date=30-04-1907&items_per_page=10&page=2&query=Grigg&snippet=true&sort_by=byDA&start_date=08-04-1907; accessed 10 October 2025).
*Evening Star*, 1903. Issue 11865, 20 April, Page 4 (https://paperspast.natlib.govt.nz/newspapers/ESD19030420.2.38?end_date=20-04-1903&items_per_page=50&query=comet&snippet=true&sort_by=byDA&start_date=20-04-1903; accessed 5 October 2025).
*Evening Star*, 1946a. Issue 25866, 9 August, Page 4 (https://paperspast.natlib.govt.nz/newspapers/ESD19460809.2.33?end_date=31-12-1946&items_per_page=100&query=comet&snippet=true&sort_by=byDA&start_date=01-08-1946; accessed 1 November 2025).
*Evening Star*, 1946b. Evening Star, Issue 25866, 9 August, Page 10 (https://paperspast.natlib.govt.nz/newspapers/ESD19460809.2.140?end_date=31-12-1946&items_per_page=100&query=comet&snippet=true&sort_by=byDA&start_date=01-08-1946#image-tab; accessed 2 November 2025).
FAU. Department of Physics, Erlangen Centre For Astroparticle Physics. Dr. Karl Remeis-Sternwarte, Astronomical Institute (https://www.sternwarte.uni-erlangen.de/remeis-start/research/plate-archive/; accessed 13 January 2026).
Find a Grave (https://www.findagrave.com/memorial/130358204/john-grigg (accessed 15 February 2026).
Fotolip. New Zealand maps (https://www.fotolip.com/new-zealand-map-960.html; accessed 18 March 2026).
Fraser, G., 2022. Two astronomers and Naval radar. *Southern Stars*, 61(4), 7–12.
Gale, W., 1921. Astronomical facts and fictions. *Transactions of the Sydney Lodge of Research*, 8, 1–14.
Geddes, M., 1932–1936. Record of astronomical activities from 1932 January 1 – 1936 August 12, by Murray Geddes. From the Carter Observatory archives 1989 May 17 sent to Norman Dickie. Sent to the authors by Ross Dickie. Unpublished
Geddes, M., and Thomsen, I.L., 1939. Surface features of Mars at the recent opposition. *Nature*, 144(3657), 944 (https://www.nature.com/articles/144944b0).
Geddes, M., 1941. Mars in 1939. *Popular Astronomy*, 49, 2–12 (https://articles.adsabs.harvard.edu/pdf/1941PA.....49....2G).
Geddes' Comet, 1932. *Waikato Times*, Volume 112, Issue 18688, 14 July 1932, Page 8 (https://paperspast.natlib.govt.nz/newspapers/WT19320714.2.82?end_date=14-07-1932&items_per_page=100&query=comet&snippet=true&sort_by=byDA&start_date=14-07-1932#image-tab; accessed 20 October 2025).
Gifford, A.C., 1931. *In Starry Skies*. Wellington, The Evening Post.

Gifford, M., 2005. *Missions, Moons & Masterpieces: The Giffords of Oamaru*. Palmerston North, Heritage Press Ltd.
Gilmore, A.C., and Jones, A.F., 2001. Comet Utsunomiya-Jones C/2000W1. *Southern Stars*, 40(1), 9–11.
Gilmore, A., and Kilmartin, P., 2004. Albert's graduation ceremony. *Southern Stars*, 43(2), 5.
Gilmore, A., 2007. Comet Gilmore P/2007 Q2. *Southern Stars*, 46(3), 5–6.
Glory of the Stars, 1932a. *NZ Herald*, Volume LXIX, Issue 21249, 1 August 1932, Page 6 (https://paperspast.natlib.govt.nz/newspapers/NZH19320801.2.27?end_date=01-08-1932&items_per_page=100&query=comet&snippet=true&sort_by=byDA&start_date=01-08-1932#image-tab; accessed 24 October 2025).
Glory of the Stars, 1932b. *Otago Daily Times*, Issue 21845, 6 January 1933, Page 9 (https://paperspast.natlib.govt.nz/newspapers/ODT19330106.2.99?end_date=06-01-1933&items_per_page=100&query=comet&snippet=true&sort_by=byDA&start_date=06-01-1933; accessed 25 October 2025).
Google Earth. https://earth.google.com/web/@0,-0.82316493,0a,22251752.77375655d,35y,0h,0t,0r/data=CgRCAggBOgMKATBCAggBSg0I____________ARAA
Green, D., 1984. IAUC Circular No. 3984: Pulsar in LMC; 1984g. *Central Bureau for Astronomical Telegrams* (http://www.cbat.eps.harvard.edu/iauc/03900/03984.html#Item2; accessed 15 January 2026).
Green, D., 1989a. IAUC Circular No. 4919. *Central Bureau for Astronomical Telegrams* (http://www.cbat.eps.harvard.edu/iauc/04900/04919.html#Item1; accessed 29 August 2025).
Green, D., 1989b. IAUC Circular No. 4921. *Central Bureau for Astronomical Telegrams* (http://www.cbat.eps.harvard.edu/iauc/04900/04921.html; accessed 29 August 2025).
Green, D., 1989c. IAUC Circular No. 4926. *Central Bureau for Astronomical Telegrams* (http://www.cbat.eps.harvard.edu/iauc/04900/04926.html; accessed 29 August 2025).
Green, D., 2007. IAUC Circular No. 8865. *Central Bureau for Astronomical Telegrams* (http://www.cbat.eps.harvard.edu/iauc/08800/08865.html#Item1; accessed 12 September 2025).
Grigg, J., 1902a. Photography with a small telescope. *Journal of the British Astronomical Association*, 12, 125–126 (https://articles.adsabs.harvard.edu/full/1902JBAA...12..125G; accessed 2 October 2025).
Grigg, J., 1902b. Letter to John Tebbutt, dated 12 August. In Letters to John Tebbutt 1860–1915, Mitchell Library, Sydney.
Grigg, J., 1902c. Über die Entdeckung eines neuen Cometen 1902 c. *Astronomische Nachrichten,* 159, 390–391 (https://articles.adsabs.harvard.edu/full/1902AN....159..389G; accessed 7 November 2025).
Grigg, J., 1903. Discovery of Grigg's Comet, 1903 *b*. *Journal of the British Astronomical Society*, 13, 320–321.
Grigg, J., 1906a. Letter to John Tebbutt, dated 21 March. In Letters to John Tebbutt 1860–1915, Mitchell Library, Sydney.
Grigg, J., 1906b. Letter to John Tebbutt, dated 20 April. In Letters to John Tebbutt 1860–1915, Mitchell Library, Sydney.
Grigg, J., 1907. Discovery of Comet 1907 *b*. *Journal of the British Astronomical Society*, 17(8), 364 (https://britastro.org/journal/journal-of-the-british-astronomical-association-vol-17-no-8: accessed 6 January 2026).
Grigg, R., 2020. John Grigg, comet discoverer and Christian. (https://creation.com/en/articles/john-grigg; accessed 27 November 2025).
GUIDE 9.1, 2020. *Project Pluto* planetarium software. Bill Gray, Maine, USA (https://www.projectpluto.com/faqnew.htm).
Hearnshaw, J., and Gilmore, A., 2015. *Mount John: The First 50 Years.* Christchurch, University of Canterbury.
Hood, P., 1951. *Observing the Heavens*. Oxford, Oxford University Press.
Hughes, D.W., 1991. J. Grigg, J.F. Skjellerup and their comet. *Vistas in Astronomy*, 34, 1–10.
Hurst, G.M., 2011. Obituary Brian Geoffrey Marsden (1937–2010). *Journal of the British Astronomical Association*, 121(1), 56–57.
Isdale, A., 1967. *History of the River Thames NZ*. Thames, printed for the author.
Jenniskens, P., Lyytinen, E., and Baggaley, J., 2020. An outburst of delta Pavonids and the orbit of parent comet C/1907 G1 (Grigg-Mellish). *Planetary and Space Science*, 189, 15 (https://www.sciencedirect.com/science/article/abs/pii/S0032063319304052?via%3Dihub; accessed 18 February 2026).
Jones, A., 1995a. The value of long-term visual monitoring of variable stars. *Australian Journal of Astronomy*, 6(3), 81–86.
Jones, A., 1995b. Variable stars and the amateur astronomer. In Orchiston, W., Carter, B., and Dodd, R., (eds.), *Astronomical Handbook for 1996*. Wellington, Carter Observatory. Pp. 103–109.
Jones, A., 2011. Seventy-five years of visual observing. *I & I News* [Newsletter of the Instruments and Imaging Section of the British Astronomical Association], New Series No. 2, 4–6.
Kammerer, A. *Analysis of past comet apparitions: C/2000 W1 (Utsunomiya-Jones)* (https://fg-kometen.vdsastro.de/koj_2000/c2000w1/00w1eaus.htm; accessed 3 February 2026).
Kelburn Observatory, 1933. *Timaru Herald*, Volume CXXXVII, Issue 19384, 7 January 1933, Page 16 (https://paperspast.natlib.govt.nz/newspapers/THD19330107.2.109?end_date=07-01-1933&items_per_page=100&query=comet&snippet=true&sort_by=byDA&start_date=07-01-1933#image-tab; accessed 26 October 2025).
King, M., 2003. *The Penguin History of NZ*. Penguin, Auckland.

Kresak, L., 1987. The 1808 apparition and the long-term physical evolution of periodic comet Grigg-Skjellerup. *Bulletin of the Astronomical Institutes of Czeckoslovakia*, 38(2), 65–75.
Kronk, G.W., 1984. *Comets, A Descriptive Catalog*. Hillside, Enslow.
Kronk, G.W., 2000. *Cometography: A Catalog of Comets. Volume 1: Ancient–1799*. Cambridge, Cambridge University Press.
Kronk, G.W., 2003. *Cometography: A Catalog of Comets. Volume 2: 1800–1899*. Cambridge, Cambridge University Press.
Kronk, G.W., 2007. *Cometography: A Catalog of Comets. Volume 3: 1900–1932*. Cambridge, Cambridge University Press.
Kronk, G.W., 2009. *Cometography: A Catalog of Comets. Volume 4: 1933–1959*. Cambridge, Cambridge University Press.
Kronk, G.W., and Meyer, M., 2010. *Cometography: A Catalog of Comets. Volume 5: 1960–1982*. Cambridge, Cambridge University Press.
Kronk, G.W., Meyer, M., and Sergeant, D.A.J., 2017. *Cometography: A Catalog of Comets. Volume 6: 1983–1993*. Cambridge, Cambridge University Press.
Kronk, G.W., and Meyer, M., 2024. *Catalog of Unconfirmed Comets - Volume 2: 1900 to the Present*. Cham, Springer.
Kronk, G., Meyer, M., Orchiston, W., and Drummond, J., 2026. Douglas C. Berry and the evolution of his cometary photography. Poster displayed at the Annual Conference of the Royal Astronomical Society of New Zealand, Blenheim, 15–17 May 2026.
Lamy, P.L., Toth, I., Weaver, H.A., A'Hearn, M.F., and Jorda, L., 2009. Properties of the nuclei and comae of 13 ecliptic comets from Hubble Space Telescope snapshot observations. *Astronomy & Astrophysics.* 508(2), 1045–1056 (https://www.aanda.org/articles/aa/full_html/2009/47/aa11462-08/aa11462-08.html; accessed 14 January 2026).
Levy, D., 1990. *Journal of the Royal Astronomical Society of Canada* Newsletter, 84, 26 (https://adsabs.harvard.edu/full/1990JRASC..84L..26G; accessed 27 January 2026).
Liller, W., 1992. *The Cambridge Guide to Astronomical Discovery*. Cambridge, Cambridge.
Local and General, 1932a. *Northern Advocate*, 29 June 1932, Page 4 (https://paperspast.natlib.govt.nz/newspapers/NA19320629.2.23?end_date=29-06-1932&items_per_page=100&query=comet&snippet=true&sort_by=byDA&start_date=29-06-1932#image-tab; accessed 17 October 2025).
Local and General, 1932b. *Taranaki Daily News*, 6 July 1932, Page 6 (https://paperspast.natlib.govt.nz/newspapers/TDN19320706.2.46?end_date=06-07-1932&items_per_page=100&query=comet&snippet=true&sort_by=byDA&start_date=06-07-1932; accessed 18 October 2025).
Local and General, 1932c. *Northern Advocate*, 5 July 1932, Page 4 (https://paperspast.natlib.govt.nz/newspapers/NA19320705.2.25?end_date=05-07-1932&items_per_page=100&query=comet&snippet=true&sort_by=byDA&start_date=05-07-1932; accessed 18 October 2025).
Local and General, 1946. *Ashburton Guardian*, Volume 66, Issue 255, 9 August 1946, Page 2 (https://paperspast.natlib.govt.nz/newspapers/AG19460809.2.8?end_date=09-08-1946&items_per_page=100&query=comet&snippet=true&sort_by=byDA&start_date=06-08-1946; accessed 5 October 2025).
McIntosh, R., 1973. Obituary - Allan Bryce. *Southern Stars*, 24(1), 19.
McLintock, A.H., 1966. *An Encyclopaedia of NZ* (https://teara.govt.nz/en/1966/comrie-leslie-john-frs; accessed 10 January 2026).
Mackrell, B., 1985. *Halley's Comet Over New Zealand*. Auckland, Reed.
Marsden, B.G., 1972a. *Central Bureau for Astronomical Telegrams*, IAUC 2432, 15 August 1972 (http://www.cbat.eps.harvard.edu/IAUCs/IAUC2432.png; accessed 15 November 2025).
Marsden. B.G., 1972b. Comets in 1971. *Quarterly Journal of the Royal Astronomical Society*, 13, 428–435.
Marsden, B.G., 1973. IAUC 2550: 1973i. *Central Bureau for Astronomical Telegrams*, IAUC 2550, 26 June 1973 (http://www.cbat.eps.harvard.edu/iauc/02500/02550.html#Item1; accessed 20 November 2025).
Marsden, B.G., 1982. *Central Bureau for Astronomical Telegrams*, IAUC 3705, 21 June 1982 (http://www.cbat.eps.harvard.edu/iauc/03700/03705.html; accessed 9 October 2025).
Marsden, B.G., 1984a. *Central Bureau for Astronomical Telegrams*, IAUC 3957, 9 July 1984 (http://www.cbat.eps.harvard.edu/iauc/03900/03957.html; accessed 23 November 2025).
Marsden, B.G., 1984b. *Central Bureau for Astronomical Telegrams*, IAUC 3958, 11 July 1984 (http://www.cbat.eps.harvard.edu/iauc/03900/03958.html; accessed 14 January 2026).
Massey, H.S.W., 1952. Leslie John Comrie. 1893–1950. *Obituary Notices of Fellows of the Royal Society*, 8(21), 96–107.
Middlehurst, B.M., and Kuiper, G.P. (eds.), 1963. *The Moon Meteorites and Comets*. Chicago, University of Chicago Press.
Minor Planet Center Search (https://minorplanetcenter.net/db_search - accessed 12 October 2025).
Mobberly, M., 2011. *Hunting and Imaging Comets* (Patrick Moore's Practical Astronomy Series). New York, Springer.
Moore, P., 1989. *The Sky at Night*.

(https://www.google.com/search?q=rod+austin+comet+discoverer&sca_esv=9bd0e26604fdbc2f&rlz=1C1GCEA_enNZ1145NZ1146&sxsrf=AE3TifM34gF8axhfUz_lrs7NifBd6OTVkQ:1763752067234&ei=g7ggaaeEDvaX0-kP_YD32Qk&start=10&sa=N&sstk=Af77f_d06ibds3oHG-oAKfO-PSLiUVmR_1LzoZQCQ6zmBMGMv_ZKH4utJBn-Bzy6SUGatiQF76JHyuPxuJ7kNU7VcEU7H_ZZF8d6zA&ved=2ahUKEwjn0cG7-IORAxX2yzQHHX3APZsQ8NMDegQIChAW&biw=1280&bih=585&dpr=1.5#fpstate=ive&vld=cid:b2bafd18,vid:UTx6-81VOAU,st:0; accessed 25 January 2026).

Mount John Find, 1973. *Press*, Volume CXIII, Issue 33257, 21 June 1973, Page 17 (https://paperspast.natlib.govt.nz/newspapers/CHP19730621.2.185?end_date=30-06-1973&items_per_page=10&query=comet&snippet=true&sort_by=byDA&start_date=01-06-1973; accessed 13 January 2026).

MPEC 2007-Q43: *Central Bureau for Astronomical Telegrams,* COMET P/2007 Q2 (GILMORE) (https://www.minorplanetcenter.net/mpec/K07/K07Q43.html; accessed 13 June 2026).

MPEC 2023-F148: *Central Bureau for Astronomical Telegrams,* COMET C/1971 M1 (Edwards) (https://www.minorplanetcenter.net/mpec/K23/K23FE8.html; accessed 3 February 2026).

Mr. Grigg's Observatory 1885. *Thames Advertiser*, Volume XVI, Issue 5122, 20 March 1885, Page 3 (https://paperspast.natlib.govt.nz/newspapers/THA18850320.2.8?end_date=31-12-1885&query=grigg&snippet=true&start_date=01-01-1885&title=HPGAZ%2CHPDG%2CKSRA%2CKCC%2CMATREC%2COG%2CPAKIOM%2CPUP%2CTAN%2CTAWC%2CTHA%2CTGMR%2CTHS%2CWHDT%2CWAIGUS%2CWAIKIN%2CWT%2CWAIPO); accessed 1 September 2025).

Nakano, S., Hale, A., Seargent, D., Biggs, J., Urata, T., Kobayashi, J., Gilmore, A.C., Jones, A.F., and Marsden, B.G., 2000. Comet C/2000 W1 (Utsunomiya-Jones). IAU Circ., No. 7526, #1 (2000). Edited by Green, D.W.E.

NASA, Astronomy Picture of the Day. (https://apod.nasa.gov/apod/astropix.html; accessed 13 September 2025).

NASA/JPL Horizons (https://ssd.jpl.nasa.gov/; accessed 21 January 2026).

*Nelson Evening Mail*, 1907a. Volume XLII, Issue XLII, 9 April, Page 2 (https://paperspast.natlib.govt.nz/newspapers/NEM19070409.2.38?end_date=09-04-1907&items_per_page=50&query=comet&snippet=true&sort_by=byDA&start_date=09-04-1907; accessed 11 October 2025).

*Nelson Evening Mail*, 1907b. Volume XLII, Issue XLII, 11 April, Page 2 (https://paperspast.natlib.govt.nz/newspapers/NEM19070411.2.14?end_date=11-04-1907&items_per_page=50&query=comet&snippet=true&sort_by=byDA&start_date=11-04-1907#image-tab; accessed 12 October 2025).

New Comet, 1932. *Waipukurau Press*, Volume XXVIII, Issue 152, 23 June 1932, Page 5 (https://paperspast.natlib.govt.nz/newspapers/WPRESS19320623.2.36?end_date=23-06-1932&items_per_page=100&query=comet&snippet=true&sort_by=byDA&start_date=23-06-1932#image-tab; accessed 16 October 2025).

New Comet, 1946. Discovery Confirmed. Photographed by Dunedin Observer, 1946. *Evening Star*, Issue 25873, 17 August 1946, Page 6 (https://paperspast.natlib.govt.nz/newspapers/ESD19460817.2.26?end_date=31-12-1946&items_per_page=100&query=comet&snippet=true&sort_by=byDA&start_date=01-08-1946; accessed 4 November 1946).

New Comet's Position, 1932. *Dominion*, Volume 25, Issue 233, 28 June 1932, Page 8 (https://paperspast.natlib.govt.nz/newspapers/DOM19320628.2.55?end_date=28-06-1932&items_per_page=100&query=comet&snippet=true&sort_by=byDA&start_date=28-06-1932#image-tab; accessed 15 September 2025).

*New Zealand Herald*, 1907. Volume XLIV, Issue 13459, 11 April, Page 4 (https://paperspast.natlib.govt.nz/newspapers/NZH19070411.2.29?end_date=11-04-1907&items_per_page=50&query=comet&snippet=true&sort_by=byDA&start_date=11-04-1907; accessed 12 October 2025).

*New Zealand Herald*, 1932. Volume LXIX, Issue 21233, 13 July, Page 6 (https://paperspast.natlib.govt.nz/newspapers/NZH19320713.2.20.6?end_date=13-07-1932&items_per_page=100&query=comet&snippet=true&sort_by=byDA&start_date=13-07-1932; accessed 21 October 2025).

NZ Map https://nz.images.search.yahoo.com/search/images;_ylt=Awrx.Ttnsbtp4YoJ3qH1Zgx.;_ylu=c2VjA3NlYXJjaARzbGsDYnV0dG9u;_ylc=X1MDMjExNDc0MjAwNQRfcgMyBGZyA21jYWZlZQRmcjIDcDpzLHY6aSxtOnNiLXRvcARncHJpZANDWE9wd0tkclJuaTZLa3hLSlhwUGpBBG5fcnNsdAMwBG5fc3VnZwMxMARvcmlnaW4DbnouaW1hZ2VzLnNlYXJjaC55YWhvby5jb20EcG9zAzAEcHFzdHIDBHBxc3RybAMwBHFzdHJsAzE1BHF1ZXJ5A25ldyUyMHplYWxhbmQlMjBtYXAEdF9zdG1wAzE3NzM5MDgzNDE-?p=new+zealand+map&fr=mcafee&fr2=p%3As%2Cv%3Ai%2Cm%3Asb-top&ei=UTF-8&x=wrt&type=E210NZ1330G0#id=3&iurl=http%3A%2F%2Fwww.fotolip.com%2Fwp-content%2Fuploads%2F2016%2F05%2FNew-Zealand-Map-6.jpg&action=click

NZ War Graves Project (https://www.nzwargraves.org.nz/casualties/murray-geddes).

Obituary, 1920. *Press*, Volume LVI, Issue 16869, 24 June 1920, Page 6 (https://paperspast.natlib.govt.nz/newspapers/CHP19200624.2.46?query=My%20own%20New%20Zealand: accessed 20 February 2026).

Orchiston, W., 1983. C.J. Westland and Comet 1914IV: a forgotten episode in New Zealand cometary astronomy. *Southern Stars*, 30, 339–345.
Orchiston, W., 1990. Albert Jones: an interview. *Astronomy Now*, 4(9), 16–17.
Orchiston, W., and Brewer, A., 1990. David Ross and the development of amateur astronomy in Victoria. *Journal of the British Astronomical Association*, 100, 173–181.
Orchiston, W., 1993. John Grigg, and the genesis of cometary astronomy in New Zealand. J*ournal of the British Astronomical Association*, 103(2), 67–76 (https://articles.adsabs.harvard.edu/full/1993JBAA..103...67O; accessed 2 November 2025).
Orchiston, W, 1997. The Waingaromia Meteorite: a IIIAB iron from New Zealand. *Meteorite*!, 3(4), 20–21 (plus Front Cover).
Orchiston, W., 1998. Mission impossible: William Scott and the first Sydney Observatory directorship. *Journal of Astronomical History and Heritage*, 1(1), 21–43.
Orchiston, W., 1999a. Comets and communication: amateur-professional tension in Australian astronomy. *Publications of the Astronomical Society of Australia*, 16, 212–221.
Orchiston, W., 1999b. Of comets and variable stars: the Afro-Australian astronomical activities of J.F. Skjellerup. *Journal of the British Astronomical Association*, 109, 328–338.
Orchiston, W., 2001. The Thames' Observatories of John Grigg. *Southern Stars*, 40(3), 14–22 (https://articles.adsabs.harvard.edu/full/2001SouSt..40c..14O; accessed 5 September 2025).
Orchiston, W., 2003a. Australia's earliest planispheres. *Journal of the British Astronomical Association*, 113, 329–332.
Orchiston, W., 2003b. J.F. Skjellerup: a forgotten name in South African cometary astronomy. *Monthly Notices of the Astronomical Society of Southern Africa*, 62, 54–73.
Orchiston, W., 2015. The amateur-turned-professional syndrome: two Australian case studies. In Orchiston, W., Greem D.A., and Strom, R. (eds.), *New Insights From Recent Studies in Historical Astronomy: Following in the Footsteps of F. Richard Stephenson. A Meeting to Honor F. Richard Stephenson on His 70th Birthday*. Cham (Switzerland), Springer. Pp. 259–350.
Orchiston, W., 2016. *Exploring the History of NZ Astronomy: Trials, Tribulations, Telescopes and Transits.* Cham (Switzerland), Springer.
Orchiston, W., 2017. *John Tebbutt: Rebuilding and Strengthening the Foundations of Australian Astronomy*. Cham (Switzerland), Springer.
Orchiston, W., and Drummond, J., 2019. The Mount Tarawera volcanic eruption in New Zealand and Maori cometary astronomy. *Journal of Astronomical History and Heritage*, 22, 521–535.
Orchiston, W., Drummond, J., and Kronk, G., 2020a. Observations of the Great September Comet of 1882 (C/1882 R1) from New Zealand. *Journal of Astronomical History and Heritage*, 23(3), 628–658.
Orchiston, W., Drummond, J., and Shylaja, B.S., 2020b. Communication issues in war-time astronomy: independent Australian, Indian, New Zealand and South African discoveries of Comet C/1941 B2 (de Kock-Paraskevopoulos). *Journal of Astronomical History and Heritage*, 23, 659–674.
Orchiston, W., Drummond, J., and Luciuk, M., 2021. Ronald McIntosh: pioneer Southern Hemisphere meteor observer. *Journal of Astronomical History and Heritage*, 24(3), 789–817.
Orchiston, W., and Drummond, J., 2022a. New Zealand observations of the Great Comet of 1858. Poster displayed at the Annual Conference of the Royal Astronomical Society of New Zealand, Whangarei, 4–6 June 2022.
Orchiston, W., and Drummond, J., 2022b. New Zealand observations of the Great Comet of 1861. Poster displayed at the Annual Conference of the Royal Astronomical Society of New Zealand, Whangarei, 4–6 June 2022.
Orchiston, W., and Drummond, J., 2022c. The 1989 Opotiki bolide: accumulated evidence for a new carbonaceous chondrite meteorite from Aotearoa/New Zealand. *Journal of Astronomical History and Heritage*, 25(2), 277–289.
Orchiston, W., and Drummond, J., 2024. Comet 26P/Grigg-Skjellerup: the Kiwi connection. *Southern Stars*, 63(4), 14–31.
Orchiston, W., and Orchiston, D.L., 2024. Celebrating the 150th anniversary of the 1875 transit of Venus: Arthur Stock and his little books. Poster displayed at the Annual Conference of the Royal Astronomical Society of New Zealand, Nelson, 24–26 May 2024 [1874 Comet].
Orchiston, W., Field-Dodgson, C., Leggott, M., Cooper, I., and Drummond, J., 2024a. The astronomical paintings and sketches of the Nelson artists Edwin and Emily Harris. Poster displayed at the Annual Conference of the Royal Astronomical Society of New Zealand, Nelson, 24–26 May 2024.
Orchiston, W., Scott, J., Rowe, J., and Wyn-Harris, S., 2024b. In search of New Zealand's missing meteorites: the role of 'Papers Past'. Poster displayed at the Annual Conference of the Royal Astronomical Society of New Zealand, Nelson, 24–26 May 2024.
Orchiston, W., Scott, J., Rowe, J., and Wyn-Harris, S., 2024c. Researching New Zealand meteorites: collaborative research by the RASNZ's Fireballs Aotearoa and Historical Sections. Poster displayed at the Annual Conference of the Royal Astronomical Society of New Zealand, Nelson, 24–26 May 2024.
Orchiston, W., and Austin, R, 2025. The historic 6-inch refractor at the New Plymouth Observatory: New Zealand's only known 'Alvan Clark telescope'. *HAD News*, 106, 15–17.
Orchiston, W., and Orchiston, D.L., 2026. Archdeacon Arthur Stock: New Zealand's first professional astronomer? *Southern Stars*, 65, in press.
Orchiston, W., Statye, J., Orchiston, D.L., and Drummond, J., 2026a. The Waingaromia Meteorite: a IIIAB iron from the East Coast of the North Island of New Zealand. Poster displayed at the Annual Conference of the Royal Astronomical Society of New Zealand, Blenheim, 15–17 May 2026.

Orchiston, W., Wyn-Harris, S., and Evans, B., 2026b. In search of New Zealand's missing meteorites. Oral presentation at the Annual Conference of the Royal Astronomical Society of New Zealand, Blenheim, 15–17 May 2026.

Orchiston, W., Wyn-Harris, S., and Evans, B., 2026c. In search of New Zealand's missing meteorites: the role of 'Papers Past' (Paper 2). Poster displayed at the Annual Conference of the Royal Astronomical Society of New Zealand, Blenheim, 15–17 May 2026.

*Otago Daily Times*, 1932. Issue 21677, 22 June, Page 1 (https://paperspast.natlib.govt.nz/newspapers/ODT19320622.2.2.4?end_date=22-06-1932&query=comet&snippet=true&start_date=22-06-1932#image-tab; accessed 17 October 2025).

Otago Institute, 1932a. *Evening Star*, Issue 21189, 24 August 1932, Page 2 (https://paperspast.natlib.govt.nz/newspapers/ESD19320824.2.10?end_date=24-08-1932&items_per_page=100&query=comet&snippet=true&sort_by=byDA&start_date=24-08-1932; accessed 25 October 2025).

Otago Institute, 1932b. *Otago Daily Times*, Issue 21761, 28 September 1932, Page 5 (https://paperspast.natlib.govt.nz/newspapers/ODT19320928.2.19?end_date=28-09-1932&items_per_page=100&query=comet&snippet=true&sort_by=byDA&start_date=28-09-1932; accessed 25 October 2025).

Papers Past (https://paperspast.natlib.govt.nz/newspapers)

Pearson, J., and Orchiston, W., 2011. The Lick Observatory solar eclipse expedition to Padang (Indonesia) in 1901. In Nakamura, T., Orchiston, W., Sôma, M., and Strom, R. (eds.), *Mapping the Oriental Sky. Proceedings of the Seventh International Conference on Oriental Astronomy*. Tokyo, National Astronomical Observatory of Japan. Pp. 207–216.

Pease, L., 2024. Murray Geddes. https://terangiaoaonunui.pukeariki.com/story-collections/taranaki-world-war-two-servicemen-and-women/geddes-murray/#:~:text=He%20died%20of%20a%20brain%20haemorrhage%20on%2023,He%20was%20buried%20at%20Cardonald%20Cemetery%2C%20Glasgow%2C%20Scotland; accessed 8 November 2025.

Personal, 1932. *Taranaki Daily News*, 25 June 1932, Page 4 (https://paperspast.natlib.govt.nz/newspapers/TDN19320625.2.25?end_date=25-06-1932&items_per_page=100&query=comet&snippet=true&sort_by=byDA&start_date=25-06-1932; accessed 16 October 2025).

Pettit, E.C., 1942. Visual magnitudes of Nova Puppis 1942. *Publications of the Astronomical Society of the Pacific*, 54(321), 259 (https://articles.adsabs.harvard.edu//full/1942PASP...54..259P/0000259.000.html; accessed 3 December 2025).

Pickering, E.C., 1902. Grigg's Comet 1902c. *Harvard College Observatory Bulletin*, 111, 1 (https://articles.adsabs.harvard.edu/full/1902BHarO.111....1P; accessed 8 January 2026).

*Poverty Bay Herald*, 1902. Volume XXIX, Issue 9488, 28 July Page 2 (https://paperspast.natlib.govt.nz/newspapers/PBH19020728.2.34?end_date=28-07-1902&items_per_page=50&query=comet&snippet=true&start_date=28-07-1902#image-tab; accessed 2 September 2025).

*Rangitikei Advocate and Manawatu Argus*, 1907. Volume XXXII, Issue 8801, 1 May, Page 3 (https://paperspast.natlib.govt.nz/newspapers/RAMA19070501.2.54.1?end_date=01-05-1907&items_per_page=50&query=comet&snippet=true&sort_by=byDA&start_date=01-05-1907: accessed 12 October 2025).

RASNZ (Royal Astronomical Society of NZ) Murray Geddes Memorial Price (https://www.rasnz.org.nz/rasnz-info/murray-geddes-memorial-prize-1; accessed 27 October 2025).

RASNZ minutes. Minutes of the Royal Astronomical Society of NZ AGM minutes, 30 November 1939.

Reinhard, R., 1987. *The Giotto Extended Mission. Proceedings of the International Symposium on the Diversity and Similarity of Comets*. Noordwijk, Netherlands.. Pp. 523–529. (Bibcode:1987ESASP.278..523R https://ui.adsabs.harvard.edu/abs/1987ESASP.278..523R/abstract; accessed 27 November 2025).

Ross, D., Grigg, J., Tebbutt, J., Maunder, E.W., 1903. Reports of the Directors of the Observing Sections - Comet Section. *Journal of the British Astronomical Society*, 14, 73–86.

Schmadel, L.D., 2007. *Dictionary of Minor Planet Names*. Springer, Heidelberg (https://link.springer.com/rwe/10.1007/978-3-540-29925-7_2538).

Siers, J., and Orchiston, W., 1998. *Annual Report July 1997 to June 98*. Wellington, Carter Observatory.

Sketch map, 1932. *Dominion*, Volume 25, Issue 233, 28 June 1932, Page 8 (https://paperspast.natlib.govt.nz/newspapers/DOM19320628.2.55.1?end_date=28-06-1932&items_per_page=100&query=comet&snippet=true&sort_by=byDA&start_date=28-06-1932; accessed 1 October 2025).

*Sky and Telescope* Magazine. February, March, April, May 1990 issues.

Steavenson, W.H., 1942. Note on the photometry of comets. *Journal of the British Astronomical Society*, 52(6), 189–191 (https://britastro.org/journal/journal-of-the-british-astronomical-association-vol-52-no-6; accessed 2 November 2025).

Steavenson, W.H., 1971. Obituary Frederick James Hargreaves. *Quarterly Journal of the Royal Astronomical Society*, 12, 336–337.

Stoy, R.H., 1942. Nova Puppis (1942). *Monthly Notes of the Astronomical Society of Southern Africa*, 1, 182 (https://ui.adsabs.harvard.edu/abs/1942MNSSA...1..182S/abstract - accessed 1 November 2025).

Success in a Comet Search, 1982: 25. *Press*, 24 June 1982, Page 25 (https://paperspast.natlib.govt.nz/newspapers/CHP19820624.2.119.3?end_date=30-09-1982&items_per_page=100&query=Comet&snippet=true&sort_by=byDA&start_date=01-06-1982#print; accessed 14 January 2026.)

Sullivan, D., 2004. Albert Jones' Honorary DSc. *Southern Stars*, 43(2),3.

Taibi, R., Morse, K., Orchiston, W., and Drummond, J., 2026. Early New Zealand meteor observers, C.P. Olivier, and the formative role of the American Meteor Society. Poster displayed at the Annual Conference of the Royal Astronomical Society of New Zealand, Blenheim, 15–17 May 2026.

Tantau, K., Comet finder and 'national anthem' composer to be remembered 100 years after death. *Stuff*, 15 March 2026 (https://www.stuff.co.nz/science/300035748/comet-finder-and-national-anthem-composer-to-be-remembered-100-years-after-death.html; accessed 15 January 2026).

Tebbutt, J., 1907. Comet Grigg, 1907, April. *The Observatory*, 30, 285–286 (https://articles.adsabs.harvard.edu//full/1907Obs....30..285T/0000286.000.html; accessed 16 October 2025).

Thames Star, 1902. *Thames Star*, Volume 11399, Issue 11399, 28 July 1902, Page 2 (https://paperspast.natlib.govt.nz/newspapers/THS19020728.2.19.5?end_date=31-07-1902&items_per_page=100&query=grigg&snippet=true&sort_by=byDA&start_date=01-07-1902; accessed 30 October 2025).

Thames Star, 1907. *Thames Star*, Volume XLIV, Issue 10535, 13 April 1907, Page 2 (https://paperspast.natlib.govt.nz/newspapers/THS19070413.2.15?end_date=13-04-1907&items_per_page=50&query=comet&snippet=true&start_date=13-04-1907; accessed 11 October 2025).

The Comet, 1907a. *Rangitikei Advocate and Manawatu Argus*, Volume XXXII, Issue 8797, 26 April 1907, Page 4 (https://paperspast.natlib.govt.nz/newspapers/RAMA19070426.2.53?end_date=26-04-1907&items_per_page=50&query=comet&snippet=true&sort_by=byDA&start_date=26-04-1907; accessed 13 October 2025).

The Comet, 1907b. *Rangitikei Advocate and Manawatu Argus*, Volume XXXI, Issue 8789, 17 April 1907, Page 3 (https://paperspast.natlib.govt.nz/newspapers/RAMA19070417.2.37?end_date=17-04-1907&items_per_page=50&query=comet&snippet=true&sort_by=byDA&start_date=17-04-1907#image-tab; accessed 14 October 2025).

The Comet, 1907c. *NZ Herald*, Volume XLIV, Issue 13465, 18 April 1907, Page 3 (https://paperspast.natlib.govt.nz/newspapers/NZH19070418.2.17.1?end_date=18-04-1907&items_per_page=50&query=comet&snippet=true&sort_by=byDA&start_date=18-04-1907; accessed 16 October 2025).

The Comet, 1907d. *Rangitikei Advocate and Manawatu Argus*, Volume XXXII, Issue 8792, 20 April 1907, Page 3 (https://paperspast.natlib.govt.nz/newspapers/RAMA19070420.2.42.1?end_date=20-04-1907&items_per_page=50&query=comet&snippet=true&sort_by=byDA&start_date=20-04-1907; accessed 16 October 2025).

The Comet, 1907e. *Rangitikei Advocate and Manawatu Argus*, Volume XXXII, Issue 8801, 1 May 1907, Page 3 (https://paperspast.natlib.govt.nz/newspapers/RAMA19070501.2.54.1?end_date=01-05-1907&items_per_page=50&query=comet&snippet=true&sort_by=byDA&start_date=01-05-1907; accessed 16 October 2025.

The Comet, 1907f. *Rangitikei Advocate and Manawatu Argus*, Volume XXXII, Issue 8804, 4 May 1907, Page 2 (https://paperspast.natlib.govt.nz/newspapers/RAMA19070504.2.29?end_date=04-05-1907&items_per_page=50&query=comet&snippet=true&sort_by=byDA&start_date=04-05-1907; accessed 16 October 2025).

The Comet Geddes, 1933a. *Otago Daily Times*, Issue 21954, 16 May 1933, Page 6 (https://paperspast.natlib.govt.nz/newspapers/ODT19330516.2.33?end_date=16-05-1933&items_per_page=100&query=comet&snippet=true&sort_by=byDA&start_date=16-05-1933; accessed 27 October 2025).

The Comet Geddes, 1933b. *Otago Daily Times*, Issue 21958, 20 May 1933, Page 19 (https://paperspast.natlib.govt.nz/newspapers/ODT19330520.2.150.7?end_date=20-05-1933&items_per_page=100&query=comet&snippet=true&sort_by=byDA&start_date=20-05-1933; accessed 27 October 2025).

The Levin Daily Chronicle, 1932. Thursday 28 July 1932. Local and General. *Levin Daily Chronicle,* 28 July 1932, Page 4 (https://paperspast.natlib.govt.nz/newspapers/LDC19320728.2.21?end_date=28-07-1932&items_per_page=100&query=comet&snippet=true&sort_by=byDA&start_date=28-07-1932; accessed 21 October 2025).

The Lost Comet Operetta, 1932. *Evening Star*, Issue 21138, 25 June 1932, Page 2 (https://paperspast.natlib.govt.nz/newspapers/ESD19320625.2.13?end_date=25-06-1932&items_per_page=100&query=comet&snippet=true&sort_by=byDA&start_date=25-06-1932; accessed 17 October 2025.

The Lost Comet, 1932. *Otago Daily Times*, Issue 21680, 25 June 1932, Page 16 (https://paperspast.natlib.govt.nz/newspapers/ODT19320625.2.131?end_date=25-06-1932&items_per_page=100&query=comet&snippet=true&sort_by=byDA&start_date=25-06-1932; accessed 15 January 2026).

The New Comet, 1932b, 1932. *Ashburton Guardian*, Volume 52, Issue 218, 28 June 1932, Page 7

(https://paperspast.natlib.govt.nz/newspapers/AG19320628.2.68?end_date=28-06-1932&items_per_page=100&query=comet&snippet=true&sort_by=byDA&start_date=28-06-1932#image-tab; accessed 17 October 2025).
The New Comet, 1932a. *NZ Herald*, Volume LXIX, Issue 21218, 25 June 1932, Page 8 (https://paperspast.natlib.govt.nz/newspapers/NZH19320625.2.45?end_date=25-06-1932&items_per_page=100&query=comet&snippet=true&sort_by=byDA&start_date=25-06-1932; accessed 16 October 2025).
The New Comet, 1932b. *Waikato Times*, Volume 111, Issue 18675, 29 June 1932, Page 6 (https://paperspast.natlib.govt.nz/newspapers/WT19320629.2.48?end_date=29-06-1932&items_per_page=100&query=comet&snippet=true&sort_by=byDA&start_date=29-06-1932#image-tab; accessed 18 October 2025).
The New Comet, 1932c. *NZ Herald*, Volume LXIX, Issue 21226, 5 July 1932, Page 8 (https://paperspast.natlib.govt.nz/newspapers/NZH19320705.2.53?end_date=05-07-1932&items_per_page=100&query=comet&snippet=true&sort_by=byDA&start_date=05-07-1932; accessed 18 October 2025).
The New Comet, 1932d. *Ashburton Guardian*, Volume 52, Issue 233, 15 July 1932, Page 5 (https://paperspast.natlib.govt.nz/newspapers/AG19320715.2.50?end_date=31-05-1933&items_per_page=100&query=comet&snippet=true&sort_by=byDA&start_date=15-07-1932#image-tab; accessed 17 October 2025).
The New Comet, 1932e. *NZ Herald*, Volume LXIX, Issue 21242, 23 July 1932, Page 12.
The New Comet, 1932f. *NZ Herald*, Volume LXIX, Issue 21274, 30 August 1932, Page 10 (https://paperspast.natlib.govt.nz/newspapers/NZH19320830.2.109?end_date=30-08-1932&items_per_page=100&query=comet&snippet=true&sort_by=byDA&start_date=30-08-1932; accessed 25 October 2025).
The New Comet, 1946. *Evening Star*, Issue 25867, 10 August 1946, Page 6 (https://paperspast.natlib.govt.nz/newspapers/ESD19460810.2.33?end_date=10-08-1946&items_per_page=100&query=comet&snippet=true&start_date=10-08-1946; accessed 12 August 2025).
The Observatory, 1932. *Evening Post*, Volume CXIV, Issue 19, 22 July 1932, Page 2 (https://paperspast.natlib.govt.nz/newspapers/EP19320722.2.7.7?end_date=22-07-1932&items_per_page=100&query=comet&snippet=true&sort_by=byDA&start_date=22-07-1932; accessed 20 October 2025).
The Thames Comet, 1903. *Gisborne Times*, Volume X, Issue 1052, 20 November 1903, Page 1 (https://paperspast.natlib.govt.nz/newspapers/GIST19031120.2.6; accessed 12 February 2026).
Thomson, I.L., 1933. The Comet Geddes. *Otago Daily Times*, Issue 21958, 20 May, Page 19 (https://paperspast.natlib.govt.nz/newspapers/ODT19330520.2.150.7?end_date=20-05-1933&items_per_page=100&query=comet&snippet=true&sort_by=byDA&start_date=20-05-1933; accessed 3 October 2025).
Thomson, I.L., 1945. Obituary notices: Murray Geddes. *Monthly Notices of the Royal Astronomical Society*, 105, 88–89.
Thomsen, I.L., 1954. The amateur in astronomy. *Southern Stars*, 16, 78–82.
*Timaru Herald*, 1907. Volume XC, Issue 13255, 9 April, Page 5 (https://paperspast.natlib.govt.nz/newspapers/THD19070409.2.34?end_date=09-04-1907&items_per_page=50&query=comet&snippet=true&sort_by=byDA&start_date=09-04-1907; accessed 10 October 2025).
Timaru Man's Observations, 1946. *Marlborough Express*, Volume 81, Issue 187, 9 August 1946, Page 4 (https://paperspast.natlib.govt.nz/newspapers/MEX19460809.2.39.1?end_date=09-08-1946&items_per_page=100&query=Comet&snippet=true&start_date=09-08-1946#image-tab; accessed 30 October 2025).
Time and Date. https://www.timeanddate.com/sun/new-zealand/new-plymouth?month=7&year=2025; accessed 14 January 2026.
Toone, J., 2005. Variable Star Section Circular No. 123, March 2005. *British Astronomical Society* (ISSN 0267-9272. https://britastro.org/vss/vssc123.pdf accessed 30 October 2025).
Toone, J., 2016. Albert Francis Arthur Lofley Jones, DSc., OBE, FRAS, FRANZ (1920-2013). *Journal of the British Astronomical Association*, 126(2), 83–94.
van Biesbroeck, G., 1946. Comet Notes: Comet Timmers; Comet Tempel (2); Comet Berry; Comet Jones. *Popular Astronomy*, 54(December), 420 (https://articles.adsabs.harvard.edu/pdf/1946PA.....54..420V; accessed 5 November 2025).
van Biesbroeck, G., 1947a. Comet Notes. *Popular Astronomy*, 55(January), 53 (https://articles.adsabs.harvard.edu/pdf/1947PA.....55...53V; accessed 5 November 2025).
van Biesbroeck, G., 1947b. Comet Notes. *Popular Astronomy*, 55(February), 110 (https://articles.adsabs.harvard.edu/pdf/1947PA.....55..110V; accessed 5 November 2025).
van Biesbroeck, G., 1947c. Comet Notes. *Popular Astronomy*, 55(December), 560 (https://articles.adsabs.harvard.edu/pdf/1947PA.....55..559V; accessed 5 November 2025).
Vaubaillon J., and Colas, F., 2005. Demonstration of gaps due to Jupiter in meteoroid streams. What happened with the 2003 Pi-Puppids? *Astronomy and Astrophysics*, 431(3), 1139–1144 (https://www.aanda.org/articles/aa/pdf/2005/09/aa1391.pdf; accessed 20 February 2026).
Volcanic theory of Venus supported, 1982. *Press*, 4 August 1982, Page 14

(https://paperspast.natlib.govt.nz/newspapers/CHP19820804.2.79.1?end_date=30-09-1982&items_per_page=100&query=Comet&snippet=true&sort_by=byDA&start_date=01-06-1982; accessed 14 January 2026).

Vsekhsvyatskii, S.K., 1964. *Physical Characteristics of Comets*. Jerusalem, Israel Program for Scientific Translations.

*Wanganui Herald*, 1903. Volume XXXVII, Issue 10951, 18 May. Page 6 (https://paperspast.natlib.govt.nz/newspapers/WH19030518.2.60?end_date=18-05-1903&items_per_page=50&query=comet&snippet=true&sort_by=byDA&start_date=18-05-1903; accessed 5 October 2025.

*Wanganui Herald*, 1907. Volume XXXXI, Issue 12138, 11 April, Page 7 (https://paperspast.natlib.govt.nz/newspapers/WH19070411.2.57?end_date=11-04-1907&items_per_page=50&query=comet&snippet=true&sort_by=byDA&start_date=11-04-1907#image-tab; accessed 12 October 2025).

Ward, B., 2022. Astronomy memories from times long ago. *Southern Stars*, 61(2), 11–13.

Warnes, P., Orchiston, W., and Englert, P., 1998. A reported tektite transported from Australasia and found at Gabriel's Gully mining camp, Central Otago, New Zealand. *Journal of the Royal Society of New Zealand*, 28, 329–331.

Weatherspark. Climate and Average Weather Year Round in Thames (https://weatherspark.com/y/144924/Average-Weather-in-Thames-New-Zealand-Year-Round; accessed 5 October 2025).

Wellington City Observatory, 1932. *Evening Post*, Volume CXIV, Issue 25, 29 July 1932, Page 9 (https://paperspast.natlib.govt.nz/newspapers/EP19320729.2.105?end_date=29-07-1932&items_per_page=100&query=comet&snippet=true&sort_by=byDA&start_date=29-07-1932; accessed 21 October 2025).

WGSBN. IAU: WG Small Bodies Nomenclature (WGSBN). (https://www.wgsbn-iau.org/; accessed 26 January 2026).

Williams, T.R., 2016. https://link.springer.com/rwe/10.1007/978-1-4419-9917-7_549 (accessed 2 January 2026).

Wikipedia. Lincoln Near-Earth Asteroid Research. (https://en.wikipedia.org/wiki/Lincoln_Near-Earth_Asteroid_Research#cite_note-NEO-STATS-1; accessed 13 June 2026).

Wright, H., 1906. Letter to John Tebbutt, dated 29 November. In Letters to John Tebbutt (1860–1915), Mitchell Library, Sydney.

**John Drummond** became fixated with astronomy at the age of ten when his mother pointed out the Pot in Orion to him. From that moment on he was hooked on the Universe. Joining the Junior Section of the local Gisborne Astronomical Society not long after, John would regularly do group meteor watches, telescope viewing and listen to astronomy talks. He also developed an interest in photography, and it was not long before he combined these two interests and began astrophotography. John's photographs have been used in many overseas books and magazines—and were used on two New Zealand stamps. He was the Director of the Royal Astronomical Society of New Zealand's Astrophotography Section for thirteen years until 2018. He is currently the Director of the Society's Comet Section.

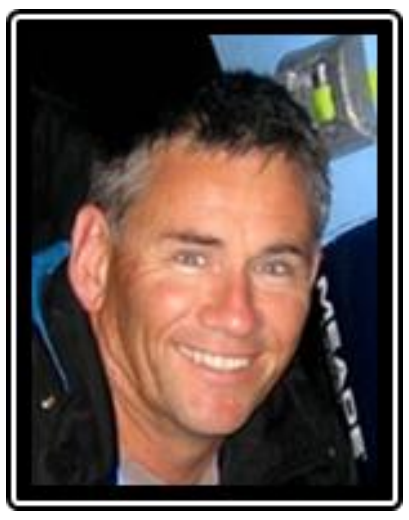

John lives about 10km west of Gisborne, on the east coast of the North Island of New Zealand, and has a range of telescopes up to 0.5 metres in aperture at his Possum Observatory. He regularly images with these telescopes and CCDs, carries out astrometry of comets, asteroids and NEOs, and sends his observations to the IAU Minor Planet Center. He has also confirmed several comets and co-discovered about 20 exoplanets in collaboration with the Ohio State University. He runs Gisborne Astro Tours from his observatories. John has authored or co-authored more than 60 research papers, many of the more recent ones being on the history of New Zealand astronomy.

John is a Past President and Past Secretary of the Royal Astronomical Society of New Zealand, and in 2019 he was made a Fellow of the Society. In 2016 he was awarded an MSc (Astronomy) by Swinburne University in Melbourne (Australia), and currently he is researching aspects of the history of cometary astronomy in New Zealand as a part-time off-campus internet-based PhD student affiliated with the Centre for Astrophysics at the University of Southern Queensland (Australia). His supervisors are the co-authors of this paper.

When he is not doing astronomy, John is a secondary school science teacher. He also enjoys surfing the great waves of Gisborne and pottering around on his small farm tending to his sheep.

**Professor Wayne Orchiston** was born in Auckland (New Zealand) in 1943, and has BA First Class Honours and PhD degrees from the University of Sydney. Currently, he is employed by the University of Science and Technology of China in Hefei as the Co-editor of the *Journal of Astronomical History and Heritage*. He is also an Adjunct Professor of Astronomy at the Centre for Astrophysics at the University of Southern Queensland (USQ) in Toowoomba, Australia. Formerly, Wayne worked at observatories, research institutes and universities in Australia, New Zealand and Thailand.

Over the past two decades Wayne has supervised more than 35 Master of Astronomy and PhD history of astronomy research projects through three different Australian universities.

Wayne has wide-ranging research interests and more than 500 publications, mainly about historic transits of Venus; historic solar eclipses; historic telescopes and observatories; the emergence of astrophysics in Asia and Oceania; the history of cometary and meteor astronomy; the astronomy of James Cook's three voyages to the Pacific; amateur astronomy and the amateur–professional interface; the history of meteoritics; Indian, Southeast Asian and Māori ethnoastronomy; and the history of radio astronomy in Australia, France, India, Japan, New Zealand and the USA.

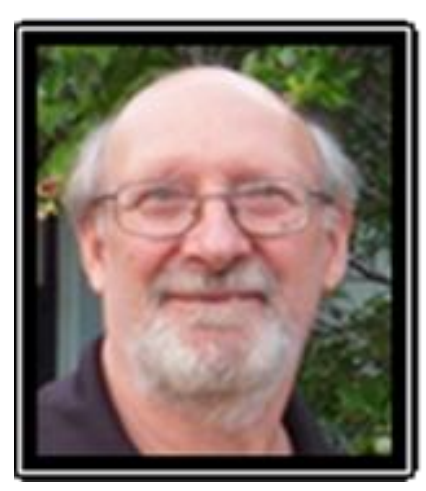

Recent books by Wayne include *Exploring the History of New Zealand Astronomy …* (2016, Springer); *John Tebbutt: Rebuilding and Strengthening the Foundations of Australian Astronomy* (2017, Springer); *The Emergence of Astrophysics in Asia …* (2017, Springer, co-edited by Tsuko Nakamura); *Exploring the History of Southeast Asian Astronomy …* (2021, Springer, co-edited by Mayank Vahia) and *Golden Years of Australian Radio Astronomy: An Illustrated History* (2021, Springer, co-authored by Peter Robertson and Woody Sullivan); and *Histoire de la Radioastronomie Française* (2025, EDP Sciences, co-authored by James Lequeux). Wayne has also edited or co-edited a succession of conference proceedings.

Since 1985 Wayne has been a member of the IAU, and he is a former President of Commission C3 (History of Astronomy). In 2003 he founded the IAU's Historical Radio Astronomy Working Group, and is the current Radio Astronomy Subject Editor for the Third Edition of Springer's *Biographical Encyclopedia of Astronomers*. He also founded the IAU Working Group on Historic Transits of Venus, and is the founding Director of the large and dynamic Historical Section of the Royal Astronomical Society of New Zealand. Wayne is also an Editor of Springer's book series on Cultural and Historical Astronomy. In 2014 he founded the History & Heritage Working Group of the Southeast Asian Astronomy Network and ran this successfully until 2024. In the process he organised three different conferences on SE Asian Astronomical History; some of the presented papers ended up in the aforementioned Southeast Asian Springer book. Since 2004 Wayne has also served on the Executive Committee of the ICOA series of Asian–Oceanic conferences, and currently he is a Councillor of the Royal Astronomical Society of New Zealand and a member of the Editorial Board of its journal, *Southern Stars*.

In 1998 Wayne Orchiston and John Perdrix co-founded the *Journal of Astronomical History and Heritage*. After John's death, Wayne was the Managing Editor until 31 July 2022 when he passed ownership of the journal to the University of Science and Technology of China. In 2013 the IAU named minor planet 48471 'Orchiston', and in 2019 former PhD student Stella Cottam, and Wayne, were awarded the Donald E. Osterbrock Prize by the American Astronomical Society for their 2015 Springer book, *Eclipses, Transits and Comets of the Nineteenth Century …* In 2023 Wayne was elected an Honorary Member of the Royal Astronomical Society of New Zealand, and two of his former doctoral students edited the following Festschrift in his honour: Gullberg, S., and Robertson, P. (eds.), 2023. *Essays in Astronomical History and Heritage: A Tribute to Wayne Orchiston on His 80th Birthday* (Springer, 2023). In January 2024 the American Astronomical Society also awarded Wayne their LeRoy E. Doggett Prize for lifetime contributions to history of astronomy, and in August 2026 he will receive an Honorary DSc from the University of Southern Queensland.

Wayne and Darunee Lingling Orchiston live in a quiet village near Chiang Mai in northern Thailand. When not involved in astronomy Wayne particularly enjoys following Australian and New Zealand athletics and Formula 1 and Indycar racing.

**Dr. Carolyn Brown** was born in Toowoomba, Queensland in Australia and has been interested in astronomy since she was four years old. After watching Stephen Hawking's "A Brief History of Time" in 1991, she proclaimed that she would become an astrophysicist. Which she did. Undertaking degrees in Physics and Astronomy at the University of Southern Queensland, Carolyn took on part-time work in the University's Physics and Astronomy Department while she completed her PhD. In 2009, Carolyn became a full-time academic in Physics and Engineering at the University of Southern Queensland while continuing to follow her passion for astronomy through research and outreach programs.

As Carolyn's career progressed, she found her love of teaching grew beyond her love of research, and she began to concentrate on educating the new generation and instilling a passion for astronomy in them, just like others had done for her. Taking on a primary teaching role at the University, Carolyn continued to foster this passion for astronomy in her students, especially her postgraduate students, until one day, convinced by one of her Doctoral students, she made the decision to move into the secondary school sector where she is hoping to make an impact on our youth to help drive them to follow their passions.

When not educating the next generation, Carolyn enjoys restoring classic Australian muscle cars where she can put her theoretical physics and engineering knowledge into a practical (and fast) application. Now all she needs is a bigger shed, with an observatory on the roof.

**Professor Jonathan (Jonti) Horner** was born in Wakefield, Yorkshire, in the United Kingdom, in 1978. At the age of five he saw part of an episode of "The Sky At Night" and became hooked on all things astronomical. He joined his local astronomical society—the West Yorkshire Astronomical Society (WYAS)—at the age of eight and began giving regular talks and writing for the society's journal, *Phobos*, by the age of ten. Thanks to the support and advice he received from both the members and the guest speakers at WYAS, Jonti moved to the University of Durham in 1996 to study for an undergraduate Masters' degree in Physics and Astronomy. After a summer project working at Armagh Observatory in 1999, with Professor Mark Bailey, Jonti moved to the University of Oxford in 2000, to begin work on his doctoral studies. His DPhil was conferred in 2004 for a thesis titled "The Behaviour of Small Bodies in the Outer Solar System." He spent time as a Postdoctoral Research Fellow at the University of Bern (Switzerland) and the Open University (UK), and a year as a teaching-only Fellow at Durham University (UK) before moving to Australia in 2010, to take up a Postdoctoral Fellowship at the University of New South Wales in Sydney. Finally, in 2014, Jonti moved to take up a position as Vice-Chancellor's Senior Research Fellow at the University of Southern Queensland in Toowoomba, where he remains to this day.

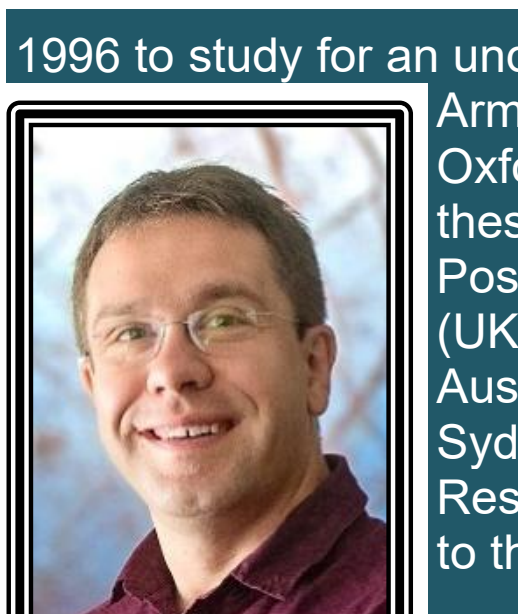

Jonti has a diverse range of research interests and has published more than 230 papers. His primary research focusses through his career have been the study of the Solar System's small bodies (particularly the Centaurs and planetary Trojans), the search for and characterisation of planets orbiting other stars (Exoplanets), and also the investigation of the various features that could render one planet more or less suitable as a target for the search for life beyond the Solar System. He is particularly proud of the work he did with his former mentor, Professor Barrie Jones, investigating the role that giant planets like Jupiter play in controlling the impact rate for terrestrial worlds—work that definitely shattered the long-held myth that Jupiter serves as Earth's celestial guardian and protector. Instead, Jupiter's role is more nuanced, with the giant planet actively throwing new objects to threaten the Earth with one hand whilst taking them away with the other. More recently, Jonti organised and led a lengthy review titled "Solar System Physics for Exoplanet Research", which has become a *de facto* textbook for undergraduate and Masters' courses around the world. His full publication list can be accessed through a library on the NASA ADS system, here: https://ui.adsabs.harvard.edu/public-libraries/YUefu-IISQSn4d6JVwKokQ

Jonti is a member of the Astronomical Society of Australia, and an ongoing Committee Member of the Astrobiology Society of the Britain. He is a past member of Australia's National Committee for Space and Radio Science, and serves as the Honorary President of the West Yorkshire Astronomical Society. He is an active and enthusiastic science communicator, regularly appearing in national and international media to discuss stories about planetary and exoplanetary science, and astrobiology. He has written more than one hundred articles for "The Conversation", and is currently serving as a guest presenter on the globally popular "SpaceNuts" podcast. Jonti's ORCID is 0000-0002-1160-7970.

When he is not working as a professional astronomer, Jonti is a keen photographer, both of wildlife and as an astrophotographer. He is also a member of the Toowoomba Philharmonic Society's chorus, and enjoys gaming with friends and family at his home in the Darling Downs.